\documentclass[12pt,letterpaper]{report}
\usepackage{amsmath}
\usepackage{amsfonts}
\usepackage{amssymb}
\usepackage{xcolor}
\usepackage{textcomp}
\usepackage{slashed}
\usepackage{accents}
\usepackage{dcolumn}
\usepackage{bm}
\usepackage{caption}
\usepackage{fancyhdr}
\usepackage{latexsym}
\usepackage{amsthm}
\usepackage{color}
\usepackage{colordvi}
\usepackage{array}
\usepackage{subfigure}
\usepackage{axodraw}
\usepackage{float}
\usepackage{calligra}
\usepackage[T1]{fontenc}

\usepackage[pdftex]{graphicx}
\usepackage{setspace}
\usepackage{hyperref}
\title{Low-energy meson phenomenology with Resonance Chiral Lagrangians}

\makeatletter
\renewcommand{\ps@plain}{
\renewcommand\@oddhead{\hfill\normalfont\textrm{\thepage}}
\renewcommand\@evenhead{}
\renewcommand\@oddfoot{}
\renewcommand\@evenfoot{}}

\def\@makechapterhead#1{%
  \vspace*{-48\p@}%
  {\parindent \z@ \raggedright \normalfont
    \ifnum \c@secnumdepth >\m@ne
        \huge\bfseries \@chapapp\space \thechapter
        \par\nobreak
        \vskip 0\p@
    \fi
    \interlinepenalty\@M
    \Huge \bfseries #1\par\nobreak
    \vskip 20\p@
  }}

\def\@makeschapterhead#1{%
  \vspace*{-48\p@}%
  {\parindent \z@ \raggedright
    \normalfont
    \interlinepenalty\@M
    \Huge \bfseries  #1\par\nobreak
    \vskip 20\p@
  }}
\makeatother

\renewcommand{\thepage}{\roman{page}}

\newcommand{\ex}{^}
\newcommand{\tu}{\tilde{u}}
\newcommand{\tU}{\tilde{U}}
\newcommand{\tr}{\tilde{r}}
\newcommand{\tl}{\tilde{\ell}}
\newcommand{\tR}{\tilde{R}}

\begin{document}
\bibliographystyle{unsrt}
\pagestyle{myheadings}

\thispagestyle{empty}

\begin{center}
\Large{
 Physics Department\\
 Centro de Investigaci\'on y de Estudios Avanzados del IPN\\\hspace*{1ex}\\
 \begin{figure}[h!]
  \centering
  \includegraphics[scale=0.20]{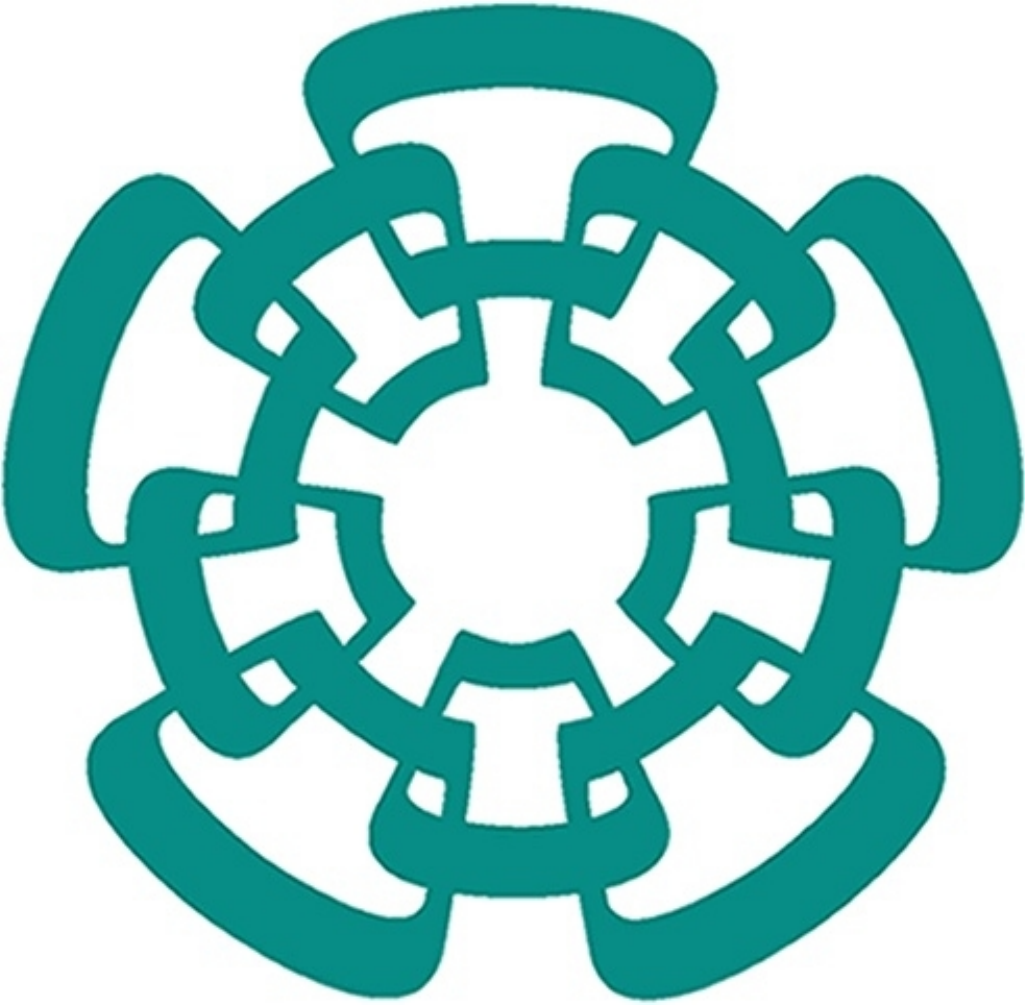}
 \end{figure}
\Huge{Low-energy meson phenomenology with Resonance Chiral Lagrangians}\\
 \vspace{58pt}
\Large{ Presented in Partial Fulfilment of the Requirements for the\\
 Degree of Doctor in Science\\}

 \vspace{58pt}
 by\\
 Adolfo Enrique Guevara Escalante\\
  \vspace{80pt} 
 Thesis advisors:\\ Dr. Gabriel L\'opez Castro and Dr. Pablo Roig Garc\'es.\\
 }
\end{center}
\pagebreak 
\hspace*{1ex}\\\hspace*{1ex}\\\hspace*{1ex}\\\hspace*{1ex}\\\hspace*{1ex}\\\hspace*{1ex}\\\hspace*{1ex}\\\hspace*{1ex}\\\hspace*{1ex}\\\hspace*{1ex}
\begin{flushright}
{\huge \calligra ``Life is like a healthy penis,\\
 it gets hard for no reason.''}
 \end{flushright}
 \pagebreak
\addcontentsline{toc}{chapter}{Table of Contents}
\tableofcontents
\pagebreak

\addcontentsline{toc}{chapter}{List of Tables}
\listoftables
\pagebreak

\addcontentsline{toc}{chapter}{List of Figures}
\listoffigures
\pagebreak

\setcounter{page}{1}
\renewcommand{\thepage}{\arabic{page}}

 {\huge \bf Agradecimientos}\\
      
      Agradezco de manera muy sincera a mis asesores, Gabriel L\'opez Castro y Pablo Roig Garc\'es, de quienes he 
      obtenido much\'isimos conocimientos y que me han ayudado de muchas maneras en mi desarrollo y formaci\'on. 
      Gracias a Sally Santiago por ser tan gran soporte para m\'i en tiempos tan dif\'iciles, en especial el tortuoso lapso de tiempo 
      que pas\'e sin beca. Agradezco much\'isimo a mis sinodales Aurore Courtoy, David Farn\'andez, Iv\'an Heredia, Omar Miranda y Genaro Toledo
      por su paciencia y ayuda durante mi periodo de formaci\'on, en especial en el periodo de revisi\'on de la tesis y el seminario.
      Agradezco especialmente todas las ayudas recibidas por parte de Eduard de la Cruz. Tambi\'en le doy las gracias al jefe del Departamento, M\'aximo L\'opez (bis) 
      por apoyarme con recursos para que pudiera mostrar mi trabajo en los Rencontres de Moriond.\\
      
      Debo agradecer tambi\'en a Jos\'e Salazar (Chepe), Blanca Ca\~nas (la doctora), Lenin Tostado, Gerardo H. Tom\'e, Alfonso Jaimes, 
      Jhovanny Mej\'ia, Idalia Sandoval, Bryan Larios y todos los que no logro recordar (y que se quedar\'an sin ser mencionados por la prisa con la que 
      tuve que escribir esta tesis y no porque no merezcan ser mencionados);
      es decir, a todos mis amigos... Y Gus (Gustavo Guti\'errez) por tantas discusiones tan 
      fruct\'iferas en el entendimiento del c\'omo funciona la naturaleza. Especiales agradecimientos a Penguin-San 
      por acompa\~narnos durante las discusiones de f\'isica y otros temas. Gracias, tambi\'en, a la naturaleza por ser 
      cu\'antica, debido a que sus fluctuaciones cu\'anticas a lo largo de la historia del universo me han permitido llegar tan lejos. 
      Tambi\'en, como becario, estoy obligado a dar las gracias a Conacyt por la beca de doctorado, cuyo sistema y personal tienen 
      grand\'isimas deficiencias que necesitan arreglo urgentemente. Agradezco el apoyo para la obtenci\'on de grado por parte del Centro (Cinvestav), 
      as\'i como los apoyos para Curso Especializado con el que asist\'i a la CERN Latin American School of High Energy Physics y 
      asistencia a Congreso que parcialmente cubri\'o gastos para asist\'ir a los Rencontres de Moriond.

 \pagebreak

{\huge \bf Abstract}\\

     It is not known how to obtain exactly transition amplitudes
     in Quantum Field Theory, so that perturbative approximation is the best we can do. Since the fundamental 
     theory of strong interactions (Quantum Chromodynamics) does not admit a perturbative approach for processes 
     with energies near or below the proton mass one needs to see how to overcome this difficulty. What common 
     sense dictates is to construct a theory that admits a perturbative description of phenomena at the energy ranges 
     in which the fundamental theory fails to be perturbative. In this thesis we present the computation of 
     some processes that cannot be obtained through an expansion of the strong coupling intensity, since nothing would 
     guarantee the convergence of such expansion, this is why we use an Effective Field Theory whose main characteristic 
     is chiral invariance.\\
     
     On the other hand, since the 1970's, the Standard Model of fundamental particles interactions has been so successful that it 
     seems very implausible to see phenomena resulting from interactions beyond this theory (with the exception of 
     everything related to neutrino masses) at leading order in perturbation theory. One then relies on precision tests, 
     for which a very good understanding of the interactions is needed. Since many experiments on the High Intensity Frontier 
     begin to take data in the very near future, in order to improve their power of prediction all possible background 
     in the search for Beyond Standard Model effects 
     must be very well understood.\\
     
     The observables we have computed are contributions within the Standard Model to processes that either need to have a very 
     well described background or that are not very well understood. Two processes are two different $\tau$ lepton decays 
     as background for processes with lepton number and lepton flavor violation such as $\tau^-\to \pi^+\ell^-\ell^-\nu_\tau$
     and background for second class currents for the decay $\tau\to\pi\eta\nu_\tau$. Another process we computed was the 
     $B^\pm\to P^\pm\ell^+\ell^-$, where $P$ is either a pion or a Kaon. This was computed in an effort to try to understand 
     the apparent lepton non-universality measured at LHCb, where we obtained a rather large CP asymmetry for the $\pi$ channel. 
     Finally, we computed the pseudoscalar light-by-light contribution to the anomalous magnetic moment of the muon, 
     giving a more robust analysis of the theoretical uncertainties and compatible with previous results.\\
     
\pagebreak

{\huge \bf Resumen}\\

   Actualmente no es posible obtener amplitudes de manera exacta usando Teor\'ia Cu\'antica de Campos, as\'i que lo mejor 
   que se puede hacer es una aproximaci\'on perturbativa. Ya que la teor\'ia fundamental de las interacciones fuertes 
   (la Cromodin\'amica Cu\'antica) no admite una descripci\'on perturbativa para procesos a escalas energ\'eticas 
   cerca o por debajo de la masa del prot\'on se vuelve necesario buscar la forma sortear esta dificultad. Lo que 
   marca la intuici\'on es construir una teor\'ia que permita una descripci\'on perturbativa de fen\'omenos a escalas
   de energ\'ia en que la teor\'ia fundamental no puede dar tal descripci\'on. En este sentido, se presenta el c\'alculo 
   de varios procesos que no pueden ser obtenidos por medio de algunos procesos que no pueden ser obtenidos por medio 
   de una expansi\'on de la intensidad de interacciones fuertes, ya que no se puede garantizar la convergencia de dicha 
   expansi\'on, por lo que hemos recurrido al uso de una Teor\'ia de Campos Efectiva cuya principal caracter\'istica 
   es la invarianza quiral.\\
   
   Por otro lado, desde la d\'ecada de 1970, el Modelo Est\'andar de part\'iculas fundamentales  ha tenido tanto \'exito 
   que parece muy poco probable encontrar alg\'un fen\'omeno resultante de interacciones m\'as all\'a de esta teor\'ia 
   (con excepci\'on de todo lo relacionado con las masas de los neutrinos) a primer orden en teor\'ia de perturbaciones. 
   Entonces se vuelve necesario recurrir a pruebas de precisi\'on, para lo cual se necesita un buen entendimiento de las 
   interacciones. Ya que muchos experimentos en la frontera de la alta intensidad empezar\'an a tomar datos en un futuro 
   muy cercano, para mejorar su poder predictivo es necesario entender muy bien cualquier posible ruido de fondo en la 
   b\'usqueda de efectos m\'as all\'a del Modelo Est\'andar.\\
   
   Las observables que calculamos son contribuciones del Modelo Est\'andar a proceso que, ya sea necesitan tener un 
   ruido de fondo muy bien descrito o no est\'an bien entendidos. Dos de los procesos son dos diferentes decaimientos 
   del lept\'on $\tau$ como ruido de fondo para procesos con violaci\'on de n\'umero y sabor lept\'onico como 
   $\tau^-\to \pi^+\ell^-\ell^-\nu_\tau$ y el ruido para el descubrimiento de corrientes de segunda clase en 
   el decaimiento $\tau\to\pi\eta\nu_\tau$. Otro proceso que calculamos fue el  decaimiento $B^\pm\to P^\pm\ell^+\ell^-$,
   donde $P=\pi,K$. Esto se calcul\'o como un esfuerzo en tratar de entender la aparente violaci\'on de universalidad 
   lept\'onica medida por LHCb, donde obtuvimos una asimetr\'ia de CP grande para el canal del $\pi$. 
   Finalmente, calculamos la contribuci\'on principal de la dispersi\'on hadr\'onica luz por luz la momento magn\'etico
   an\'omalo del mu\'on, dando un an\'alisis m\'as robusto de la incertidumbre te\'orica y que es compatible con resultados previos.
 \ \ \ \ \ 

\graphicspath{{Figs/}} 
\newcommand{\ds}{\slashed{\partial}}
\newcommand{\pip}{{\pi^+}}
\newcommand{\pim}{{\pi^-}}
\newcommand{\pz}{{\pi^0}}
\newcommand{\kp}{{K^+}}
\newcommand{\km}{{K^-}}
\newcommand{\kz}{{K^0}}
\newcommand{\kzb}{{\bar{K^0}}}

\chapter{Theoretical Framework}
 \section{Introduction}
 
  In this chapter we show the theoretical framework within quantum field theory needed to compute the observables in subsequent chapters. 
  First we give an introduction using a historical approach of the development of the Standard Model of elementary particles. 
  In section \ref{ChiPeeTee} we develop Chiral Perurbation Theory from the chiral symmetry of the QCD Lagrangian and its Spontaneous 
  Symmetry Breaking into vectorial $SU(3)$. In section \ref{ResChiTh} we give the main features of Resonance Chiral Theory, lying the 
  foundations to further enlarge the theory to higher chiral orders.
 
 \section{Standard Model}
  
  \subsection{Introduction}
  
  In this section we give a summary of the historical development of the now called Standard Model of elementary particles. In subsection \ref{EWw}
  we follow the development of the electroweak unification starting with the chiral symmetry of neutrinos up to the Glashow-Weinberg-Salam model 
  of electroweak interactions. In subsection \ref{qcd} we show the historical development of strong interactions until Ne'eman and Gell-Mann's 
  extension of the isospin model, then introduce the concept of partons and the color charge to conclude with the Lagrangian of strong interactions. 
  In subsection \ref{SMc} we briefly summarize $CP$ violation, the Kobayashi-Maskawa scenario and the dates in which the remaining particles of the 
  Standard Model were discovered. In subsection \ref{Limitations} we discuss the limitations of QCD and define the concept of Effective Field Theory. 
  
  \subsection{Electroweak Standard Model}\label{EWw}
  
  Since Ernest Rutherford's discovery in 1909 that protons were confined in atomic nuclei positively charged \cite{Rutherf}, the question of how same 
  charge particles can remain together without repelling each other arose. After James Chadwick's discovery of the neutron in 1932 \cite{Chadw}, 
  a {\it strong} interaction was hypothesized to explain why the nucleus (a bounded state of protons and neutrons, as suggested by Dmitri 
  Ivanenko \cite{Ivanenko}) remain bounded, where Werner Heisenberg proposed the isospin model \cite{isospin}. 
  The next year, Enrico Fermi proposed the existence of a new interaction to explain $\beta$-decay \cite{Ferm}, later known  as {\it weak} 
  interaction, where the interacting term came as products of fermion currents 
  \begin{equation}
  \mathcal{L}_{Fermi}=g\left(\bar{\psi}_p\gamma^\mu\psi_n\right)\left(\bar{\psi}_e\gamma_\mu\psi_\nu\right), 
  \end{equation}
  where the subindex in each fermion operator $\psi$ denotes the physical field referred to.
  It also was the first attempt of including the neutrino as
  a fundamental field. With this and except for gravity, all now known fundamental interactions had been postulated by then at a quantum level. \\
  
  Fermi's theory of beta decay only included the proton, neutron (both within the isospin model), electron and neutrino fields as fundamental, 
  but could be very easily extended to include muons (earlier called $\mu$-mesons), heavier baryons and spin zero fields. Also the particles 
  with strangeness (earlier called $\eta$-charge) were able to be allocated in a Fermi-like theory. \\
  
  Since the Fermi theory was not able to predict some nuclear processes involving $\Delta J=0$ between nuclei, a generalization of Fermi's theory 
  was sought by considering all linearly independent combinations of Dirac matrices \cite{Gamow}, namely
  {\bf 1}, $\gamma_5$, $\gamma_\mu$, $\gamma_\mu\gamma_5$ and $\sigma_{\mu\nu}=\frac{i}{2}[\gamma_\mu,\gamma_\nu]$, where the squared brackets 
  denotes the commutator. The Lagrangian reads
  
  \begin{equation}\label{FermiGen}
   \mathcal{L}=g_i\left(\bar{\psi}_1\Gamma_i\psi_2\right)\left(\bar{\psi}_3\Gamma_i\psi_4\right), \hspace*{10ex} i=S,P,V,A,T\hspace*{1ex},
  \end{equation}
  
  where $\Gamma_i$ is one of the linearly independent operators and $S,P,V,A,T$ stands for scalar, pseudoscalar, vector, axial-vector and tensor operators respectively.
  This implied an effort to experimentally determine the coupling constants $g_i$.\\
  
  The fact that Tsung-Dao Lee and Chen-Ning Yang \cite{LeeYang} suggested the non conservation of parity in $\beta$-decays (or $P$ violation,
  confirmed experimentally some months later in Co$^{60}$ decays \cite{WuAmbler} and in $\pi^+\to\mu^++\nu$ and $\mu\to e+2\nu$ decays \cite{GarwinLederman}), 
  lead Abdus Salam to propose what was known as chirality or $\gamma_5$ invariance \cite{g5inv}. The argument is as follows: since the neutrino is a massless field, 
  no term mixing chiralities exists in its free Lagrangian, this means that under the substitution $\nu\to-\gamma_5\nu$ the free Lagrangian remains 
  invariant\footnote{Our convention of $\gamma_5$ is different from the convention followed in the cited papers, this is $$\gamma_5^{us}=-i
  \gamma_5^\text{Salam},$$ where $\gamma_5^\text{us}$ is our convention which is used in this thesis and $\gamma_5^\text{Salam}$ is the convention 
  used in the cited papers. However, the expressions in all cited papers will be adjusted to fit our convention.}. Then, it is postulated that no
  neutrino interaction can generate a self-mass term, {\it i.e.}, all interactions must respect this non-mixing chirality of neutrino terms. The way 
  of fulfilling this idea is by imposing the $\gamma_5$ invariance to all the neutrino interaction terms. Therefore, to the lepton\footnote{A lepton is 
  defined as a field which undergoes no strong interactions at tree level.} current in eq.(\ref{FermiGen}) must be added a term violating parity conservation. This is 
  accomplished by taking 

  \begin{equation}
   \mathcal{L}=g_i\left(\bar{\psi}_1\Gamma_i\psi_2\right)\left[\bar{e}\Gamma_i(1-\gamma_5)\nu\right],   
  \end{equation}
  where this added term must have the same coupling constant due to the same $\gamma_5$ invariance. Since $\frac{1}{2}(1-\gamma_5)$ is a projection 
  operator, meaning that it is hermitian and that any power of such operator gives the same operator, it can be noticed that the lepton accompanying the neutrino in the 
  current must have a determined chirality depending on the operator $\Gamma_i$ in the interaction. In other words, the lepton current can be divided 
  into scalar, pseudoscalar and tensor operators for an electron with opposite chirality than that of the neutrino and into vector and axial for same 
  chiralities.\\
  
  By making use of a Fierz identity for the muon decay, one is able to detect that two kind of processes may take place. One occurs with 
  the emission of two neutrinos and the other with the emission of a neutrino and an antineutrino
  
  \begin{subequations}
  \begin{align}
  \mathcal{L}_A=&g_i\left(\bar{\mu}\Gamma_ie\right)\left[\bar{\nu}\Gamma_i(1-\gamma_5)\nu\right], \hspace*{13ex} i=V,A,\\
  \mathcal{L}_B=&g_i\left(\bar{\mu}\Gamma_ie^*\right)\left[\nu^T\gamma_0\Gamma_i(1-\gamma_5)\nu\right], \hspace*{8ex} i=S,P,T.
  \end{align}
  \end{subequations}

  In the previous equations we can see that $\gamma_5$ invariance would require $g_V=-g_A$ for vector and pseudo-vector interactions, 
  while it requires $g_S=g_P$ for scalar and pseudoscalar interactions. These two interaction Lagrangian densities give different values for
  the Michel parameter\footnote{The Michel parameters \cite{MICHEL} in a three body decay give the energy and angular distributions, $\frac{d^2\Gamma}{x^2dxd\cos(\theta)}$,
  where $x=E/E_{max}$ is the normalized energy of a final state particle and $\theta$ is the angle between two final state particle three-momenta, which in the case of muon 
  decay can be given as a function of the angle between the final state charged lepton three-momentum and the spin of the decaying muon.} $\rho$, namely 
  $\rho_A=\frac{3}{4}$ and $\rho_B=0$ for $\mathcal{L}_A$ and $\mathcal{L}_B$ respectively. This was the first prediction of the correct $\rho$ value for the 
  $\mu$ decay, however the coupling constants in the generalized Fermi theory were not known and there was doubt if all operators would really contribute.\\
  
  At the moment there was not any certainty in which operators participated in the interactions since some experiments gave inconsistent results 
  among them. However, Richard Feynman \cite{FeynmanGellmann} showed an inconformity in describing the fundamental fermion field as a four component spinor, 
  arguing that for a spin 0 field (Klein-Gordon) we only need a wave function of one component and therefore the electron field should be described by 
  a two component field. Thus, he showed that the fermion field in the Dirac equation $(i\slashed{\partial} - \slashed{A})\psi=m\psi$ can be substituted by another 
  fermion field 
  \begin{equation}
   \psi=\frac{1}{m}(i\slashed{\partial}-\slashed{A}+m)\chi,
  \end{equation}
  which, as in the case of the Klein-Gordon field, is described by a second order equation
  \begin{equation}
   (i\slashed{\partial}-\slashed{A})^2\chi=\left[(i\partial^\mu-A^\mu)(i\partial_\mu-A_\mu)-\frac{1}{2}\sigma^{\mu\nu}F_{\mu\nu}\right]=m^2\chi.
  \end{equation}
  However, $\chi$ is still a four component spinor, but since $\sigma^{\mu\nu}$ commutes with $\gamma_5$ the $\chi$ field can be splitted 
  into $\gamma_5$ eigenvectors of two components, which are $\gamma_5\chi_-=-\chi_-$ and $\gamma_5\chi_+=\chi_+$. The connection with the 
  original spinor field is given by 
  \begin{equation}
   \chi_\mp=\frac{1}{2}(1\mp\gamma_5)\psi.
  \end{equation}
  Feynman also states that it is these two-component fields that should be treated as fundamental,
  and therefore, it is this field which should enter the weak current interaction. Then, connecting with $\gamma_5$ invariance Feynman postulates 
  that all fermions in the generalized Fermi theory should be inserted with the {\it left} projection operator. This can only lead to 
  currents of the type $V-A$, for the rest must vanish and, therefore, having a universal weak coupling strength. Worth is to mention that also Robert Marshak 
  and George Sudarshan in an independent work \cite{Marshak} showed that a universal Fermi interaction together with $\gamma_5$ invariance 
  can be achieved only through $V-A$ currents.\\
  
  Great success was achieved with this description of weak interactions, however one problem still remained: the Fermi interaction was not renormalizable. 
  A more fundamental approach could be achieved by merging both, weak and electromagnetic interactions in a more general theory. This idea has its origin 
  in some shared characteristics:
  \begin{itemize}
   \item {Both forces affect equally all forms of hadrons and charged leptons.
   }
   \item{Both are vector in character.
   }
   \item{Both (individually) possess universal couplings.
   }
  \end{itemize}
  
  Since universality and vector character are features of a gauge theory, these shared characteristics suggested that weak forces, just as electromagnetic 
  interactions arise from a gauge principle.\\
  
  In an attempt to give a more fundamental description and a unification of all particle interactions (except for gravity), Julian Schwinger suggested 
  \cite{SuperSchwinger} 
  that all intrinsic degrees of freedom are dynamically exhibited by specific interactions, each interaction with its characteristic symmetry properties, and 
  that the final effect of interactions with successively lower symmetry is to produce a spectrum of physically different particles from an initially degenerate 
  state. He also postulated that only unitary groups should be taken into account for internal symmetries, and by assuming a $SO(6)$ group for describing 
  electromagnetic, weak and strong interactions he made the suggestion that electroweak interactions among leptons could be unified in a {\it local} $SO(3)$ 
  subgroup. The fact that the electromagnetic field must be a realization of this symmetry assumed to be a component of a $SO(3)$ iso-triplet encouraged 
  Schwinger to say that the other two components of the iso-triple responsible for weak interactions must also be vector particles.\\
  
  This was the first step towards a consistent perturbative description of electroweak interactions, however there was a problem with Schwinger's theory and 
  all theories that tried to unify electromagnetic and weak interactions as a group $SU(2)$. By describing the Maxwell field as the component of such iso-triplet 
  one ends up with an interaction term with charged fermions that {\it do not} conserve parity, which is undeniably incompatible with electrodynamics. The 
  electromagnetic current found in this way is
  \begin{equation}\label{GlaShow}
   j^{3}_\mu=\bar{\psi}\gamma_\mu O_3\psi=\bar{\psi}\gamma_\mu\frac{1}{4}\left[t_3-\gamma_5(t_3^2-2)\right]\psi,
  \end{equation}
   where $t_i$ are generators of the $SO(3)$ electroweak subgroup.\\
  
  This was first noticed by Sheldon Glashow \cite{Glashow}, who introduced the concept of partial symmetries. This concept states that there might be a symmetry 
  under which part of the Lagrangian density is invariant; more precisely, it is only the mass terms in the Lagrangian density that break the conservation 
  under a determined symmetry transformation. Then, by proving that an $SO(3)$ subgroup give the inconsistent results shown in eq (\ref{GlaShow}), he arrives to 
  the conclusion that the only way to give a consistent description of a unified theory of electroweak interactions is by adding more vector fields, 
  where the minimal amount of added fields in this case is 1. This field is assumed to be a singlet under the $SO(3)$ group, meaning that it does not interact 
  with neither the charged weak fields nor with the neutral field. He then introduced the lepton current associated with this boson 
  \begin{equation}
   j^{B}_\mu=\bar{\psi}\gamma_\mu S\psi=\bar{\psi}\gamma_\mu\frac{3}{4}\left[t_3+\gamma_5(t_3^2-\frac{2}{3})\right]\psi.
  \end{equation}
  Thus, it is found that the operator $S$ satisfies the following relations 
  \begin{subequations}
   \begin{align}
    [\vec{O},S]&=0,\label{OS}\\
    O_1^2+O_2^2+O_3^2&+S^2=\boldsymbol{1},\label{remix}\\
    Q:=t_3=&O_3+S\label{Nakano}
   \end{align}
  \end{subequations}
  
  Now, eq. (\ref{OS}) means that the field $B$ associated with the operator $S$ must be a scalar under $SO(3)$ transformations, so that $j^B_\mu$ must remain 
  invariant under such transformations as expected. However, eq. (\ref{remix}) shows that $S$ is not independent from all the other operators. So, one can 
  find a $SO(2)$ symmetry to the $B$ field and the $W_3$ field associated with the $O_3$ operator in their own {\it neutral bosons space}. 
  \begin{equation}
   \left(\begin{array}{c} A\\Z \end{array}\right)=\left(\begin{array}{cc} \cos\theta_W&\sin\theta_W\\-\sin\theta_W&\cos\theta_W\end{array}\right)
  \left(\begin{array}{c} W_3\\B \end{array}\right).
  \end{equation}
 The relation of 
  eq. (\ref{Nakano}) give a relation resembling the Nakano--Nishijima--Gell-Mann (NNG) relation \cite{NN, Nishijima, GM}.\\
  
  This rotation between the neutral fields ends up with one that is associated with a parity conserving current, and therefore is identified as the Maxwell field.
  The mixing of these fields was first done to permit an arbitrary choice of strengths of the triplet and singlet interactions, which would explain 
  the difference between the weak and electromagnetic coupling strengths.\\
  
  Glashow had successfully unified the electromagnetic and weak interactions by relying on a $SO(3)\otimes U(1)$ symmetry of the interaction Lagrangian density and 
  the kinetic free Lagrangian density, but, as Glashow said, the mass terms do not preserve any weak interaction symmetry. This was a problem, since by adding a 
  mass term to the free Lagrangian density for the intermediate weak bosons one has to add the term $k^\mu k^\nu/m^2(k^2-m^2)$ to the propagator of the gauge bosons. 
  \begin{equation}
   D_W(k)=\frac{-g_{\mu\nu}}{k^2-m_W^2}\to -\frac{g_{\mu\nu}+\frac{k_\mu k_\nu}{m_W^2}}{k^2-m_W^2}
  \end{equation}
 The problem with this term 
  in the propagator is that it made loop corrections not renormalizable. So, the problem of having a perturbatively consistent theory was not solved yet.\\
  
  The solution thought was that the masses of the vector bosons should be generated dynamically. It was proven by Abdus Salam and John Clive Ward \cite{DeltaIRule} 
  that a non-zero vacuum expectation value of a scalar field interacting with other fields may give mass to the latter, however with a non-zero vacuum 
  expectation value breaking a global symmetry scalar bosons with zero mass must come about. This was first conjectured \cite{Goldstone1} and the proven by 
  Jeffrey Goldstone, Abdus Salam and Steven Weinberg \cite{Goldstone2} by using the K\"all\'en-Lehmann spectral representation. Then, Fran\c{c}ois Englert and 
  Robert H. Brout \cite{BE} and Peter Higgs \cite{Higgs} showed that one may be able to {\it exorcise out} the Goldstone bosons by choosing a gauge in which this 
  scalar particles could be transformed into the longitudinal mode of some of the gauge bosons propagators.\\
  
  Then, Weinberg \cite{Weinberg} gave the correct description of the complete unification of electromagnetic and weak interactions with the correct 
  mechanism that gives mass to all fields in the model (except for neutrinos and the photon fields) by taking the weak interaction symmetry to be
  $SU(2)_L\otimes U(1)_Y$ with a spontaneous symmetry breaking (SSB) of the $SU(2)$ symmetry given by the introduction of a weak doublet of scalar fields
  $\phi=\left(\begin{array}{c}\phi^-\\\phi^0\end{array}\right)$ that interact with the gauge bosons. Weinberg's Lagrangian is given by
  \begin{multline}
   \mathcal{L}=-\frac{1}{4}W_{\mu\nu}W^{\mu\nu}-\frac{1}{4}B_{\mu\nu}B^{\mu\nu}-\bar{R}\slashed{D}_RR-\bar{L}\slashed{D}_LL
   -\frac{1}{2}\left|(\partial-D_L)\phi\right|^2\\-Y(\bar{L}\phi R+\bar{R} \phi^\dagger L)-\mu^2\phi^\dagger\phi+\frac{\lambda}{24}|\phi^\dagger\phi|^2,
  \end{multline}
   where $W_{\mu\nu}=\partial_\mu W_\nu-\partial_\nu W_\mu+g\varepsilon_{abc}W_\mu^aW_\nu^b$, $B_{\mu\nu}=\partial_\mu B_\nu-\partial_\nu B_\mu$, 
   $L=\frac{1}{2}(1-\gamma_5)\left(\begin{array}{c}\nu_e\\e\end{array}\right)$, $R=\frac{1}{2}(1+\gamma_5)e$, $D_R=\partial-ig B$, 
   $D_L=\partial+ig\vec{t}\cdot\vec{W}-i\frac{1}{2}g'B$ and $t$ are the generators of the $W$ fields algebra. As in the case of the $K$ meson isospin doublet 
   (see subsection \ref{SMc} below), both scalar form a charge doublet, so that the $\bar{\phi^0}$ should be differentiated from the $\phi^0$, 
   therefore three of the states are absorbed (two charged and one neutral), the remaining one is called the Brout-Englert-Higgs (BEH) boson. \\
   
   There are completely analogous terms to those 
   consisting of the electron field and a neutrino with same leptonic charge as the electron field but for 
   muon and the neutrino related to muon production. This neutrino was found to be different to the one produced in beta decay by Leon Lederman, Melvin 
   Schwartz and Jack Steinberger at Brookhaven \cite{numu}, which was used by Weinberg to construct a $\mu$ $SU(2)_L$ doublet analogous to the electron one.\\
   
   Salam also arrived at these expressions from a
   more general symmetry principle by stating that it should be $SU(2)_L\otimes U(1)_Y$ the symmetry that unifies electromagnetic and weak gauge 
   bosons\cite{Salam-Ward,Salam}. Following Higgs' idea of {\it gauging out} the Goldstone bosons, Salam showed that with the most general model for a 
   scalar autointeracting field the symmetry breaking would then absorb three of the scalar fields into the $SU(2)$ gauge bosons in order to give them mass. 
   The Glashow angle now played the role of choosing the interaction strength $g\sin\theta_W$ such that one can find the appropriate massless vector boson 
   to identify it as the Maxwell field. He also noticed that since the theory before spontaneous symmetry breaking is renormalizable, it should keep like 
   this after breaking the symmetry, solving thus the problem of finding a perturbative description of a unified theory of electromagnetic and weak interactions 
   later known as the Standard Model of Electroweak interactions. But the problem still remained for particles that undergo strong interactions, which led 
   Glashow to the conclusion that this was just an academic exercise if this model is not general enough so that it also applies for fields that can 
   interact via strong interactions. So, a theory of strong interactions should be developed which was compatible with the electroweak Standard Model.
   
   \subsection{Strong Interactions and Quantum Chromodynamics}\label{qcd}
   
   The eta-charge, now known as strangeness, in particle physics was conjectured by Toshiyuki Nakano, Kazuhiko Nishijima \cite{NN, Nishijima} and Murray 
   Gell-Mann \cite{GM} where they proposed the relation known as the Nakano--Nishijima--Gell-Mann (NNG) relation 
   \begin{equation}
    Q=I_3+\frac{1}{2}(B+S),
   \end{equation}
   where $Q$ is the electric charge of the particle, $I_3$ is the isotopic spin third component, $B$ is the baryon number and $S$ is the strangeness (or $\eta$-charge) 
   of the particle.\\
   
   The NNG relation used to identify new baryons had been very successful describing newfound particles, which seemed like there should be 
   a more fundamental principle behind the relation. This was the thought followed by Soichi Sakata \cite{Sakata}, making an analogy with the coincidence between 
   the mass number of atomic nuclei and its spin: when the spin is integer the mass number is even, while if the spin is half-integer the mass number should 
   be odd. After Chadwick's discovery of the neutron, this puzzle was solved by developing the isotopic spin model. So this even-odd rule for nuclei was explained 
   by means of the sub-atomic particles. Therefore, Sakata showed us that, in analogy with the even-odd rule the NNG relation may be explained if one assumes the 
   existence of new particles from which all newfound hadrons should be made of\footnote{In his paper, Sakata gives credit for the first composite model to 
   Markov (Rep. Acad. Sci. USSR, 1955), although we were not able to track down such paper.}. He also stresses out the lack of interaction laws between 
   such fundamental particles.\\
   
   Then, Yuval Ne'eman \cite{Neeman} and Murray Gell-Mann \cite{8fold} proposed both in 1961 a $SU(3)$ symmetry between the currents generated with the fundamental 
   Sakata fields (also called sakatons) called the eight-fold way or symmetric Sakata model. This was intended as an extension to $SU(2)$ isospin to include strangeness. 
   These sakatons were $p$, $n$ and $\Lambda$, with the same quantum numbers as the proton, neutron and $\Lambda$ baryon. The statement was that 
   sakatons should interact via a massive vector boson, and that these interactions should be invariant under the $SU(3)$ symmetry. Now, for weak 
   currents of sakatons, one should arrive at the expressions
   \begin{equation}
    J_\mu=i\bar{p}\gamma_\mu(1-\gamma_5)n+i\bar{p}\gamma_\mu(1-\gamma_5)\Lambda,
   \end{equation}
   which can be achieved by taking the combinations of currents 
   \begin{equation}\label{WeakJ}
    J_\mu=i\bar{b}\gamma_\mu\frac{1}{2}(\lambda_1+i\lambda_2+\lambda_4+i\lambda_5)b,
   \end{equation}
   where $b=\left(\begin{array}{c}p\\n\\\Lambda\end{array}\right)$ is the baryon triplet and $\lambda_i$ is the $i$th Gell-Mann matrix. It should be noticed the lack of 
   $(\lambda_6\pm i\lambda_7)$ currents that would couple the $n$ and $\Lambda$ fields giving rise Flavor Changing Neutral Current (FCNC) 
   which were not be seen by the time this theory was postulated. (The combinations of $\lambda_3$ and $\lambda_8$ will give the diagonal charged current.)\\
   
   By making use of the $V-A$ model of weak interactions \cite{FeynmanGellmann,Marshak} and the eight-fold way model, Nicola Cabibbo found in 1963 that 
   weak currents of strongly interacting fields should have some additional symmetry\cite{Cabibbo}; those belonging to a $SU(3)$ representation of the 
   symmetric Sakata model with $\Delta S=0$, and $\Delta Q=1$, $j_\mu^{(0)}$, should be related to currents $j_\mu^{(1)}$ with $\Delta S=\Delta Q=1$, the 
   former with selection rule $\Delta I=1$, the latter with $\Delta I=\frac{1}{2}$. In this model, strangeness changing weak current should be blended with $\Delta S=0$ 
   currents since a $SU(3)$ transformation would necessarily mix $N$ and $\Lambda$ in the general weak current for sakatons in eq. (\ref{WeakJ}). Thus, the 
   total weak current for strongly interacting fields is
   \begin{equation}
    J_\mu=aj_\mu^{(0)}+bj_\mu^{(1)},
   \end{equation}
   where $a$ and $b$ should have some universality constraint stemming from the mix between strangeness conserving and changing currents. A naive universality 
   relation would be $a=b=1$; however this might give rise to uncoupled currents. Therefore, Cabibbo assumes a weaker form of universality, namely that 
   $J_\mu$ should be of `unit-length', {\it i.e.}, $a^2+b^2=1$. Hence, $J_\mu$ can be re-expressed as\footnote{A similar expression following the 
   same universality statement was obtained first by Gell-Mann and L\`evy\cite{GMLevy}, but using the relation $$\bar{p}\gamma_\mu(n+\epsilon\Lambda)
   (1+\epsilon^2)^{-\frac{1}{2}}$$ which is a very good approximation to the expression given by Cabibbo when expanded near $\theta\approx0$. 
   Gell-Mann and L\`evy take $\epsilon^2\sim0.06$, which gives $\sin\theta\sim\epsilon\approx 0.26$ a very good approximation.}
   \begin{equation}
    J_\mu=\cos(\theta)j_\mu^{(0)}+\sin(\theta)j_\mu^{(1)}.
   \end{equation}
   This weaker form of universality solved several experimental discrepancies between different processes that implied the use of weak currents of strongly 
   interacting particles.\\
   
   The symmetric Sakata model was successful explaining the octuplet allocation of pseudoscalar mesons, however it failed in constructing the nucleons 
   from the sakatons. This is why in 1964 George Zweig \cite{Zweig} proposed some fundamental fields called aces instead of sakatons. (Gell-Mann made 
   the same proposition also in 1964, calling the fundamental fields quarks \cite{quarks}.) This fields should have 
   fractional electric charge with spin 1/2 and should form a triplet of the same $SU(3)$ symmetry
   \begin{equation}
    q=\left(\begin{array}{c}u^{\frac{2}{3}}\\d^{-\frac{1}{3}}\\s^{-\frac{1}{3}}\end{array}\right),
   \end{equation}
   where the super-index denotes the electric charge of each ace or quark in units of the proton electric charge $e$. Since quarks have spin $1/2$, baryons should be composed from an odd number of 
   quarks and mesons by an even number of quarks. Then, by taking products of quarks and anti-quarks baryons should be 
   represented as the product of three quark fields since $(qqq)=\boldsymbol{1}\oplus\boldsymbol{8}\oplus\boldsymbol{8}\oplus\boldsymbol{10}$, 
   and mesons as the product of a quark and an anti-quark since $(q\bar{q})=\boldsymbol{1}\oplus\boldsymbol{8}$, where $\boldsymbol{1}$, 
   $\boldsymbol{8}$ and $\boldsymbol{10}$ are one, eight and ten dimensional representations of $SU(3)$. For example, 
   the meson octet would be composed by $\pi$, $K$ and $\eta$ mesons. Further more, a 27 dimensional representation that must be 
   considered in the symmetric Sakata model not found experimentally should be absent in the quark model.\\
   
   Unsatisfied with having a fractional electric charge (in units of the proton charge), Moo-Young Han and Yoichiro Nambu postulated that the fundamental 
   fields should be each one a triplet of an $SU(3)$ symmetry \cite{Han-Nambu} which was not the one proposed by Ne'eman. These should be (as Schwinger had proposed 
   \cite{SuperSchwinger}) a local gauge group, leading to eight neutral vector bosons, named {\it gluons}, that mediate the strong interactions and does not mix the triplets. Since it 
   was derived as a subgroup of an embedding $SU(6)$ symmetry, the fundamental triplets should interact through other $SU(3)$ symmetry which, as the weak 
   interaction does, would blend the three interacting triplets. In this case, similar to the Salam $SU(6)$ model where the correct symmetry was  
   the subgroup $SU(2)\otimes U(1)$, the correct group seemed to be the one with the eight vector bosons which did not mixed the triplets of fermions.
   This necessarily implies a new charge for these fields. 
   This new charge would explain the fact that the newfound 
   $\Omega^-$ had $S=-3$ and $J=\frac{3}{2}$ \cite{BarnesOmega-} and the existence of the $\Delta^{++}$ baryon with same spin as the $\Omega^-$ but with isospin
   $I=\frac{3}{2}$ found in 1951 \cite{KeithBruecknerLambda}, since without this extra charge, the Pauli exclusion principle would forbid the existence 
   of these baryons. The fact the $\Delta^{++}$ and $\Omega^-$ baryons should be each made of three identical fermions made the fractional charge 
   model of quarks the most viable model discarding the integer charge model since in the Han-Nambu model it is not possible to reproduce the $\Omega^-$.\\
   
   However, quarks were still fundamental fields understood mainly as mathematical objects with no physical evidence of their existence. Then, in 1968 
   James D. Bjorken studying the inelastic lepton-proton scattering, demonstrated that in the limit of infinite energy transfer the structure functions 
   upon which the cross section depends remain finite \cite{Bjorken}, furthermore, he showed that these structure functions can be expressed as dependent of the ratio 
   of the virtual photon four-momentum squared $q^2$ and the initial energy of the proton $P_0$, which is taken as constant as one takes the limit
   $q_0,P_0\to\infty$. So, the structure functions have no dependence on the scaling of the energy of the process, meaning that no structure can be 
   discerned as the energy of the process is augmented. Since by reducing the de Broglie wavelength one has a higher energy state, 
   by taking a higher center of mass energy one will have a greater resolution scale, probing smaller space regions. Therefore, the electrons 
   scattered must be interacting with free point-like particles inside the proton.\\
   
   With only three kind of quarks, the theory of weak interactions seemed to have some problems with experimental selection rules of weak processes. 
   By using the Pauli-Villars regularization technique, a cut-off energy remarkably small of $\Lambda\sim$ 3 GeV seemed necessary. Also 
   there should be amplitudes with $\Delta S=2$ contributing to $K$ decays that were not observed since a $\bar{s}\gamma^\mu d $ hadronic neutral current was not 
   prohibited by the electroweak model with three quarks. In 1970, Sheldon Glashow, Jean ($I\omega\acute{\alpha}\nu\nu\eta\varsigma$) 
   Iliopoulos and Luciano Maiani proposed the existence of a 
   new quark \cite{GIM} in analogy with the electroweak model for four leptons ($e,\nu_e,\mu$ and $\nu_\mu$) leading to the hadronic current 
   \begin{equation}
    J^{H}_\mu=\bar{q}C_H\gamma_\mu(1-\gamma_5)q,
   \end{equation}
   where $q=(c,u,d,s)$, and in order for $J^H_\mu$ to be unit charge current, $C_H$ must have the following form
   \begin{equation}\label{CKM2flav}
    C_H=\left(\begin{array}{cc}\boldsymbol{0}&U\\\boldsymbol{0}&\boldsymbol{0}\end{array}\right),
   \end{equation}
   where $\boldsymbol{0}$ is a $2\times2$ zero matrix and $U$ is a matrix that must be unitary in analogy with the leptonic weak current. By rephasing 
   the quark fields one gets the most general form
   \begin{equation}
    U=\left(\begin{array}{cc}-\sin\theta&\cos\theta\\\cos\theta&\sin\theta\end{array}\right).
   \end{equation}
   This proposed quark should have $Q=\frac{2}{3}$, $Y=-\frac{2}{3}$ and, since it must be an isospin singlet, a new quantum number called {\it charm}\footnote{
   Although Glashow had already proposed a charm number to describe a different quantum number, Iliopoulos affirmed that the name 
   {\it charm} used to baptize the new quark was thought of as a good-luck charm for them so that it would exist and be detected soon \cite{CLASHEP15}.}. 
   In this model, the weak symmetry group for quarks $SU(2)\otimes U(1)$ is a partial symmetry (in the sense of Glashow's paper \cite{Glashow}) 
   of the gluonic $SU(3)$ symmetry of strong interactions. With this, Glashow, Iliopoulos and Maiani gave the term complementary to the Glashow-Salam-Weinberg 
   theory of leptonic weak interactions, completing the theory of weak interactions. On the other hand, the 3 GeV cut needed without the charm quark can 
   be qualitatively explained with the model including the charm quark since $\Lambda\sim m_c$, where $m_c$ is the charm quark mass, this is, the charm 
   quark becomes an active degree of freedom of the theory only above these energies.\\
   
   In 1972, William Bardeen, Harald Fritzsch and Murray Gell-Mann proposed that the quarks should have the extra charge mentioned above and coined the 
   term {\it color} for this new quantum number. Each quark should exist with one of three possible color values, `say red, white and blue' \cite{Gell-Mann:QCD}. 
   Also, all physical states and all observable quantities must be color $SU(3)$ singlets. This theory solved immediately the tension between the quark 
   model predicted $BR(\pi^0\to\gamma\gamma)$ and the experimental result that gave a factor $9$ greater. Eventually, this became the standard theory 
   for strong interactions that has been used since then to describe all strong interactions. 
   
   \subsection{Standard Model of Particle Physics}\label{SMc}
   
   There was still one problem with the Lagrangian of elementary particles. As Nakano and Nishijima showed, the $K^0$ must be different from $\bar{K^0}$,
   then, they must belong to two different isospin doublets and have different strangeness. These states can be expressed as a linear combination of 
   $CP$ symmetry eigenstates, $K_1$ and $K_2$ which makes them identifiable by their decay products into pions. Since the $\pi$ has intrinsic parity 
   $P=-1$ and $C=1$, the $K_1$ being $CP$ even should decay into two pions and the $K_2$ into three, so each state would have definite mass and lifetime. 
   Since one has a longer lifetime, they were called $K$ short ($K_S=K_1$) and $K$ long ($K_L=K_2$).
   However, as it was shown by James Christenson, James Cronin, Val Logsdon Fitch and Ren\'e Turlay\footnote{Only Cronin and Fitch were awarded with the 
   Physics Nobel Prize despite the four of them contributed to the same work.} at the Alternating Gradient Synchrotron at Brookhaven \cite{K0NL}, 
   there were a small probability that $K_L$ would decay in the channel that $K_S$ does, giving a clear indication of $CP$ symmetry violation.\\
   
   Within the models of electroweak and strong interactions there is no $CP$ violation, meaning that something should be still missing. In 1972, Makoto Kobayashi 
   and Toshihide Maskawa suggested four possible scenarios within the electroweak model to include $CP$ violation \cite{Kobayashi-Maskawa}. One of them was 
   the scenario where two more quarks should be included in a left $SU(2)_L$ doublet and two right handed singlets, so that including these extra quarks 
   the matrix in eq. (\ref{CKM2flav}) must now have a non-factorisable phase responsible for $CP$ violation.\\
   
   Then, in 1977 the E288 experimental team lead by Leon Lederman at Fermilab discovered a meson resonance which should have a different content 
   of quarks \cite{LLederman}. This new quark was named {\it bottom} and should be part of a new $SU(2)_L$ doublet. two years earlier, Martin Lewis Perl 
   discovered the tau lepton with the SLAC-LBL group \cite{tau}. The top quark, which was the left weak doublet partner of the bottom was discovered in 
   1995 in the CDF and D$\slashed{0}$ experiments at Fermilab \cite{topquark}. The tau neutrino was discovered in the year 2000 by the DONUT collaboration
   \cite{DONUT} completing the Kobayashi-Maskawa frame (also generalized to leptons).\\
   
   Thus, the Standard Model (SM) is the theory that describes the interactions between charged leptons, neutrinos, quarks and gauge bosons, which are the mediators 
   of the electromagnetic, weak and strong interactions. The internal symmetry that generates these gauge bosons is thus $SU(3)_C\otimes SU(2)_L\otimes U(1)_Y$,
   with a weak gauge spontaneous symmetry breaking field experimentally discovered in 2012 at the Large Hadron Collider experiments \cite{LHCBEH}. The electroweak 
   Standard Model (EWSM) Lagrangian is thus expressed as the sum of four terms,
   
   \begin{equation}\label{EWSM}
    \mathcal{L}_{EWSM}=\mathcal{L}_{fermion}+\mathcal{L}_{gauge}+\mathcal{L}_{\phi}+\mathcal{L}_{Yukawa},
   \end{equation}
   the kinetic term of the fermions and their interaction with the gauge bosons ($\mathcal{L}_{fermion}$), the pure gauge bosons contributions 
   (kinetic and interactions) ($\mathcal{L}_{gauge}$), the BEH field and interaction with gauge bosons ($\mathcal{L}_{\phi}$) and the Yukawa 
   interaction between the BEH field and the SM fermions ($\mathcal{L}_{Yukawa}$). Each term is invariant under the $SU(2)_L\otimes U(1)_Y$ 
   symmetry group. Since the electroweak symmetry is non abelian, there will be an interaction term among gauge bosons that is included in 
   
   \begin{equation}\label{gauge}
   \mathcal{L}_{gauge}=W_{\mu\nu}^aW^{\mu\nu}_a+B_{\mu\nu}B^{\mu\nu},
   \end{equation}
   where $W_a^{\mu\nu}=(\partial^\mu W^\nu_a-\partial^\nu W_a^\mu-gf_{abc}W^\mu_bW^\nu_c)$, $f$ is the structure constant of the group and 
   $B_{\mu\nu}=\partial_\mu B_\nu-\partial_\nu B_\mu$. The correct interaction term of the gauge bosons with the fermions is obtained by the 
   standard method of making an arbitrary local gauge transformation, obtaining thus
   \begin{equation}
    \mathcal{L}_{fermion}=\sum_{i}\bar{\chi}^i_L(i\slashed{\partial}-g\tau_a \slashed{W}_a-g'\slashed{B})\chi^i_L,
   \end{equation}
   where $i$ runs through all the lepton flavors $l$ and quark doublets $q$, $\tau_a$ are the $2\times2$ Pauli matrices and $g$ and $g'$ are the $SU(2)_L$ 
   and $U(1)_Y$ coupling constants respectively. As Weinberg showed us \cite{Weinberg}, the correct representation of the SM fermions is through doublets 
   defined as 
   \begin{equation}
    \chi_L^l=\left(\begin{array}{c}\psi_\ell\\\nu_\ell\end{array}\right)_L\text{\hspace*{5ex} and \hspace*{5ex}}
    \chi_L^q=\left(\begin{array}{c}\psi_u\\\psi_d\end{array}\right)_L,
   \end{equation}
   where $\psi_{u}$ is an up-type quark and $\psi_d$ its corresponding down-type quark ($u\leftrightarrow d,c\leftrightarrow s,t\leftrightarrow b$). The 
   subindex $L$ denotes the left projection of the fermion field, {\it i.e.}, $\psi_L=\frac{1}{2}(1-\gamma_5)\psi$.
   
   The scalar fields term in the Lagrangian is obtained in an analogous way to the fermion case, and reads
   \begin{equation}
    \mathcal{L}_{\phi}=(D^\mu\phi)^\dagger D_\mu\phi-V(\phi)
   \end{equation}
   The field $\phi$ is realized as a $SU(2)_L$ doublet $\phi=\left(\begin{array}{c}\phi^+\\\phi^0\end{array}\right)=
   \frac{1}{\sqrt{2}}\left(\begin{array}{c}\phi_1-i\phi_2\\\phi_3-i\phi_4\end{array}\right)$, where the $\phi_i$ are all real fields.
   The covariant derivative is $D^\mu=\partial^\mu+ig\frac{\tau_a}{2}W^\mu_a-i\frac{g'}{2}B^\mu$. 
   The $V(\phi)$ term is made of the self-interacting terms of the introduced scalar fields 
   \begin{equation}
    V(\phi)= \mu^2|\phi|^2+\frac{\lambda}{4!}|\phi|^4,
   \end{equation}
   where by taking $\mu^2<0$ one obtains a non-zero vacuum expectation value $vev$, which gives the SSB of the electroweak symmetry into the 
   electromagnetic symmetry $SU(2)_L\otimes U(1)_Y\to U(1)_{EM}$. As was said above, three of the scalar fields are absorbed by the weak bosons 
   as longitudinal component of these fields leaving only one physical scalar field, the BEH field. Since the remaining scalar has non-zero $vev$,
   we can take the covariant derivative term acting on the $vev$ of the BEH field to get
   \begin{multline}
    \left|\left(ig\frac{\tau_a}{2}W_a^\mu+i\frac{g}{2}B^\mu\right)\left(\begin{array}{c}0\\v\end{array}\right)\right|^2=\frac{1}{8}v^2g^2\left(
    W_1^\mu {W_1}_\mu+W_2^\mu {W_2}_\mu\right)\hspace*{25ex}\\
    \hspace*{35ex}+\frac{1}{8}v^2(g'B_\mu-g{W_3}_\mu)(g'B^\mu-g{W_3}^\mu)\\
    \hspace*{19ex}=\left(\frac{1}{2}vg\right)^2{W^+}_\mu {W^-}^\mu+\frac{1}{8}v^2\left(W^3_\mu,B_\mu\right)\left(\begin{array}{cc}g^2&-gg'\\-gg'&g'^2\end{array}\right)
    \left(\begin{array}{c}{W^3}^\mu\\B^\mu\end{array}\right),
   \end{multline}
   where $W^\pm=\frac{1}{\sqrt{2}}\left(W^1\mp iW^2\right)$. All terms in the previous expression are mass terms since they are quadratic forms of the fields,  
   to see this we need to express $W^3$ and $B$ in a base which has no terms of the kind $W^3\cdot B$, this is, no gauge boson mix terms. In order to obtain 
   these mass eigenstates a rotation of the neutral current bosons $W^3$ and $B$ is done
   \begin{equation}
    \left(\begin{array}{c}A_\mu\\Z_\mu\end{array}\right)=\left(\begin{array}{cc}\cos\theta_W&\sin\theta_W\\-\sin\theta_W&\cos\theta_W\end{array}\right)
     \left(\begin{array}{c}B_\mu\\W^3_\mu\end{array}\right).
   \end{equation}
   To obtain such states one finds that the relation $\tan\theta_W=g/g'$ must be fulfilled so that one of the states remains massless. This field is
   identified, as Glashow, Weinberg and Salam did, as the Maxwell (electromagnetic) field. Therefore, the mass of the other fields are $m_W =vg/2$ and
   $m_Z=\frac{1}{2}\sqrt{g^2+g'^2}$. With this, the interaction of the fermions with the electroweak gauge bosons can be written as follows
   \begin{equation}
    {\cal L}_{CC}=\frac{g}{2\sqrt{2}}W^{-}_{\mu}\left\{ \sum_\ell\bar{\psi}_\ell \gamma^{\mu}(1-\gamma_5) \psi_{\nu_\ell} + \sum_i\bar{\psi_d}_i\gamma^{\mu}(1-\gamma_5) {\psi_u}_i\right\} + {\rm h. c.}\ ,
   \end{equation}
   where $\ell$ runs on the lepton number, $i$ on the quark family and h.c. is the hermitian conjugate of the previous terms. For the weak neutral 
   one has 
   \begin{multline}
    {\cal L}_{NC} = \frac{g}{2\cos\theta _W}Z_\mu\sum_{i,\ell}\left[\frac{}{}{\overline{\psi}_{u_i}}_L\gamma^\mu{\psi_{u_i}}_L -{\overline{\psi}_{d_i}}_L\gamma^\mu{\psi_{d_i}}_L +
    {\overline{\psi}_{\nu_\ell}}_L\gamma^\mu{\psi_{\nu_\ell}}_L -{\overline{\psi}_\ell}_L\gamma^\mu{\psi_\ell}_L\right.\\
     \left. -2\sin^2\theta_W\left(\frac{2}{3}\overline{\psi}_{u_i}\gamma^\mu\psi_{u_i} -
     \frac{1}{3}\overline{\psi}_{d_i}\gamma^\mu\psi_{d_i} -\overline{\psi}_\ell\gamma^\mu\psi_\ell\right)\right].
   \end{multline}
   The remaining piece of the Lagrangian is the interactions of the fermions with the scalar doublet introduced to induce the SSB of the EW gauge. This is given by
   \begin{equation}\label{Yukawa}
    \mathcal{L}_{Yukawa}=-\sum_{j,k}\left[\Gamma^u_{jk}\overline{\chi}^{q_j}_L\tilde{\phi}\left(\psi_{u_k}\right)_R+
    \Gamma^d_{jk}\overline{\chi}^{q_j}_L\phi\left(\psi_{d_k}\right)_R + \Gamma^l_{jk}\overline{\chi}^{l_j}_L\phi\left(\psi_{l_k}\right)_R +\ {\rm h.c.}
    \right],
   \end{equation}
   where a different convention to that of Weinberg has been used for the terms with right-handed up-type quarks, here $\tilde{\phi}=i\tau_2\phi^\dagger$. 
   Since in the broken EW gauge the scalar doublet field is just $\phi=\left(\begin{array}{c}0\\v+H\end{array}\right)$, one finds for the $u$ type quarks 
   (for the down-type quarks and the leptons a completely analogous procedure follows) the Lagrangian density
   \begin{equation}\label{yuk}
    \mathcal{L}_{Yuk}=\sum_{jk}\left(\overline{\psi}_{u_j}\right)_L{Y}_{jk}^u\left(\psi_{u_k}\right)_R+\ {\rm h.c.},
   \end{equation}
   where the matrix $Y$ is in general neither hermitian, nor diagonal. Since the physical states are the mass eigenstates we need to express the previous 
   equation in terms of mass eigenstates. Since any matrix can be diagonalized by multiplying it by the left and by the right with the adequate unitary 
   matrices, the left and right fermion fields are transformed with unitary matrices mixing all the $u$ type quarks. Therefore, by making 
   ${\psi_u}_L\to A_L^u{\psi_u}_L$ and ${\psi_u}_R\to A_R^u{\psi_u}_R$ one finds
   \begin{equation}
    A^{u\dagger}_LM^uA^u_R=\left(\begin{array}{ccc}m_u&0&0\\0&m_c&0\\0&0&m_t\end{array}\right).
   \end{equation}
   There are analogous terms for the down-type quarks and for the charged leptons. Notice that since in the SM neutrinos are assumed massless, there is 
   no such term for these fields. If one forbids the right-handed projection of the neutrino field, no mass term can be generated in eq. (\ref{Yukawa}).\\

   It is trivially verified that this unitary transformation does not affect the neutral currents of the standard model, since they include a $\bar{\psi_X}$ and 
   a $\psi_X$ factor and all SM operators are diagonal in flavor space\footnote{Flavor is defined as the attributes that distinguish quarks and charged leptons,
   namely the $u$, $d$, $s$, $c$, $b$, $t$ for quarks and the $e$, $\mu$ and $\tau$ for leptons.}. However, charged currents are modified by these rotations since 
   they involve fields of two different kinds (e.g. $\bar{\psi_e}$ and $\psi_{\nu_e}$). These matrices will give one unitary matrix in the interaction Lagrangian since the product of two unitary matrices 
   is a unitary matrix. The parameters of this unitary matrix can be diminished by rephasing the fermion fields. So, for the hadronic charged currents one has
   
   \begin{equation}
    {\cal L}_{cc}^q= \frac{g}{2\sqrt{2}}W^+_{\mu} \left\{ \left(\bar{d},\bar{s},\bar{b}\right)V^{\dagger}_{CKM}\gamma^{\mu}(1-\gamma_5) \left(\begin{array}{c}u\\c\\t\end{array}\right) \right\}+\ {\rm h.c.}\ ,
   \end{equation}
   where $V_{CKM}$ is the unitary matrix generated by the transformation done to obtain the mass eigenstates of the quarks, which after rephasing the quark fields 
   has three real parameters and one phase. This phase is the responsible for $CP$ violation which was detected in $K^0$ and $\bar{K^0}$ decays. Notice that since 
   no right-handed projection of the neutrino field exist in the SM, no Yukawa term can be generated for these fields, so that there is no special unitary 
   transformation in the flavor space of the neutrinos to diagonalize a $Y$ matrix as in eq. (\ref{yuk}). Therefore, when one transforms the charged lepton 
   fields, the same unitary transformation can be applied to the neutrino field so that the resulting matrix is the identity. Thus, no mixing matrix is 
   obtained for leptons as it is for quarks.\\
   
   \subsection{QCD, limitations and Effective Field Theories}\label{Limitations}

    Quantum Chromodynamics (QCD), as was previously stated, is described by a local $SU(3)$ symmetry, with a Lagrangian density 
   \begin{equation}\label{QCDLd}
    \mathcal{L}_{QCD}=\mathcal{L}_{quark}+\mathcal{L}_{gluon},
   \end{equation}
   where $\mathcal{L}_{gluon}$ is analogous to $\mathcal{L}_{gauge}$ in eq. (\ref{EWSM}). It can be obtained by taking the first term in eq. (\ref{gauge})
   and changing the weak ($W$) for the gluon ($G$) fields and by taking $f$ as the structure constant of $SU(3)$. The quark term is 
   \begin{equation}
    \mathcal{L}_{quark}=\sum_r\bar{q}_{r\alpha} i\slashed{D}^\alpha_\beta q^\beta_r,
   \end{equation}
   where $r$ runs over all flavors and $\alpha$ and $\beta$ are color indices. The covariant derivative is found in an analogous way to the 
   EW case, and reads
   \begin{equation}
    D^\mu_{\alpha\beta}=\delta_{\alpha\beta}\partial^\mu +i\frac{g_s}{2}G_i^\mu\lambda^i_{\alpha\beta},
   \end{equation}
   where $g_s$ is the strong interaction constant, $\delta$ is the Kronecker $\delta$ and $\lambda$ are the generators of the $su(3)$ algebra.
    With these all the terms of the SM have been described. \\

    As has been said, all the processes computed within the SM give outstandingly precise predictions of a very vast amount of processes. This relies on the 
    renormalizability of the whole model, as a result the coupling constant varies with the energy of the studied process. Nevertheless, when QCD is 
    renormalized it is found that the coupling constant diverges as one approaches the 1 GeV energy region (where the energy at which it diverges is called the 
    Landau pole) and the theory, despite being correct becomes 
    not perturbative. Since there is no way known of computing exactly amplitudes in Quantum Field Theory, QCD is of no use without a perturbative 
    description of phenomena. A way to overcome this problem is using an Effective Field Theory (EFT) which relies on some symmetries of the original 
    QCD Lagrangian.\\
    
    There are two kinds of EFTs \cite{Ecker95}, namely
    \begin{itemize}
     \item {Decoupling EFTs: These are characterized by an energy scale $\Lambda$ below which only light degrees of freedom are left and the heavy ones are frozen 
     and so can be {\it integrated out}. Such theories are described by Lagrangians such as
     \begin{equation}
      \mathcal{L}_{eff}=\mathcal{L}_{d\le4}+\sum_{d>4}\frac{1}{\Lambda^{d-4}}\sum_{i_d}g_{i_d}\mathcal{O}_{i_d},
     \end{equation}
     where $d$ is the dimension of the operator $\mathcal{O}$ in natural units.
     }
     \item{Non-decoupling EFTs: The transition from fundamental to effective theory is made through a phase transition via a SSB generating Goldstone bosons. 
     Processes with different number of Goldstone bosons relate $d\le4$ with $d>4$ terms, so that one cannot distinguish between these terms.
     }
    \end{itemize}

    \subsection{Chiral symmetry of the QCD Lagrangian density}
    
    One needs to rely on EFTs to compute processes involving strong interaction at energies near or below the Landau pole, so the question of how to 
    construct such EFT comes about. First of all, as previously stated a symmetry of the underlying theory is needed to develop the EFT. By reviewing 
    the historical development of the theories of strong interactions in the previous section in becomes appealing to rely on some kind of flavor symmetry due to 
    its great success in the Symmetric Sakata model and the quark model in describing phenomena below the lepton-nucleon inelastic scattering, $\sim1$ 
    GeV, where the structure of the nucleons becomes apparent. Since we are interested in regions around and below the Landau pole, the QCD Lagrangian 
    density (\ref{QCDLd}) can be split into light and heavy degrees of freedom, 
    \begin{equation}
     \mathcal{L}_{quark}=\sum_{q=u,d,s}\bar{q}i\slashed{D}q+\mathcal{L}_{heavy\hspace*{1ex} quarks}
     =\sum_{q=u,d,s}\left(\bar{q_L}i\slashed{D}q_L+\bar{q_R}i\slashed{D}q_R\right)+\mathcal{L}_{heavy\hspace*{1ex} quarks}.
    \end{equation}
    By separating the light quarks in their right and left chiral parts it becomes apparent a $SU(3)_L\otimes SU(3)_R\otimes U(1)_V\otimes U(1)_A$ symmetry. 
    The $U(1)_V$ is just the baryon number conservation and the $U(1)_A$ is not conserved at the quantum level. The remaining $G=SU(3)_L\otimes SU(3)_R$ is 
    called the chiral group. 
    Notice that only the weak symmetry $SU(2)_L$ is explicitly broken by including a mass term for the quarks, meaning that QCD does not forbid a mass term 
    for the quarks. This mass term breaks explicitly the chiral group symmetry. However one can exclude the quark masses and include them later in a 
    consistent way.\\
    
    The Noether currents of the chiral symmetry $G$ are $\bar{q}\gamma^\mu\frac{1}{2}(1\pm\gamma_5)\frac{\lambda^a}{2}q$ for right and left quark currents,
    with which one can construct the vector and axial quark currents 
    \begin{subequations}
     \begin{align}\centering
     V^{\mu,\alpha}&=R^{\mu,\alpha}+ L^{\mu,\alpha}=\bar{q}\gamma^\mu\frac{\lambda^a}{2}q,\\
     A^{\mu,\alpha}&=R^{\mu,\alpha}- L^{\mu,\alpha}=\bar{q}\gamma^\mu\gamma_5\frac{\lambda^a}{2}q.
     \end{align}
    \end{subequations}
    In this way one arrives to a $SU(3)_V\otimes SU(3)_A$ symmetry. This basis is chosen since theoretically and experimentally there is evidence that the chiral 
    group must be spontaneously broken to $SU(3)_V$ \cite{Ecker95, Pich94}. A way to see this is with a similar argument 
    to that used by Scherer \cite{Scherer02}: If $SU(3)_V\otimes SU(3)_A$ was the symmetry of the meson spectrum, to the lowest lying octet of vector mesons 
    would correspond an octet of opposite parity and with same spin and mass, {\it i.e.}, the lowest lying axial-vector mesons. The fact that empirically both octets 
    have significantly different masses (eg. $m_{a_1}\approx 2m_\rho$) means that $SU(3)_A$ must be broken. Since the Lagrangian density is invariant under 
    the complete chiral group, this means that $SU(3)_A$ must be spontaneously broken, generating eight Goldstone bosons. Since the Goldstone bosons inherit the 
    properties of the generators of the broken symmetry, they must be pseudoscalar, with zero baryon number.
    
  \subsection{Inclusion of external currents}
    
    Aiming to construct a theory which can be obtained through a generating functional, external currents are introduced in order to generate Green functions of 
    quark currents. This external fields do not propagate and can be introduced by extending the Lagrangian density adding quark currents coupled to 
    some external hermitian fields
    \begin{equation}
     \mathcal{L}=\mathcal{L}_{quarks}+\bar{q}\gamma^\mu(v_\mu+a_\mu\gamma_5)q-\bar{q}(s-ip\gamma_5)q.
    \end{equation}
    The major advantages of this method is that one can include the electroweak gauge boson interactions by making
    \begin{subequations}
     \begin{align}
      \centering
      r_\mu&=v_\mu+a_\mu=-eQA_\mu^{ext}\\
      \ell_\mu&=v_\mu-a_\mu=-eQA_\mu^{ext}-\frac{e}{\sqrt{2}\sin\theta_W}\left(W^{ext,+}_\mu T_+\right)
     \end{align}
    \end{subequations}
    Where $Q=\frac{1}{3}\text{diag}(2,-1,-1)$, $(T_+)_{ij}=\delta_{i1}(\delta_{j2}V_{ud}+\delta_{j3}V_{us})$ and $V_{ij}$ are the Cabibbo-Kobayashi-Maskawa 
    mixing matrix. Also, one can include (as previously said) the quark masses by means of the scalar external current, this is, 
    symmetry breaking terms can be introduced by means of the external fields. If one takes $v=a=p=0$ and 
    \begin{equation}
     s=\left(\begin{array}{ccc}m_u&0&0\\0&m_d&0\\0&0&m_s\end{array}\right),
    \end{equation}
    one can include symmetry breaking terms in a manifestly chiral invariant way. The inclusion of external fields promotes chiral symmetry to a local one. 
    The transformation rules of the external fields are 
    \begin{subequations}
     \begin{align}
      \centering
      r^\mu&\to g_Rr^\mu g_R^\dagger+ig_R\partial^\mu g_R^\dagger,\\
      \ell^\mu&\to g_L\ell^\mu g_L^\dagger+ig_L\partial^\mu g_L^\dagger,\\
      s+ip&\to g_R(s+ip)g_L^\dagger\quad\quad\\ 
      s-ip&\to g_L(s-ip)g_R^\dagger\ .
     \end{align}     
    \end{subequations}
    Since now chiral symmetry has been promoted to a local one the derivative of the quark fields must be modified in order to keep the chiral invariance
    also, by introducing a field strength tensor that transforms like $F^{\mu\nu}_x\to g_x F^{\mu\nu}_x g_x^\dagger$ for $x=r,\ell$. Thus, one has
    \begin{subequations}
     \begin{align}
      \centering
      F^{\mu\nu}_R&=\partial^\mu r^\nu - \partial^\nu r^\mu-i[r^\mu,r^\nu],\\
      F^{\mu\nu}_L&=\partial^\mu \ell^\nu - \partial^\nu \ell^\mu-i[\ell^\mu,\ell^\nu].
     \end{align}
    \end{subequations}
    We have now all the elements needed to construct the EFT. The covariant derivative will be defined in the next section once the 
    the symmetry has been realized.
    
 \section{Chiral Perturbation Theory}\label{ChiPeeTee}
 
%

    \subsection{Construction of Chiral Perturbation Theory ($\chi$PT)}
  
     Weinberg showed that in order to get consistent results, one must use a non-linear realization of the chiral symmetry, so that soft pions cannot be 
     emitted from virtual particles in hard scattering processes \cite{Weinberg68NonLinearSU2}. A year later, Callan, Coleman, Wess and Zumino developed 
     a generalized way to construct non-linear realizations of arbitrary symmetry groups \cite{ColemanCallan69}. For the chiral symmetry $G$, SSB 
     generates eight pseudoscalar bosons which are identified with the lightest octet of pseudoscalar mesons. So, these are the fundamental fields upon 
     which the theory is constructed. These meson fields are collected in a unitary matrix using the Gell-Mann matrices
     \begin{equation}
      U(\varphi)=\exp(i\sqrt{2}\lambda_a\varphi^a/F),\hspace*{6ex}\lambda_a\varphi^a=\left(\begin{array}{ccc}
      \frac{\pi^0}{\sqrt{2}}+\frac{\eta_8}{\sqrt{6}}&\pi^+&K^+\\\pi^-&-\frac{\pi^0}{\sqrt{2}}+\frac{\eta_8}{\sqrt{6}}&K^0\\
      K^-&\bar{K^0}&-2\frac{\eta_8}{\sqrt{6}}\end{array}\right),
     \end{equation}
     where $\lambda_a$ are the eight Gell-Mann matrices and $F_\pi=F[1+\mathcal{O}(m_q)]\sim$ 92.4 MeV is the pion decay constant. As stated at the end 
     of the previous subsection, the promotion of $G$ to a local symmetry lead us to define a covariant derivative, given by
     \begin{equation}
       D_\mu U=\partial_\mu U-ir_\mu U +iU\ell_\mu,
     \end{equation}
     and which transforms as $D_\mu U\xrightarrow{\hspace*{1ex}G\hspace*{1ex}}g_RD_\mu Ug_L^{-1}$.\\
     
     Since one is able to construct a mass term for quarks stemming from the interaction with a constant external scalar current, a mass term of the 
     mesons might arise by making them interact with an external scalar current $\chi=2B(s+ip)$, the constant $B$ is related to the quark condensate 
     $\langle0|\bar{q}q|0\rangle=-F^2B[1+\mathcal{O}(m_q)]$. The lowest dimension operator which is also chiral invariant that can be constructed is 
     \begin{equation}
      \mathcal{L}_{mass}=\frac{F^2}{4}\langle \chi U^\dagger +\chi^\dagger U\rangle,
     \end{equation}
     where $\langle A\rangle=\text{Tr}(A)$. One can obtain, by expanding the $U$ fields to $\mathcal{O}(\varphi^2)$ the relations
     \begin{subequations}
      \begin{align}\centering
       M_\pip^2&=2mB\\
       M_\pz^2&=2mB+\mathcal{O}\left[\frac{(m_u-m_d)^2}{m_s-m}\right]\\
       M_\kp^2&=(m_u+m_s)B\\
       M_\kz^2&=(m_d+m_s)B\\
       M_{\eta_8}^2&=\frac{2}{3}(m+2m_s)B+\mathcal{O}\left[\frac{(m_u-m_d)^2}{m_s-m}\right],
      \end{align}
     \end{subequations}
     where $m=\frac{1}{2}(m_u+m_d)$. A relation between quark and pseudoscalar meson masses has been now constructed.\\
     
     Given the realization of the chiral symmetry through the matrix $U(\varphi)$, the general Lagrangian density can be constructed by means of this 
     matrix and the external fields. However, the question of how to do this in a systematic way including only the relevant terms comes naturally.
     An answer to this question is given by Weinberg \cite{Weinberg79}, where he conjectures that, since Quantum Field Theory by itself has no content 
     beyond analyticity, unitarity, cluster decomposition\footnote{In a QFT having the property of Cluster decomposition means that the vacuum-to-vacuum expectation value 
     of a product of many operators defined in two disjoint small space-time regions $A$ and $B$ with very large separation equals the vacuum-to vacuum
     expectation value of the products of operators in region $A$ times the vacuum-to-vacuum expectation value of the product of operators in region $B$,
     meaning that the effects of operators in $A$ cannot affect what happens in $B$ and vice-versa.} and symmetry, by giving all the terms consistent 
     with the assumed symmetry principles one gets the most general Lagrangian density, and calculating the matrix elements with this Lagrangian density 
     to any given order in perturbation theory the result is the most general possible S-matrix consistent with analyticity, perturbative unitarity, 
     cluster decomposition and the assumed symmetry principles. This, however, is of no use by itself since an infinite number of terms should be 
     considered in the Lagrangian to give the most general S-matrix.\\
     
     Nevertheless, one can rescale the moments and the meson masses, taking $p\to tp$ and $m_P\to t m_P$. Then, the chiral dimension is defined as
     \begin{equation}
      \mathcal{M}(tp_i,tm_P)=t^D\mathcal{M}(p_i,m_P)
     \end{equation}
     with
     \begin{equation}
      D=2+\sum_{n=1}^\infty(d-2)N_{d}+2N_L,
     \end{equation}
     where $N_L$ is the number of loops, $N_{d}$ is the number of vertices formed from interactions with $d$ derivatives. Thus, by stating that amplitudes 
     with the lowest chiral counting $D$ will give a dominant contribution and with aid of Weinberg's conjecture one is able to construct, order by order 
     in chiral expansion the most general Lagrangian consistent with all features of QFT and chiral symmetry. \\
     
     The elements to construct Chiral Perturbation Theory are shown with their chiral order
     \begin{equation}\label{ChCount}
      U=\mathcal{O}(p^0),\hspace*{1.5ex} D_\mu U=\mathcal{O}(p),\hspace*{1.5ex} r_\mu,l_\mu=\mathcal{O}(p),\hspace*{1.5ex} F^{L/R}_{\mu\nu}=\mathcal{O}(p^2),
        \hspace*{1.5ex}\chi=\mathcal{O}(p^2).
     \end{equation}
     Now, the most general chiral invariant Lagrangian density that can be constructed at lowest chiral order $D=2$ is
     \begin{equation}
      \boxed{\mathcal{L}_2=\frac{F^2}{4}\text{Tr}[D_\mu U(D^\mu U)^\dagger]+\frac{F^2}{4}\text{Tr}(\chi U^\dagger+U\chi^\dagger)}.
     \end{equation}
     
     In the same way as it was done for $\mathcal{L}_2$, the chiral Lagrangian of order $D=4$ can be constructed using the operators which 
     give the most general lagrangian at this order invariant under $G$. However, its associated functional will only give Green functions with even powers of Goldstone 
     bosons (even intrinsic parity sector). From the $U(1)_A$ anomaly one can construct the most general Lagrangian with chiral dimension 
     $D=4$ that will give Green functions that involve only odd numbers of Goldstone bosons (odd intrinsic parity sector),
     this is the Wess-Zumino-Witten Lagrangian $\mathcal{L}_{WZW}$ \cite{WZW}.
     The corresponding functional is given by 
     
     \begin{multline}\label{WZW}
 Z[U,l,r]=-\frac{i N_C}{240\pi^2}\int_{M^5}d^5x\varepsilon^{ijklm}\langle\Sigma^L_i\Sigma^L_j\Sigma^L_k\Sigma^L_l\Sigma^L_m\rangle\\
 -\frac{i N_C}{48\pi^2}\int d^4x\varepsilon_{\mu\nu\rho\sigma}(W(U,l,r)^{\mu\nu\rho\sigma}-W({\bf 1},l,r)^{\mu\nu\rho\sigma})
 \end{multline}
 where $\langle A\rangle=\text{Tr}(A)$, 
 \begin{multline}
 W(U,l,r)_{\mu\nu\rho\sigma}=\langle U\ell_\mu\ell_\nu\ell_\rho U^\dagger r_\sigma+\frac{1}{4}U\ell_\mu U^\dagger r_\nu U\ell_\rho U^\dagger r_\sigma+
i U\partial_\mu\ell_\nu\ell_\rho U^\dagger r_\sigma \\\hspace*{5ex}+i\partial_\mu r_\nu U\ell_\rho U^\dagger r_\sigma-i\Sigma^L_\mu\ell_\nu U^\dagger r_\rho U\ell_\sigma 
+\Sigma^L_\mu U^\dagger\partial_\nu r_\rho U\ell_\sigma\\\hspace*{5ex}-\Sigma^L_\mu\Sigma^L_\nu U^\dagger r_\rho U\ell_\sigma+\Sigma^L_\mu\ell_\nu\partial_\rho\ell_\sigma
+\Sigma_\mu^L\partial_\nu\ell_\rho\ell_\sigma-i\Sigma_\mu^L\ell_\nu\ell_\rho\ell_\sigma\\ + \frac{1}{2}\Sigma^L_\mu\ell_\nu\Sigma^L_\rho\ell_\sigma
-i\Sigma_\mu^L\Sigma_\nu^L\Sigma_\rho^L\ell_\sigma - (L\leftrightarrow R)\rangle,\hspace*{5ex}
\end{multline}
    and $\Sigma^L_\mu=U^\dagger\partial_\mu U$, $\Sigma^R_\mu=U\partial_\mu U^\dagger$.

 \section{Resonance Chiral Theory (R$\chi$T)}\label{ResChiTh}
 
     Chiral Perturbation Theory \cite{Weinberg79, Gasser:1983yg,Gasser:1984gg,Bijnens:1999sh,Bijnens:2001bb} has been very successful, 
     however it has some serious limitations. Chiral Perturbation Theory is only reliable at 
     energies below $\sim$ 500 MeV. Above this energy, other meson resonances become active degrees of freedom. So, to push Chiral Perturbation Theory
     to higher energies one needs to include resonances as active degrees of freedom. The way to do this is relying not only in chiral symmetry, but 
     in $1/N_C$ expansion\cite{LargeN-QCD} from which Chiral Perturbation Theory can be obtained with the restoration of the $U(1)_A$ anomaly \cite{Manohar}. 
     Therefore, $1/N_C$ is completely compatible with chiral symmetry. It has been shown that at low energies $(<500$ MeV) $\mathcal{L}_4$ is completely 
     described by the lowest-lying resonance exchange \cite{Ecker89-1}, so that the most general Lagrangian will be given by $\mathcal{L}_2+\mathcal{L}_{WZW}$ and all possible 
     contributions from resonance exchange. For the sake of simplicity, a different representation is used for the realization of the Goldstone 
     bosons, namely the matrix $u$ defined by the relation $U=u\ex2$, and which transforms under $G$ as
     \begin{equation}
      u(\varphi)\xrightarrow{\hspace*{1ex}G\hspace*{1ex}}g_Ru(\varphi)h(\varphi)^\dagger=h(\varphi)u(\varphi)g_L^\dagger,
     \end{equation}
     where $h\in SU(3)_V$. Also, the fields $\chi_\pm$ and $f^{\mu\nu}_\pm$ are used instead of $\chi$ and $F^{\mu\nu}_{R/L}$, which are defined as
     \begin{subequations}
     \begin{align}
      f^{\mu\nu}_\pm&=uF^{\mu\nu}_Lu^\dagger\pm u^\dagger F^{\mu\nu}_Ru,\\
      \chi_\pm&=u^\dagger\chi u^\dagger\pm u\chi^\dagger u
     \end{align}
     \end{subequations}
     Depending on the nature of the resonance field one can write down a chiral invariant
     interaction term to couple the resonance field with the chiral objects in eq. (\ref{ChCount}) \cite{Ecker89-1,Ecker89-2}. The realization of 
     $G$ on resonance fields in the antisymmetric tensor formalism is given by
     \begin{equation}
      R_{\mu\nu}\xrightarrow{\hspace*{1ex}G\hspace*{1ex}}h(\varphi)R_{\mu\nu}h(\varphi)^\dagger,
     \end{equation}
     with covariant derivative 
     \begin{equation}
      \nabla_\mu R=\partial_\mu R+[\Gamma_\mu,R],\end{equation}\begin{equation}
      \Gamma_\mu=\frac{1}{2}[u^\dagger(\partial_\mu-ir_\mu)u+u(\partial_\mu-i\ell_\mu)u^\dagger].
     \end{equation}
     By using the covariant derivative, the kinetic Lagrangian density can now be written down
     \begin{equation}\label{KinLagr}
      \mathcal{L}_{kin}(R_{\mu\nu})=-\frac{1}{2}\langle\nabla^\lambda R_{\lambda\mu}\nabla_\nu R^{\nu\mu}-\frac{1}{2}M_R^2R_{\mu\nu}R^{\mu\nu}\rangle.
     \end{equation}
     Similarly, one can write the interaction terms at the leading $1/N_C$ order in the following way
     \begin{subequations}\label{L2ResCHiTee}
      \begin{align}\centering
       \mathcal{L}_2(V)&=\frac{F_V}{2\sqrt{2}}\langle V_{\mu\nu}f^{\mu\nu}_+\rangle+\frac{iG_V}{\sqrt{2}}\langle V_{\mu\nu}u^\mu u^\nu\rangle,\\
       \mathcal{L}_2(A)&=\frac{F_A}{2\sqrt{2}}\langle A_{\mu\nu}f^{\mu\nu}_-\rangle,\\
       \mathcal{L}_2(S)&=c_d\langle Su_\mu u^\mu\rangle+c_m\langle S\chi_+\rangle,\\
       \mathcal{L}_2(P)&=id_m\langle P\chi_-\rangle.
      \end{align}
     \end{subequations}
     In this antisymmetric tensor formalism, the propagator of the meson resonance is given by the following relation \cite{Ecker89-1}
     \begin{multline}
      \langle 0|T\{R_{\mu\nu}R_{\rho\sigma}\}|0\rangle=i\frac{1}{M_R^2}\int \frac{d^4ke^{-ik(x-y)}}{(2\pi)^4(M_R^2-k^2-i\epsilon)}\left[\frac{}{}
      M^2g_{\mu\rho}g_{\nu\sigma}+g_{\mu\rho}k_\nu k_\sigma\right.\\
      \left.\frac{}{}-g_{\mu\sigma}k_\nu k_\rho-k^2g_{\mu\rho}g_{\nu\sigma}-(\mu\leftrightarrow\nu)\right],
     \end{multline}
      corresponding to the normalization
      \begin{equation}
       \langle0|R_{\mu\nu}|R,p\rangle=i\frac{1}{M_R}[p_\mu\epsilon_\nu(p)-p_\nu\epsilon_\mu(p)].
      \end{equation}
      The final ingredient of this theory is a match between QCD and this EFT. The way to do this is by taking the Green function due to some 
      process and then take the high energy limit, by comparing this result with the exact QCD prediction one is able to find constrictions 
      in some of the coupling constants. This reduces significantly the number of free parameters in the theory.
 \pagebreak
 

\def\N{N_C}
\def\RCT{R\chi T}
\def\t{\tau}
\def\n{\nu}
\def\g{\gamma}
\def\p{\pi}
\def\lmas{\ell^+}
\def\lmenos{\ell^-}
\def\pt{p_\tau}
\def\CPT{\chi PT}

\chapter{Lepton universality violation and new sources of CP violation}

  \section{Introduction}
  
  In the Standard Model (SM) all lepton currents couple with the same strength to the weak gauge bosons. This is an interesting 
  property of the SM, since any deviation from this prediction would be a clear signal of Beyond SM (BSM) interactions, therefore 
  all processes in which a false signal of lepton universality might come about due to kinematical or dynamical effects stemming 
  from processes unrelated to real lepton universality. Since High Intensity Frontier experiments such as Belle-II will be 
  able to look up for processes with lepton universality violation it becomes mandatory to know well all the processes that may 
  be an important background for the observation of BSM effects in these decays.\\
  
  In order to get reliable predictions one needs to have under control all the hadronic effects in the processes studied. 
  Therefore, predictions of background in the search for BSM effects have to be taken into account so that this background
  can improve the predictions of the experiments. Some experimental results show effects that cannot be 
  explained by means of the SM (athough, not so far from the SM $\lesssim4\sigma$). Therefore, the hadronic effects on such 
  processes need to be under very good control.\\
  
  Since CP asymmetries have been measured recently in $B$ meson decays, the hadronic effects in these decays need to be very well 
  studied to increase the precision in the search for BSM phenomena and determine whether these measurements agree with the 
  Standard Model or needs an explanation from effects beyond it. Therefore, the hadronic parts in potential background in 
  the search for BSM phenomena needs to be very well known and under control.

  \section{The $\tau^-\to\pi^-\nu_\tau\ell^+\ell^-$ decays as background for BSM interactions}\label{taupillnu MoFo}
  
  \subsection{Introduction}
  
  Being the heaviest of the leptons, the $\tau$ offers an opportunity that neither of its lighter copies can provide, 
  namely the chance to study in a clean way the production of hadrons. This unique characteristic makes it appealing for the analysis of all 
  kind of processes involving leptons and hadrons. One such effect is Lepton Flavor Violation (LFV) and Lepton Number 
  Violation (LNV), as well as lepton universality violation in semileptonic $\tau$ decays \cite{leptUV}. Lepton universality violation 
  has been suggested in heavy meson decays including $\tau$'s in the final state, therefore to ensure that no spurious
  violation of this universality is been measured due to $\tau$ decays one has to compute the possible configurations 
  in which such effect could be induced. The hadronic effects in the studies of LFV in $\tau$ decays has to be very well 
  understood, since otherwise this effects can give different orders of magnitude\cite{Arganda:2008jj,Celis:2013xja, Celis:2014asa, Lami:2016vrs}\\
  
  In this section we study the $\tau^-\to\pi^-\nu_\tau\ell^+\ell^-$ decays, 
  where the effective vertex $W\gamma^*\pi$ comes about. The presence of this effective vertex makes the processes 
  appealing for one more reason, which is the fixing of parameters for the pion transition form factor by means of a 
  weak isospin rotation (and its correction factor) where both gauge bosons are off-shell and also with high virtualities.
  In subsection \ref{Mat elem} we show the different contributions to the total amplitude and their complete expressions.
  In subsection \ref{FFpinull} the expressions of the form factors are given, as well as the Feynman diagrams needed 
  for computing them. In subsection \ref{SDpillnu} we show the form in which the couplings in the form factors are 
  constrained through the short distance behavior of QCD. In subsection \ref{BR pillnu} we show our results for the 
  total branching fraction and the different contributions for the process, also, the invariant mass spectra for 
  both decay channels are shown.
  
  \subsection{Matrix element of the process}\label{Mat elem}
  
  The study of the $\tau^-(p_\tau)\to\pi^-(p)\nu_\tau(q)\ell^+(p_+)\ell^-(p_-)$ decays is a generalization of the 
  computation in ref \cite{Guo:2010dv}, where the photon is real. The opportunity of studying this process for virtual 
  photon turns out to be interesting since by a weak isospin rotation one can determine the principal contribution to the Hadronic 
  Light by Light scattering to the anomalous magnetic moment of the muon $a_\mu$. So, if this process could be measured one 
  could in principle constraint the relevant parameters for the relevant contribution in the $a_\mu$. It also 
  opens the possibility of studying lepton universality violation by comparing the $\gamma^*\to e^+e^-$ and
  $\gamma^*\to \mu^+\mu^-$ processes. Also, a Majorana neutrino exchange could give place to LNV for which this process 
  might become a very important background \cite{Nestor}.\\
  
  \begin{figure}[ht!]
   \centering\includegraphics[scale=0.5]{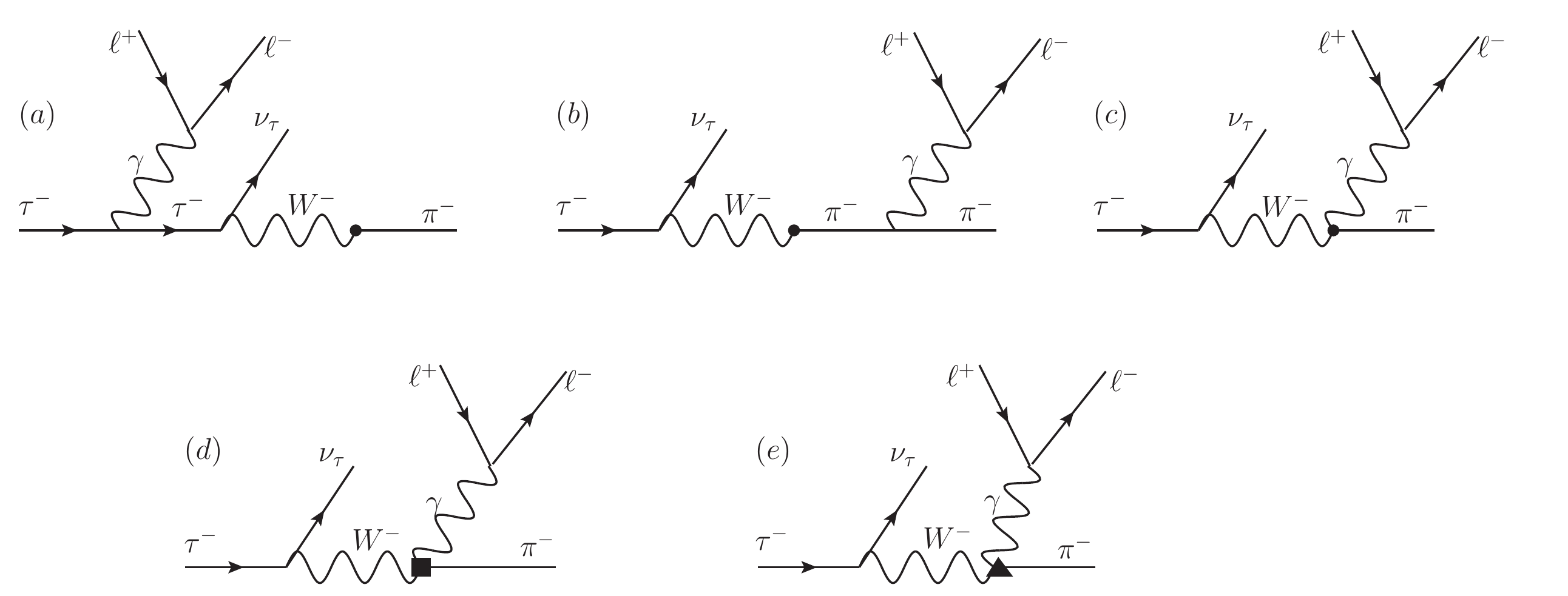}\caption{Feynman diagrams of the different contributions to the 
   $\tau\to\pi\ell^+\ell^-\nu_\tau$ decay. Diagrams (a) to (c) give the model independent contribution, while the 
   structure dependent has been separated into two contributions for convenience}\label{Feyn diags}
  \end{figure}

  To describe this process we need to compute the relevant amplitude, which is given by one model independent
  and two structure dependent contributions. The model independent contribution has the bremsstrahlung off the 
  $\tau$ (a), off the $\pi$ (b) and the diagram with the local vertex $W\gamma^*\pi$ (c) shown in figure \ref{Feyn diags}
  which are given by pure QED assuming a point-like pion. The contributions to the amplitude are thus
  \begin{eqnarray}\label{explicit expressions matrix element}
   \mathcal{M}_{IB} & = & -i G_F V_{ud}\frac{e^2}{k^2}F M_\tau \bar{u}(p_-)\gamma_\mu v(p_+)\bar{u}(q)(1+\gamma_5)\left[\frac{2p^\mu}{2p\cdot k+k^2} + 
   \frac{2\pt^\mu-\slashed{k}\gamma^\mu}{-2\pt\cdot k+k^2}\right]u(\pt)\,,\nonumber\\
   \mathcal{M}_{V} & = & - G_F V_{ud} \frac{e^2}{k^2} \bar{u}(p_-)\gamma^\nu v(p_+) F_V(p\cdot k,k^2)\epsilon_{\mu\nu\rho\sigma}k^\rho p^\sigma \bar{u}(q)\gamma^\mu(1-\gamma_5)u(\pt)\,,\\
   \mathcal{M}_{A} & = & i G_F V_{ud} \frac{2e^2}{k^2}\bar{u}(p_-)\gamma_\nu v(p_+) \left\lbrace F_A(p\cdot k,k^2)\left[(k^2+p\cdot k)g^{\mu\nu}-k^\mu p^\nu\right] - \frac{1}{2} A_2(k^2) 
   k^2g^{\mu\nu} \right. \nonumber \\ && \ \ \ \ \left. + \frac{1}{2} A_4(k^2) k^2(p+k)^\mu p^\nu\right\rbrace\bar{u}(q)\gamma_\mu(1-\gamma_5)u(\pt)\nonumber\, .
  \end{eqnarray}
  Here $G_F$ is the Fermi constant, $V_{ud}$ is the first entry of the CKM matrix, $k=p_++p_-$ is the virtual photon four-momentum and all the values 
  for the SM parameters were taken from the 
  Particle Data Group Review on Particle Physics \cite{PDG16}. $\mathcal{M}_{IB}$ is the total model independent contribution and 
  $\mathcal{M}_{V}$ and $\mathcal{M}_{A}$ are the vector and axial-vector structure dependent contributions, respectively.\\
  
  By inspection of the
  previous expressions it can be seen that the decay amplitudes for the real photon are obtained by making $\frac{e}{k^2}\bar{u}(p_-)\gamma_\mu v(p_+)\to\epsilon(k)_\mu^*$, 
  for $\epsilon^*_\mu(k)$ the polarization of the real photon, then setting $k^2\to0$. Our computation of the form factors $F_A$ and $F_V$ agrees in the $k^2\to0$ limit
  with the expressions from ref \cite{Guo:2010dv}. In addition, we provide for the first time their the dependence on the photon virtuality. 
  The additional axial form factors $A_2(k^2)$ and $A_4(k^2)$ can be seen in ref \cite{Bijnens:1992en}. 
  These can be expressed in terms of one form factor at the order we are interested in computing the process (further details can be seen on ref \cite{We:2012}). By defining $B(k^2):=-\frac{1}{2}A_2(k^2)$ one
  gets $\frac{1}{2}A_4(k^2)=-B(k^2)/(k^2+2p\cdot k)$, so that the amplitude simplifies to
  
\begin{eqnarray}\label{A matrix element}
  \mathcal{M}_{A} & = & i G_F V_{ud} \frac{2e^2}{k^2}\bar{u}(p_-)\gamma_\nu v(p_+) \left\lbrace F_A(p\cdot k,k^2)\left[(k^2+p\cdot k)g^{\mu\nu}-k^\mu p^\nu\right]\right.\nonumber\\
& & \left. +B(k^2) k^2 \left[g^{\mu\nu}-\frac{(p+k)^\mu p^\nu}{k^2+2p\cdot k}\right]\right\rbrace\bar{u}(q)\gamma_\mu(1-\gamma_5)u(\pt)\,.
\end{eqnarray}
  
  The $k$ dependence of the amplitudes gives us an idea of what to expect. All the amplitudes have a factor $1/k^2$, nevertheless the structure independent 
  amplitude will give the dominant contribution when $k$ is small since it has extra $1/k$ factors compared to the structure dependent part, which has an $\mathcal{O}(k)$ 
  dependence. Instead, for $k^0\sim m_\tau-m_\pi$ the structure dependent terms will give the main contribution to the process. Since the mass of the electron 
  is 200 times smaller than that of the muon one would expect to see this decay channel dominated by the pure QED contribution. Being the muon nearly as heavy as 
  the pion, one would expect that hadronic effects would be more important since the structure dependent amplitude $\sim \mathcal{O}(k)$.
  
  It should be noticed that ref \cite{PDG16} neglects the $A_4(k^2)$ in the $\pi\to\mu\nu_\mu e^+e^-$ for kinematic reasons, however, given the different phase 
  space of our process  we will keep this form factor. \\
  
  The decay rate will be conveniently separated in six terms which will be given by the squared moduli of the three amplitudes and three interference terms. 
  \begin{equation}
   \Gamma_{total}=\Gamma_{IB}+\Gamma_{VV}+\Gamma_{AA}+\Gamma_{IVB}+\Gamma_{IVA}+\Gamma_{VA}
  \end{equation}
  where the integrals for $d\Gamma(\tau^-\to\pi^-\ell^+\ell^-\nu_\tau)$ were obtained using the phase space configuration of reference \cite{AlainGabriel}.
  All the squared amplitudes sum over final and averaged over initial polarizations are shown in appendix \ref{LongExpressions}.
  
\subsection{Form Factors}\label{FFpinull}
  
\begin{figure}[ht!]
   \centering\includegraphics[scale=0.6]{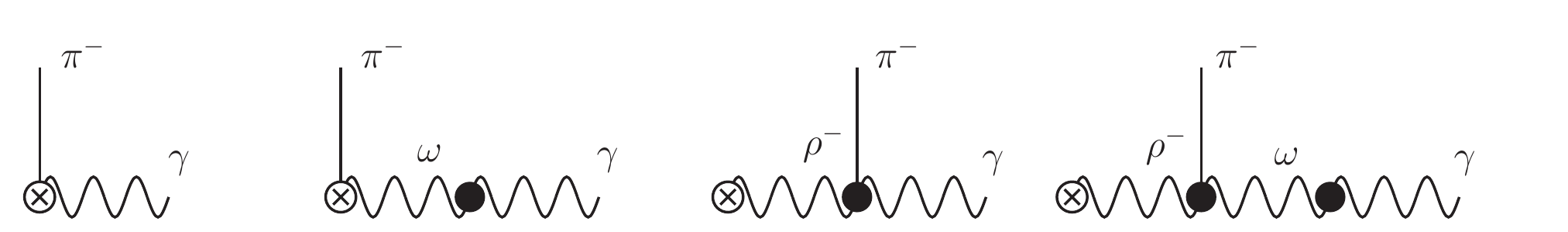}\caption{Contribution to the vector form factor in eq (\ref{explicit expressions matrix element}),
   where the circle with cross denotes the weak vertex.}\label{VFF}
  \end{figure}

\begin{figure}[ht!]
   \centering\includegraphics[scale=0.6]{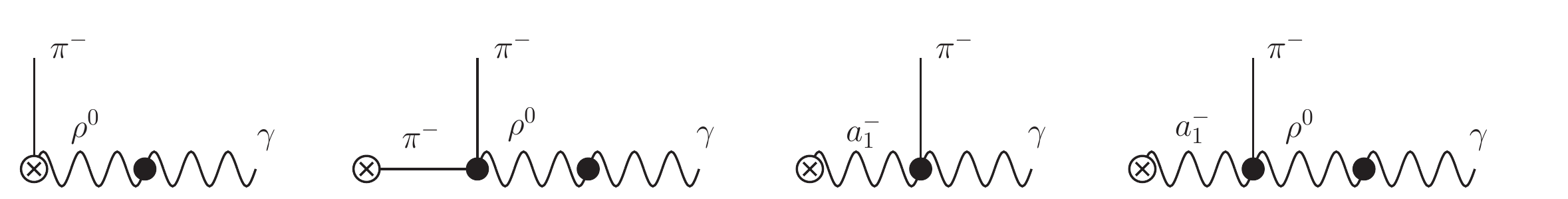}\caption{Contribution to the axial form factor in eq (\ref{explicit expressions matrix element}),
   where the circle with cross denotes the weak vertex.}\label{AFF}
  \end{figure}

  The expressions for the form factors are obtained by computing the Feynman diagrams in figure \ref{VFF} for the vector and in figure \ref{AFF} for axial 
  form factors. These contributions are computed by means of R$\chi$T since at the energies probed in the $\tau$ decay, the resonances region is available. 
  However, near the mass of the $\tau$ the process is kinematically suppressed, therefore contributions of higher multiplets of resonances can be safely neglected.
  
  Since both gauge bosons in the $W\gamma\pi$ vertex are off-shell, the effective vertex will depend upon two Lorentz invariants, which we choose as $t:=(p+k)^2$ and $k^2$. 
  In ref \cite{Guo:2010dv} the same diagrams were obtained, however the one with the contribution to the electromagnetic form factor (the second in fig \ref{AFF}), 
  is zero for on-shell photon, so the extra form factor is expected to be proportional to the electromagnetic form factor of the pion for $\rho-\gamma$ mixing. 
  
  Therefore, the expression of the vector form factor is expressed in the following way
  
  {\footnotesize\begin{eqnarray}\label{F_V}
 F_V(t,k^2) &=& -\frac{N_C}{24\pi^2 F}+ \frac{2\sqrt2 F_V}{3 F M_V
}\bigg[ (c_2-c_1-c_5) t +
(c_5-c_1-c_2-8c_3) m_\pi^2 + 2 (c_6-c_5) k^2\bigg]\times\nonumber \\
& &  \left[ \frac{\mathrm{cos}^2\theta}{M_\phi^2-k^2-iM_\phi\Gamma_\phi}\left(1-\sqrt{2} \mathrm{tg}\theta \right)
+ \frac{\mathrm{sin}^2\theta}{M_\omega^2-k^2-iM_\omega\Gamma_\omega}\left(1+\sqrt{2} \mathrm{cotg}\theta \right)\right]
\nonumber \\
& & + \frac{2\sqrt2 F_V}{3 F M_V }\, D_\rho(t)\,  \bigg[ ( c_1-c_2-c_5+2c_6) t +
(c_5-c_1-c_2-8c_3) m_\pi^2 + (c_2-c_1-c_5)k^2\bigg] \nonumber \\
& & + \frac{4 F_V^2}{3 F }\, D_\rho(t)\,  \bigg[ d_3 (t+4k^2) +
(d_1+8d_2-d_3) m_\pi^2 \bigg]\times\nonumber \\
& & \left[ \frac{\mathrm{cos}^2\theta}{M_\phi^2-k^2-iM_\phi\Gamma_\phi}\left(1-\sqrt{2} \mathrm{tg}\theta \right)
+ \frac{\mathrm{sin}^2\theta}{M_\omega^2-k^2-iM_\omega\Gamma_\omega}\left(1+\sqrt{2} \mathrm{cotg}\theta \right)\right]\,, 
\end{eqnarray}}
where $\theta$ is the mixing of $\omega-\phi$ vector mesons and
\begin{equation}
D_R(t) = \frac{1}{M_R^2 - t - i M_R \Gamma_R(t)}\,\,,
\end{equation}
 and $\Gamma_R(t)$ is the off-shell width of the resonance meson. The $c_i$ and $d_k$ are couplings of the VJP and VVP operators respectively, given 
 in section \ref{sec:RChT}. The off-shell width can be read from appendix \ref{OffShellWidthApp} \cite{GomezDumm:2000fz}. If the ideal mixing of vector mesons $\omega-\phi$ 
 is considered, 
 \begin{equation}
  \left(\begin{array}{c}\omega_1\\\omega_8\end{array}\right)=\left(\begin{array}{cc}\sqrt{\frac{2}{3}}&-\frac{1}{\sqrt{3}}
  \\\frac{1}{\sqrt{3}}&\sqrt{\frac{2}{3}}\end{array}\right)\left(\begin{array}{c}\omega\\\phi\end{array}\right),
 \end{equation}
 the contribution from the $\phi$ meson vanishes. Since these are rather narrow resonances, their energy dependent widths will be taken constant 
 and equal to their total widths.\\
 
 In the case of the axial form factors one finds
 \begin{eqnarray} \label{F_A}
F_A(t, k^2) &=& \frac{F_V^2}{F}\left(1-\frac{2G_V}{F_V}\right)\,D_\rho(k^2) - \frac{F_A^2}{F} D_{\mathrm{a}_1}(t)
+ \frac{F_A F_V}{\sqrt{2} F}\,D_\rho(k^2)\, D_{\mathrm{a}_1}(t)\,  \bigg( - \lambda'' t +
\lambda_0 m_\pi^2 \bigg)\,,\nonumber\\
\end{eqnarray}
where we have used the notation 
\begin{eqnarray}
\sqrt{2}\lambda_0 &=&-4\lambda_1- \lambda_2-\frac{\lambda_4}{2}-\lambda_5\,, \nonumber \\
\sqrt{2} \lambda''  &=& \lambda_2-\frac{\lambda_4}{2}-\lambda_5\,,
\end{eqnarray}
for the relevant combinations of couplings of the VAP operators shown in section \ref{sec:RChT}. Again, 
the off-shell width of the $a_1$ meson is shown in appendix \ref{OffShellWidthApp}.\\

   What one gets for the $B(k^2)$ form factor is, as previously mentioned, completely related to the electromagnetic 
   form factor of the pion for $\rho-\gamma$ mixing.
   \begin{equation} \label{B}
    B(k^2) = F \left[\frac{F_V^{\pi^+\pi^-}|_\rho(k^2)-1}{k^2}\,\right],
   \end{equation}
   
   The form factor $F_V^{\pi^+\pi^-}|_\rho(k^2)$ has been obtained by means of dispersion relations in ref. \cite{Pich:2001pj, De Troconiz:2001wt, Ananthanarayan:2011xt, Hanhart:2012wi},
   although, we follow the approach in ref. \cite{Dumm:2013zh} and use a dispersive representation of the form factor at low energies matched  to a 
   phenomenological description at intermediate energies, including the excited resonance contribution. So, the form factor is obtained through a three 
   times subtracted dispersion relation 
   
   \begin{equation}\label{FV_3_subtractions}
 F_V^\pi(s) \,=\,\exp \Biggl[ \alpha_1\, s\,+\,\frac{\alpha_2}{2}\,
s^2\,+\,\frac{s^3}{\pi}\! \int^\infty_{s_{\rm thr}}\!\!ds'\,
\frac{\delta_1^1(s')} {(s')^3(s'-s-i\epsilon)}\Biggr] \, ,
\end{equation}
where \cite{Boito:2008fq}
\begin{equation}
\label{delta}
 \tan \delta_1^1(s) = \frac{\Im m F_V^{\pi(0)}(s)}{\Re e
F_V^{\pi(0)}(s)} \ ,
\end{equation}
with 
\begin{eqnarray} \label{SU2formula}
\hspace{-.5cm} F_V^{\pi\,(0)}(s) & = & \frac{M_\rho^2}{M_\rho^2
\left[1+\frac{s}{96\pi^2 F^2}\left(A_\pi(s)+
\frac12 A_K(s)\right)\right]-s}\nonumber \\
& = & \frac{M_\rho^2}{M_\rho^2 \left[1+\frac{s}{96\pi^2 F^2}\Re e
\left(A_\pi(s) + \frac12 A_K(s)\right)\right]-s-i M_\rho
\Gamma_\rho(s)}\ .
\end{eqnarray}
The loop function is ($\mu$ can be taken as $M_\rho$)
\begin{equation}\label{loopfun_2pi}
A_{P}(k^2) \, = \ln{\left( \frac{m^2_P}{\mu^2}\right)} + {8 \frac{m^2_P}{k^2}} -
\frac{5}{3}  + \sigma_P^3(k^2) \,\ln{\left(\frac{\sigma_P(k^2)+1}{\sigma_P(k^2)-1}\right)}\,,
\end{equation}
and the phase--space factor $\sigma_P(k^2)$ is defined after the $\rho$ width in appendix \ref{OffShellWidthApp}.\\

The parameters $\alpha_1,\,\alpha_2$ and the $\rho(770)$ resonance parameters entering $B(k^2)$ will be extracted \cite{Dumm:2013zh} from fits to BaBar 
$\sigma(e^+e^-\to\pi^+\pi^-)$ data \cite{Aubert:2009ad} excluding the $\omega(782)$ contribution. We have used the fitted values with corresponding errors, 
as discussed in ref \cite{Hanhart:2012wi}
$\alpha_1=1.87,\,\alpha_2=4.26$.

\subsection{Short distance constraints}\label{SDpillnu}

   One of the main advantages of R$\chi$T is the fact that the couplings from operators in the theory can be constrained by means of 
   the QCD behavior at infinite energies, specifically the Green functions behavior. The Large $N_C$ description of QCD gives us a way 
   to relate the quark Green functions to those generated by means of meson exchange, therefore, by matching 
   consistently the leading order Operator Product Expansion result to a given order in the $1/N_C$ expansion Green functions in both descriptions, the relations among them will
   constrain the couplings of the effective theory. When the energy in the quark current in the Green function is very large 
   ($E\to\infty$, this is, when the distance between the points of the Green function $\to0$) the quarks Green function 
   must be equal to the meson exchange Green function, this is called quark-hadron duality. Assuming quark-hadron duality, the study of two point spin-one Green functions 
   in pQCD shows that the imaginary part of the quark loop must be constant at infinite momentum transfer \cite{Floratos:1978jb} and 
   can be understood as a sum of infinitely many positive contributions from intermediate hadron states. Since all these infinite positive 
   contributions must add up to a constant value at high energies, each of these contributions must vanish in this limit. \\
   
   In principle all resonance excitations must enter the short distance relations, however most experimental processes can be very 
   well described by considering the lowest-lying (vector and axial) resonances approximation \cite{Peris}. The effect of the excited resonances usually 
   give very small corrections to the short distance relations (see some examples in \cite{Guerrero:1997ku,Jamin:2006tk,We:2012}). 
   Therefore, by assuming this lowest-lying resonances dominance and the sum rules given by Weinberg \cite{Weinberg:1967kj} we get the results
   
   \begin{eqnarray}\label{Couplings}
    c_1-c_2+c_5&=&0\,,\nonumber \\ 2(c_6-c_5)&=&\frac{-N_C M_V}{64\pi^2F}\,,\nonumber \\ c_1-c_2-8c_3+c_5&=&0\,, \nonumber \\ d_1+8d_2-d_3&=& \frac{1}{16}\,, \\
    d_3&=&\frac{-N_C M_V^2}{128\pi^2F^2}+\frac{1}{16}\,,\nonumber \\ G_V&=& \frac{F}{\sqrt{2}}\,,\nonumber \\ F_V&=&\sqrt{2}F\,,\nonumber \\ F_A&=& F\,,\nonumber \\ \lambda'&=& \frac{1}{2}\,,\nonumber \\ \lambda''&=&0 \,,\nonumber \\ \lambda_0&=&\frac{1}{8}\,. \nonumber
   \end{eqnarray} 
   Given the corrected short distance relations of reference \cite{Roig:2013baa} a reanalysis of the process should be done, however this effect would give 
   a small correction (even for the $\ell=\mu$ channel), so that the conclusion in this thesis and in ref. \cite{We:2012} stands. For the branching fraction of the process we will take $M_V=775$ MeV.

 \subsection{Branching ratio and invariant mass spectrum}\label{BR pillnu}

  The form factors given in eqs. (\ref{F_V}), (\ref{F_A}) and (\ref{B}) parametrize the structure dependent contribution. With the information 
  on the couplings of the form factors, we are now ready to compute all the contributions to the branching fraction and the invariant mass 
  spectrum. The branching fractions are, therefore, predicted with the values \cite{We:2012} of table \ref{Tab:1 pillnu}, where we let a variation of 
  a 20\% around the short distance prediction of couplings is allowed in order to estimate a theoretical uncertainty. These are the errors 
  of the structure dependent contributions in table \ref{Tab:1 pillnu}, the errors in the completely structure independent contribution 
  comes from the numerical integration of this contribution. The error ranges are basically given by the uncertainties in $F_V,G_V.F_A,d_3$ and $c_5-c_6$.
  
  \begin{table}[!ht]
 \begin{center}
\begin{tabular}{|c||c|c||c|c|}
\hline
 & $\ell=e$ & $\ell=\mu$& $\ell=e$ & $\ell=\mu$\\
\hline
IB& $1.461\cdot10^{-5}$ & $1.600\cdot10^{-7}$ & $\pm 0.006\cdot10^{-5}$& $\pm 0.007\cdot10^{-7}$\\
IB-V& $-2\cdot10^{-8}$ & $1.4\cdot10^{-8}$ & $\left[-1\cdot10^{-7},1\cdot10^{-7}\right]$ & $\left[-4\cdot10^{-9},4\cdot10^{-8}\right]$\\
IB-A& $-9\cdot10^{-7}$ & $1.01\cdot10^{-7}$ & $\left[-3\cdot10^{-6},2\cdot10^{-6}\right]$ & $\left[-2\cdot10^{-7},6\cdot10^{-7}\right]$\\
VV & $1.16\cdot10^{-6}$ & $6.30\cdot10^{-7}$ & $\left[4\cdot10^{-7},4\cdot10^{-6}\right]$ & $\left[1\cdot10^{-7},3\cdot10^{-6}\right]$\\
AA& $2.20\cdot10^{-6}$ & $1.033\cdot10^{-6}$ & $\left[1\cdot10^{-6},9\cdot10^{-6}\right]$ & $\left[2\cdot10^{-7},6\cdot10^{-6}\right]$\\
V-A& $2\cdot10^{-10}$ & $-5\cdot10^{-11}$ & $\sim10^{-10}$  & $\sim10^{-10}$\\
\hline
TOTAL& $1.710\cdot10^{-5}$& $1.938\cdot10^{-6}$ & $\left(1.7^{+1.1}_{-0.3}\right)\cdot 10^{-5}$& $\left[3\cdot10^{-7},1\cdot10^{-5}\right]$\\
\hline
\end{tabular}
\caption{\small{The central values of the different contributions to the branching ratio of the $\tau^-\to\pi^-\nu_\tau\ell^+\ell^-$ decays ($\ell=e,\,\mu$) are displayed 
on the left-hand side of the table. The error bands of these branching fractions are given in the right-hand side of the table. The error bar of the IB contribution stems 
from the uncertainties on the pion decay constant $F$ and $\tau_\ell$ lepton lifetime \cite{PDG16}.}} \label{Tab:1 pillnu}
\end{center}
\end{table}

   We computed also the invariant (di-lepton) mass spectrum normalized to the $\tau$ total width for both channels
   \begin{equation}\label{spectrum}
    \frac{1}{\Gamma_\tau}\cdot\frac{d\Gamma(\tau^-\to\pi^-\nu_\tau \ell^+\ell^-)}{dk^2}.
   \end{equation}

\begin{figure}[!ht]
\centering
\vspace{1.3cm}
\includegraphics[scale=0.55,angle=-90]{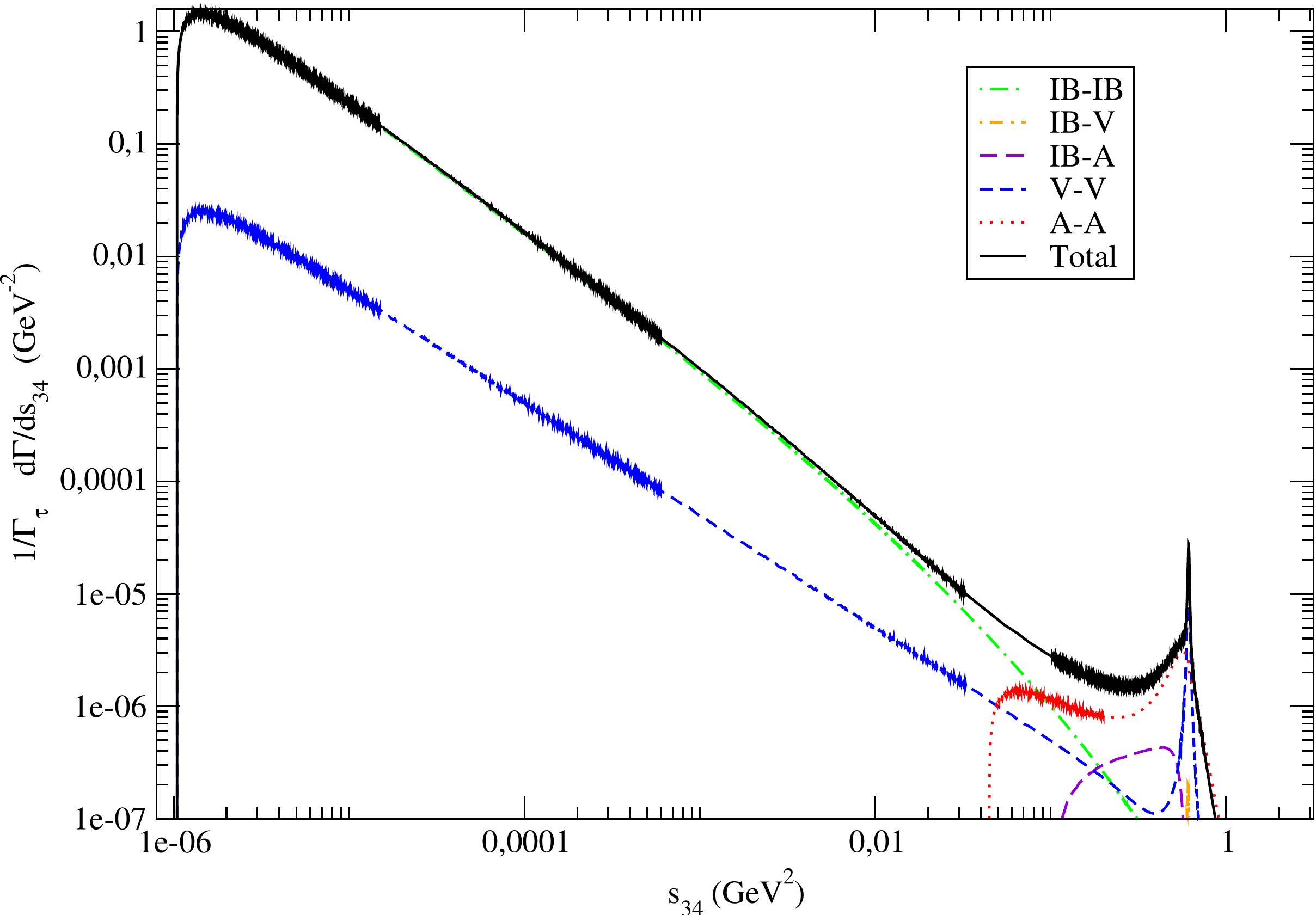}
\caption{\small{The different contributions to the normalized $e^+e^-$ invariant mass distribution defined in Eq. (\ref{spectrum}) are plotted. A double logarithmic scale was 
needed.} \label{Fig:4}}
\end{figure}

   As it was noticed in subsection \ref{Mat elem}, since the electron can reach very low invariant mass values, the decay width for this channel is mainly given by the 
   model independent part of the amplitude and the structure dependent part gives a negligible contribution. This behavior of the invariant mass spectrum is 
   shown in figure \ref{Fig:4}, where $s_{34}=k^2$ is the invariant mass of the 
   di-lepton \cite{We:2012}.\\

   Thus, we can see that for the muon channel the the opposite of the electron case happens. Since the mass of the muon is nearly that of the pion, the energy region 
   probed in this decay has a great contribution from the hadronic processes (structure dependent).\\

\begin{figure}[!ht]
\includegraphics[scale=0.55,angle=-90]{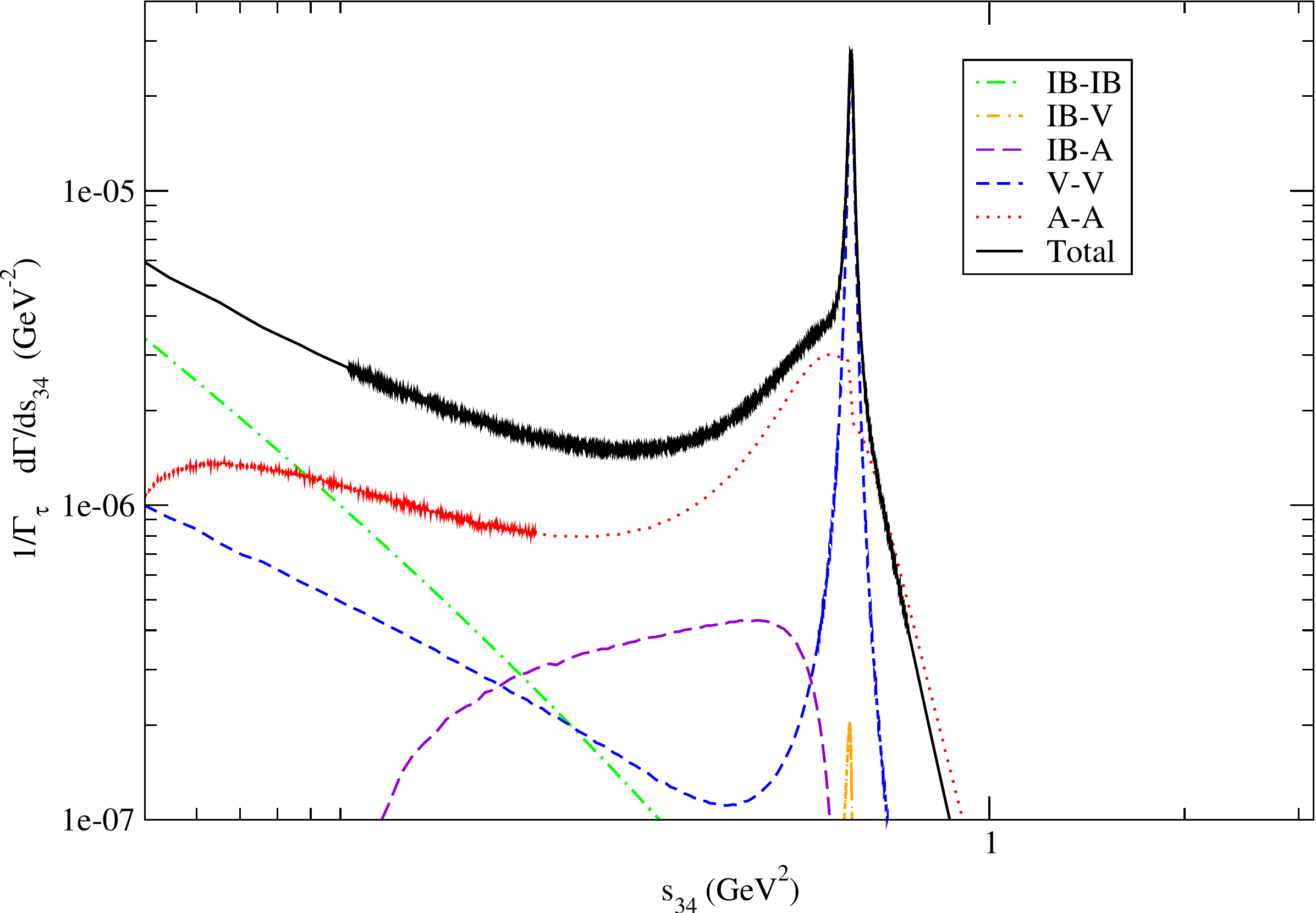}
\caption{\small{The different contributions to the normalized $e^+e^-$ invariant mass distribution defined in Eq. (\ref{spectrum}) are plotted in a magnification for 
$s_{34}\gtrsim 0.1$ GeV$^2$ intended to better appreciate the $SD$ contributions. A double logarithmic scale was needed.} \label{Fig:5}}
\end{figure}

\begin{figure}[!ht]
\begin{center}
\vspace*{1.2cm}
\includegraphics[scale=0.50,angle=-90]{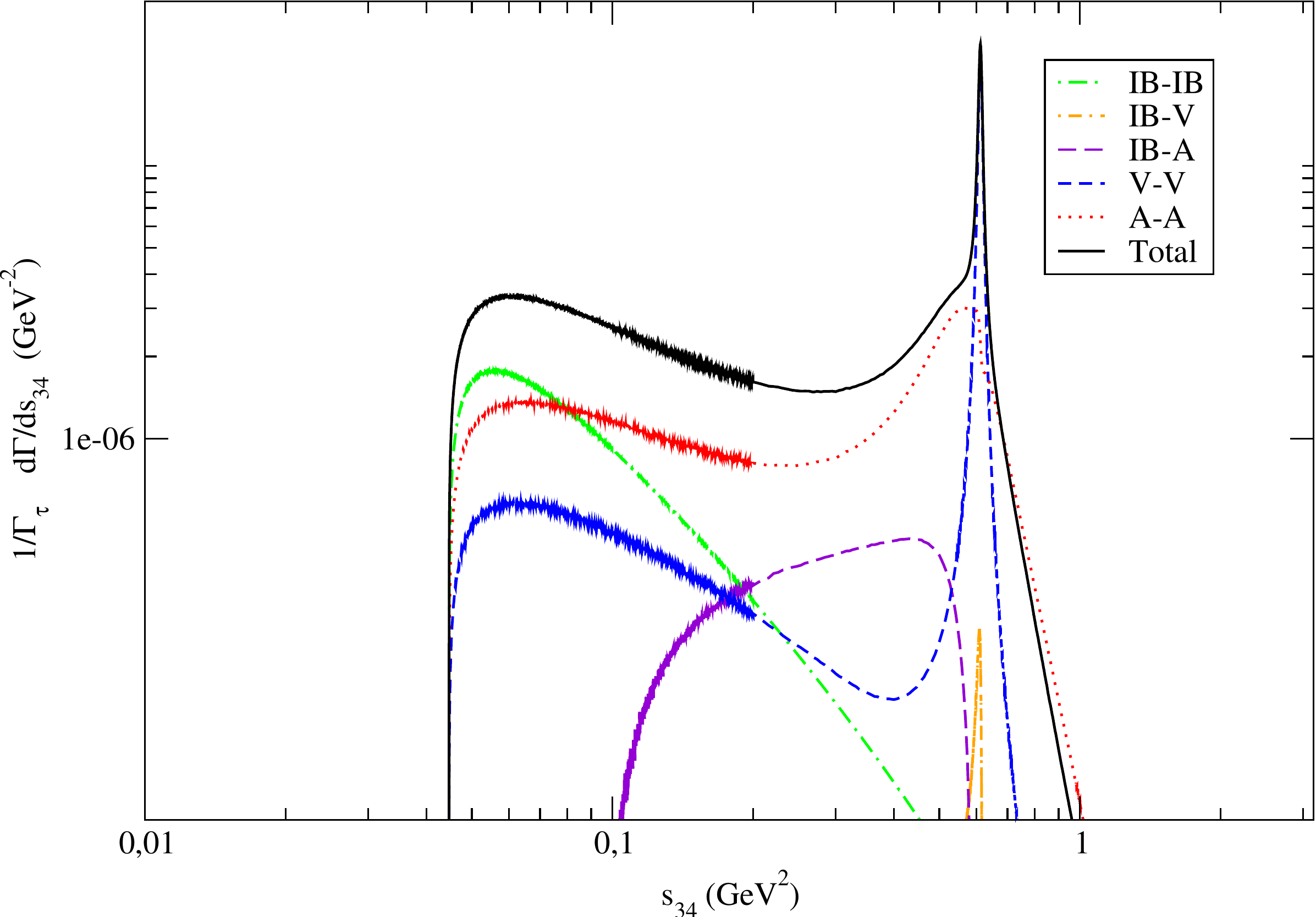}
\caption[]{\small{The different contributions to the normalized $\mu^+\mu^-$ invariant mass distribution are plotted. A double 
logarithmic scale allows to display the different contributions more clearly.}} \label{Fig:6}
\end{center}
\end{figure}

In Figs.~\ref{Fig:4}-\ref{Fig:6} vertical fluctuations can be appreciated in certain energy regions of the normalized invariant-mass distributions. In order to compute these 
distributions in the $s_{34}$ variable, we have integrated numerically the decay probability over the remaining four independent kinematical variables. 
The observed fluctuations  arise from the Monte Carlo evaluation over the four-body phase space integration. The branching fractions shown in Table 
\ref{Tab:1 pillnu} were obtained by integrating numerically these invariant-mass distributions and checked from a direct integration over the five independent kinematical variables.\\

The fact that, in both decays, the contribution to the decay width of the $s_{34}>1$ GeV$^2$ region is negligible justifies our assumption of including only the lightest 
multiplet of vector and axial-vector resonances. This result is not trivial in the axial-vector case and in the vector case it is not modified even if the 
$\rho(1450)$ exchange is included phenomenologically \cite{Dumm:2009va}.\\

We can see the effect of including the $B(k^2)$ form factor into the axial amplitude by neglecting its contribution and then comparing it to the full axial contribution
(squared $A$ modulus plus the interference with $IB$). The effect in both of the parts are 33 and 25\% respectively of the values in table \ref{Tab:1 pillnu}. 
It becomes essential to include this contribution since the muon channel is dominated by the axial amplitude. Thus, it drops a 44\% of the value in table 
\ref{Tab:1 pillnu} when one neglects this contribution. This explains the peak in the axial amplitude around the rho mass, since $B(k^2)$ is proportional to 
the electromagnetic form factor of the pion with only $\rho-\gamma$ mixing. \\

It is worth to notice the difference between the total branching fractions for the 
electron and (for its central value) the muon channels. If this effect is not taken into account in the search for lepton universality violation  might lead 
to spurious signal of beyond Standard Model effects. 

\subsection{Conclusions}

We have studied for the first time the branching ratio of the $\tau\to\pi\ell^+\ell^-\nu_\tau$ decays as well as the normalized 
invariant di-lepton mass spectrum. The analysis of this decays lead to a lepton universality violation induced by dynamic and 
kinematical effects. If this effect is not taken into account, the study of lepton flavor violation in heavy hadrons involving 
$\tau$ leptons in the final state might lead to a non-universality induced effect which might be confused with a genuine BSM 
effect.\\

As previously mentioned, the observables in this process are needed in order to give a good estimate of the background for 
the search of lepton flavor violation processes as $\tau\to\ell'\ell^+\ell^-$ or lepton number violation as in the process 
$\tau^-\to\pi^+\ell^-\ell^-\nu_\tau$ \cite{Nestor}. This form factors were coded in the R$\chi$T based version of TAUOLA, 
which is the standard Monte Carlo generator for $\tau$ decays \cite{TAUOLA2, Shekhovtsova:2012ra}.
  

 \section{Long-distance contribution to $B^\pm\to(\pi\ex\pm,K\ex\pm)\ell\ex+\ell\ex-$ decays.}
 \subsection{Introduction}
 
 \hspace*{3ex} Being the lightest element of the third generation quark doublet, all $b$-flavored hadrons occur through generation changing processes. 
 This means that all amplitudes will be $\mathcal{O}(\lambda)$ suppressed, within the Wolfenstein parameterization 
 \cite{Wolfenstein} of the CKM matrix. 
 Thus, several processes involving $b$-flavored mesons will be highly suppressed in the Standard Model yielding an excellent ground to 
 search for Beyond Standard Model Physics.\\
 
 In 2014 several interesting precision tests of the Standard Model were made at the LHCb experiment, in particular a test of lepton
 universality in $B\ex\pm\to K\ex\pm\ell\ex+\ell\ex-$ decays, 
 \begin{equation}
  R_K:=BR(B\to K\mu^+\mu^-)/BR(B\to Ke^+e^-)=0.745^{+0.090}_{-0.074}\pm0.036
 \end{equation}
 within the squared invariant mass of the lepton pair in the range $[1,6]$ GeV$^2$ \cite{AaijRK}, where the errors shown are the first statistic and the second systematic. 
 This very interesting result contrasts with current SM prediction, which is $R^{SM}_K=1+(3.0^{+1.0}_{-0.7})\cdot10^{-4}$
 \cite{BobethHillerPiranishvili}.
 We saw previously (in $\tau\to\pi\ex-\ell\ex+\ell\ex-\nu_\tau$ decays) \cite{UsBDecay}
 that the weak radiative pion vertex gives different values
 by taking either $\ell=\mu$ or $\ell=e$ for the model dependent terms, especially close to 1 GeV, where they became relevant.
 This analysis hinted the possibility that such violation of lepton universality might be due to hadronic effects at the GeV scale.\\
 
 In this chapter we compute a Long-Distance (LD) term of the Weak Annihilation contribution of $B^\pm\to P^\pm\ell\ex+\ell\ex-$
 decays which, by extending R$\chi$T to include $b$-flavored mesons and after computing all possible diagrams, we find that the only non-negligible
 contribution comes from the $P$ meson electromagnetic form factor for $P=\pi,K$. Also, to have reference point, we used the Gounaris-Sakurai 
 (GS) parametrization used by the BaBar collaboration in the fit of their data to the mentioned form factors
 \cite{GounarisSakurai}. The observables are computed for invariant dilepton mass 
 below the charmonium threshold $q^2\le8$ GeV$^2$, overlapping the experimental $q^2$ range in the measurement of 
 \begin{equation}
  R_P:=BR(B^\pm\to P^\pm\mu^+\mu^-)/BR(B^\pm\to P^\pm e^+e^-),
 \end{equation}
 where $P$ is either $\pi$ or $K$.
 Our short distance (SD) analysis of the amplitude was based in the results given in ref \cite{BobethHillerPiranishvili}.
 
 \subsection{R$\chi$T contribution to the Weak Annihilation amplitude}
\begin{figure}[!h]
  \centering
  \includegraphics[scale=0.45]{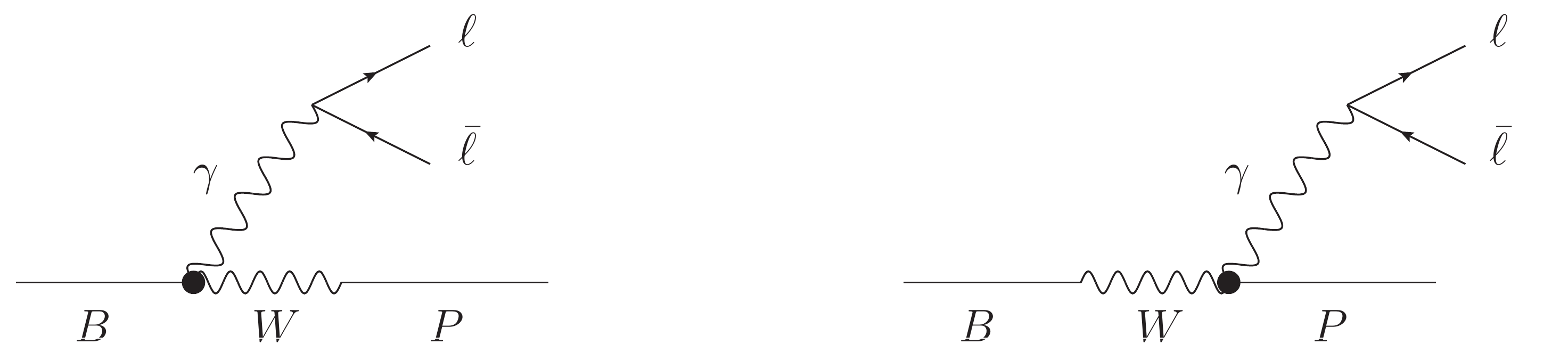}\caption{All possible contributions to the WA amplitude at leading order in $1/N_C$. The thick dot
  denotes interactions between resonances and the fields coupled to the vertex.}\label{generalWA}
 \end{figure}

 The QCD factorization (QCDf) \cite{QCDf1,QCDf2} contribution to the $B^\pm\to(\pi\ex\pm,K\ex\pm)\ell\ex+\ell\ex-$ decays \cite{BobethHillerPiranishvili,UsBDecay,Khodj}
 is taken at next to leading order, where a comparable uncertainty to that obtained using R$\chi$T with only the leading terms is expected. 
 In this approximation the small 
 contribution of the Weak Annihilation (WA) of the valence quarks in the $B$ meson becomes relevant.
  At this order in $1/N_C$, the quark currents of both mesons can be {\it naively  factorized},
 since a gluon exchange would mean a higher order term in $1/N_C$; this means that the hadronic current can be written as 
 products of decay constants and/or form factors \cite{QCDf2} which is described by the diagrams in Figure \ref{generalWA}.\\
 
 Now, at leading order in QED the leptonic current can be factorized 
 from the hadronic current since no photon exchange exists except for the one connecting the lepton current with the hadronic one. Thus, 
 the LD WA amplitude can be written as the coupling of the leptonic 
 current and an effective hadronic electromagnetic current $(\mathcal{M}_\mu^{WA})$
 \begin{equation}
 \mathcal{M}^{WA}_{LD}=\frac{e^2}{q^2}\bar{\ell}\gamma^\mu\ell\mathcal{M}_\mu^{WA},
 \end{equation}
 where $q^2$ is the dilepton invariant mass squared and, by conservation of the electromagnetic current, the effective hadronic 
 electromagnetic current takes the form
 \begin{equation}\label{LDWASI}
  \mathcal{M}_\mu^{WA}=\left[\left(p_B+p_P\right)_\mu-\frac{m_B^2-m_P^2}{q^2}q_\mu\right]F(q^2),
 \end{equation}
 
 where $m_{B/P}$ is the $B$ meson (pseudo-Goldstone boson) mass and $p_{B/P}$ is the $B$ meson (pseudo-Goldstone boson) four-momentum. 
 All strong, weak and electromagnetic interactions happening within the $B\to P\gamma^*$ transition are embedded in the form factor $F(q^2)$.
 Since $q_\mu\bar{\ell}\gamma^\mu\ell=0$, only the first term in the previous equation contributes where, by the same argument $p_B+p_P$ can
 be replaced by $2p_B$. The next section shows the framework to compute the form factor of the $B^\pm\to\gamma^* {W^*}^\pm$ effective vertex.
 
 \subsection{Extending R$\chi$T for heavy flavor mesons}
 
 To obtain the form factor of the interactions of the effective vertex shown in Figure \ref{generalWA}, we proceed in analogy with 
 the extension of Resonance Chiral Theory to include $c$-flavored quarks, $c$-flavored resonances and the interactions between them, the 
 Goldstone bosons and the light resonances \cite{B-RChT}. 
 This model {\it does not} rely on the heaviness of the charm quark and therefore cannot ensure a better description depending on the mass of 
 the heavy quark in the meson, nevertheless it can be extended in a straightforward way to $b$-flavored mesons.\\
 
 The $b$-flavored mesons are included by constructing a flavor triplet in a convenient realization and demanding it 
 to transform linearly under the chiral group $G$, 
 
 \begin{equation}
  B:=\left(\begin{array}{c} B^-\\B^0_d\\B^0_s \end{array}\right),\hspace*{20ex} B\xrightarrow{\hspace*{1ex} G\hspace*{1ex}}h(\varphi)B,
 \end{equation}

 where $h(\varphi)\in SU(3)_V$. In other words, $B$ mesons are taken to be the components of a $SU(3)_V$ triplet which is 
 a linear realization of $SU(3)_V$. The same procedure is followed for $b$-flavored resonances where, as SSB of the chiral group 
 dictates, $B^V$ and $B^A$ will have different masses.\\
 
 Analogously to the chiral theory for light mesons, the electroweak interactions are introduced as non-propagating external fields where
 they have to be extended to $SU(4)$ hermitian fields coupling to the two quark doublets of weak $SU(2)_L$ via
 \begin{equation}
  \mathcal{L}=\mathcal{L}^{0}_{QCD}-m_b\bar{b}b+\bar{q}\gamma^\mu\left[\frac{1}{2}(1-\gamma_5)\tilde{\ell}_\mu+\frac{1}{2}(1+\gamma_5)\tilde{r}_\mu
  \right]q-i\bar{q}\left(\tilde{s}-i\gamma_5\tilde{p}\right)q.\label{WALDamp}
 \end{equation}
 Even when all external fields vanish, the mass term for the $b$ quark still remains. Since the fields have to be extended, at the meson level 
 a new realization is needed in order to couple the pseudoscalar mesons to the external currents. This means extending the $U$ matrix in the 
 chiral group to a $4\times4$ operator $\tilde{U}$. The way to realize this is by constructing a matrix $4\times4$ including the $B$ triplet
 \begin{equation}
  \tilde{U}=\tilde{u}^\dagger_R\tilde{u}_L,
 \end{equation}
 where
 \begin{equation}
  \tilde{u}^\dagger_R=\left(\begin{array}{cc} u(\varphi)&\frac{i}{\sqrt{2}f_B}u(\varphi)B\\
  \frac{i}{\sqrt{2}f_B}B^\dagger&f_B/F \end{array}\right),
  \hspace*{5ex}\tilde{u}_L=\left(\begin{array}{cc} u(\varphi)&\frac{i}{\sqrt{2}f_B}B\\
  \frac{i}{\sqrt{2}f_B}u(\varphi)B^\dagger&f_B/F \end{array}\right),
 \end{equation}
 with $f_X$ the decay constant of meson $X$. In a similar way, the resonance fields can be extended to $SU(4)$
 \begin{equation}
  \tilde{R}=\left(\begin{array}{cc}R&B^R\\{B^R}^\dagger&0\end{array}\right).
 \end{equation}
 The important thing to notice here is that in this model, the Goldstone bosons and the light resonances enter as non-linear realizations and 
 $b$-flavored mesons as linear realizations.
 That is to say, the $SU(4)$ realization is only introduced to see how the external currents have to be implemented, meaning that this is 
 {\it not} an implementation of a chiral realization with four flavors whatsoever.\\
 \hspace*{1ex}\\
   \begin{figure}[!h]
  \centering
  \includegraphics[scale=0.302]{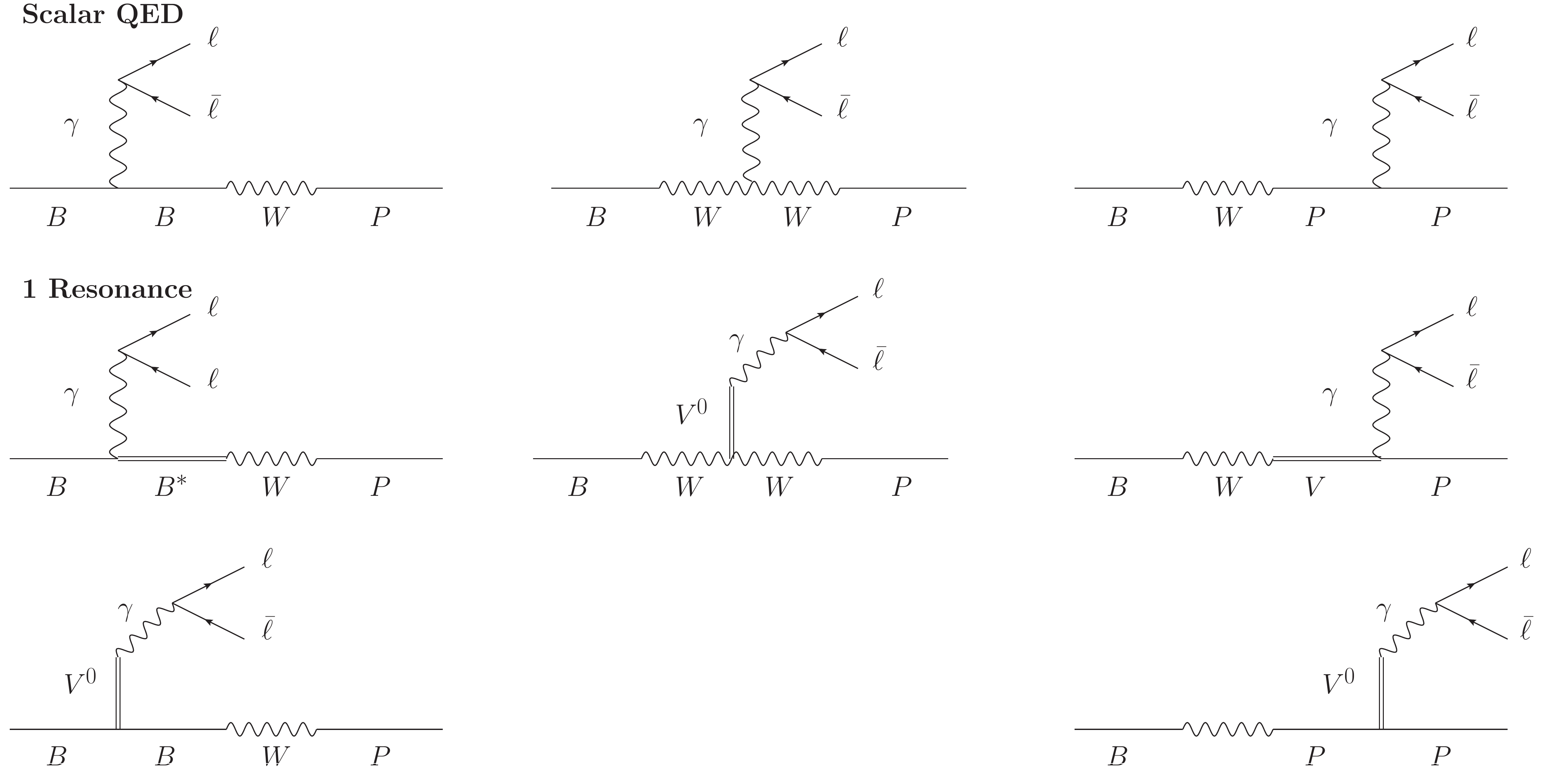}
  \includegraphics[scale=0.302]{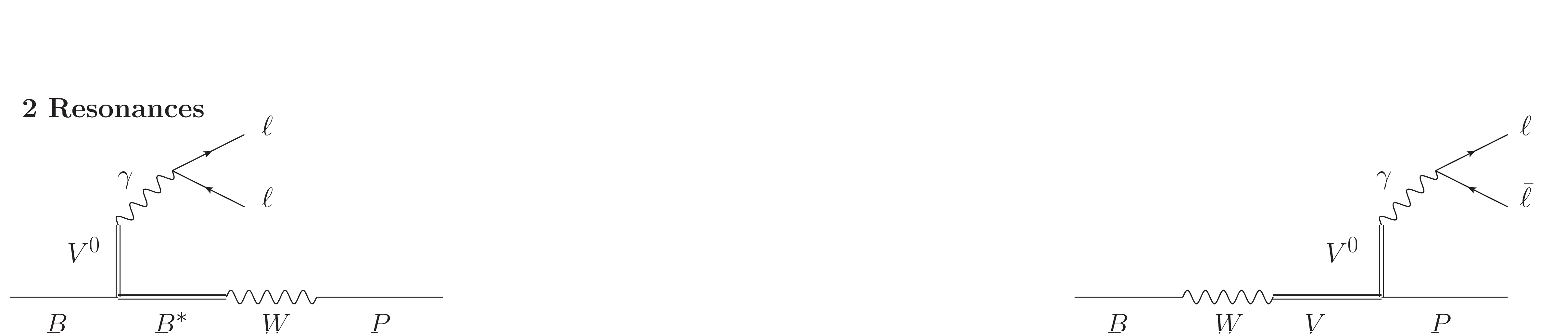}\caption{All LD WA Feynman diagrams at leading order in $1/N_C$. The first row shows the contribution 
  from model independent interactions, while the second and third shows contributions from diagrams with one 
  and two resonances respectively. $V^{(0)}$ stands for light charged (neutral) vector resonances.}\label{fullWA}
 \end{figure}
   The coupling to weak interactions through external currents comes from the definition of the covariant derivatives on the relevant objects
  \begin{subequations}
  \begin{align}
   D_\mu\tU&=\partial_\mu\tU-i\tr\tU+i\tU\tl,\\
   \nabla_\mu\tR&=\partial_\mu\tR+\left[\tilde{\Gamma}_\mu,\tR\right],
  \end{align}
 \end{subequations}
 where
 \begin{equation}
  \tilde{\Gamma}_\mu=\frac{1}{2}\left\{\tu_R\left[\partial_\mu-i\tr\right]\tu^\dagger_R+\tu_L\left[\partial_\mu-i\tl\right]\tu^\dagger_L\right\}.
 \end{equation}
 The extended right ($\tr$) and left ($\tl$) external fields are
 \begin{equation}
  \tr_\mu=\left(\begin{array}{cc}r_\mu&0\\0&\gamma_\mu\end{array}\right)\hspace*{5ex}
  \tl_\mu=\left(\begin{array}{cc}l_\mu&\omega_\mu\\\omega_\mu^\dagger&\delta_\mu\end{array}\right)
 \end{equation}
 where 
 \begin{subequations}
  \begin{align}
   \gamma_\mu&=-\frac{1}{3}e\left[A_\mu-\tan\theta_WZ_\mu\right],\\
   \delta_\mu&=-\frac{1}{3}eA_\mu+e\left[-\frac{1}{\sin2\theta_W}+\frac{1}{3}\tan\theta_W\right]Z_\mu,\\
   \omega_\mu&=2m_W\sqrt{\frac{G_F}{\sqrt{2}}}\left(\begin{array}{c}V_{ub}\\0\\0\end{array}\right)W_\mu.
  \end{align}
 \end{subequations}
 Now, the covariant derivative acting on the $B$ resonances is defined in the following way
 \begin{equation}
  \nabla_\mu B^{(R)}=\left[\partial_\mu+\Gamma_\mu+\frac{i}{2}(\gamma_\mu+\delta_\mu)\right]B^{(R)}.
 \end{equation}
 At the lowest chiral order, the whole set of operators that can be constructed regarding the relevant objects 
 defined previously can be obtained from ref \cite{B-RChT} by making the substitutions $B\to C$ and $F_C\to f_B$
 and using the definitions given above for all the operators.\\
 
\begin{figure}[!h]
  \centering
  \includegraphics[scale=0.45]{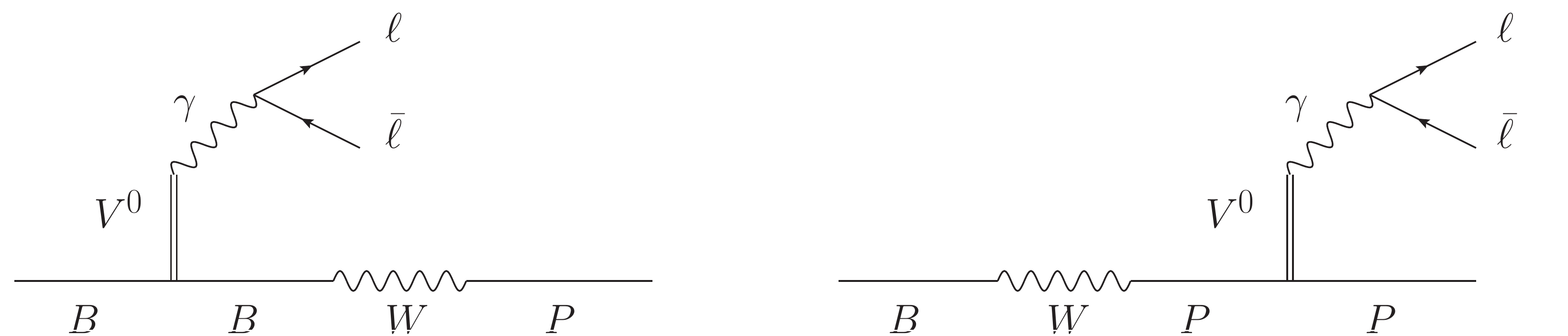}\caption{Only non-vanishing structure dependent contribution to the WA LD amplitude.}\label{EFFWA}
 \end{figure}

 Once all operators have been constructed, the Feynman diagrams that need to be computed are those shown in 
 Figure \ref{fullWA}. All $b$-flavored resonances in the Feynman diagrams will have a 
 propagator $D_{B^*}$ (left-hand side diagrams of Figure \ref{fullWA}) with invariant mass $k^2=m_P^2$ equals to the 
 pseudo-Goldstone mass, such contributions are not taken into account due to the suppression of the heavy
 resonance,
 \begin{equation}
  D_{B^*}(m_P^2)=\frac{1}{m_{B^*}^2-m_P^2}.  
 \end{equation}
 By electromagnetic gauge invariance, the structure independent terms computed with scalar QED, are found to vanish. 
 The diagram in the middle of the first line in Figure \ref{fullWA} is obtained by computing the bremsstrahlung off the $B$ and the 
 corresponding diagram for $P$ and then applying the Ward-Takahashi identity, which by gauge invariance gives a relation
 between the sum of the bremsstrahlung diagrams and the remaining one.
 Also, after computing the remaining diagrams it is found that several vanish too. The only non-vanishing diagrams 
 are those described only by the electromagnetic form factor of the mesons, namely those of Figure \ref{EFFWA}.
 \subsection{The electromagnetic form factor $F_P(q^2)$}
 Therefore, the amplitude of eq. (\ref{LDWASI}) will be fully described by the electromagnetic form factors of the pseudo-Goldstone and 
 the $B$ meson. Thus, this amplitude can be written in the following way\\
 \begin{equation}\label{LDamp}
  \mathcal{M}^{WA}_{LD}=\frac{\sqrt{2}G_F(4\pi\alpha)V_{ub}V_{uD}^*f_Bf_P}{q^2(m_B^2-m_P^2)}\left[m_B^2\left(F_P(q^2)-1\right)
  -m_P^2\left(F_B(q^2)-1\right)\right]{p_B}_\mu\bar{\ell}\gamma^\mu\ell.
 \end{equation}

 Since the second term is smaller by a factor $m_P^2/m_B^2$, it can be neglected once it can be claimed that $F_B(q^2)$ is not too large 
 compared to $F_P(q^2)$. By using the formalism developed in the previous section, the electromagnetic form factor of the $B$ meson is
 found to be
 \begin{equation}
 F_B(q^2)= 1 + \frac{3}{2}q^2\left(\frac{1}{M_\rho^2-q^2-iM_\rho\Gamma_\rho(q^2)}-\frac{1}{3\left[M_\omega^2-q^2-iM_\omega\Gamma_\omega(q^2)\right]}\right),  
 \end{equation}
 where $\Gamma_X(s)$ is the off-shell width of the $X$ meson resonance \cite{GomezDumm:2000fz} with invariant mass $\sqrt{s}$. 
 So that it becomes apparent that $F_B(q^2)$ will not surpass the suppression factor $m_P^2/m_B^2$ so that it can be neglected, only considering the 
 electromagnetic form factor of the light pseudoscalar. \\
 
 On the other hand, QCDf gives the following amplitude \cite{BobethHillerPiranishvili,Khodj,UsBDecay}
 \begin{equation}\label{SDamp}
  \mathcal{M}_{QCDf}=\frac{G_F\alpha}{\sqrt{2}\pi}V_{tb}V_{tD}^*\xi_P(q^2)p^\mu_B\left[F_V(q^2)\bar{\ell}\gamma_\mu\ell
  +F_A(q^2)\bar{\ell}\gamma_\mu\gamma_5\ell\right],
 \end{equation}
 where $V_{tx}$ is the CKM mixing term between the top quark and the down-type quark $x$, $\xi$ is a long-distance form factor obtained through 
 Light Cone Sum Rules (LCSM) \cite{BallZwicky} and $F_V$ and $F_A$ are functions of the Wilson Coefficients of the Operator Product Expansion. 
 (All the details about the computation of the QCDf amplitude can be seen in Sergio Lennin Tostado Robledo's Ph. D. thesis.)
 By comparing this expression with eq. (\ref{LDamp}) it can be seen that the LD WA amplitude\footnote{Although the $B$ meson form factor is neglected,
 in the rest of the chapter, it was included in the numerical results. No difference is noticed in the observables computed by including it.} can be absorbed in 
 the vector form factor of the QCDf amplitude by doing the replacement
 \begin{equation}
  \xi_P(q^2)F_V\to\xi_P(q^2)F_V+\kappa_Pm_B^2\left[\frac{F_P(q^2)-1}{q^2}\right],
 \end{equation}
 where 
 \begin{equation}
  \kappa_P=-8\pi^2\frac{V_{ub}V^*_{uD}}{V_{tb}V^*_{tD}}\frac{f_Bf_P}{m_B^2-m_P^2}
 \end{equation}
 
 is a dimensionless constant $\sim\mathcal{O}(10^{-2})\times \frac{V_{ub}V^*_{uD}}{V_{tb}V^*_{tD}}$. This means that for $P=K$ ({\it i.e.} $D=s$) there will be 
 an extra suppression $\sim\mathcal{O}(\lambda^{2})$ compared to the case when $P=\pi$ ({\it i.e.} $D=d$), where the CKM factors are $\mathcal{O}(\lambda^0)$.
 Resonance Chiral Theory with only the lightest vector multiplet gives the functional form of the light pseudoscalar form factor
 \begin{equation}
  F_P(q^2)=1+\frac{F_VG_V}{F^2}\frac{q^2}{M_V^2-q^2}.
 \end{equation}

  \begin{figure}[!t]
  \centering
  \includegraphics[scale=0.48]{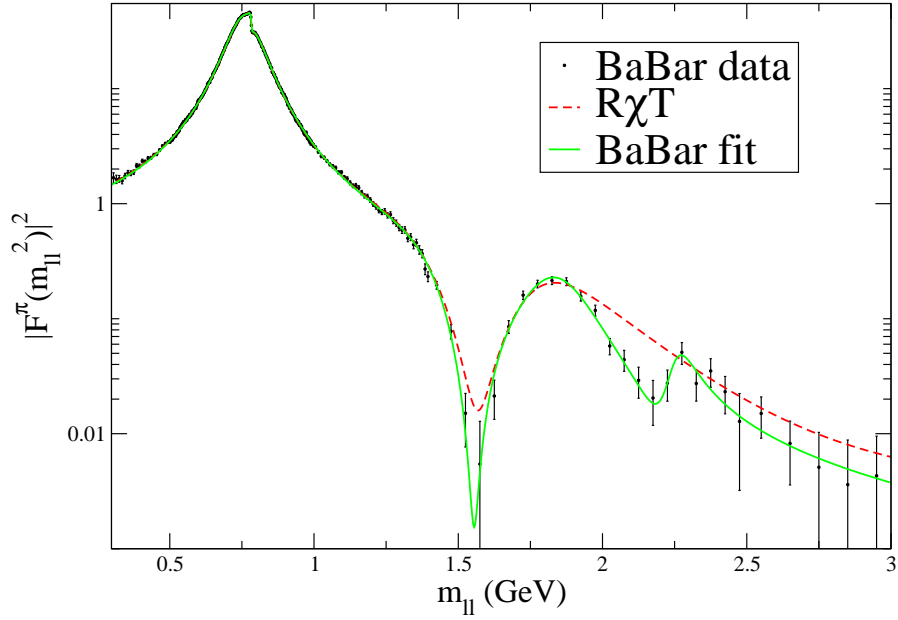}\caption{BaBar parametrization and our form factor compared 
  with data from BaBar. Here m$_\text{ll}=\sqrt{q^2}$. Both form factors overlap below 1.4 GeV, which is the 
  dominant region of the form factor in the observables of the studied decays.}\label{BaBarPiWA}
 \end{figure}

 In the previous expression, by demanding a Brodsky Lepage behavior \cite{BrodskyLepage} 
 of the form factor one gets the relation $F_VG_V=F^2$ (also at LO in $1/N_C$). This expression for the 
 form factor must be improved to obtain a more precise determination of the amplitude through the off-shell width of the resonances, 
 (in the $P=\pi$ case) heavier vector multiplets and the dominant isospin breaking effect in the $\rho-\omega$ mixing giving the factor
 \cite{ThCirigliano}
 \begin{equation}
  1-\theta_{\rho\omega}\frac{q^2}{3m_\rho^2}\frac{1}{m_\omega^2-q^2-im_\omega\Gamma_\omega},
 \end{equation}
 
 where $\theta_{\rho\omega}=(-3.3\pm0.5)\cdot10\ex{-3}$ GeV$^2$ \cite{ThRhoOmega}. 
 For the $F_\pi(q^2)$ we used the parametrization given in ref \cite{Parametr}
 including three vector resonances with same quantum numbers ($\rho(770)$, $\rho(1450)$ and $\rho(1700)$) and a resummation of final state 
 interactions encoded through chiral loop functions.

  \begin{figure}[!h]
  \centering
  \includegraphics[scale=0.48]{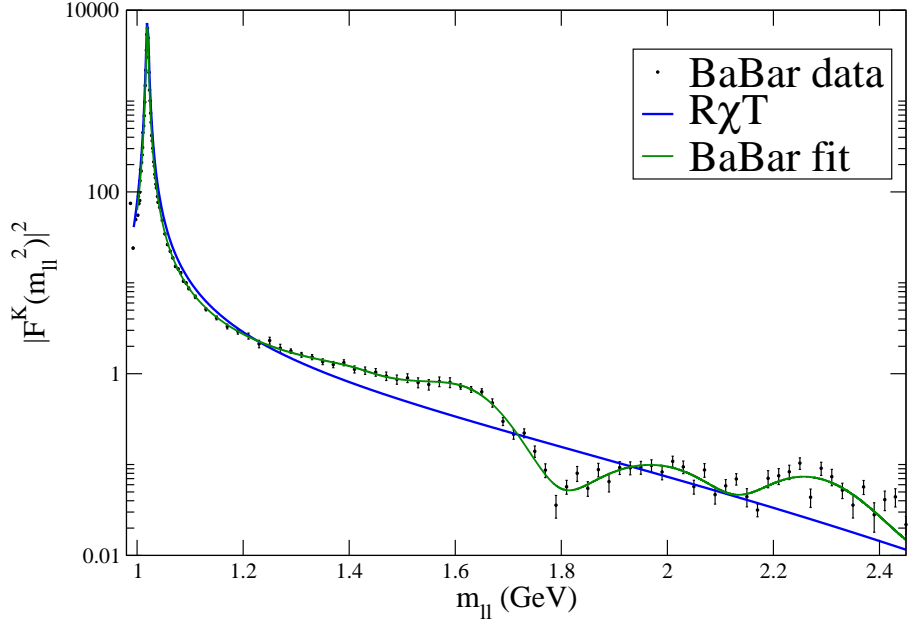}\caption{Electromagnetic form factor of the $K$ meson with the BaBar parametrization and our form factor compared 
  with data from BaBar. Here m$_\text{ll}=\sqrt{q^2}$.}\label{BaBarKWA}
 \end{figure}

 The comparison between our form factors and those used by BaBar for fitting their data \cite{BaBarPiFF} are shown in Figures \ref{BaBarPiWA} for $P=\pi$ with
 energy region $2m_\pi\le \sqrt{q^2}\le$ 3GeV, where the GS parametrization includes an extra iso-vector resonance. In these plots it is made 
 evident the lack of the latter resonance ($\rho(2250)$), also small differences can be seen in the region where the $\rho(1450)$ interferes 
 with the $\rho(1700)$. All these differences are taken into account to obtain the uncertainty induced by $F_\pi(q^2)$. \\
 \begin{figure}[!h]
  \centering
  \includegraphics[scale=0.48]{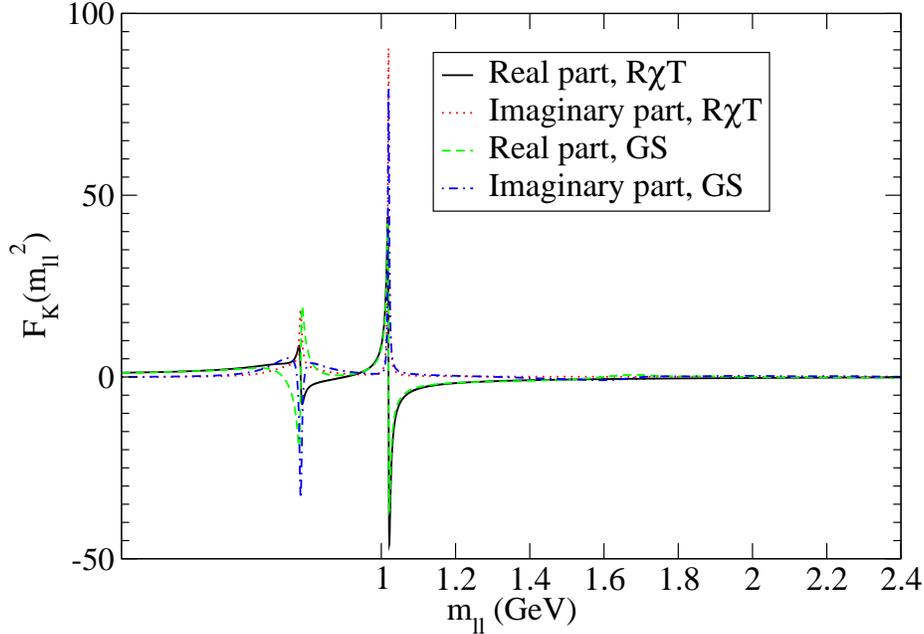}\caption{Real and imaginary parts of $F_K(q^2)$ using R$\chi$T and GS.}
  \end{figure}

 For the $P=K$ case, we also made use of the BaBar parametrization of the form factor \cite{BaBarKFF} to compare it with our form factor as can be seen in Figure 
 \ref{BaBarKWA}. In this case, the $\phi(1020)$ resonance peak is so large that no other multiplet of resonances need to be considered to improve 
 the precision of the LD WA amplitude for $P=K$, since there is a very good match between the data and our form factor around the peak of this 
 resonance and, in addition, the squared modulus of the form factor drops 4 orders of magnitude just outside the peak of the resonance.
 That is to say, R$\chi$T gives a very good description from threshold to around 1.3 GeV and deviations at higher energies will have a
 negligible effect on the integrated observables. The remaining structure shown in the BaBar fit (above 1.3 GeV) is due to heavier vector resonances
 (two $\phi$, three $\rho$ and three $\omega$), which also includes the lightest iso-vector multiplet included in the R$\chi$T description.\\

  \begin{figure}
  \centering
  \includegraphics[scale=0.4]{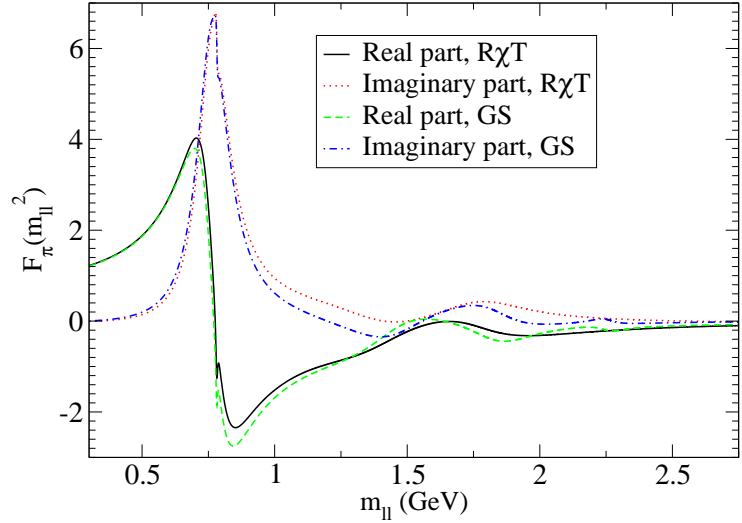}\caption{Real and imaginary parts of $F_\pi(q^2)$ using R$\chi$T and GS.}\label{GSvsRChT}
 \end{figure}

 \begin{figure}[!h]
  \centering
  \includegraphics[scale=.54]{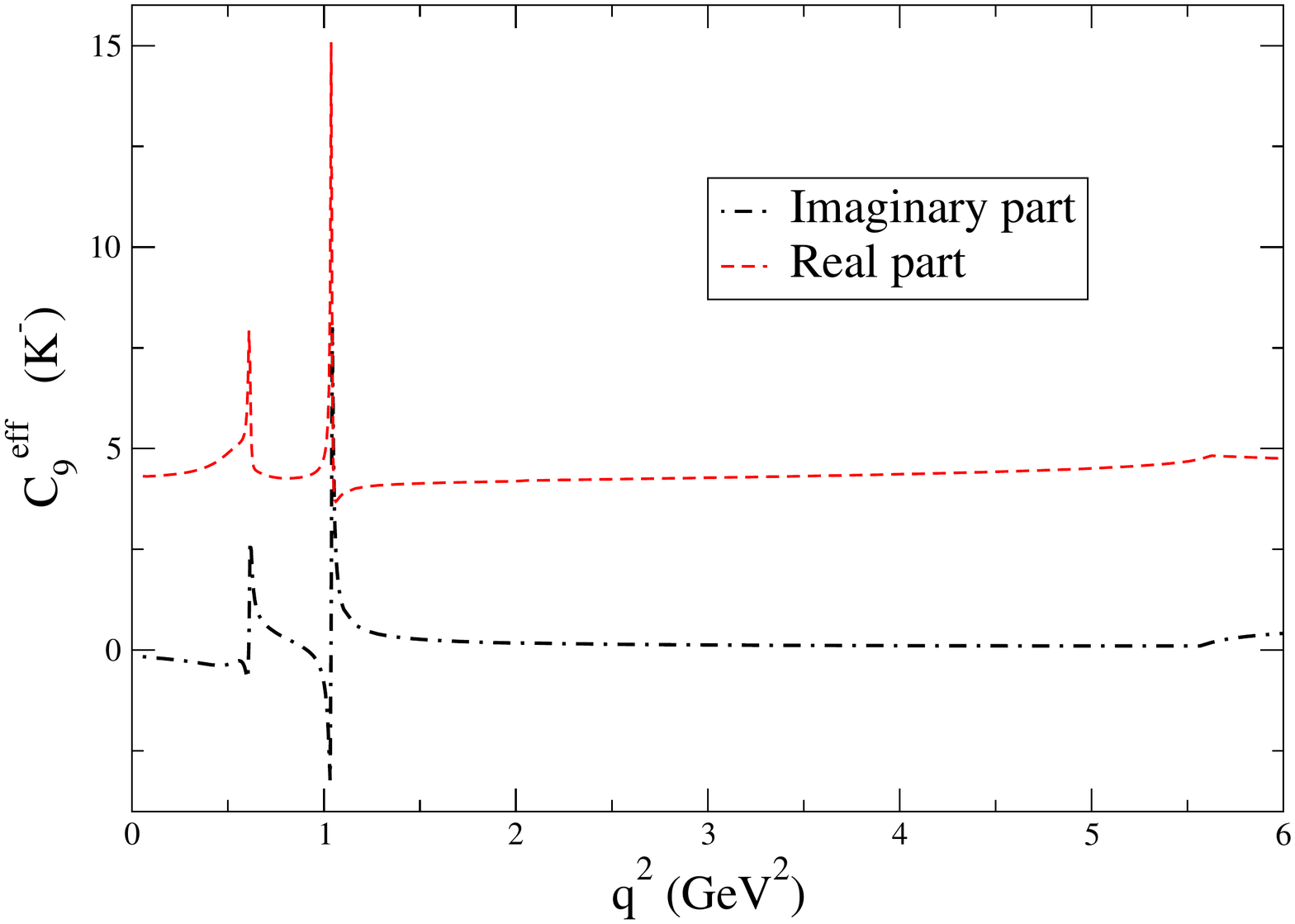}
  \caption{The smooth match between LD and QCDf description of $F_K$ at 2 GeV$^2$ is shown.}\label{SDLDKMatch}
  \end{figure}
  
 \begin{figure}[!h]
 \centering
  \includegraphics[scale=.54]{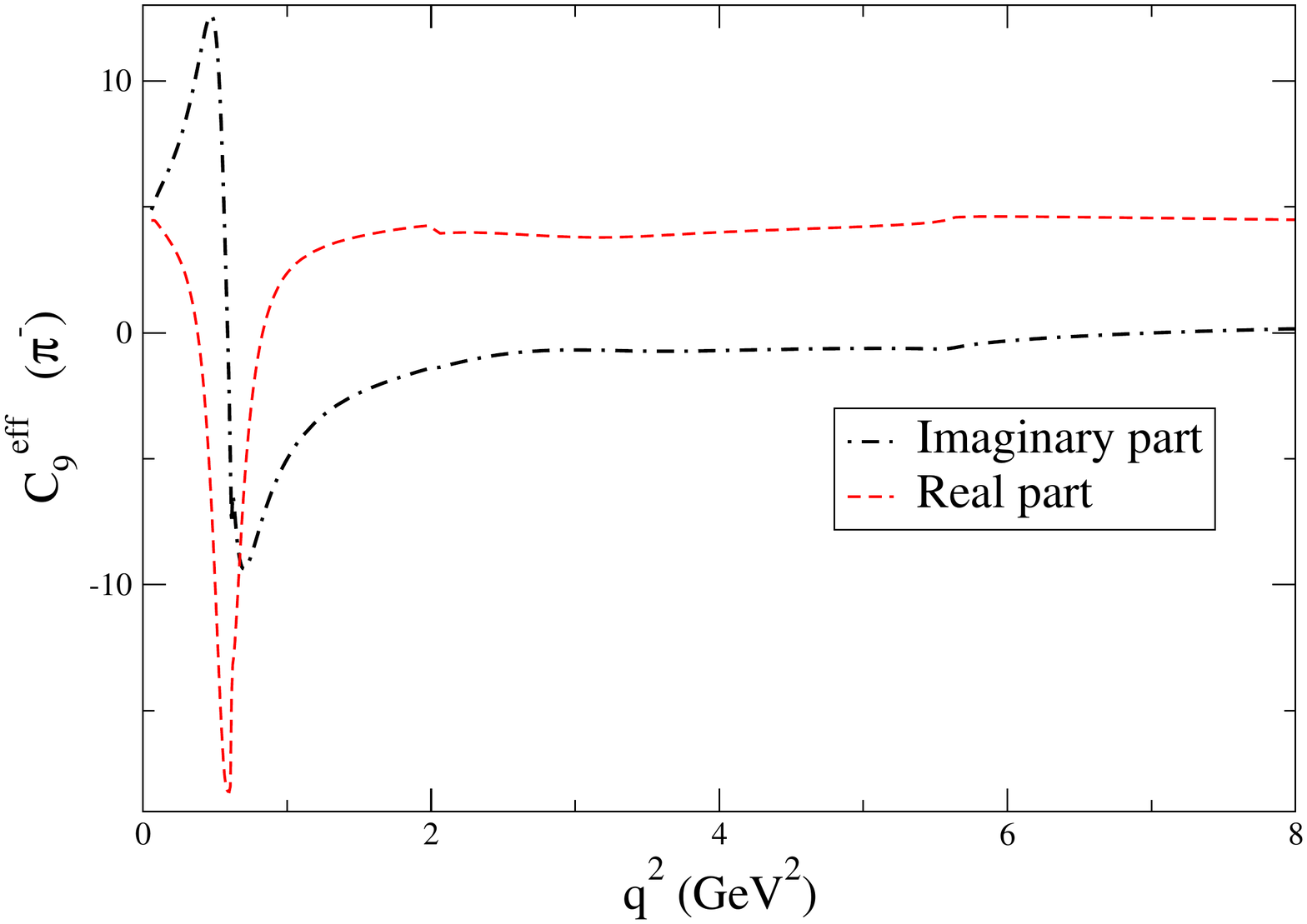}\caption{The smooth match between LD and QCDf description of $F_\pi$ at 2 GeV$^2$ is shown.}\label{SDLDPiMatch}
 \end{figure}

 As previously mentioned, the WA is also considered in the QCDf amplitude, thus, at some point this would mean making a double counting of the same process. 
 Therefore, a check on whether this is happening is needed. To do so, it must be remarked that QCDf is expected to give a more precise description of 
 phenomena for high $q^2$ region of energy. Also, since R$\chi$T is expected to give adequate description of phenomena at small $q^2$, we assume there 
 is an intermediate energy scale where both models describe adequately this process. This match must be done at the form factor level, meaning that 
 the real (imaginary) part of one model must match the real (imaginary) part of the other model if our assumption is correct. So, a comparison between our 
 model and the fit done 
 by the BaBar collaboration has to be made before trying to match the QCDf and the chiral Lagrangian descriptions. This is shown in Figure \ref{GSvsRChT}, 
 where we can see that both parametrizations give a similar description of the form factor for $q^2\ge1$ GeV, {\it i.e.}, in the $q^2$ region of the $R_P$ 
 observable. The fact that both parametrizations are significantly different at low $q^2$ (specifically around the $\rho(770)$ resonance) for $P=K$ is expected, 
 since chiral Lagrangians are more precise the lower the energy of the process is. \\
 
 At the precision order of the QCDf amplitude $F_V(q^2)=C_9^\text{eff}$
 is found to make a very smooth match with the chiral Lagrangians description at around 2 GeV$^2$. This match is shown in 
 Figure \ref{SDLDKMatch} and Figure \ref{SDLDPiMatch}. This shows that it is meaningless to compute the observables as we proposed, 
 where the chiral Lagrangian amplitude must be taken (for $q^2<1$ GeV$^2$, otherwise GS also gives a good parameterization) up to 2 GeV$^2$ and the QCDf 
 amplitude from $q^2=2$ GeV$^2$ up to the $c\bar{c}$ threshold. Our branching ratio for the $P=\pi$ case can be compared with 
 that of ref \cite{Ali}, however in their analysis of QCDf  
 parameters, these were fitted to reproduce $B^+\to\pi^0\ell^+\nu_\ell$ data. The branching fractions obtained are shown in 
 table \ref{SDLDTab}, and since no dedicated study of the errors stemming from the QCDf contribution was made, the errors shown 
 were obtained by rescaling the errors in ref \cite{Ali} for $P=\pi$ and in \cite{BobethHillerPiranishvili}
 for $P=K$ according to the different central values obtained by them and us. Also, by using different quark mixing values (\cite{PDG14}
 and \cite{CKMFitter})
 the branching ratio is $\sim$5\% larger when the parameters from the CKM fitter group is used compared to the result using the PDG values. 
 And thus, the $R_P$ ratios are $R_K=1.0003(1)$ in the (1,6) GeV$^2$ range and $R_\pi=1.0006(1)$ in the (1,8) GeV$^2$ range. Finally, to compare 
 our result with that of reference \cite{Ali} we computed the branching ratio in the whole kinematical domain, which gives 
 $BR(B^-\to\pi^-\ell^+\ell^-)=(2.6_{-0.3}^{+0.4})\cdot10^{-8}$. For the $K$ channel we find $BR(B^-\to K^-\ell^+\ell^-)=(1.92_{-0.41}^{+0.69})\cdot10^{-8}$
 for $q^2\in[1,6]$ GeV$^2$. Comparing our results with the measurements done at BaBar \cite{AaijRK,AaijPi}
 
 \begin{subequations}
  \centering
  \begin{align}
   BR(B^-\to\pi^-\mu^+\mu^-)=&(2.3\pm0.6\pm0.1)\cdot10^{-8},\\
   BR(B^-\to K^-e^+e^-)=&(1.56_{-0.16}^{+0.20})\cdot10^{-7},\hspace*{5ex} \text{for }1<q^2<6\text{ GeV}^2,
  \end{align}
 \end{subequations}
 we see there is a very good agreement within errors.  
  
 \begin{table}
  \centering
  \begin{tabular}{|c||c|c||c|}\hline
  &$P=\pi$&$P=\pi$&$P=K$\\ \hline \hline
  &$0.05 \leq q^2 \leq 8$ GeV$^2$&\ \ \ $1 \leq q^2 \leq 8$ GeV$^2$&$1\leq q^2\leq 6\text{ GeV}^2$\\ \hline
  $LD$&$(9.16\pm0.15)\cdot 10^{-9}$&$(5.47\pm0.05)\cdot 10^{-10}$&$(1.70\pm0.21)\cdot10^{-9}$\\ 
  Interf&$(-2.62\pm 0.13)\cdot 10^{-9}$& $(-2^{+2}_{-1})\cdot 10^{-10}$&$(-6\pm2)\cdot10^{-11}$\\ 
  $SD$&$(9.83^{+1.49}_{-1.04})\cdot 10^{-9}$& $(8.71^{+1.35}_{-0.90})\cdot10^{-9}$&$(1.90^{+0.69}_{-0.41})\cdot10^{-7}$\\ \hline%
 \end{tabular}\caption{LD, SD and their interference contributions to the branching ratio for both channels.}\label{SDLDTab}
 \end{table}

 \subsection{CP Asymmetry}
 
 We can analyze further the decays proposed by computing the CP asymmetry in the dilepton invariant mass region where our description works. 
 The interest in this observable stems from the proposal made in ref \cite{CpAsymmetry}
  (within QCDf) that, in these decays a large CP asymmetry might come about. By inspecting the behavior of the form factor in 
 Figure \ref{SDLDPiMatch} a measurable CP asymmetry seems possible, due to the large values of the real and imaginary parts of $P_\pi(q^2)$. 
 Since the off-shell width of the vector resonances describes the imaginary 
 part of the electromagnetic form factor of the pseudo-Goldstone bosons, it will be responsible for the strong phase required to generate 
 the CP asymmetry. The CP asymmetry is defined as follows,
 \begin{equation}
  A_{CP}(P)=\frac{\Gamma(B^+\to P^+\ell^+\ell^-)-\Gamma(B^-\to P^-\ell^+\ell^-)}{\Gamma(B^+\to P^+\ell^+\ell^-)+\Gamma(B^-\to P^-\ell^+\ell^-)}.
 \end{equation}

 As just mentioned above, the CP asymmetry will be mainly consequence of the off-shell width of vector resonances, but for $q^2\ge2$ GeV$^2$ it
 must be verified that QCDf also gives a measurable asymmetry. There are, in fact two sources of such asymmetry. The first stems from on-shell
 radiating light quarks, since heavy quarks bremsstrahlung is suppressed by a factor $m_q/m_b$, where $m_{q(b)}$ is the mass of the light ($b$) quark.
 The second source arises from light $q\bar{q}$ loops ref \cite{CpAsymmetry}. The CP asymmetry taking into account all the previously 
 mentioned effects for different $q^2$ ranges is shown in Table \ref{CPas}.

 \begin{table}[!h]
  \centering
  \begin{tabular}{|c||c|c||c|}\hline
  $(q^2_\text{min},q^2_\text{max})$&Ref \cite{CpAsymmetry}&$P=\pi$&$P=K$\\ \hline \hline
  $(1,8)\text{ GeV}^2$&$13\pm2$&$7.8\pm2.9$&------\\ 
  $(1,6)\text{ GeV}^2$&$16\pm2$&$9.2\pm1.7$&$-1.0\pm0.3$\\ 
  $(2,6)\text{ GeV}^2$& $13^{+2}_{-3}$&$7.7\pm0.5$&------\\ 
  $(0.05,8)\text{ GeV}^2$& ------&$16.1\pm1.9$&------\\ \hline%
 \end{tabular}\caption{CP asymmetry computed for different $q^2$ ranges, all values are given as percentages.}\label{CPas}
 \end{table}

 For $P=\pi$, in the (0.05,8) GeV$^2$ range, 83\% of the $A_{CP}$ has a LD WA origin, while  in the (1,8) GeV$^2$ this contribution is 
 reduced to a 31\%. For $P=K$, 70\% of the asymmetry is due to LD WA. As seen in Table \ref{CPas},in ref \cite{CpAsymmetry} the $A_{CP}(\pi)$ was 
 predicted to be larger than ours, while in ref \cite{KhodjCP}
 reports a result that lies between both predictions (for (1,6) GeV$^2$ they predict $A_{CP}(\pi)=(14.3^{+2.9}_{-3.5})$\%).\\
 
 A remark regarding the $F_B(q^2)$ needs to be done. By computing the CP asymmetry from muon threshold we obtained a contribution 
 to this observable that is completely negligible, since its contribution is smaller than the uncertainty, namely $\mathcal{O}(10^{-4})$.
 Also note that the asymmetry becomes significantly larger as $q^2$ becomes smaller. That is to say, the CP asymmetry becomes larger as 
 $q^2_{min}$ moves towards the region where the predictions based on QCDf are not reliable.\\
 
 Currently, experimental data gives $A_{CP}$ values consistent with zero ($A_{CP}(K)=0.011\pm0.017$ \cite{PDG14} and $A_{CP}(\pi)=-0.11\pm0.12$), but 
 within errors are also compatible with the different theoretical predictions.\\
 
 Despite the fact that one-photon exchange diagrams at leading order give a small contribution to the decay rates, within the Standard Model 
 they can generate a non-negligible CP asymmetry. This CP asymmetry together with measurements of the decay rates can be used as test 
 in the decays studied in this chapter. This makes us emphasize the need of a dedicated measurement of these observables at LHCb in the next run
 and in the forthcoming Belle-II experiment. 

 \subsection{Conclusions}
   We computed the $\mathcal{M}(B\to P\ell^+\ell^-)$ amplitude for $P=\pi,K$ obtaining a contribution that had not been considered before in the 
   analysis of this process. Our contribution for the $K$ channel is a correction $\sim$ 1\%, which we believe cannot be measured in the forthcoming 
   experiments. 
   
   However, for $P=\pi$ the contribution becomes very significant when $q^2_{min}$ is lowered near the threshold. This suggest that the $[1,8]$ GeV$^2$
   range is free from hadronic pollution (within current experimental errors).On the other hand, more refined measurements of the fully integrated 
   branching fraction for this decay could be sensitive to our contribution once the error is reduced below half of the current uncertainty.
   
   Interestingly, the different weak and strong phases of the QCDf and LD WA one photon exchange contributions are capable to generate a CP asymmetry. 
   Again, this CP asymmetry is large in the case of a pion in the initial state for $q$ taken from threshold, but also sizable and worth to measure 
   in the experimental range for the leptons squared invariant mass. For the $K$, the range $1\le q^2\le6$ GeV$^2$ is optimal for such search. 
   Our CP asymmetry result and the magnitudes of the decay rates at LHCb and future Belle II can provide another non-trivial test of the SM.
  

\newcommand{\GP}{$G$-parity}
\newcommand{\etap}{\eta^{(\prime)}}
\newcommand{\invm}{m_{\pi\eta}}
\newcommand{\invmp}{m_{\pi\eta'}}
\newcommand{\MeV}{\text{ MeV}}
\newcommand{\GeV}{\text{ GeV}}
\newcommand{\deceta}{\tau\to\pi\eta\gamma\nu_\tau}
\newcommand{\decetap}{\tau\to\pi\eta'\gamma\nu_\tau}
\newcommand{\both}{\tau\to\pi\etap\gamma\nu_\tau}

\chapter{New charged current structures}

   \section{Introduction}
   
    The question of whether weak charged currents are different to those coupled with the $W$ boson has a great relevance 
    in nowadays High Intensity Frontier experiments, as it offers an excellent place for the search of Beyond Standard Model (BSM) effects. 
    Several extensions to the SM predict new charged currents whose origins might stem, for example, on quark-lepton symmetries 
    (leptoquarks) or additional copies of the BEH scalar doublet (charged Higgs), which would induce a scalar charged current 
    instead of the $V -A$ weak current described in chapter 1.\\
    
    In semileptonic processes, these currents can be classified by means of $G$-parity into first and Second Class Currents (SCC). 
    This parity taken as an extension of the charge conjugation $C$ is defined by $G:=Ce^{i\pi I_2}$, where $I_2$ is the 
    generator of isospin rotations \cite{weinbergscc}. 
    If {\GP} was an exact symmetry, there would only be first class currents in the SM, however, it is not since isospin violating 
    processes will induce SCC. Although SCC are present within the SM, they are highly suppressed 
    with respect to the first class ones, which keeps the feature of a good test of the SM.\\
    
    In this chapter we compute the $\tau\to\pi\etap\gamma\nu_\tau$ decay as a background for the search of SCC. 
    We compare our results with the $\tau\to\pi\etap\nu_\tau$ decay, which is the cleanest channel of those proposed for the 
    search of these currents \cite{Leroy:1977pq} with predicted $BR\lesssim10^{-5}$ \cite{estimates-scc}. The latter can be induced by {\GP} breaking, while the former 
    will receive contributions from first class currents and a very suppressed contribution from second class ones. Experimental limits 
    on these processes are near the thoretical estimates based on isospin breaking \cite{babar2011, Aubert:2008nj, Hayasaka:2009zz}. Since Belle-II increase 
    considerably the luminosity compared to previous B-factories, these isospin breaking decays are likely to be measured. We, therefore, 
    look for a sufficiently low cut in the photon energy such that the radiative process (above such cut) can be neglected as background in the search for 
    BSM SCC.\\
    
     In section \ref{sec:General} we show the relevant 
    
\section{Matrix Element and Form Factors}\label{sec:General}

   The $\tau\to\pi\etap\gamma\nu_\tau$ process does not present {\GP} suppression, since the photon is not an isospin eigenstate. It also  
   will have, as stated before, bremsstrahlung contributions stemming from the $G$ suppressed process which will be further suppressed 
   by a factor $\alpha$. Contributions from the effective vertex $W\pi\eta\gamma$, which does not present the $G$ 
   suppression, is expected to give an effect comparable to the non radiative process.\\
   
\begin{figure}[ht!]
 \centering
 \includegraphics[scale=0.75]{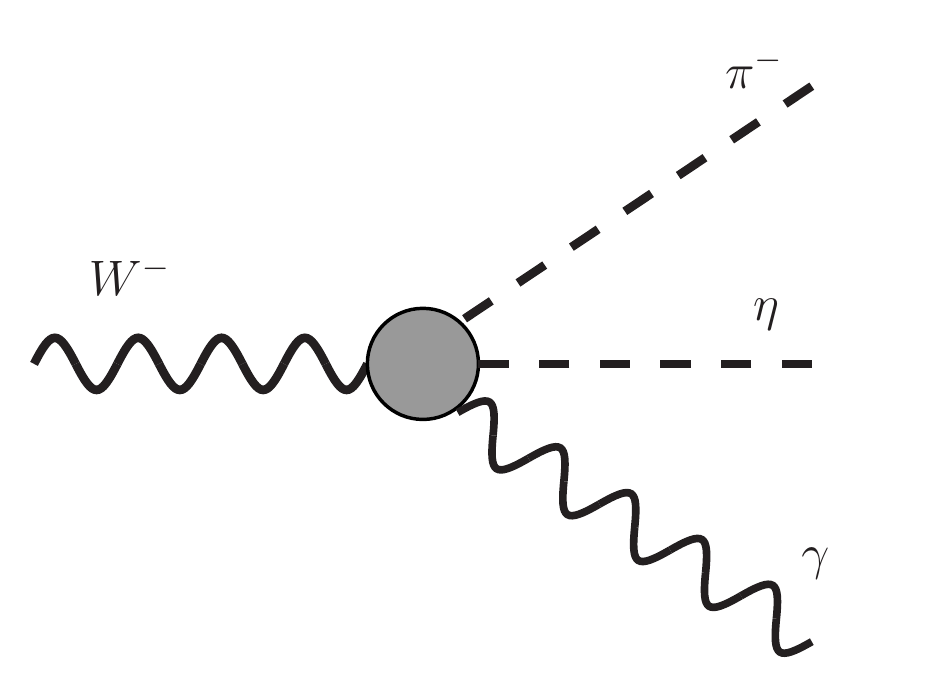}\caption{Effective hadronic vertex (grey blob) that defines the $V_{\mu\nu}$ and $A_{\mu\nu}$ tensors.}\label{fig:hadvertex}
\end{figure}

   To check this assertion we compute the amplitude with the convention $\tau^-(P)\to \pi^-(p)\eta^{(\prime)}(p_0)\nu(p')\gamma(k,\epsilon)$,
   given by
   \begin{equation}\label{m.e.}
    \mathcal{M}=\frac{e G_F V_{ud}^*}{\sqrt{2}}\epsilon^{*\mu}\left\lbrace (V_{\mu\nu}-A_{\mu\nu})\bar{u}(p')\gamma^\nu(1-\gamma_5)u(P)\right\rbrace\,.
   \end{equation}
   where the bremsstrahlung contribution has been neglected, {\it i. e.}, eq. (\ref{m.e.}) gives the leading contribution to our process 
   which originates from the $W\pi\eta\gamma$ effective vertex. Here, the hadronic tensors $V_{\mu\nu}$ and $A_{\mu\nu}$ are directly 
   related to the vector and axial-vector contributions to the effective vertex in fig. \ref{fig:hadvertex} and are parametrized as follows \cite{Cirigliano:2002pv, BEG}
   
   \begin{eqnarray} \label{decomp}
V_{\mu\nu}&=&v_1(p.kg_{\mu\nu}-p_\mu k_{\nu})+v_2\left(g_{\mu\nu}p_0.k-p_{0\mu}k_{\nu}\right) \nonumber \\
&&+ v_3(p_\mu p_0.k-p_{0\mu}p.k)p_\nu+v_4(p_\mu p_0.k-p_{0\mu}p.k)p_{0\nu} \nonumber \\
A_{\mu\nu}&=&i\varepsilon_{\mu\nu\rho\sigma}\left(a_1p_0^{\rho}k^{\sigma}+
a_2k^{\rho}W^{\sigma}\right)+
i\varepsilon_{\mu\rho\sigma\tau}k^{\rho}p^{\sigma}p_0^{\tau}\left(a_3W_{\nu}+
a_4(p_0+k)_{\nu}\right)\,,
\end{eqnarray}
where $W=P-p'=p+p_0+k$. These tensors depend on four vector $(v_i)$ and four axial $(a_i)$ form factors, respectively, depending on three 
Lorentz scalars. \\

\begin{figure}[!ht]
 \centering\includegraphics[angle=-90,scale=0.32]{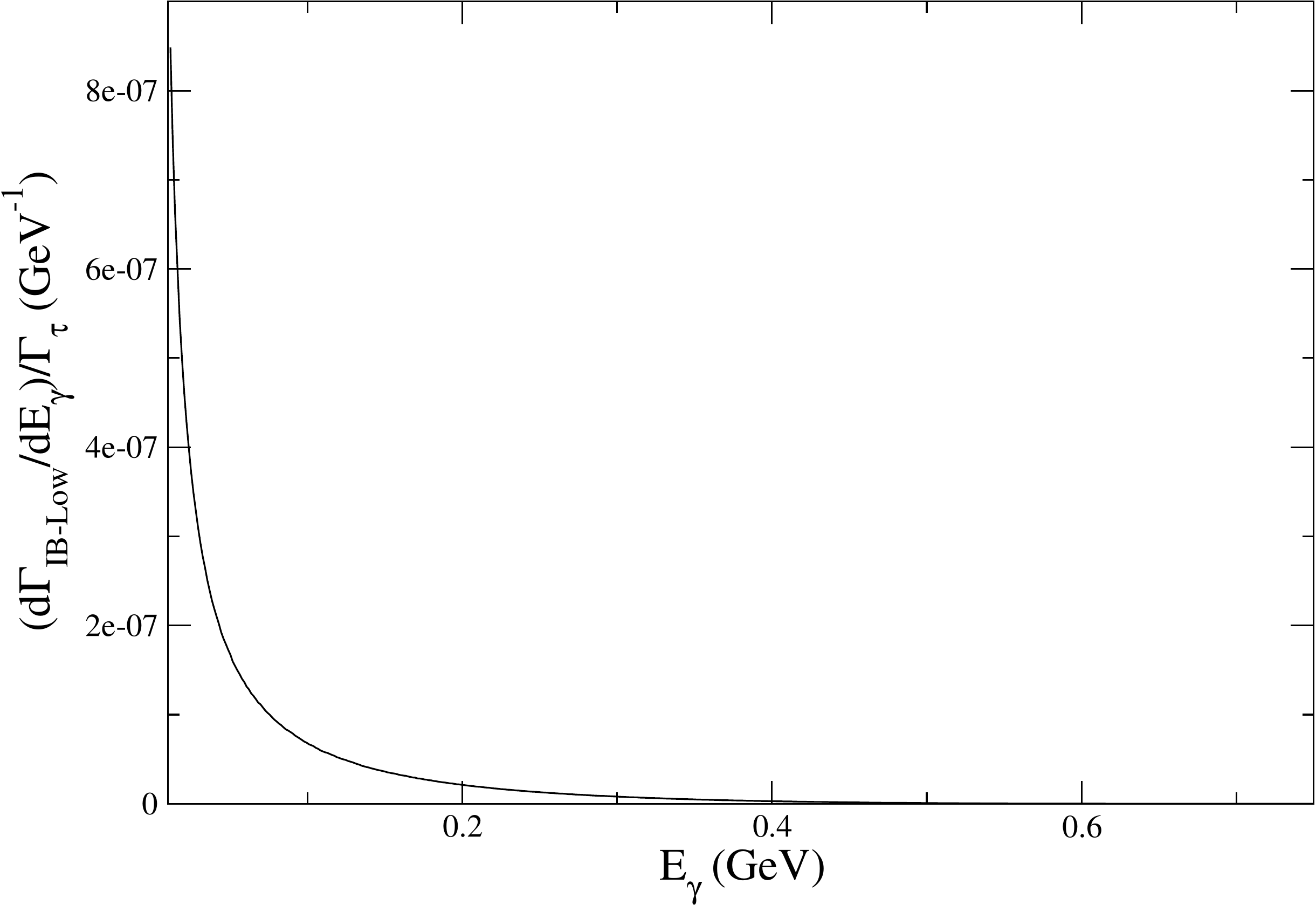}\includegraphics[angle=-90,scale=0.32]{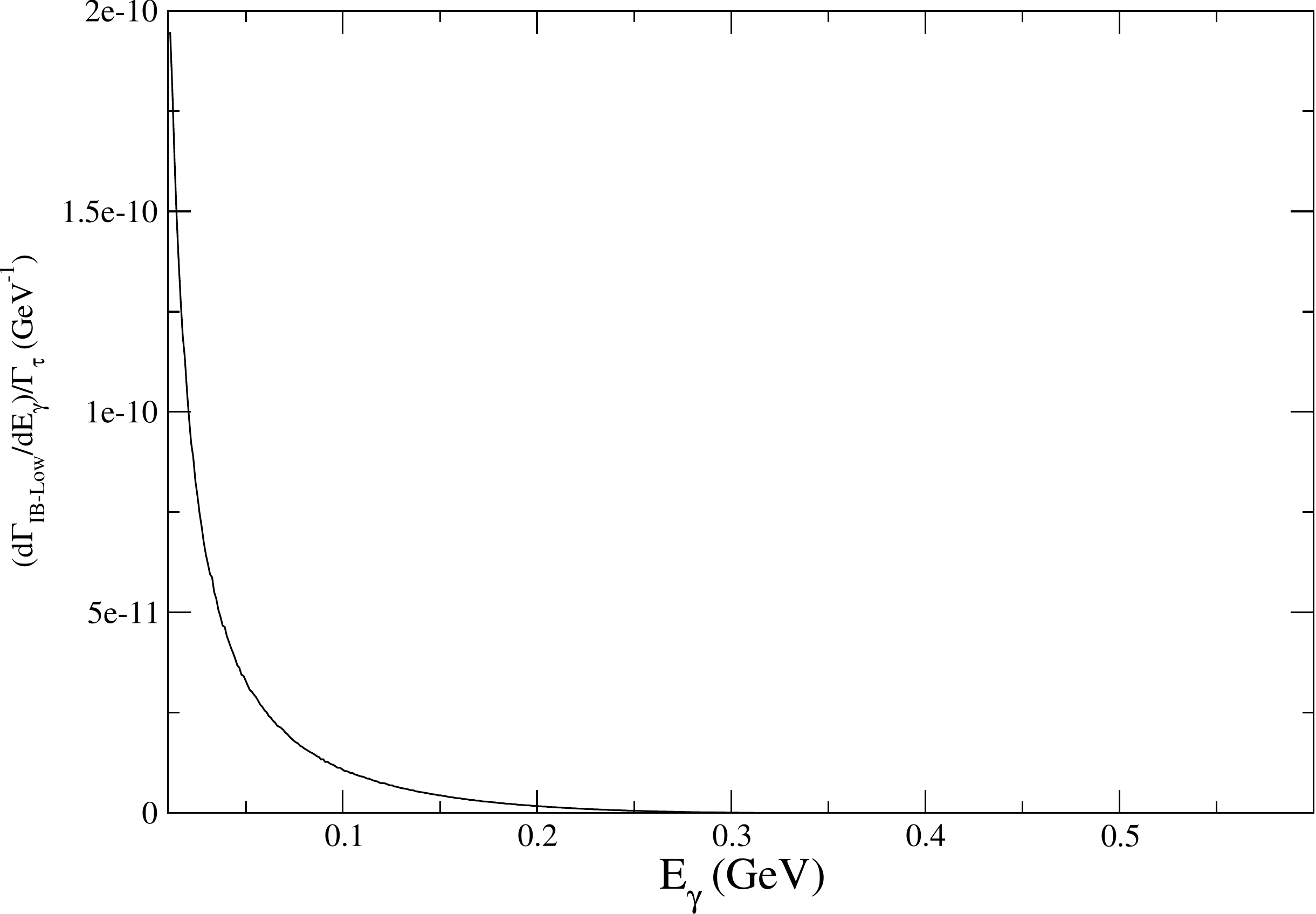}\caption{Photon energy spectra for the 
 leading bremsstrahlung terms in $BR(\tau\to\pi\etap\gamma\nu_\tau)$}\label{IBscc}
\end{figure}

We have argued that bremsstrahlung amplitudes will give a negligible contribution to the process, however we can go further and try to give 
an estimate of whether this is true for a reasonable experimental cut in the energy of the photon. The way to do this is by using Low's theorem \cite{Low:1958sn}, 
which tells that an amplitude with a real photon will have the dependence 
\begin{equation}\label{LowThm}
 \mathcal{M}_\gamma=\frac{A}{k}+B+\mathcal{O}(k),
\end{equation}
where $A$ and $B$ are given in terms of the non radiative amplitude $\mathcal{M}_0$. Since the four momentum is taken to be very small, one can expand the 
radiative amplitude as
\begin{equation}
 \mathcal{M}_\gamma=-e\mathcal{M}_0\left(\frac{P\cdot \epsilon}{P\cdot k}-\frac{p\cdot \epsilon}{p\cdot k}\right)+\cdots,
\end{equation}
and the amplitude $\mathcal{M}_0$ can be taken to be independent of $k$. Then, the $\mathcal{M}_0$ amplitude can be calculated with the expressions of 
ref. \cite{Escribano:2016ntp}. The infrared divergence might surpass the suppression and give a comparable contribution, therefore one 
needs to establish a minimum energy in order to determine the kinematical region of the phase space to be probed. By choosing this threshold as 
10 MeV, photons with smaller energies will not be included. Thus, we obtain the bremsstrahlung contribution to the branching fractions $BR(\deceta)\sim2.5\times10^{-8}$ and for $BR(\decetap)\sim4.6\times10^{-12}$
for an energy cut $E_\gamma>10 \MeV$, and the corresponding photon energy spectra is shown in fig. \ref{IBscc}. Now we can safely neglect the bremsstrahlung 
contribution to our process.

\section{Meson dominance model prediction}\label{sec:MDM}

 Before trying to give the very complex description of the problem within R$\chi$T, we will estimate the form factors of the process in the 
 Meson Dominance Model (MDM) \cite{AlainGabriel}, which are given by a considerably smaller amount of diagrams. In this model weak and 
 electromagnetic couplings are dominated by the exchange of a few light mesons and their excitations. The determination of the relevant 
 couplings is done phenomenologically and by symmetry relations of the theory. \\
 
\begin{figure}[ht!]
 \includegraphics[scale=0.6]{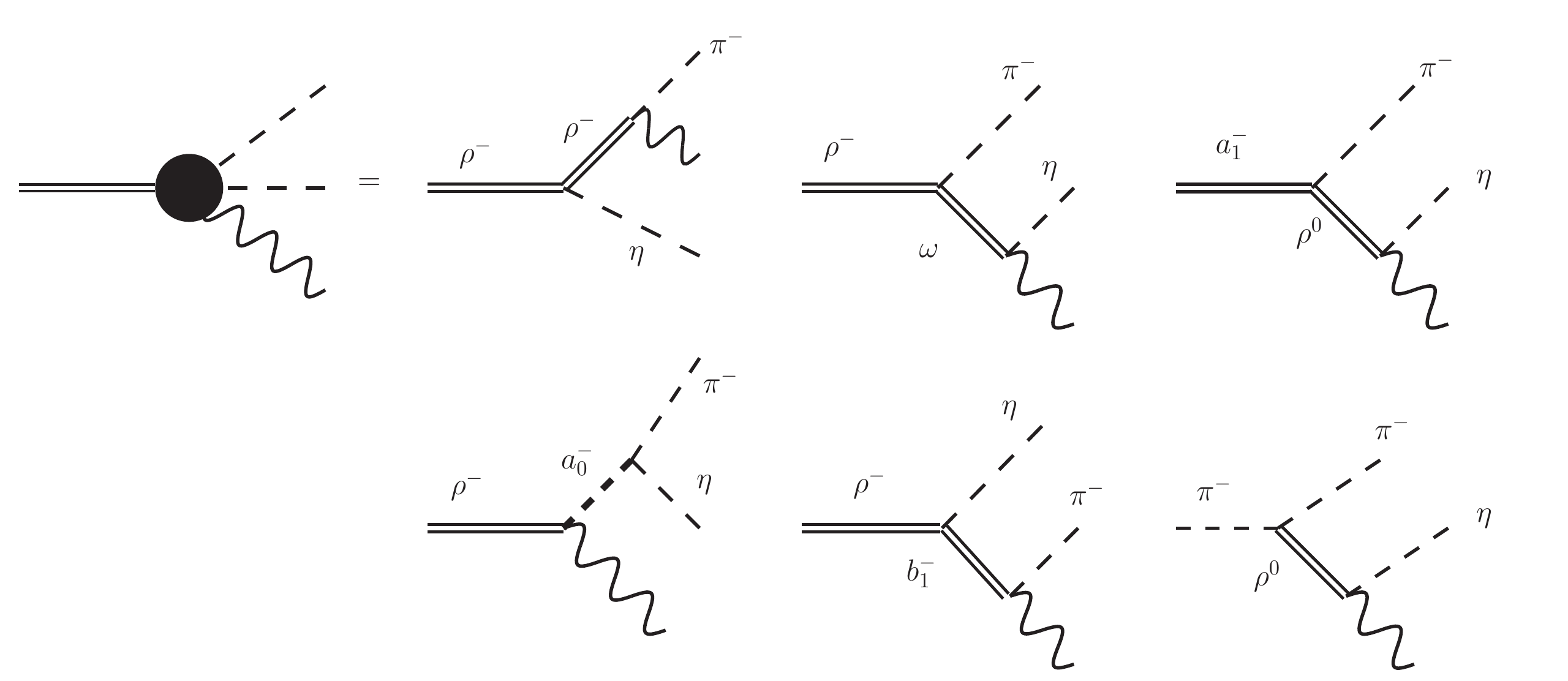}\caption{Contributions to the effective weak vertex in the MDM model. The wavy line denotes the photon.}\label{fig:vertex}
\end{figure}

 Therefore one needs to compute all the processes in fig. \ref{fig:vertex} to obtain the MDM contribution to the form factors, where the diagram with the $b_1(1235)$ 
 exchange can be neglected given $BR(b_1\to\pi\gamma)=(1.6\pm0.4)10^{-3}$ and $BR(b_1\to\rho\eta)<10\%$. Also, the contribution from the pion pole in the 
 last diagram in figure \ref{fig:vertex} will be suppressed since the pion will be far off its mass-shell.\\
 
 The Feynman rules for 
 the different vertices in the diagrams are 

 \begin{eqnarray}\label{FeynmanrulesMDM}
 V'^\mu(r) \to V^\alpha(s) P(t) &:& i g_{V'VP}\epsilon^{\mu\alpha\rho\sigma}s_\rho t_\sigma\,,\label{FRVVP}\\ 
 V^\mu(r) \to \gamma^\alpha(s) P(t) &:& i g_{VP\gamma}\epsilon^{\mu\alpha\rho\sigma}s_\rho t_\sigma\,,\label{FRVAgamma}\\ 
 A^\mu(r) \to V^\alpha(s) P(t) &:& i g_{VAP} (r\cdot s g_{\mu\alpha}-r_\alpha s_\mu)\,,\label{FRVAP}\\
 V^\mu(r) \to \gamma^\alpha(s) S(t) &:& i g_{VS\gamma} (r\cdot s g_{\mu\alpha}-r_\alpha s_\mu)\,,\label{FRVSgamma}\\ 
 S(r) \to P(s) P'(t) &:& i g_{SPP'}\,. \label{FRSPP}
\end{eqnarray}

The following contributions to the effective weak vertex are found (the superscripts denote the ordering of diagrams in the right-hand side of figure \ref{fig:vertex}), from left to right and from top to 
bottom; we have used the following definition ${\cal H}_\nu=(V_{\mu\nu}^{SD}-A_{\mu\nu})\epsilon^{*\mu}$:

\begin{eqnarray}
{\cal H}_{\nu}^a&=&\frac{i\sqrt{2}m^2_{\rho}}{g_{\rho}} g_{\rho^-\rho^-\eta}g_{\rho^-\pi^-\gamma}\frac{1}{D_{\rho}(W^2)}\frac{1}{D_{\rho}((p+k)^2)}\nonumber\\
&&\hspace*{33ex}\times\varepsilon_{\nu\alpha\rho\sigma}(p+k)^{\rho}p_0^{\sigma} \varepsilon^{\alpha\mu\gamma\delta}k_{\gamma}p_{\delta}\epsilon^*_{\mu} \,, \label{hme-mdmfirst}\\
{\cal H}_{\nu}^b&=&\frac{i\sqrt{2}m^2_{\rho}}{g_{\rho}} g_{\rho^-\omega\pi^-}g_{\omega\eta\gamma}\frac{1}{D_{\rho}(W^2)}\frac{1}{D_{\omega}((p_0+k)^2)}\nonumber\\
&&\hspace*{31.5ex}\times\varepsilon_{\nu\alpha\rho\sigma}(p_0+k)^{\rho}p^{\sigma} \varepsilon^{\alpha\mu\gamma\delta}k_{\gamma}p_{0\delta}\epsilon^*_{\mu} \,, \\
{\cal H}_{\nu}^c &=& \frac{i\sqrt{2}m_{a_1}^2}{g_{a_1}}g_{\rho^0 a_1^-\pi^-}g_{\rho^0\eta\gamma} \left((p_0+k).Wg_{\nu\alpha}-W_{\alpha}(p_0+k)_{\nu} \right) \nonumber \\ 
&&\ \ \ \ \hspace*{22ex} \times \frac{1}{D_{a_1}(W^2)D_{\rho}((p_0+k)^2)} \varepsilon^{\alpha\mu\gamma\delta}k_{\gamma}p_{0\delta}\epsilon^*_{\mu} \label{AconinMDM} \,, \\
{\cal H}_{\nu}^d &=& \frac{i\sqrt{2}m_{\rho}^2}{g_{\rho}}g_{\rho^-a_0^-\gamma}g_{a_0^-\pi^-\eta} \left(W.kg_{\mu\nu}-k_{\nu}W_{\mu} \right) \epsilon^{*\mu} \frac{1}{D_{\rho}(W^2)D_{a_0}((p+p_0)^2)}\, .\label{hme-mdmlast}
\end{eqnarray}
In the above expressions, we have defined $D_{X}(Q^2)$ as the denominator of the meson propagator, which may 
(or not) have an energy-dependent width; $g_X$ represents the weak couplings of spin-one mesons, 
defined here as $\langle X| J_{\mu} |0\rangle =i\sqrt{2}m^2_X/g_X\eta_{\mu}$ ($\eta_{\mu}$ is the polarization 
four-vector of meson $X$) and $g_{XYZ}$ denotes the trilinear coupling among mesons $XYZ$. The effects of the $\rho$ meson excitations can be taken into account through the following replacement
\begin{equation}
\frac{\sqrt{2}m_\rho^2}{g_\rho}\frac{1}{D_\rho(W^2)}\to\frac{\sqrt{2}}{g_{\rho\pi\pi}}\frac{1}{1+\beta_\rho}\left[BW_\rho(W^2)+\beta_\rho BW_{\rho'}(W^2)\right]\,,
\end{equation}
where 
\begin{equation}
BW_\rho(W^2)\,=\,\frac{m_\rho^2}{m_\rho^2-W^2-im_\rho\Gamma_\rho(W^2)} \,,
\end{equation}
with $BW_\rho(0)=1$ and $\beta_\rho$ encodes the strength of the $\rho'=\rho(1450)$ meson contribution. The $\rho\to\pi\pi$ coupling is denoted $g_{\rho\pi\pi}$ and $BW_{a_0}(X^2)$, $BW_{a_1}(X^2)$ and $BW_\omega(X^2)$ are defined in analogy to 
$BW_\rho(W^2)$.\\

Note that all the amplitudes in eqs.~(\ref{hme-mdmfirst}) to (\ref{hme-mdmlast}) are of $\mathcal{O}(k)$ in agreement with Low's theorem. All of them correspond to contributions to the vector current, except 
eq.~(\ref{AconinMDM}), which is due to the axial-vector current.\\

The MDM leads to the following form factors:
\begin{eqnarray}\label{FFsMDM}
v_1^{MDM}&=&iC_{\rho}(W^2)\left[-\frac{g_{\rho^-\rho^-\eta}g_{\rho^-\pi^-\gamma}}{D_{\rho}[(p+k)^2]}p.p_0 +\frac{g_{\rho^-\omega\pi^-}g_{\omega\eta\gamma}}{D_{\omega}[(p_0+k)^2]}p_0.(p_0+k)\right.\nonumber\\
&&\hspace*{40ex}\left.+\frac{g_{\rho^-a_0^-\gamma}g_{a_0^-\pi^-\eta}}{D_{a_0}[(p+p_0)^2]}\right]\,, \\
v_2^{MDM}&=& iC_{\rho}(W^2)\left[\frac{g_{\rho^-\rho^-\eta}g_{\rho^-\pi^-\gamma}}{D_{\rho}[(p+k)^2]}p.(p+k) -\frac{g_{\rho^-\omega\pi^-}g_{\omega\eta\gamma}}{D_{\omega}[(p_0+k)^2]}p.p_0\right.\nonumber\\
&&\hspace*{40ex}\left.+\frac{g_{\rho^-a_0^-\gamma}g_{a_0^-\pi^-\eta}}{D_{a_0}[(p+p_0)^2]}\right] \,,\\
v_3^{MDM}&=& iC_{\rho}(W^2)\left[-\frac{g_{\rho^-\rho^-\eta}g_{\rho^-\pi^-\gamma}}{D_{\rho}[(p+k)^2]} \right]\,, \\
v_4^{MDM}&=& iC_{\rho}(W^2)\left[ \frac{g_{\rho^-\omega\pi^-}g_{\omega\eta\gamma}}{D_{\omega}[(p_0+k)^2]}\right] \,, \\
a_1^{MDM}&=& C_A(W^2) \left[\frac{g_{\rho^0a_1^-\pi^-}g_{\rho^0\eta\gamma}}{D_{\rho}[(p_0+k)^2]}\right] (p_0+k).W\,, \\ 
a_2^{MDM} &=& 0\,, \\
a_3^{MDM} &=& 0\,, \\
a_{4}^{MDM} &=& - \frac{a_1^{MDM}}{(p_0+k).W}\, .
\end{eqnarray}
In the above equations the shorthand notation $C_X(W^2)=\sqrt{2}m_X^2/[g_XD_X(W^2)]$ has been used.\\

The coupling constants required in MDM are defined in equations (\ref{FRVVP})-(\ref{FRSPP}). Comparisons of the calculated and measured rates 
allows to determine the relevant coupling constants assuming they are real and positive as indicated in the following.\\

 \begin{itemize}
  \item We can use the $\tau^-\to(\rho,a_1)^-\nu_\tau$ decays to extract the (axial-)vector weak coupling constants defined as indicated before. 
%
We use the decay width for $\tau^- \to X^-\nu_{\tau}$
 \begin{equation}
  \Gamma(\tau^-\to\nu_\tau X^-)\,=\,\frac{G_F^2|V_{ud}|^2}{8\pi M_\tau^3}\frac{M_X^2}{g_X^2} \left(M_\tau^2-M_X^2\right)^2 (M_\tau^2+2M_X^2)\,.
 \end{equation}
 
%
For the $a_1(1260)$ we assume 
$  BR(\tau^-\to a_1^-\nu_\tau)=0.1861\pm0.0013$ \cite{PDG16}.
  Similarly, we can extract $g_\rho$ from $\tau^-\to\rho^-\nu_\tau$ decays; instead, we compare the measured value of the $\rho^0\to\ell^+\ell^-$ decay width with
   \begin{equation}
   \Gamma(\rho^0\to\ell^+\ell^-)\,=\,\frac{4\pi}{3}\left(\frac{\alpha}{g_\rho}\right)^2\left(1+\frac{2m_\ell^2}{M_V^2}\right)\sqrt{M_V^2-4m_\ell^2}\,.
  \end{equation}
  
  \item We extract the coupling constants $g_{VP\gamma}$ from the $V^\mu\to\gamma^\alpha(s) P(t)$ decays, 
using the decay width
 \begin{equation}
  \Gamma(V\to P\gamma)\,=\,\frac{|g_{VP\gamma}|^2}{96\pi M_V^3}(M_V^2-M_P^2)^3\,.
 \end{equation}
 This expression, together with $\Gamma(\rho/\omega \to \pi/\eta\, \gamma)$ \cite{PDG16} allows to determine four of the required coupling constants. 
 
 \item In order to fix the $\rho a_1\pi$ coupling we consider the decay amplitude 
$ \mathcal{M}\,=\,i g_{\rho a_1\pi}(r\cdot s g_{\mu\alpha}-r_\alpha s_\mu)\eta_{a_1}^\mu \eta_\rho^{*\alpha}\,,$ for $a_1^\mu (r,\eta_{a_1}) \to\rho^\alpha(s,\eta_\rho)\pi(t)$ decays. This gives the
decay rate ($\lambda(a,b,c)$ is the ordinary K\"al\'en function)
\begin{equation}
 \Gamma(a_1\to\rho\pi)\,=\,\frac{|g_{\rho a_1\pi}|^2}{96\pi M_{a_1}^3}\left[\lambda(M_{a_1}^2,M_\rho^2,m_\pi^2)+6M_\rho^2M_{a_1}^2\right] \lambda^{1/2}(M_{a_1}^2,M_\rho^2,m_\pi^2)\,.
\end{equation}
According to the PDG16 \cite{PDG16} $a_1\to\rho\pi$ decays make up $61.5\%$ \cite{Asner:1999kj} of the total decay width of $a_1(1260)$, which we take as  $\Gamma_{a_1}=(475\pm175)$ MeV \cite{PDG16}. Using isospin symmetry to relate 
the two decay modes of charged $a_1$ mesons lead us to the result in  Table \ref{ParsMDM}. 

\item
The following partial widths of $a_0(980)$ meson
\begin{eqnarray}
 \Gamma(a_0\to\gamma\gamma)\,&=&\,\frac{|g_{a_0\gamma\gamma}|^2}{32\pi}M_{a_0}^3\,, \\
 \Gamma(a_0\to\pi\eta)&=&\frac{|g_{a_0\pi\eta}|^2}{16\pi M_{a_0}^3}\lambda^{1/2}(M_{a_0}^2,m_\eta^2,m_\pi^2)\,,
\end{eqnarray}
can be used to extract the required coupling constants involving the $a_0$ meson.
%
Neither of these individual $a_0$ decay rates have been measured separately. Instead, measurements of their product have been reported by several groups with good agreement among them. The average value reported in PDG16~\cite{PDG16} is
\begin{equation}
 \Gamma(a_0\to\gamma\gamma)\times \frac{\Gamma(a_0\to\pi\eta)}{\Gamma_{a_0}}\,=\,\left(0.21^{+0.08}_{-0.04}\right) \mathrm{keV}\,.
\end{equation}
We can extract the product of coupling constants of the $a_0$ by comparing the previous equations and using  $\Gamma_{a_0}\,=\,\left(75.6\pm1.6^{+17.4}_{-10.0}\right)$ MeV~\cite{Uehara:2009cf} for the total decay width.

\item The coupling $g_{\rho\omega\pi}$ was fixed using the relation

\begin{equation}
 g_{\rho\omega\pi}\,=\,\frac{G^8}{\sqrt{3}}\left[\mathrm{sin}\theta_V+\sqrt{2}r\mathrm{cos}\theta_V\right]\,,
\end{equation}
where $G^8\ (G^0)$ is the $SU(3)$ invariant coupling of one pseudoscalar meson with two octets (one octet and one singlet) of vector mesons, and $r\equiv G^0/G^8$. Using the rates of $V \to P\gamma$ decays and assuming ideal 
$\omega-\phi$ mixing, $\theta_V=$tan$^{-1}\left(\frac{1}{\sqrt{2}}\right)$, one gets $G^8\,=\,(1.052\pm0.032)\cdot 10^{-2}$ MeV$^{-1}$ and $r=1.088\pm0.018$ \cite{AlainGabriel}.
 
\item 
The following MDM relations between strong and electromagnetic couplings
\begin{equation}
 g_{\rho\rho\eta}=\frac{g_\rho}{e}g_{\rho\eta\gamma}\,,
 \quad g_{a_0\rho\gamma}=\frac{g_\rho}{e}g_{a_0\gamma\gamma}\,.\label{MDMrelations}
\end{equation}
can be used to extract other relevant coupling constants.

\item 
Finally for the decays involving the $\eta^\prime$ meson, the couplings $g_{a_0\pi\eta'}$, $g_{\rho\rho\eta'}$, $g_{\omega\eta'\gamma}$ and $g_{\rho\eta'\gamma}$ need to be determined. Employing 
the above formulas it is straightforward to obtain the last two from the measured $\Gamma(\eta'\to\omega\gamma)$ and $\Gamma(\eta'\to\rho\gamma)$ decays \cite{PDG16}. $g_{\rho\rho\eta'}$ is fixed in terms 
of $g_{\rho\eta'\gamma}$ in analogy to eq.~(\ref{MDMrelations}). It is not possible to determine $g_{a_0\pi\eta'}$ easily, because the involved masses forbid all possible one-to-two body decays. However, 
according to \cite{Guo:2012yt}, $g_{a_0\pi\eta'}<<g_{a_0\pi\eta}$. We will take $g_{a_0\pi\eta'}/g_{a_0\pi\eta}\leq 0.1$ as a conservative estimate.\\
\end{itemize}

 In table \ref{ParsMDM} we show the values of the coupling constants obtained using the above procedure. The errors are propagated from the experimental ones adding them in quadrature. In section \ref{sec:radbkgMDM} we 
 will present the MDM predictions for the $\tau^-\to\pi^-\eta^{(\prime)}\gamma\nu_\tau$ decays using these inputs.\\

  \begin{table}
   \centering
   \begin{tabular}{|c|c|} \hline
   Coupling constant&Fitted value\\ \hline
   $g_{\rho}$&$5.0\pm0.1$ \\ \hline
   $g_{a_1}$&$7.43\pm0.03$\\ \hline
   $eg_{\rho\eta\gamma}$&$(4.80\pm0.16)\times10^{-1}$ GeV$^{-1}$\\ \hline
   $g_{\rho\rho\eta}$&$(7.9\pm0.3)$ GeV$^{-1}$\\ \hline
   $eg_{\omega\eta\gamma}$&$(1.36\pm0.06)\times10^{-1}$ GeV$^{-1}$ \\ \hline
   $eg_{\rho\pi\gamma}$&$(2.19\pm0.12)\times10^{-1}$ GeV$^{-1}$\\ \hline
   $g_{\rho\omega\pi}$&$(11.1\pm0.5)$ GeV$^{-1}$\\ \hline
   $g_{a_1\rho\pi}$&$(3.9\pm1.0)$ GeV$^{-1}$\\ \hline
   $eg_{\rho a_0\gamma}$&$(9.2\pm1.6)\times10^{-2}$ GeV$^{-1}$\\ \hline
   $g_{a_0\pi\eta}$&$(2.2\pm0.9)$ GeV\\ \hline \hline
   $eg_{\rho\eta'\gamma}$& $(4.01\pm0.13)\times10^{-1}$ GeV$^{-1}$\\ \hline
   $eg_{\omega\eta'\gamma}$& $(1.30\pm0.08)\times10^{-1}$ GeV$^{-1}$\\ \hline
   $g_{\rho\rho\eta'}$& $(6.6\pm0.2)$ GeV$^{-1}$\\ \hline
   $g_{a_0\pi\eta'}/g_{a_0\pi\eta}$& $\leq 0.1$ \\ \hline
   \end{tabular}\caption{Our fitted values of the coupling parameters. Those involving a photon are given multiplied by the unit of electric charge.}\label{ParsMDM}

  \end{table}

 \section{Resonance Chiral Theory}\label{sec:RChT}
 
 \subsection{Resonance Lagrangian operators}\label{sec:theoRChT}
  The interaction terms linear in resonance fields which -upon their integration out- contribute to the low-energy constants of the $\chi PT$ 
Lagrangian at $\mathcal{O}(p^4)$ were also derived in refs.~\cite{Ecker89-1,Ecker89-2,Donoghue:1988ed}. These are
\begin{eqnarray}\label{linear_interactions}
{\cal L}
^R & = & c_d \left\langle S u^\mu u_\mu\right\rangle + c_m \left\langle S\chi_+\right\rangle + i d_m \left\langle P \chi_-\right\rangle+i \frac{d_{m0}}{N_F}\left\langle P\right\rangle\left\langle \chi_-\right\rangle\nonumber\\
& & + \frac{F_V}{2\sqrt{2}} \langle V_{\mu\nu} f_+^{\mu\nu}\rangle + i\,\frac{G_V}{\sqrt{2}} \langle V_{\mu\nu} u^\mu u^\nu\rangle+\frac{F_A}{2\sqrt{2}}\left\langle A_{\mu\nu}f_-^{\mu\nu}\right\rangle\,\,.
\end{eqnarray}
The last two operators on the first line involving pseudoscalar resonances do not play any role in our study \footnote{Although it may seem that the operator with coefficient $d_{m0}$ is suppressed 
with respect to the others in 
eq.~(\ref{linear_interactions}) because of its additional trace, this is not the case since it is enhanced due to $\eta^\prime$ exchange \cite{Kampf:2011ty}.} because they couple the pseudoscalar 
resonances to spin-zero sources 
instead of to the weak $V-A$ current.\\

Resonant operators contributing at $\mathcal{O}(p^6)$ in the chiral expansion (in the low-energy limit) were studied systematically in refs.~\cite{Cirigliano:2006hb} and \cite{Kampf:2011ty} for the even- and odd-intrinsic parity sectors, 
respectively. We will be discussing those entering our study of $\tau^-\to\pi^-\eta^{(\prime)}\gamma\nu_\tau$ decays in the following.\\

We will consider first the even-intrinsic parity sector and start with the operators containing one resonance field. There, only one of the operators involving a scalar resonance matters to our analysis: 
$\mathcal{O}_{15}^S=\langle Sf^{\mu\nu}_+{f_{+}}_{\mu\nu}\rangle$ \cite{Cirigliano:2006hb}, while again no operators including pseudoscalar resonances contribute (in either intrinsic parity sector).\\

The corresponding Lagrangian with one vector resonance field was derived in ref.~\cite{Cirigliano:2006hb}:
\begin{equation}\label{Lagrangian_V_Towards}
 {\cal L}_{(4)}^{V} = \sum_{i=1}^{22} \, \lambda_i^{V} \, {\cal O}^V_i\,,
\end{equation}
with the operators
\begin{eqnarray}\label{Operators_V_Towards}
 {\cal O}^V_1\,=\, i \, \langle \, V_{\mu \nu} \,  u^{\mu} u_{\alpha} u^{\alpha} u^{\nu} \, \rangle & , & {\cal O}^V_2\,=\, i \, \langle \,  V_{\mu \nu} \,  u^{\alpha} u^{\mu} u^{\nu} u_{\alpha} \, \rangle\,,\nonumber\\
 {\cal O}^V_3\,=\, i \, \langle \, V_{\mu \nu} \, \{ \,  u^{\alpha} , u^{\mu} u_{\alpha} u^{\nu} \, \} \,\rangle & , & {\cal O}^V_4\,=\, i \, \langle \, V_{\mu \nu} \, \{ \, u^{\mu} u^{\nu},  u^{\alpha} u_{\alpha} \, \} \, \rangle\,,\nonumber\\
 {\cal O}^V_5\,=\, i \, \langle \, V_{\mu \nu} \, f_{-}^{\mu \alpha} \, f_{-}^{\nu \beta} \, \rangle \, g_{\alpha \beta} & , & {\cal O}^V_6\,=\, \langle \, V_{\mu \nu} \, \{ \, f_{+}^{\mu \nu} \, , \, \chi_{+} \, \} \, \rangle\,,\nonumber\\
 {\cal O}^V_7\,=\, i \, \langle \, V_{\mu \nu} \, f_{+}^{\mu \alpha} \, f_{+}^{\nu \beta}  \, \rangle \, g_{\alpha \beta} & , & {\cal O}^V_8\,=\, i \, \langle \, V_{\mu \nu} \, \{ \, \chi_{+} \, , \, u^{\mu} u^{\nu} \, \} \, \rangle\,,\nonumber\\
 {\cal O}^V_9\,=\, i \, \langle \, V_{\mu \nu} \, u^{\mu} \, \chi_{+} \, u^{\nu} \, \rangle & , & {\cal O}^V_{10}\,=\, \langle \, V_{\mu \nu} \, [ \, u^{\mu} \, , \, \nabla^{\nu} \chi_{-} \, ] \, \rangle\,,\nonumber\\
 {\cal O}^V_{11}\,=\, \langle \, V_{\mu \nu} \, \{ \, f_{+}^{\mu \nu} \, , \, u^{\alpha} u_{\alpha} \, \} \, \rangle & , & {\cal O}^V_{12}\,=\, \langle \, V_{\mu \nu} \, u_{\alpha} \,  f_{+}^{\mu \nu} \, u^{\alpha} \, \rangle\,,\nonumber\\
 {\cal O}^V_{13}\,=\, \langle \, V_{\mu \nu} \, ( \, u^{\mu} \, f_{+}^{\nu \alpha} \, u_{\alpha} \, + \, u_{\alpha} \, f_{+}^{\nu \alpha} \, u^{\mu} \, ) \, \rangle & ,
 & {\cal O}^V_{14}\,=\, \langle \, V_{\mu \nu} \, ( \, u^{\mu} u_{\alpha} \, f_{+}^{\alpha \nu} \, + \, f_{+}^{\alpha \nu} \, u_{\alpha} u^{\mu} \, ) \,\rangle\,,\nonumber\\
 {\cal O}^V_{15}\,=\, \langle \, V_{\mu \nu} \, ( \, u_{\alpha} u^{\mu} \, f_{+}^{\alpha \nu} \, + \, f_{+}^{\alpha \nu} \, u^{\mu} u_{\alpha} \, ) \, \rangle & ,
 & {\cal O}^V_{16}\,=\,i \, \langle \, V_{\mu \nu} \, [ \, \nabla^{\mu} f_{-}^{\nu \alpha} \, , \, u_{\alpha} \, ] \, \rangle \,,\nonumber\\
 {\cal O}^V_{17}\,=\, i \, \langle \, V_{\mu \nu} \, [ \, \nabla_{\alpha} f_{-}^{\mu \nu } \, , \, u^{\alpha} \, ] \, \rangle & , & {\cal O}^V_{18}\,=\, i \, \langle \, V_{\mu \nu} \, [ \, \nabla_{\alpha} f_{-}^{\alpha \mu} \, , \, u^{\nu} \, ] \, \rangle\,,\nonumber\\
 {\cal O}^V_{19}\,=\, i \, \langle \, V_{\mu \nu} \, [ \, f_{-}^{\mu \alpha} \, , \, h^{\nu}_{\alpha} \, ] \, \rangle & , & {\cal O}^V_{20}\,=\, \langle \, V_{\mu \nu} \, [ \, f_{-}^{\mu \nu} \, , \, \chi_{-} \, ] \, \rangle\,,\nonumber\\
 {\cal O}^V_{21}\,=\, i \, \left\langle \, V_{\mu\nu} \, \nabla_{\alpha} \nabla^{\alpha} \, \left( u^{\mu} \, u^{\nu} \right) \, \right\rangle & , & 
 {\cal O}^V_{22}\,=\, \langle \, V_{\mu \nu} \, \nabla_{\alpha} \nabla^{\alpha} \, f_{+}^{\mu \nu} \, \rangle\,.
\end{eqnarray}

Two-resonance operators which conserve intrinsic parity are discussed in the following. We begin with the basis of operators for vertices with one $V$ and one $A$ resonances and a pseudoscalar meson \cite{GomezDumm:2003ku} (here denoted $P$ in the 
operators indices, like in the quoted reference) in the normal parity sector. This is
 \begin{equation}\label{VAP}
  \mathcal{L}^{VAP}
  =\sum_{i=1}^5\lambda^i\mathcal{O}^i_{VAP},
 \end{equation}
 where the operators are
\begin{eqnarray}\label{operatorsVAP}
  \mathcal{O}_{VAP}^1&=&\langle[V^{\mu\nu},A_{\mu\nu}]\chi_-\rangle,\nonumber\\
  \mathcal{O}_{VAP}^2&=&i\langle[V^{\mu\nu},A_{\nu\alpha}]h_\mu^{\hspace*{0.7ex}\alpha}\rangle,\nonumber\\
  \mathcal{O}_{VAP}^3&=&i\langle[\nabla^\mu V_{\mu\nu},A^{\nu\alpha}]u_\alpha\rangle,\nonumber\\
  \mathcal{O}_{VAP}^4&=&i\langle[\nabla^\alpha V_{\mu\nu},A_\alpha^{\hspace*{0.7ex}\nu}]u^\mu\rangle,\nonumber\\
  \mathcal{O}_{VAP}^5&=&i\langle[\nabla^\alpha V_{\mu\nu},A^{\mu\nu}]u_\alpha\rangle.
\end{eqnarray}

There is only one relevant operator with both a $V$ and a $S$ field, $O_3^{SV}=\left\langle \, \{ \, S \, , \, V_{\mu \nu} \, \} \, f_{+}^{\mu \nu} \, \right\rangle $, with coupling $\lambda_3^{SV}$ \cite{Cirigliano:2006hb}.\\

Finally, we include the relevant operators with two $V$ resonances in this even-intrinsic parity sector \cite{Cirigliano:2006hb}
\begin{equation}
 \mathcal{L}^{VV}\,=\,\sum_{i=1}^{\;\;\;\;18} \lambda_i^{VV} \mathcal{O}_i^{VV}\,,
\end{equation}
where
\begin{eqnarray}
 \mathcal{O}_1^{VV}&=&\langle V_{\mu\nu}V^{\mu\nu}u^\alpha u_\alpha\rangle\,,\nonumber\\
 \mathcal{O}_2^{VV}&=&\langle V_{\mu\nu} u^\alpha V^{\mu\nu} u_\alpha\rangle\,,\nonumber\\
 \mathcal{O}_3^{VV}&=&\langle V_{\mu\alpha}V^{\nu\alpha}u^\mu u_\nu\rangle\,,\nonumber\\
 \mathcal{O}_4^{VV}&=&\langle V_{\mu\alpha}V^{\nu\alpha}u_\nu u^\mu\rangle\,,\nonumber\\
 \mathcal{O}_5^{VV}&=&\langle V_{\mu\alpha}(u^\alpha V^{\mu\beta}u_\beta+u_\beta V^{\mu\beta}u^\alpha)\rangle\,,\nonumber\\
 \mathcal{O}_6^{VV}&=&\langle V_{\mu\nu}V^{\mu\nu}\chi_+\rangle\,,\nonumber\\
 \mathcal{O}_7^{VV}&=&i\langle V_{\mu\alpha}V^{\alpha\nu}f_{+\beta\nu}\rangle g^{\beta\mu}\,.
\end{eqnarray}

Next we turn to the odd-intrinsic parity sector, where the two terms involving a scalar and an axial-vector resonance \cite{Kampf:2011ty} are
\begin{equation}
 O_1^{SA}\,=\,i\epsilon_{\mu\nu\alpha\beta}\left\langle \left[A^{\mu\nu}, S\right]f_+^{\alpha\beta}\right\rangle\,,\quad O_2^{SA}\,=\,\epsilon_{\mu\nu\alpha\beta}\left\langle A^{\mu\nu} \left[S,u^\alpha u^\beta\right]\right\rangle\,. 
\end{equation}

In this intrinsic parity sector, operators with only vector resonances and sources and at most one pseudoscalar (again denoted $P$ in the naming of the operators) were derived in reference 
\cite{RuizFemenia:2003hm}
 \begin{equation}\label{VJP&VVP}
 \mathcal{L}^{V,odd}
 =\sum_{a=1}^7\frac{c_a}{M_V}\mathcal{O}^a_{VJP}+\sum_{a=1}^4d_a\mathcal{O}^a_{VVP},
 \end{equation}
 where the operators are 
 \begin{eqnarray}\label{operatorsVJP&VVP}
  \mathcal{O}_{VJP}^1&=&\varepsilon_{\mu\nu\rho\sigma}\langle\{V^{\mu\nu},f^{\rho\alpha}_+\}\nabla_\alpha u^\sigma\rangle\,,\nonumber\\
  \mathcal{O}_{VJP}^2&=&\varepsilon_{\mu\nu\rho\sigma}\langle\{V^{\mu\alpha},f^{\rho\sigma}_+\}\nabla_\alpha u^\nu\rangle\,,\nonumber\\
  \mathcal{O}_{VJP}^3&=&i\varepsilon_{\mu\nu\rho\sigma}\langle\{V^{\mu\nu},f^{\rho\sigma}_+\}\chi_-\rangle\,,\nonumber\\
  \mathcal{O}_{VJP}^4&=&i\varepsilon_{\mu\nu\rho\sigma}\langle V^{\mu\nu}[f^{\rho\sigma}_-,\chi_+]\rangle\,,\nonumber\\
  \mathcal{O}_{VJP}^5&=&\varepsilon_{\mu\nu\rho\sigma}\langle\{\nabla_\alpha V^{\mu\nu},f^{\rho\alpha}_+\}u^\sigma\rangle\,,\nonumber\\
  \mathcal{O}_{VJP}^6&=&\varepsilon_{\mu\nu\rho\sigma}\langle\{\nabla_\alpha V^{\mu\alpha},f^{\rho\sigma}_+\}u^\nu\rangle\,,\nonumber\\
  \mathcal{O}_{VJP}^7&=&\varepsilon_{\mu\nu\rho\sigma}\langle\{\nabla^\sigma V^{\mu\nu},f^{\rho\alpha}_+\}u_\alpha\rangle\,;\\ \nonumber\\
  \mathcal{O}_{VVP}^1&=&\varepsilon_{\mu\nu\rho\sigma}\langle\{V^{\mu\nu},V^{\rho\alpha}\}\nabla_\alpha u^\sigma\rangle\,,\nonumber\\
  \mathcal{O}_{VVP}^2&=&i\varepsilon_{\mu\nu\rho\sigma}\langle\{V^{\mu\nu},V^{\rho\sigma}\}\chi_-\rangle\,,\nonumber\\
  \mathcal{O}_{VVP}^3&=&\varepsilon_{\mu\nu\rho\sigma}\langle\{\nabla_\alpha V^{\mu\nu},V^{\rho\alpha}\}u^\sigma\rangle\,,\nonumber\\
  \mathcal{O}_{VVP}^4&=&\varepsilon_{\mu\nu\rho\sigma}\langle\{\nabla^\sigma V^{\mu\nu},V^{\rho\alpha}\}u_\alpha\rangle\,.
\end{eqnarray}

In our case, however, we will not only need odd-intrinsic parity couplings of a $V$ resonance, a $J$ source and a pseudoGoldstone; but also such vertices with two pseudoscalars \footnote{Obviously, in 
this case $J$ has opposite parity than in the 
case with one pseudoGoldstone since both vertices are of odd-intrinsic parity.}. In this case, as warned in ref.~\cite{RuizFemenia:2003hm}, the set $\left\lbrace \mathcal{O}^a_{VJP}\right\rbrace_{a=1}^{\;\;\;\;7}$ is no longer a basis 
\footnote{Analogous comment applies to eq.~(\ref{VAP}), as pointed out in ref~\cite{GomezDumm:2003ku}.} and one needs to use the operator basis with a $V$ resonance derived in ref.~\cite{Kampf:2011ty}; 
i. e.
\begin{equation}\label{V_KN}
\widetilde{\mathcal{L}^{V,odd}}
    =\varepsilon^{\mu\nu\alpha\beta}\sum_i\kappa_i^V{\mathcal{O}_i^V}_{\mu\nu\alpha\beta},
  \end{equation}
with the operators
 \begin{eqnarray}\label{operatorsV_KN}
&& ({\mathcal{O}^{V}_1})^{\mu\nu\alpha\beta}=i\langle V^{\mu\nu}(h^{\alpha\sigma}u_\sigma u^\beta-u^\beta u_\sigma h^{\alpha\sigma}) \rangle\,,\nonumber\\
&& ({\mathcal{O}^{V}_2})^{\mu\nu\alpha\beta}=i\langle V^{\mu\nu}(u_\sigma h^{\alpha\sigma} u^\beta-u^\beta h^{\alpha\sigma}u_\sigma)\rangle\,,\nonumber\\
&& ({\mathcal{O}^{V}_3})^{\mu\nu\alpha\beta}=i\langle V^{\mu\nu}(u_\sigma u^\beta h^{\alpha\sigma}-h^{\alpha\sigma}u^\beta u_\sigma)\rangle\,,\nonumber\\
&& ({\mathcal{O}^{V}_4})^{\mu\nu\alpha\beta}=i\langle [V^{\mu\nu},\nabla^\alpha\chi_+]u^\beta\rangle\,,\nonumber\\
&& ({\mathcal{O}^{V}_5)}^{\mu\nu\alpha\beta}=i\langle V^{\mu\nu}[f_-^{\alpha\beta},u_\sigma u^\sigma]\rangle\,,\nonumber\\
&& ({\mathcal{O}^{V}_6})^{\mu\nu\alpha\beta}=i\langle V^{\mu\nu}(f_-^{\alpha\sigma}u^\beta u_\sigma-u_\sigma u^\beta f_-^{\alpha\sigma})\rangle\,,\nonumber\\
&& ({\mathcal{O}^{V}_7})^{\mu\nu\alpha\beta}=i\langle V^{\mu\nu}(u_\sigma f_-^{\alpha\sigma}u^\beta-u^\beta f_-^{\alpha\sigma}u_\sigma)\rangle\,,\nonumber\\
&& ({\mathcal{O}^{V}_8})^{\mu\nu\alpha\beta}=i\langle V^{\mu\nu}(f_-^{\alpha\sigma}u_\sigma u^\beta-u^\beta u_\sigma f_-^{\alpha\sigma})\rangle\,,\nonumber\\
&& ({\mathcal{O}^{V}_9})^{\mu\nu\alpha\beta}=\langle V^{\mu\nu}\lbrace\chi_-,u^\alpha u^\beta\rbrace\rangle\,,\nonumber\\
&& ({\mathcal{O}^{V}_{10}})^{\mu\nu\alpha\beta}=\langle V^{\mu\nu} u^\alpha \chi_- u^\beta\rangle\,,\nonumber\\
&& ({\mathcal{O}^{V}_{11}})^{\mu\nu\alpha\beta}=\langle V^{\mu\nu} \lbrace f_+^{\alpha\rho},f_-^{\beta\sigma}\rbrace\rangle g_{\rho\sigma}\,,\nonumber\\
&& ({\mathcal{O}^{V}_{12}})^{\mu\nu\alpha\beta}=\langle V^{\mu\nu} \lbrace f_+^{\alpha\rho},h^{\beta\sigma}\rbrace\rangle g_{\rho\sigma}\,,\nonumber\\
&& ({\mathcal{O}^{V}_{13}})^{\mu\nu\alpha\beta}=i\langle V^{\mu\nu} f_+^{\alpha\beta}\rangle\langle\chi_-\rangle\,,\nonumber\\
&& ({\mathcal{O}^{V}_{14}})^{\mu\nu\alpha\beta}=i\langle V^{\mu\nu} \lbrace f_+^{\alpha\beta},\chi_-\rbrace\rangle\,,\nonumber\\
&& ({\mathcal{O}^{V}_{15}})^{\mu\nu\alpha\beta}=i\langle V^{\mu\nu} [f_-^{\alpha\beta},\chi_+]\rangle\,,\nonumber\\
&& ({\mathcal{O}^{V}_{16}})^{\mu\nu\alpha\beta}=\langle V^{\mu\nu} \lbrace \nabla^\alpha f_+^{\beta\sigma},u_\sigma\rbrace\rangle\,,\nonumber\\
&& ({\mathcal{O}^{V}_{17}})^{\mu\nu\alpha\beta}=\langle V^{\mu\nu} \lbrace \nabla_\sigma f_+^{\alpha\sigma},u^\beta\rbrace\rangle\,,\nonumber\\
&& ({\mathcal{O}^{V}_{18}})^{\mu\nu\alpha\beta}=\langle V^{\mu\nu} u^\alpha u^\beta\rangle\langle\chi_-\rangle\,.
  \end{eqnarray}

The operators in eq.~(\ref{operatorsVJP&VVP}) can be written in terms of those in eq.~(\ref{operatorsV_KN}). This yields the following identities among the corresponding couplings \cite{Roig:2013baa}

\begin{eqnarray}\label{eq: relation different basis}
\kappa_1^{VV}&=& \frac{-d_1}{8 n_f}\,\nonumber\\\qquad \kappa_2^{VV}
&=& \frac{d_1}{8} + d_2\,,\nonumber\\\qquad \kappa_3^{VV}&=&d_3\,,\qquad\nonumber\\
\kappa_4^{VV}&=&d_4\,,
\nonumber\\
-2 M_V \kappa_5^V \, &=& \,  M_V \kappa_6^V \, =\, M_V \kappa_7^V = \frac{c_6}{2}\,,\nonumber\\
M_V \kappa_{11}^V &=& \frac{c_1 - c_2 - c_5 + c_6 + c_7}{2}\,,
\nonumber\\
M_V \kappa_{12}^V &=& \frac{c_1 - c_2 - c_5 + c_6 - c_7}{2}\,,\\
\qquad n_f M_V \kappa_{13}^V &=& \frac{- c_2 + c_6}{4}\,,\nonumber\\
\qquad M_V \kappa_{14}^V &=& \frac{c_2 + 4 c_3 - c_6}{4}\,,\nonumber\\
 M_V \kappa_{15}^V &=& c_4\,,\nonumber\\
M_V \kappa_{16}^V &=& c_6 + c_7\,,\nonumber\\
M_V \kappa_{17}^V &=& -c_5 + c_6\,.\nonumber\\
\end{eqnarray}

The analogous Lagrangian to eq.~(\ref{V_KN}) involving an $A$ resonance \cite{Kampf:2011ty} is the last missing piece needed for our computations. This is
\begin{equation}\label{A_KN}
\mathcal{L}^{A,odd}
    =\varepsilon^{\mu\nu\alpha\beta}\sum_i\kappa_i^A{\mathcal{O}_i^A}_{\mu\nu\alpha\beta},
  \end{equation}
with the operators
 \begin{eqnarray}\label{operatorsA_KN}
&& ({\mathcal{O}^{A}_1})^{\mu\nu\alpha\beta}=\langle A^{\mu\nu}[u^\alpha u^\beta,u_\sigma u^\sigma]\rangle,\nonumber\\
&& ({\mathcal{O}^{A}_2})^{\mu\nu\alpha\beta}=\langle A^{\mu\nu}[u^\alpha u^\sigma u^\beta, u_\sigma]\rangle,\nonumber\\
&& ({\mathcal{O}^{A}_3})^{\mu\nu\alpha\beta}=\langle A^{\mu\nu}\{\nabla^\alpha h^{\beta\sigma},u_\sigma\}\rangle,\nonumber\\
&& ({\mathcal{O}^{A}_4})^{\mu\nu\alpha\beta}=i\langle A^{\mu\nu}[f_+^{\alpha\beta},u^\sigma u_\sigma]\rangle,\nonumber\\
&& ({\mathcal{O}^{A}_5)}^{\mu\nu\alpha\beta}=i\langle A^{\mu\nu}(f_+^{\alpha\sigma}u_\sigma u^\beta-u^\beta u_\sigma f_+^{\alpha\sigma})\rangle,\nonumber\\
&& ({\mathcal{O}^{A}_6})^{\mu\nu\alpha\beta}=i\langle A^{\mu\nu}(f_+^{\alpha\sigma} u^\beta u_\sigma-u_\sigma u^\beta f_+^{\alpha\sigma})\rangle,\nonumber\\
&& ({\mathcal{O}^{A}_7})^{\mu\nu\alpha\beta}=i\langle A^{\mu\nu}(u_\sigma f_+^{\alpha\sigma} u^\beta -u^\beta f_+^{\alpha\sigma}u_\sigma)\rangle,\nonumber\\
&& ({\mathcal{O}^{A}_8})^{\mu\nu\alpha\beta}=\langle A^{\mu\nu}\{f_-^{\alpha\sigma},h^{\beta}_{\sigma}\}\rangle,\nonumber\\
&& ({\mathcal{O}^{A}_9})^{\mu\nu\alpha\beta}=i\langle A^{\mu\nu}f_-^{\alpha\beta}\rangle\langle \chi_-\rangle,\nonumber\\
&& ({\mathcal{O}^{A}_{10}})^{\mu\nu\alpha\beta}=i\langle A^{\mu\nu} u^{\alpha}\rangle \langle\nabla^\beta\chi_-\rangle ,\nonumber\\
&& ({\mathcal{O}^{A}_{11}})^{\mu\nu\alpha\beta}=i\langle A^{\mu\nu}\{f_-^{\alpha\beta},\chi_-\}\rangle,\nonumber\\
&& ({\mathcal{O}^{A}_{12}})^{\mu\nu\alpha\beta}=i\langle A^{\mu\nu}\{\nabla^{\alpha}\chi_-,u^{\beta}\}\rangle,\nonumber\\
&& ({\mathcal{O}^{A}_{13}})^{\mu\nu\alpha\beta}=\langle A^{\mu\nu}[\chi_+,u^{\alpha}u^{\beta}]\rangle,\nonumber\\
&& ({\mathcal{O}^{A}_{14}})^{\mu\nu\alpha\beta}=i\langle A^{\mu\nu}\{f_+^{\alpha\beta},\chi_+\}\rangle,\nonumber\\
&& ({\mathcal{O}^{A}_{15}})^{\mu\nu\alpha\beta}=\langle A^{\mu\nu}\{\nabla^{\alpha}f_-^{\beta\sigma},u_{\sigma}\}\rangle,\nonumber\\
&& ({\mathcal{O}^{A}_{16}})^{\mu\nu\alpha\beta}=\langle A^{\mu\nu}\{\nabla_{\sigma}f_-^{\alpha\sigma},u^{\beta}\}\rangle.
  \end{eqnarray}
  
 We recall that the basis for odd-intrinsic parity operators with two vector resonances and a pseudoscalar meson was given in eq.~(\ref{VJP&VVP}).\\
 
  \subsubsection{Short-distance QCD constraints on the $R\chi L$ couplings}\label{sec:QCDconstraints}
  We have discussed in the previous section how symmetry determines the structure of the operators in the $R\chi L$ though it leaves, however, the corresponding couplings undetermined (as in $\chi PT$ or any other effective field 
  theory with a corresponding fundamental theory in the strongly coupled regime). It was soon 
  observed \cite{Weinberg:1967kj, Ecker89-1,Ecker89-2} that demanding that the Green functions (and related form factors) computed in the meson theory to match their known asymptotic behavior according to the operator product expansion \cite{Wilson:1969zs} 
  of QCD relates some of the $R\chi L$ couplings and thus increases the predictive power of the theory. We will quote in the following the results of this programme interesting to our study.\\
  
  In the odd-intrinsic parity sector, the analysis of three-point $VVP$ Green function and associated form factors yields \cite{RuizFemenia:2003hm, Kampf:2011ty, Roig:2013baa}
  \begin{eqnarray}
M_V( 2 \kappa_{12}^V + 4 \kappa_{14}^V +\kappa_{16}^V -\kappa_{17}^V) \quad =&
4 \,c_3 \,+\, c_1 & = \quad  0\,,
\nonumber
\\[3mm]
M_V (2\kappa_{12}^V + \kappa_{16}^V -2\kappa_{17}^V) \quad =&
c_1 \, - \, c_2 \, + \, c_5 \, & = \quad  0 \, ,
\nonumber
\\[3mm]
-\, M_V \kappa_{17}^V \quad = &  c_5 \, - \, c_6 \, & =
\quad  \frac{N_C \,M_V}{64 \,\sqrt{2}\, \pi^2\,F_V} \, ,
\nonumber
\\[3mm]
8\kappa_2^{VV} \quad =&  d_1 \, + \, 8 \,d_2 & = \quad
\frac{F^2}{8\,F_V^2} - \frac{N_C \,M_V^2}{64\, \pi^2\, F_V^2}\,,
\nonumber
\\[3mm]
\kappa_3^{VV}\quad = & d_3& =\quad  -\, \frac{N_C}{64\pi^2} \frac{M_V^2}{F_V^2} \,,
\nonumber
\\[3mm]
& 1 \, + \, \frac{32 \,\sqrt{2} \, F_V \,d_m\, \kappa_3^{PV}}{F^2} & = \quad  0\,,
\nonumber
\\[3mm]
& F_V^2 & = \quad  3\, F^2\,.
\label{eq: Consistent set of relations}
\end{eqnarray}
It is remarkable that the last of eqs.~(\ref{eq: Consistent set of relations}) involves couplings belonging to the even-intrinsic parity $R\chi L$, despite it was obtained demanding consistency to the high-energy constraints derived in the 
odd-intrinsic parity sector \cite{RuizFemenia:2003hm, Kampf:2011ty, Roig:2013baa, Guo:2008sh, Dumm:2009kj, Guo:2010dv, We:2012}. Let us also mention that the short-distance QCD constraint $\kappa_2^S\,=\,0$ \cite{Kampf:2011ty} forbids a diagram similar to the third one in fig.~\ref{fig:1R-FV} where this time the coupling to the 
current would conserve intrinsic parity (it would be thus a contribution to the axial-vector form factors, since $a_0^-\to\pi^-\eta$ belongs to the unnatural intrinsic parity sector)~\footnote{For completeness we quote the corresponding 
operator, $O_2^S\,=\,\epsilon_{\mu\nu\alpha\beta}\langle i S\left[f_+^{\mu\nu},f_-^{\alpha\beta}\right]\rangle$.}. Another relevant short-distance constraint in the odd-intrinsic parity sector which is derived from the study of the 
$VAS$ Green function \cite{Kampf:2011ty} is $\kappa^{14}_A=0$. Interestingly, this same analysis also yields the relation $\kappa^V_4=2\kappa_{15}^V$, where $\kappa^V_4$ does not enter the relations (\ref{eq: relation different 
basis}). Other high-energy constraints derived in the quoted study are not relevant to our computation.\\

 In the even-intrinsic parity sector, the study of $VAP$ and $SPP$ Green functions~\footnote{Four-point functions have been studied in ref.~\cite{Ananthanarayan:2004qk}.} and their form factors allowed to derive the 
 following restrictions \cite{Cirigliano:2004ue, Cirigliano:2005xn, Cirigliano:2006hb}
 \begin{eqnarray}\label{Relations_even}
  \lambda'&\equiv&\frac{1}{\sqrt{2}}\left(\lambda_2-\lambda_3+\frac{\lambda_4}{2}+\lambda_5\right)=\frac{F^2}{2\sqrt{2}F_AG_V}\,,\nonumber\\
  \lambda''&\equiv&\frac{1}{\sqrt{2}}\left(\lambda_2-\frac{\lambda_4}{2}-\lambda_5\right)=\frac{2G_V-F_V}{2\sqrt{2}F_A}\,,\nonumber\\
  \lambda_0&\equiv&-\frac{1}{\sqrt{2}}\left(4\lambda_1+\lambda_2+\frac{\lambda_4}{2}+\lambda_5\right)=\frac{\lambda'+\lambda''}{4}\,,\nonumber\\
  \kappa_{1}^{SA}&\equiv&\frac{F^2}{32\sqrt{2}c_mF_A},\quad
 \end{eqnarray}
supplemented by $ F_V G_V=F^2\,,\; F_A=\sqrt{2}F$ and $F_V=\sqrt{3}F$ (this one in accord with the result found in the odd-intrinsic parity sector)~\cite{Weinberg:1967kj, Ecker89-1, Peris}. Since $\lambda^V_{21}=0=\lambda^V_{22}$ 
\cite{Cirigliano:2006hb}, we will not consider the contribution of the corresponding operators. The well-known relation $c_d c_m\,=\,F^2/4$ \cite{JOP} arising in the study of strangeness-changing scalar form factors will 
also be employed.\\
  
  Although not all the operators appearing in section \ref{sec:theoRChT} do actually contribute to the considered decays, the number of asymptotic relations looks too small compared to the number of free couplings to allow a meaningful general 
  phenomenological study of the $\tau^-\to\pi^-\eta^{(\prime)}\gamma \nu_\tau$ decays within $R\chi L$. Also there is not enough phenomenological information on the couplings of eqs.~(\ref{operatorsV_KN}) and (\ref{operatorsA_KN}), for instance. 
  Due to that we will first consider only the diagrams with at most one resonance and then comment on the possible extension to include two-resonance diagrams in section \ref{sec:radbkgRChT}.\\
  
  \subsubsection{Form factors according to Resonance Chiral Lagrangians}\label{sec:FFsRChT}
\begin{figure}[ht!]\centering
  \includegraphics[scale=0.8]{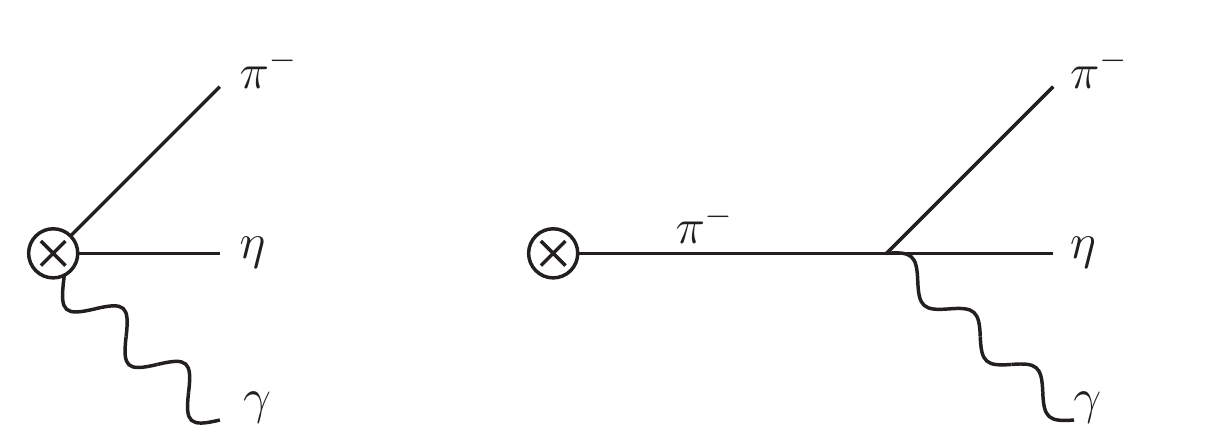}\caption{Contributions from the Wess-Zumino-Witten functional \cite{WZW} to $\tau^-\to\pi^-\eta\gamma\nu_\tau$ decays. The cross circle indicates the 
  insertion of the charged weak current.}\label{fig:ChPT}
 \end{figure}

\begin{figure}[ht!]
  \centering
  \includegraphics[scale=0.5]{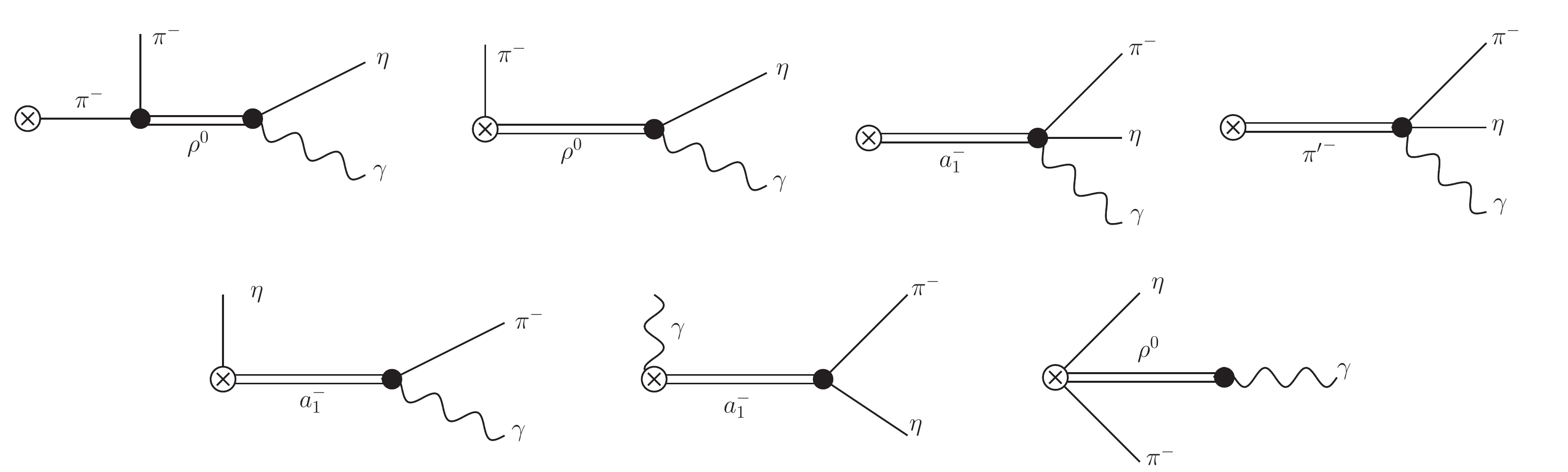}\caption{One-resonance exchange contributions from the $R\chi L$ to the axial-vector form factors of the $\tau^-\to\pi^-\eta\gamma\nu_\tau$ decays. Vertices involving 
  resonances are highlighted with a thick dot.}\label{fig:1R-FA}
 \end{figure}

\begin{figure}[ht!]
  \centering
  \includegraphics[scale=0.5]{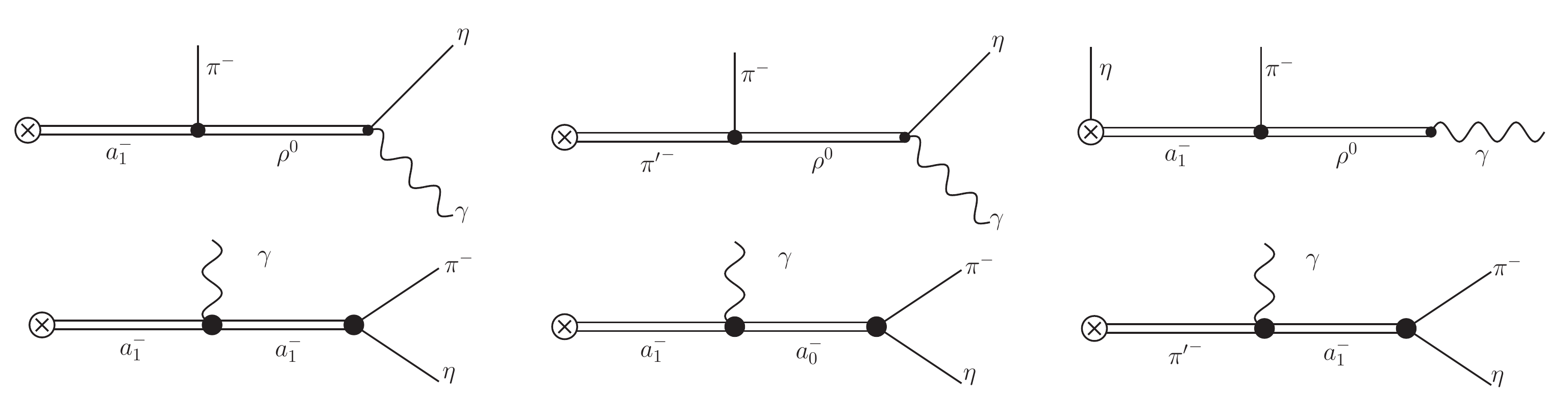}\caption{Two-resonance exchange contributions from the $R\chi L$ to the axial-vector form factors of the $\tau^-\to\pi^-\eta\gamma\nu_\tau$ decays. Vertices involving 
  resonances are highlighted with a thick dot.}\label{fig:2R-FA}
 \end{figure}

\begin{figure}[ht!]
  \centering
  \includegraphics[scale=0.45]{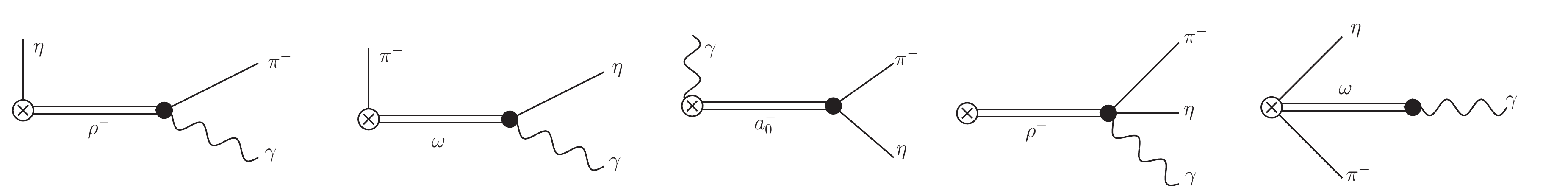}\caption{One-resonance exchange contributions from the $R\chi L$ to the vector form factors of the $\tau^-\to\pi^-\eta\gamma\nu_\tau$ decays. Vertices involving 
  resonances are highlighted with a thick dot.}\label{fig:1R-FV}
 \end{figure}

\begin{figure}[ht!]
  \centering
  \includegraphics[scale=0.35]{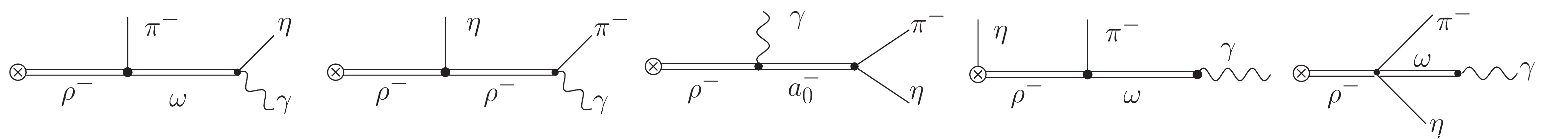}\caption{Two-resonance exchange contributions from the $R\chi L$ to the vector form factors of the $\tau^-\to\pi^-\eta\gamma\nu_\tau$ decays. Vertices involving 
  resonances are highlighted with a thick dot.}\label{fig:2R-FV}
 \end{figure}

  The relevant Feynman diagrams are shown in figures \ref{fig:ChPT} to \ref{fig:2R-FA}~\footnote{We remind that only diagrams which do not violate G-parity are considered.}. Fig.~\ref{fig:ChPT} corresponds to the model-independent 
  contribution given by the chiral $U(1)$ anomaly, fixed by QCD~\footnote{We note that this contribution is absent in the MDM approach.}. The left-hand side diagram is the purely local contribution while, in the one on the right, 
  the Wess-Zumino-Witten functional provides the $\pi\pi\eta\gamma$ vertex (and all hadronic information corresponding to the coupling of the pion to the axial-vector current is encoded in the pion decay constant). The anomalous 
  vertices violate intrinsic parity, as these two diagrams do. Figs.~\ref{fig:1R-FA} to \ref{fig:2R-FV} are, on the contrary, model-dependent. Figs.~\ref{fig:1R-FA} and \ref{fig:2R-FA} (\ref{fig:1R-FV} and \ref{fig:2R-FV}) correspond 
  to the one- and two-resonance mediated contributions to the axial-vector (vector) form factors in eqs.~(\ref{m.e.}) to (\ref{decomp}), respectively.\\

\vspace*{1.5cm}
  
   As a general fact, the axial-vector form factors in radiative tau decays to two pseudoscalars violate intrinsic parity as it can be checked for all contributing diagrams in figs.~\ref{fig:1R-FA} and \ref{fig:2R-FA}. The last vertex 
   in all diagrams in the first line of fig.~\ref{fig:1R-FA} is of odd-intrinsic parity (as well as it happens with the second diagram in the second line of this figure). In the first and third diagrams of the second line of 
   fig.~\ref{fig:1R-FA} intrinsic parity is violated in the coupling to the weak (thus axial-vector) current. The odd-intrinsic parity violating vertices appearing in the diagrams in fig.~\ref{fig:2R-FA} are $\rho^0\to\eta\gamma$, 
   $a^{-\mu}\to a_1^-\eta$ ($a^\mu$ stands for the axial-vector current), $a_1^-\to\pi^-\eta$ and $a_1^-\to a_0^-\gamma$.\\

\vspace*{1.5cm}

 \vspace*{1.5cm}
     
  We note that the first two diagrams of figs.~\ref{fig:1R-FV} contain only odd-intrinsic parity violating vertices while the last three diagrams in this figure contain only even-intrinsic 
  parity vertices in such a way that intrinsic-parity is not violated in neither of them (as it corresponds to the vector form factors). Similarly, in fig.~\ref{fig:2R-FV}, the first, second and fourth diagram contain two 
  intrinsic parity violating vertices and the third and fifth diagram contain only even-intrinsic parity vertices. Thus, again intrinsic parity is conserved in these diagrams as well.\\
     
\vspace*{1.5cm}

 \vspace*{1.5cm}

%
%

 Using the $R\chi L$ introduced in section \ref{sec:theoRChT}, it is straightforward to verify that all three diagrams involving the $\pi'$ resonance vanish (in figs.~\ref{fig:1R-FA} and \ref{fig:2R-FA}). Also the last 
 diagram of fig.~\ref{fig:1R-FV} is null but all other diagrams in figures \ref{fig:ChPT} to \ref{fig:2R-FV} contribute non trivially to the considered $\tau^-\to\pi^-\eta^{(\prime)}\gamma\nu_\tau$ decays. Since the left-handed 
 weak current has both vector and axial-vector components, one could expect to have two different contributions per given topology, with intrinsic parity conserving and violating coupling to the weak charged current, 
 respectively. However, we point out that using the Lagrangian introduced in section \ref{sec:theoRChT} this only happens for the last diagrams in figs.~\ref{fig:1R-FA} and \ref{fig:1R-FV}. In our computation we have 
 neglected subleading contributions in the chiral counting, namely the coupling to the weak current in the second diagram of fig.~\ref{fig:1R-FA} receives contributions from the piece of the  Lagrangian in eq.~(\ref{linear_interactions}). 
 Correspondingly, we are not considering the contributions given by the Lagrangian in eq.~(\ref{Lagrangian_V_Towards}), which are suppressed by one chiral order.\\
 
 Comparing the $R\chi L$ diagrams in figs.~\ref{fig:ChPT} to \ref{fig:2R-FV} with the MDM diagrams in fig.~\ref{fig:hadvertex}, we see first that the model-independent contribution of both diagrams in fig.~\ref{fig:ChPT} (axial-form factors at 
 lowest order in the chiral expansion) is not included in the MDM approach. Among the 13 contributions in figs.~\ref{fig:1R-FA} and ~\ref{fig:2R-FA} (which are subleading in the chiral regime) only one is considered in MDM 
 \footnote{The diagram with the pion pole also appears in fig.~\ref{fig:vertex}, but it is neglected.} (the first diagram in figure \ref{fig:2R-FA}). Finally, 10 diagrams appear in figs.~\ref{fig:1R-FV} and \ref{fig:2R-FV} but only 
 three of them (those including the vertices $\rho-\omega-\pi$, $\rho-a_0-\gamma$ and $\rho-\rho-\eta$) enter the MDM description.\\
 
 We would like to make a final comment regarding gauge invariance before quoting our form factor results using $R\chi L$. It can be checked that the contribution of $\mathcal{O}_{10}^A$ to the third diagram in fig.~\ref{fig:1R-FA} is not gauge 
invariant by itself. However, for this particular operator, the cancellation of gauge-dependent pieces involves the diagrams with radiation off the $a_1$ and off the weak vertex in figs.~\ref{fig:1R-FA} and \ref{fig:2R-FA}. As a result of this 
mechanism, we note the presence of $D_{a_1}(W^2)$ and $D_{a_1}[(p+k)^2]$ factors and the absence of $D_{a_1}[(p+p_0)^2]$ terms in the corresponding contributions to the axial-vector form factors~\footnote{We note that, among the $\mathcal{O}_i^A$ 
operators, only $\mathcal{O}_{10}^A$ couples to $\pi^-\eta^{(\prime)}$. This vertex does not contribute to the corresponding non-radiative decays because at least an additional independent momentum is needed for a non-vanishing contraction with 
the Levi-Civita symbol.}.\\

For convenience, we will quote the individual contributions to each form factor figure by figure (following the order of the diagrams in a given figure). We will start with the axial-vector form factors. The diagrams in fig.~\ref{fig:ChPT} give 
\begin{equation}
 a_1^{\chi PT}\,=\,\frac{N_C C_q}{6 \sqrt{2} \pi^2 F^2}\,,\quad a_3^{\chi PT}\,=\,\frac{a_1^{\chi PT}}{D_\pi\left[W^2\right]}\,,\label{FFsRChT_ChPT}
\end{equation}
which is a model-independent result coming from the QCD anomaly.\\

The contribution of the remaining diagrams (figures \ref{fig:1R-FA} and \ref{fig:2R-FA} for the axial-vector form factors and \ref{fig:1R-FV} and \ref{fig:2R-FV} for the vector form factors) is collected in appendix A. The corresponding off-shell 
width of meson resonances used in our numerical analysis can be found in appendix B. We will discuss in the next section if further insight can be gained on the $R\chi L$ couplings values restoring to phenomenology and using the expected scaling 
of the low-energy constants of the $\chi PT$ Lagrangian.\\
\subsubsection{Phenomenological estimation of $R\chi L$ couplings}
Although the relations in section \ref{sec:QCDconstraints} only reduce the number of unknowns in eqs.~(\ref{FFsRChT_ChPT}) and (\ref{a1-1R-RChL}) to (\ref{v4-2R-RChL}), some of the remaining free couplings can still be estimated phenomenologically. The 
high-energy constraint $c_dc_m=F^2/4$ leaves either $c_d$ or $c_m$ as independent. We will use $c_d = \left(19.8^{+2.0}_{-5.2}\right)$ MeV \cite{Guo:2012yt}. In this way all relevant couplings in eq.(\ref{linear_interactions}) have been determined.\\
$\lambda_{15}^S$ is the only leading operator contributing to $a_0\to\gamma\gamma$. From $\Gamma(a_0\to\gamma\gamma)=(0.30\pm0.10)$ keV$=\frac{64\pi\alpha^2}{9}M_{a_0}^3|\lambda_{15}^S|^2$ we can estimate $|\lambda_{15}^S|=(1.6\pm0.3)\cdot 10^{-2}$ 
GeV$^{-1}$. We note that the coupling relevant for the $a_1-a_0-\gamma$ vertex, $\kappa_1^{SA}$ is fixed by a short-distance constraint in eqs.~(\ref{Relations_even}).\\

We turn now to the $\lambda_i$ couplings in eq.~(\ref{VAP}). Short-distance constraints leave two such couplings undetermined. The three combinations of them that are predicted by high-energy conditions have the following numerical values:
\begin{equation}\label{values_lambdasVAP}
 \lambda^\prime\,\sim\,0.4\,,\quad \lambda^{\prime\prime}\,\sim\,0.04\,,\quad \lambda_0\,\sim\,0.12\,.
\end{equation}
The same linear combination of $\lambda_4$ and $\lambda_5$ enters all couplings in eq.~(\ref{values_lambdasVAP}). Therefore we can take one them as independent ($\lambda_4$ for us). We will choose as the other independent coupling $\lambda_2$, which 
enters all couplings in eq.~(\ref{values_lambdasVAP}). A conservative estimate would be $|\lambda_2|\sim|\lambda_4|\leq0.4$, to which we will stick in our numerical analysis.\\

According to ref.~\cite{Cirigliano:2006hb} the $\lambda_i^V$ couplings can be estimated from the expected scaling of the NNLO low-energy constants of the $\chi PT$ Lagrangian (we also employ short-distance QCD constraints on the $R\chi L$ couplings 
to write the following expression conveniently) as
\begin{equation}\label{estimate_lambda_i^V}
 \lambda_i^V\,\sim\,3C_i^R\frac{M_V^2}{F}\sim 0.05\; \mathrm{GeV}^{-1}\,,
\end{equation}
that can be considered an upper bound on $|\lambda_i^V|$ because the employed relation $C_i^R\sim\frac{1}{F^2(4\pi)^4}$ is linked to $L_i^R\sim\frac{1}{(4\pi)^2}\sim5\cdot10^{-3}$, which is basically the size of $L_9^R$ and $|L_{10}^R|$ but clearly 
larger than the remaining eight $L_i^R$ \cite{Ecker89-1, Bijnens:2011tb}. There is not that much information on the values of the $C_i^R$ (see, however ref.~\cite{Jiang:2009uf}). We will take $|\lambda_i^V|\leq 0.04$ GeV$^{-1}$ for the variation of these 
couplings ($i=6,11,12,13,14,15$ are relevant to our analysis), although it may be expected that only one or two of them (if any) are close to that (upper) limit. Proceeding similarly we can estimate $\lambda_i^{VV}\sim \frac{M_V^4}{2F^2}C_i^R$ and 
$\lambda_i^{SV}\sim\sqrt{2}\frac{M_S^2 M_V^2}{c_m F}C_i^R$. This sets a reasonable upper bound $|\lambda_i^{SV}|\sim|\lambda_i^{VV}|\lesssim 0.1$ that we will assume in the numerics.\\

We discuss next the values of the $c_i$ ($\kappa_i^V$) couplings in eqs.~(\ref{VJP&VVP}) and (\ref{V_KN}). Eqs.~(\ref{eq: Consistent set of relations}) predict the vanishing of two linear combinations of $c_i$'s. The numerical value for the predicted 
$c_6-c_5$ is $-0.017$. There are some determinations of $c_3$. It was estimated (although with a sign ambiguity) studying $\tau^-\to\eta\pi^-\pi ^0\nu_\tau$ decays \cite{Dumm:2012vb}. Taking into account the determinations by Y.~H.~Chen \textit{et. al.} 
\cite{Chen:2012vw,Chen:2013nna,Chen:2014yta} as well, we will use $c_3=0.007^{+0.020}_{-0.012}$. $c_4$ was first determined studying $\sigma(e^+e^-\to KK\pi)$ in ref.~\cite{Dumm:2009va}, although with a value yielding inconsistent results in 
$\tau^-\to K^-\gamma\nu_\tau$ \cite{Guo:2010dv}. We will take the determination $c_4=-0.0024\pm0.0006$ \cite{Chen:2013nna} as the most reliable one. Two other independent $c_i$ combinations appear in our form factors. We will take them as $c_5$ and 
$c_7$ whose modulus we will vary in the range $[0,0.03]$. Using eqs.~(\ref{eq: relation different basis}) to relate the $c_i$ and $\kappa_i^V$ couplings we can find reasonable guesses on the latter from $|c_i|\lesssim 0.03$. Thus, we will take 
$|\kappa_i^V|\leq 0.04$ GeV$^{-1}$ for their variation.\\

There is very little information on the $\kappa_i^A$ couplings. As a reasonable estimate we will make them vary in the same interval as the $\lambda_i^V$ and $\kappa_i^V$ couplings.\\

The numerical values of the two $d_i$ couplings (VVP operators) which were determined in eq.~(\ref{eq: Consistent set of relations}) are $d_1+8d_2\sim0.15$ and $d_3\sim-0.11$. $d_2$ has been determined jointly with $c_3$ (discussed above). According 
to the quoted references we will employ $d_2=0.08\pm0.08$. Then only $d_4$ would remain free. Given the previous values for the other $d_i$'s we will assume $|d_4|<0.15$.\\

  We will discuss in the next section the phenomenology of $\tau^-\to\pi^-\eta^{(\prime)}\gamma\nu_\tau$ decays, focusing on the background they constitute to the searches 
 of SCC in their corresponding non-radiative decays. We will start discussing the simplified case of $MDM$, according to eqs.(\ref{FFsMDM}), to turn next to the $R\chi L$ prediction corresponding to eqs.~(\ref{FFsRChT_ChPT}) and (\ref{a1-1R-RChL}) to 
 (\ref{v4-2R-RChL}).\\

 \section{$\tau^-\to\pi^-\eta^{(\prime)}\gamma\nu_\tau$ as background in the searches for $\tau^-\to\pi^-\eta^{(\prime)}\nu_\tau$}\label{sec:radbkg}

\subsection{Meson dominance predictions}\label{sec:radbkgMDM}

 The SM values for the non radiative process $\tau^-\to\eta^{(\prime)}\pi^-\nu_\tau$ mark the threshold below which no SCC stemming from BSM interactions will be detected 
 in these decays, therefore, one needs to provide a clean scenario for the experimental study of this process. Since the radiative decay can be a considerable background 
 in measuring the non radiative one, the former must be determined in such a way that its effect can be discerned from the latter. The way to do this is by imposing an 
 energy cut on the photon above which we discard all the contributions to the observables.\\
 
 One has to choose the cut in order to fairly reduce the background, but leaving a photon energy range that can be explored in the experiment. As explained in ref. 
 \cite{We:2016}, 50 MeV can be a too stringent cut for the process and, therefore, we take 100 MeV as the upper limit on photon energy. Notice that since no bremsstrahlung of on-shell particles will enter the 
 structure dependent description (see section \ref{sec:General}), our computation of the relevant observables is free of infrared divergences, however to be consistent 
 with the disregarding of the bremsstrahlung contribution a lower energy bound should be taken experimentally. This gives us the photon energy range $10\text{ MeV}\le E_\gamma\le100
 \text{ MeV}$. To verify this assertion we must analyze the photon energy spectrum. The statistical uncertainties in both models are given assuming a Gaussian distribution of the parameters 
 and letting them to randomly vary within such distribution. \\
 
 \begin{figure}[!ht]
  \centering
  \includegraphics[scale=0.3,angle=-90]{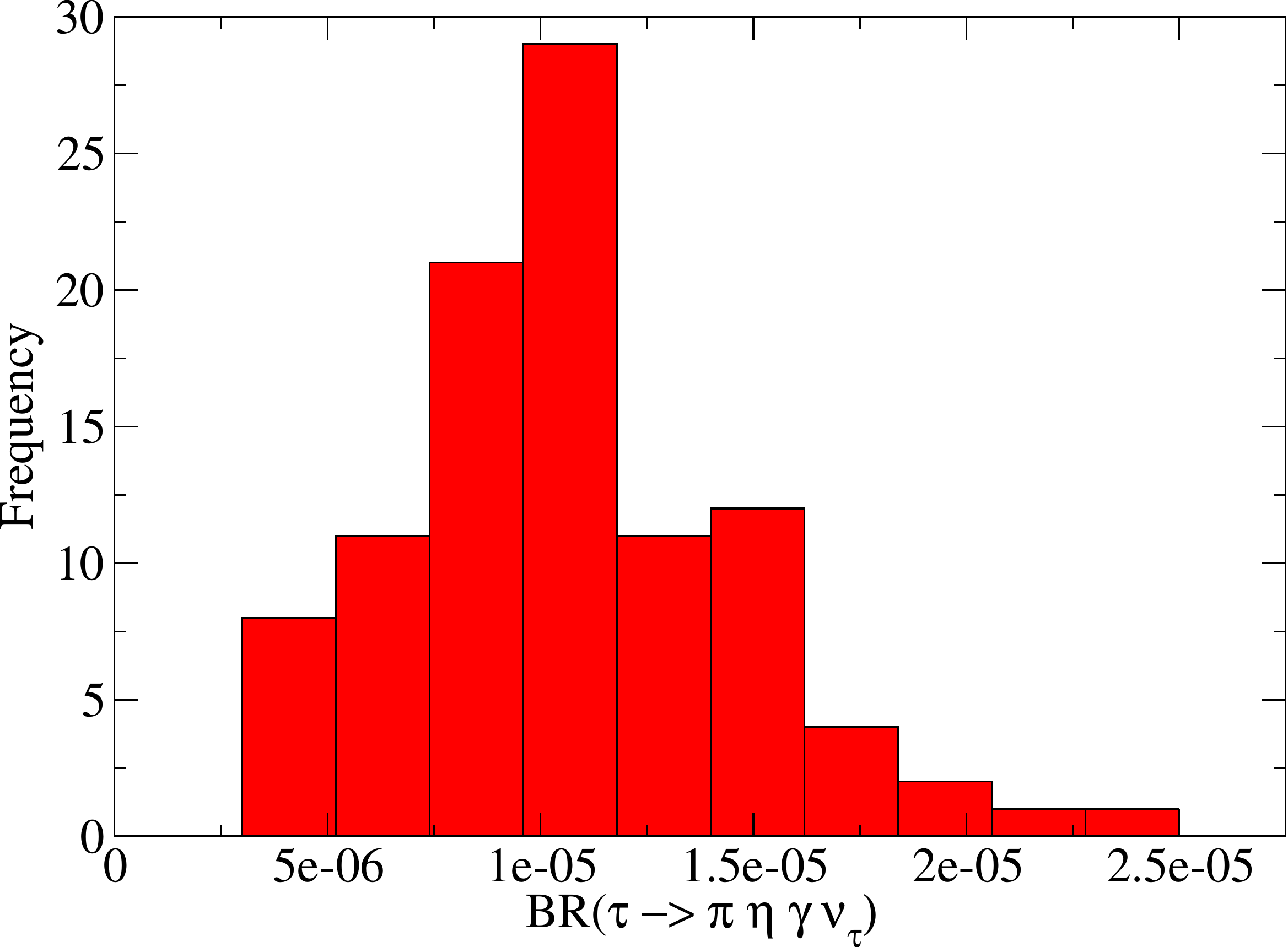}\includegraphics[scale=0.3,angle=-90]{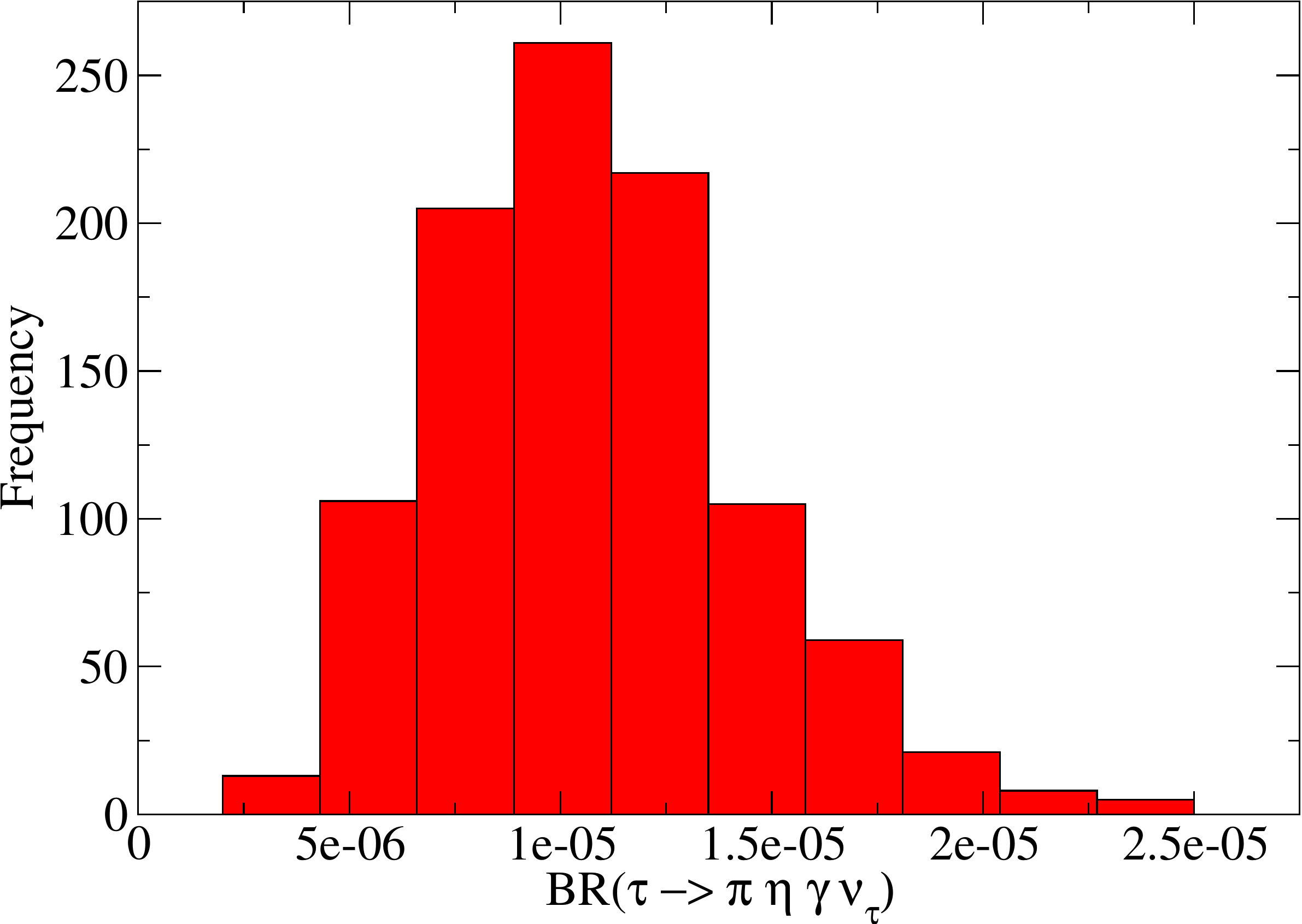}\caption{Histogram of $BR(\tau^-\to\pi^-\eta\gamma\nu_\tau )$ for 
  $100$ (left) and 1000 (right) random points in the MDM parameter space are plotted.}\label{fig:VMD100/1000}
 \end{figure}

 The purpose of computing the observables in the MDM is to have an estimation of the magnitude of the process and to compare it with the R$\chi$T, therefore no 
 errors are including due to model uncertainties. We use the values of the couplings in table \ref{ParsMDM}, where the error in the prediction of MDM will come 
 by their variation, as told in the previous paragraph.  We will first plot the predicted branching ratios when sampling these 10 parameters within one-sigma 
 uncertainties (using normal distributions). This information is collected in figures \ref{fig:VMD100/1000}, where the branching ratio for the variation of these 
 10 parameters is shown using 100 points (left) and 1000 (right) in the parameter space scan. \\
 
 \begin{figure}[!ht]
  \centering
  \includegraphics[scale=0.3,angle=-90]{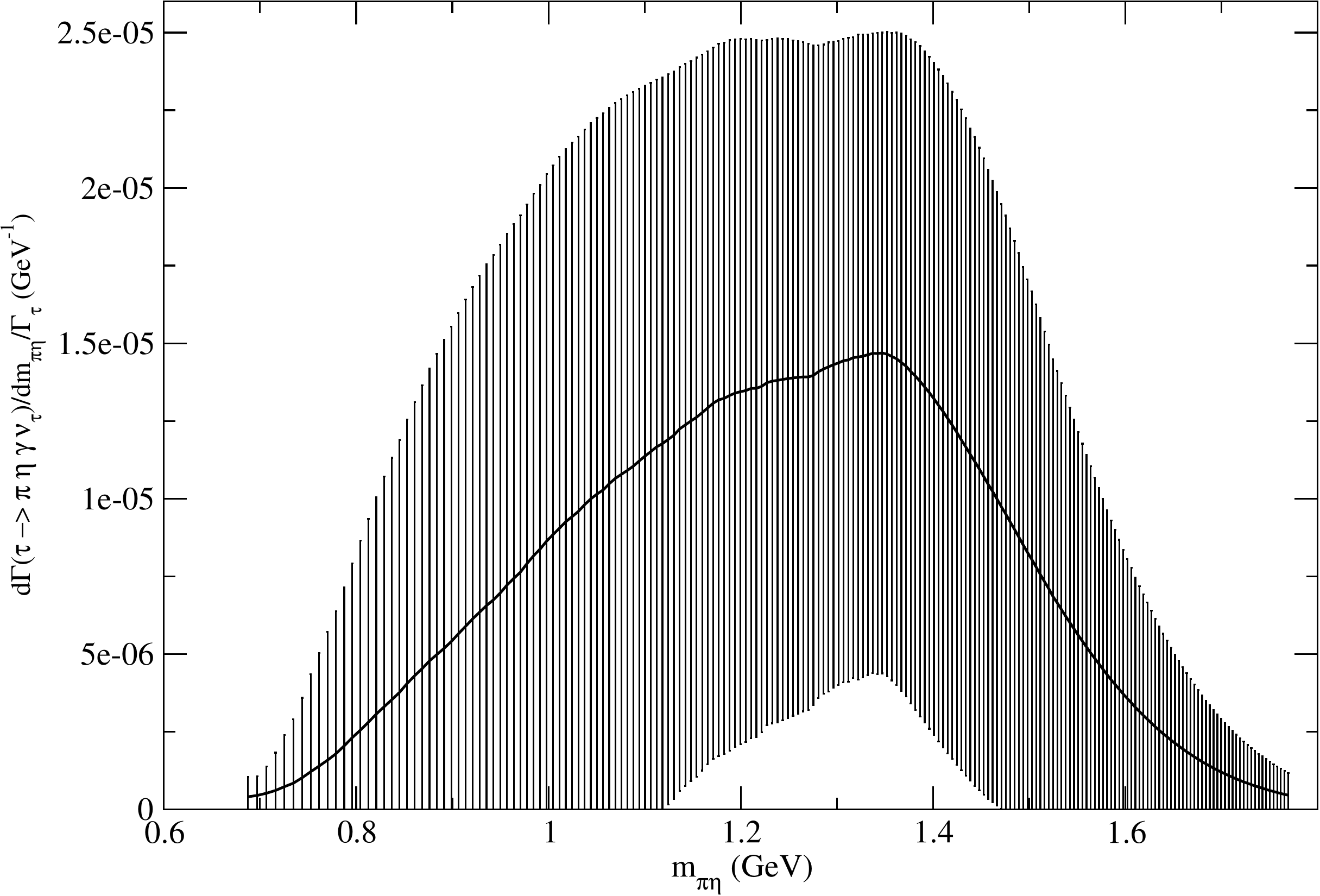}\includegraphics[scale=0.3,angle=-90]{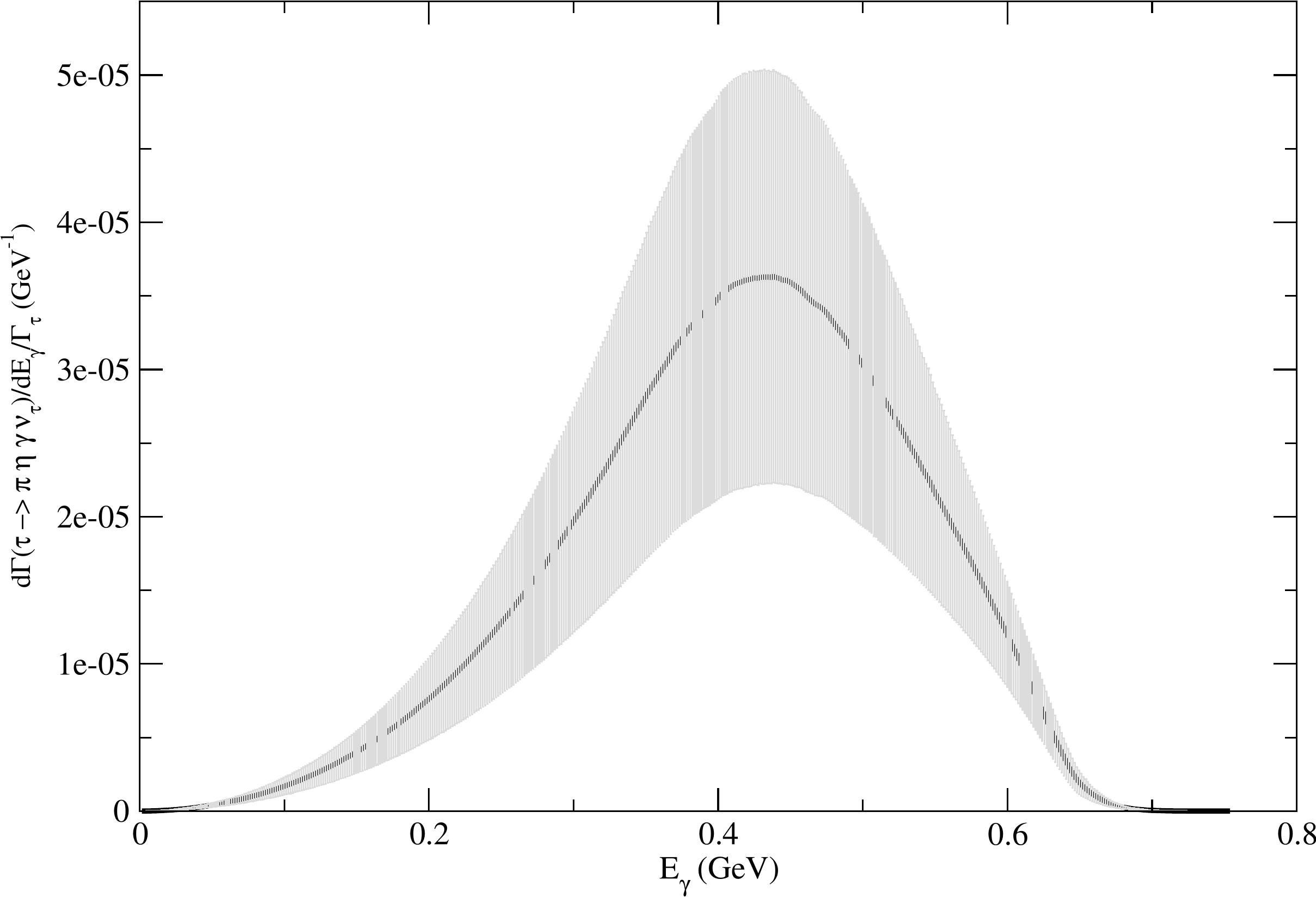}\caption{$\tau^-\to\pi^-\eta\gamma\nu_\tau$ 
  normalized spectra according to MDM in the invariant mass of the $\eta\pi^-$ system (left) and in the photon energy (right) are plotted for
  some characteristic points in fig.~\ref{fig:VMD100/1000}}\label{fig:VMDSpectra_eta}
 \end{figure}

 We find the Branching ratios $BR_{100}=(1\pm1)\times10^{-5}$ for a hundred points and $BR_{1000}=(1.1\pm0.3)\times10^{-5}$ using a thousand points. Then, by using 
 the same phase space integrals given in ref \cite{AlainGabriel} and taking 100 random sets of points in parameter space we computed the $\pi\eta$ invariant 
 mass $\invm$ spectrum dividing the domain length in 200 steps (left) and the photon energy dividing its domain length into 500 steps $E_\gamma$ spectrum (right), 
 both spectra normalized to the $\tau$ lepton lifetime are shown in figure \ref{fig:VMDSpectra_eta}. \\
 
 \begin{figure}[!ht]
  \centering\includegraphics[scale=0.35,angle=-90]{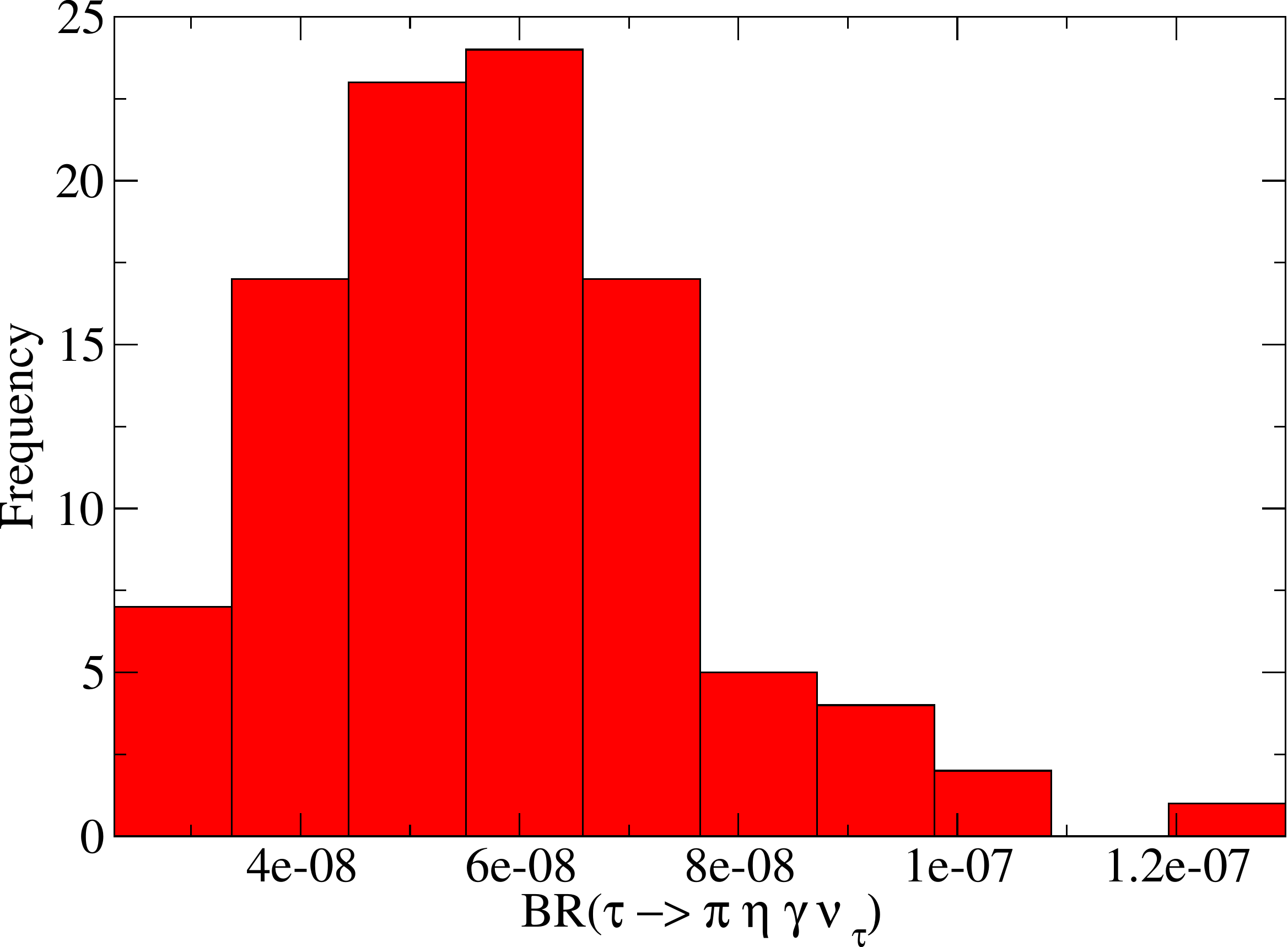}\caption{Histogram of $BR(\deceta)$ where photons with $E_\gamma>100$ MeV are rejected.}\label{fig:VMDcuts}
 \end{figure}
 
 It can be noticed the peak in the $1.15\GeV\le \invm\le1.35\GeV$ region for the $\invm$ spectrum, however there is not any marked dynamics responsible for 
 this effect. As was discussed previously, the effect of bremsstrahlung will become greater at lower energies, while by Low's theorem \cite{Low:1958sn} 
 the structure dependent amplitude will give greater contributions at large photon energies stemming from its dependence on $k$ ($\sim\mathcal{O}(k)$). 
 Thus, the photon energy spectrum gives the possibility to analyze the effect of the lowest upper bound imposed for the energy of the photon.
 Since, as already discussed (ref \cite{We:2016}), a 50 MeV cut is too restrictive we will take the 100 MeV cut. Using this cut, we reevaluated the 
 branching fraction with 100 parameter space points obtaining the plot in fig. \ref{fig:VMDcuts}. Thus, the upper bound obtained for the branching 
 ratio with a larger simulation sample (not shown in ) is $BR\le0.6\times10^{-7}$, two orders of magnitude smaller than the non radiative decay 
 \cite{Escribano:2016ntp}. \\
 
 \begin{figure}[!ht]
  \centering
    \includegraphics[scale=0.3,angle=-90]{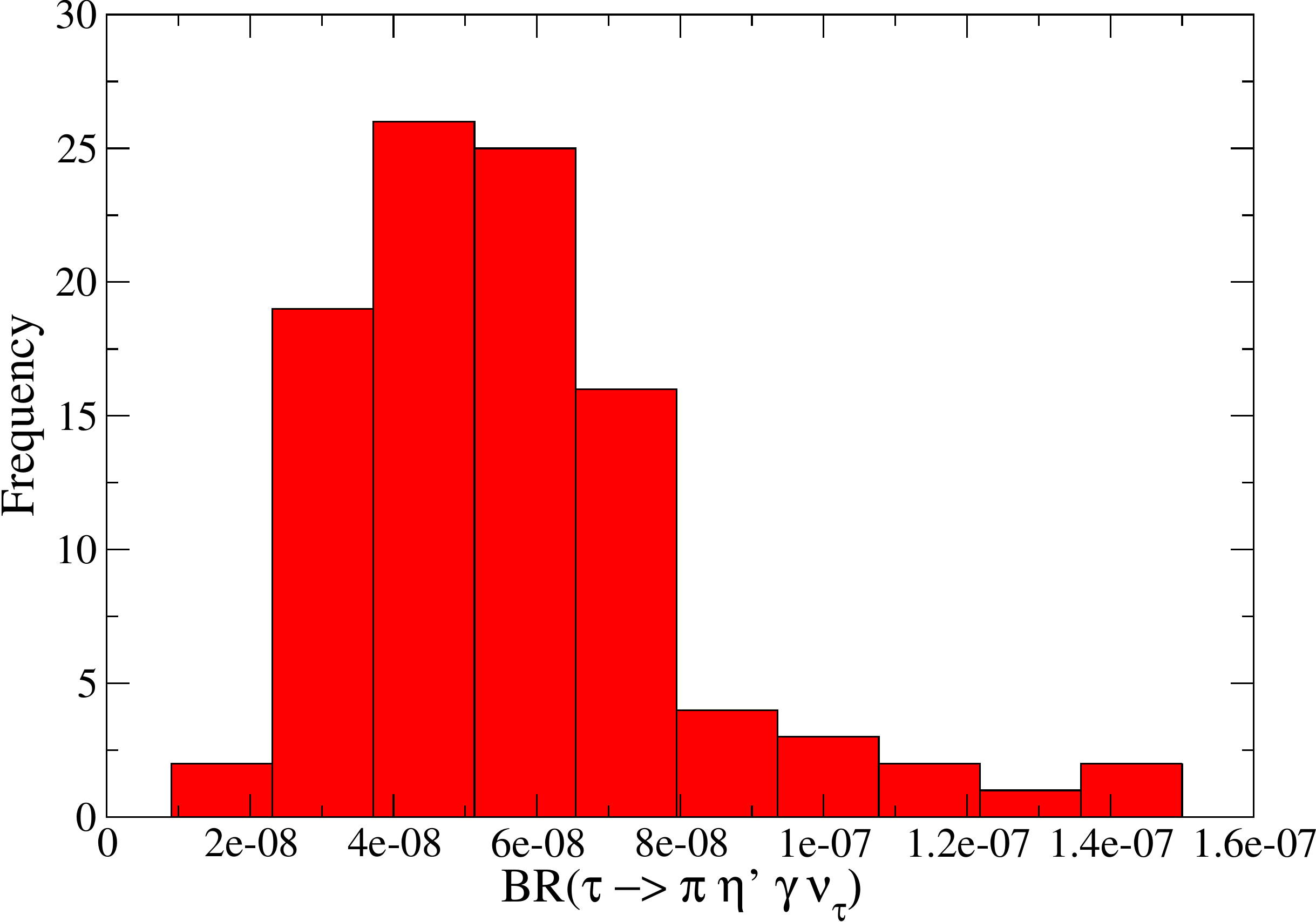}\includegraphics[scale=0.3,angle=-90]{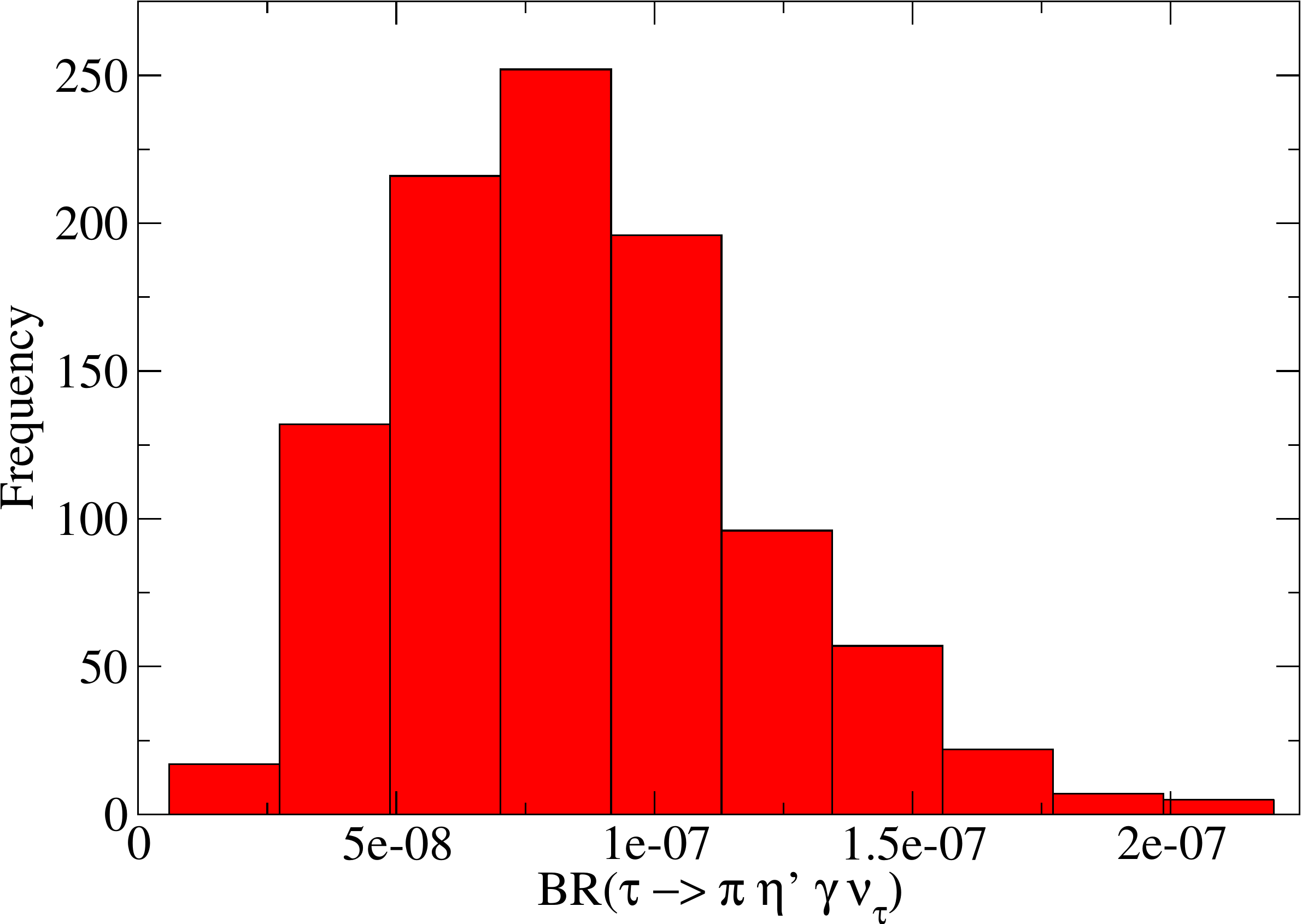}\caption{Histogram of 
    $BR(\tau^-\to\pi^-\eta'\gamma\nu_\tau)$ for $100$ (left) and 1000 (right) random points in the MDM parameter space are plotted.}\label{fig:VMD100/1000p}
 \end{figure}

\begin{figure}[!ht]
  \centering
    \includegraphics[scale=0.3,angle=-90]{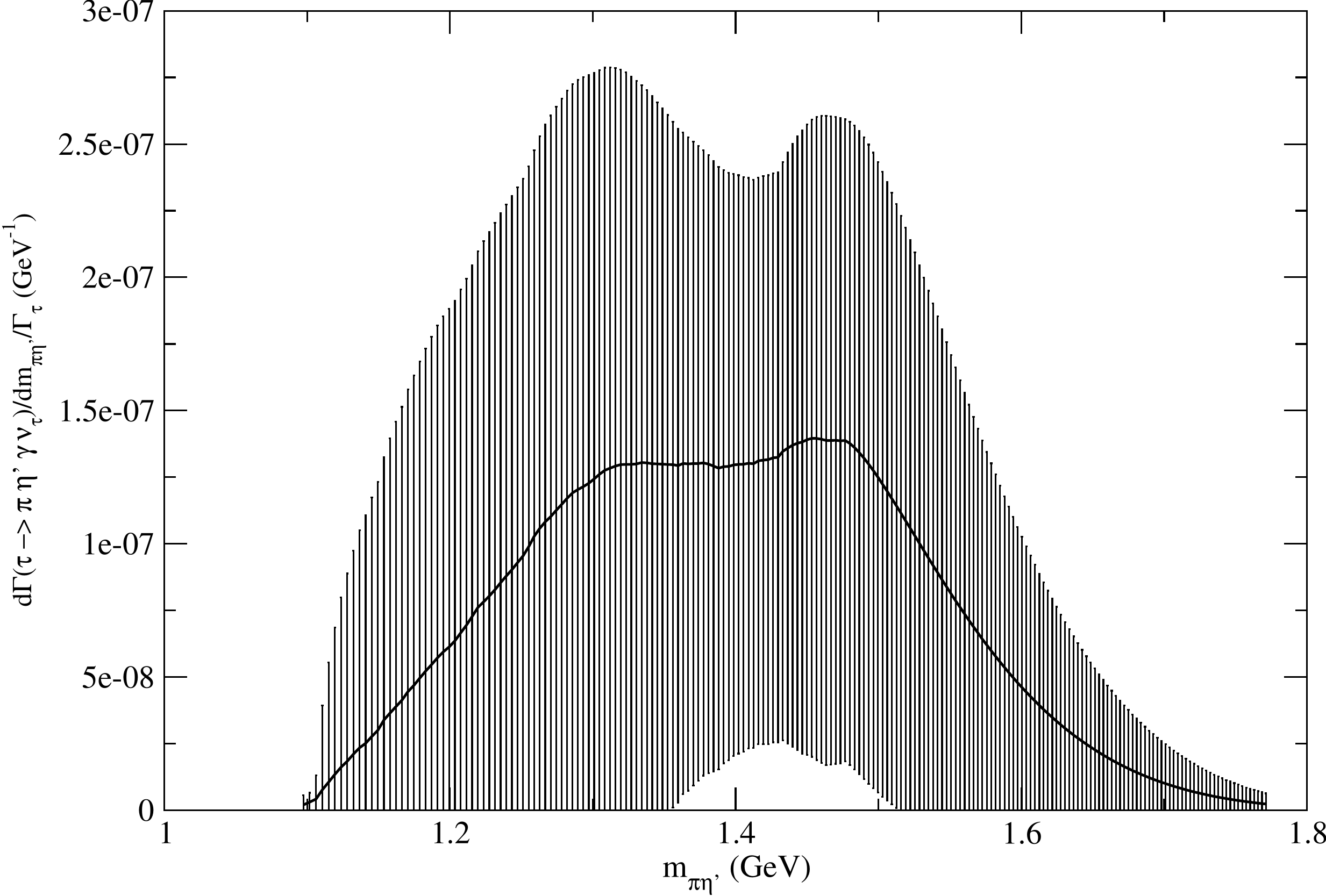}\includegraphics[scale=0.3,angle=-90]{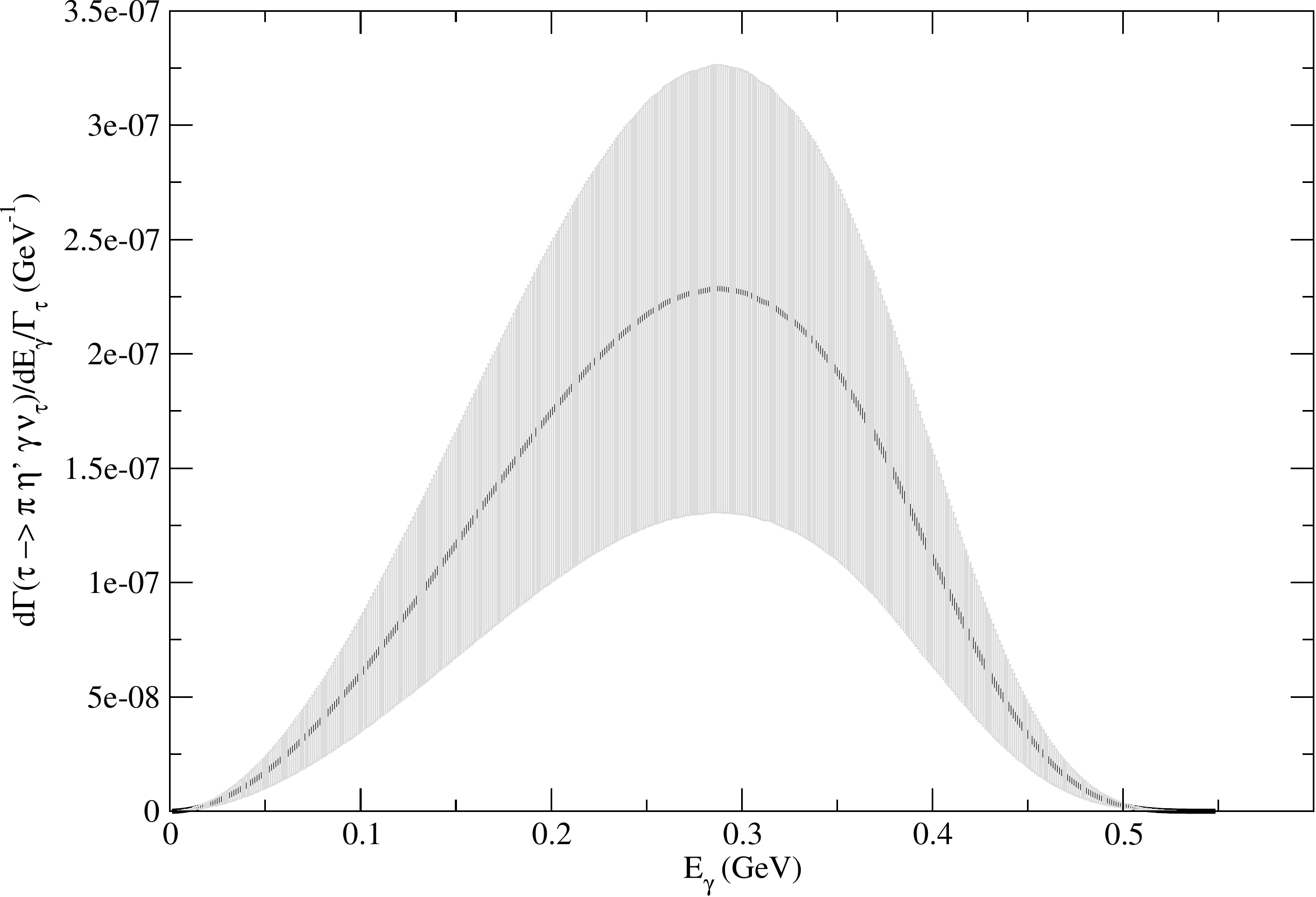}\caption{$\tau^-\to\pi^-\eta'\gamma\nu_\tau$ 
  normalized spectra according to MDM in the invariant mass of the $\pi^-\eta'$ system (left) and in the photon energy (right) are plotted for
  some characteristic points in fig.~\ref{fig:VMD100/1000p}}\label{fig:VMDSpectra_etap}
 \end{figure}

 For the $\eta'$ channel the procedure is completely analogous. We first plot the branching ratio for 100 (1000) normally sampled points in the 
 parameter space in fig \ref{fig:VMD100/1000p}, where the corresponding branching fractions are $BR_{100}\sim6\times10^{-8}$ and 
 $BR_{1000}=(0.8\pm0.8)\times10^{-8}$. Just as in the case of the $\eta$, we obtained the $\invmp$ spectrum shown in figure \ref{fig:VMDSpectra_etap} 
 dividing the domain length in 200 steps (left) and the photon energy dividing its domain length into 500 steps $E_\gamma$ spectrum (right) 
 both for a 100 random points in parameter space. Since the limited phase space does not allow an on-shell $a_0$ meson exchange, no possible 
 related substructure arises.\\
 
\begin{figure}[!ht]
  \centering\includegraphics[scale=0.32,angle=-90]{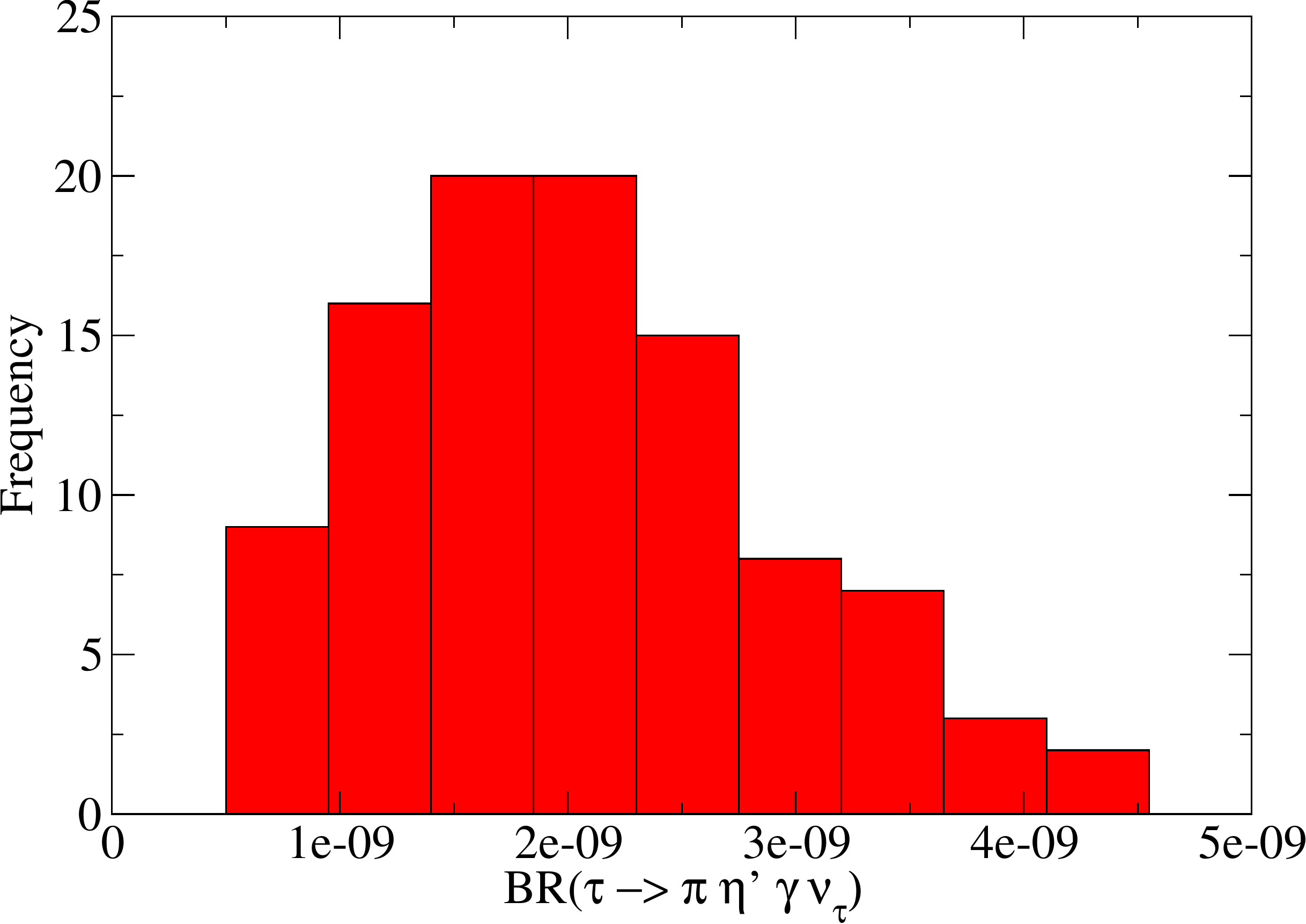}\caption{Histogram of $BR(\decetap)$ where photons with $E_\gamma>100$ MeV are rejected.}\label{fig:VMDcutsp}
 \end{figure}

 Completely analogous to the $\eta$ decay, the branching fraction is reanalyzed imposing the cut in photon energy of $100\MeV$ and shown in fig \ref{fig:VMDcutsp} using 
 100 points of parameter space. The branching fraction obtained thus is $\le0.2\times10^{-8}$, which was obtained by using a thousand parameter space 
 points. This is suppressed by a factor of 50 with respect to the non radiative decay.  It must be noticed that the MDM contribution to this process is 
 mainly given by the last diagram in figure \ref{fig:hadvertex}, when all the other contributions are neglected we find that $\sim80\%$ of the process 
 is given by this contribution in the $\eta$ channel, while for the $\eta'$ it is essentially saturated by it \cite{We:2016}. All the 
 results will be compared to R$\chi$T predictions in the next section.
 
 \subsection{R$\chi$L predictions}\label{sec:radbkgRChT}

 Contrary to the MDM case which has only contributions from diagrams involving two resonances, in R$\chi$T we have point interaction and one Goldstone exchange contributions 
 arising from the WZW functional along with diagrams including one and two resonances. All the observables obtained with MDM are computed using R$\chi$T and 
 then compared using all contributions to those without the two resonances exchange diagrams. \\
 
 \begin{figure}[!ht]
  \centering
  \includegraphics[scale=0.3,angle=-90]{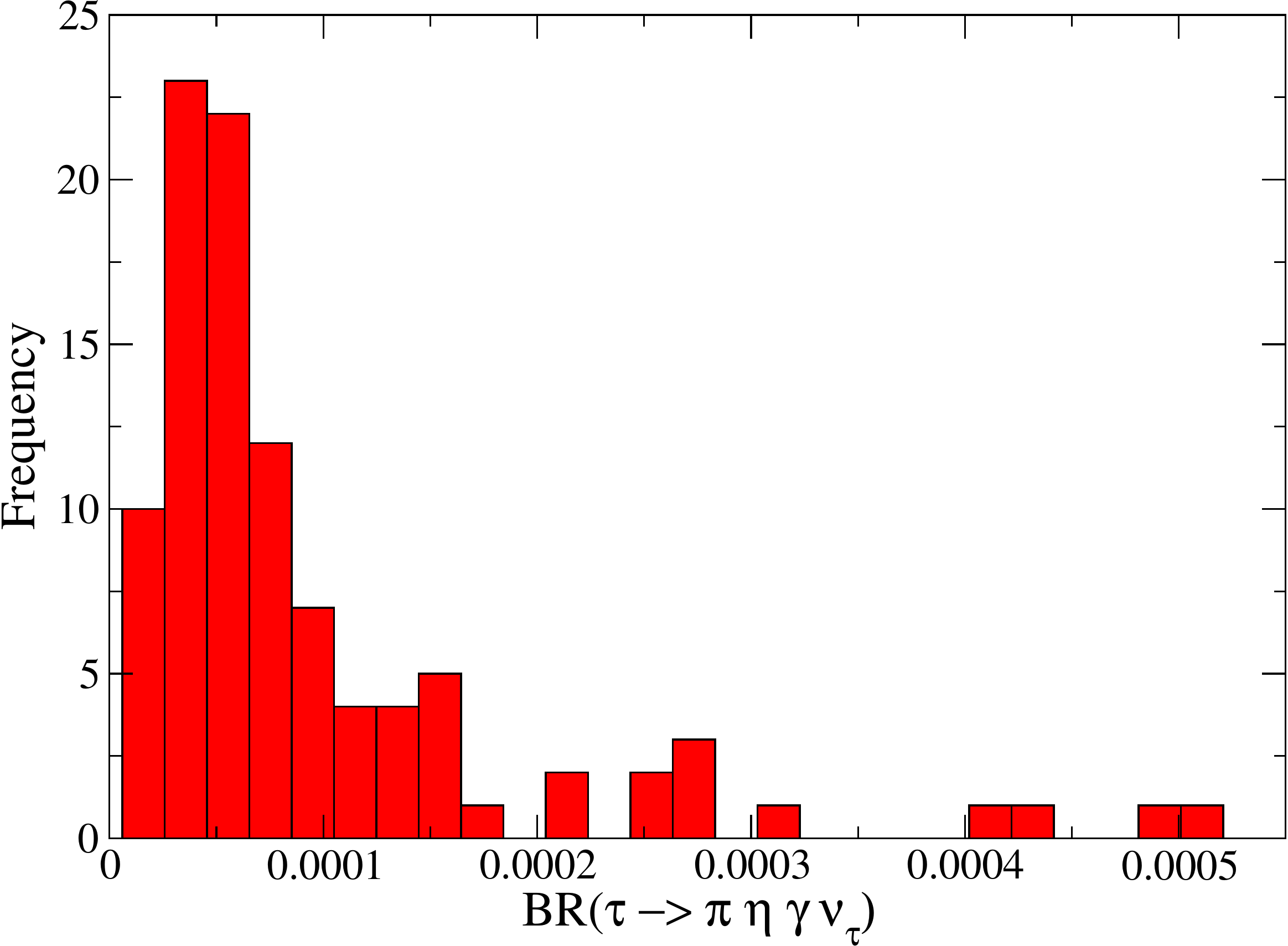}\includegraphics[scale=0.3,angle=-90]{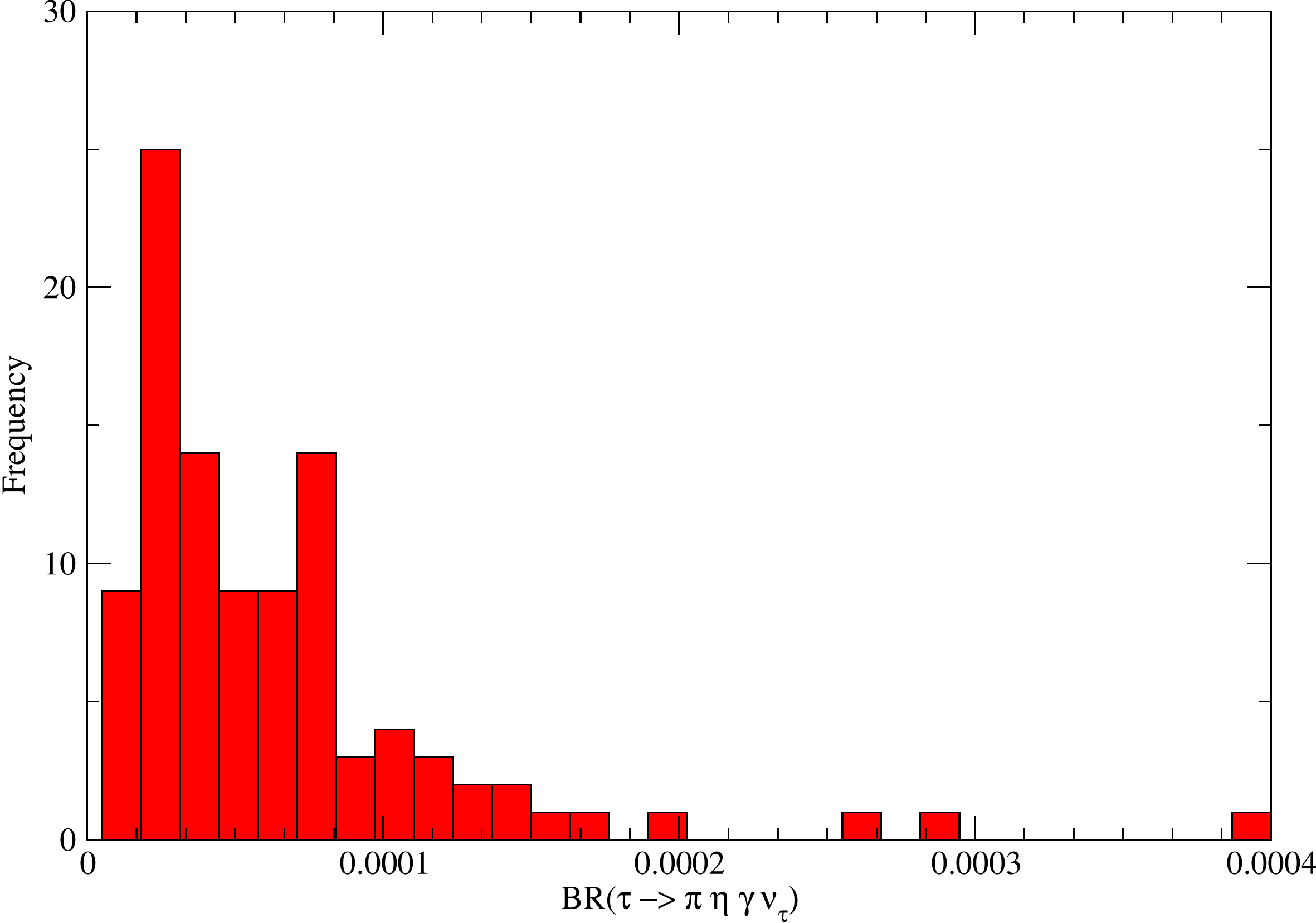}\caption{Histogram of $BR(\tau^-\to\pi^-\eta\gamma\nu_\tau )$ 
  with a sample of $100$ R$\chi$T parameter space points for the complete (left) and neglecting 2R diagrams (right) branching fractions.}\label{fig:RChT100/1000}
 \end{figure}

 As was stated above in subsection \ref{sec:radbkgMDM}, a Gaussian distribution of the parameters is assumed. With this, a 
 sample of 100 points in parameter space is studied, which throws a result of $BR_{100}=(1.0\pm0.2)\times10^{-4}$, where the error is statistical. This 
 error can be largely reduced by sampling a larger region of the parameter space, however the error will now be saturated by systematic theoretical error. 
 Thus we find a $BR_{1000}=(0.98\pm0.15)\times10^{-4}$, where the error is statistical. To give a more reliable result one has to consider theoretical uncertainties from the model, where one 
 can estimate the error by assigning a $1/N_C$ uncertainty to the amplitude (which is leading order in $1/N_C$), which gives a $1/N_C^2$ error that 
 becomes comparable to the statistical error. A conservative estimation of the theoretical uncertainty, which accounts an uncertainty twice larger,
 in this branching fraction is $\sim0.22\times10^{-4}$. So, by adding in quadratures the statistical and systematic errors we find $BR(\deceta)=(0.98\pm0.27)
 \times10^{-4}$ including all contributions. One should be careful when comparing the results from MDM and R$\chi$T, since they are different by an order 
 of magnitude. This will be contrasted to the result using a less general statistical error analysis, discussed in section \ref{sec:statistics}, 
 where the results agree with the VMD ones. 
 The result for 1000 points is plotted in fig \ref{fig:RChT100/1000} (left), 
 along with the contribution neglecting two resonance (2R) diagrams (right). These last one gives a reduced branching ratio $BR(1R)=(0.65\pm0.17)\times10^{-4}$.\\
  
 \begin{figure}[!ht]
  \centering
    \includegraphics[scale=0.3,angle=-90]{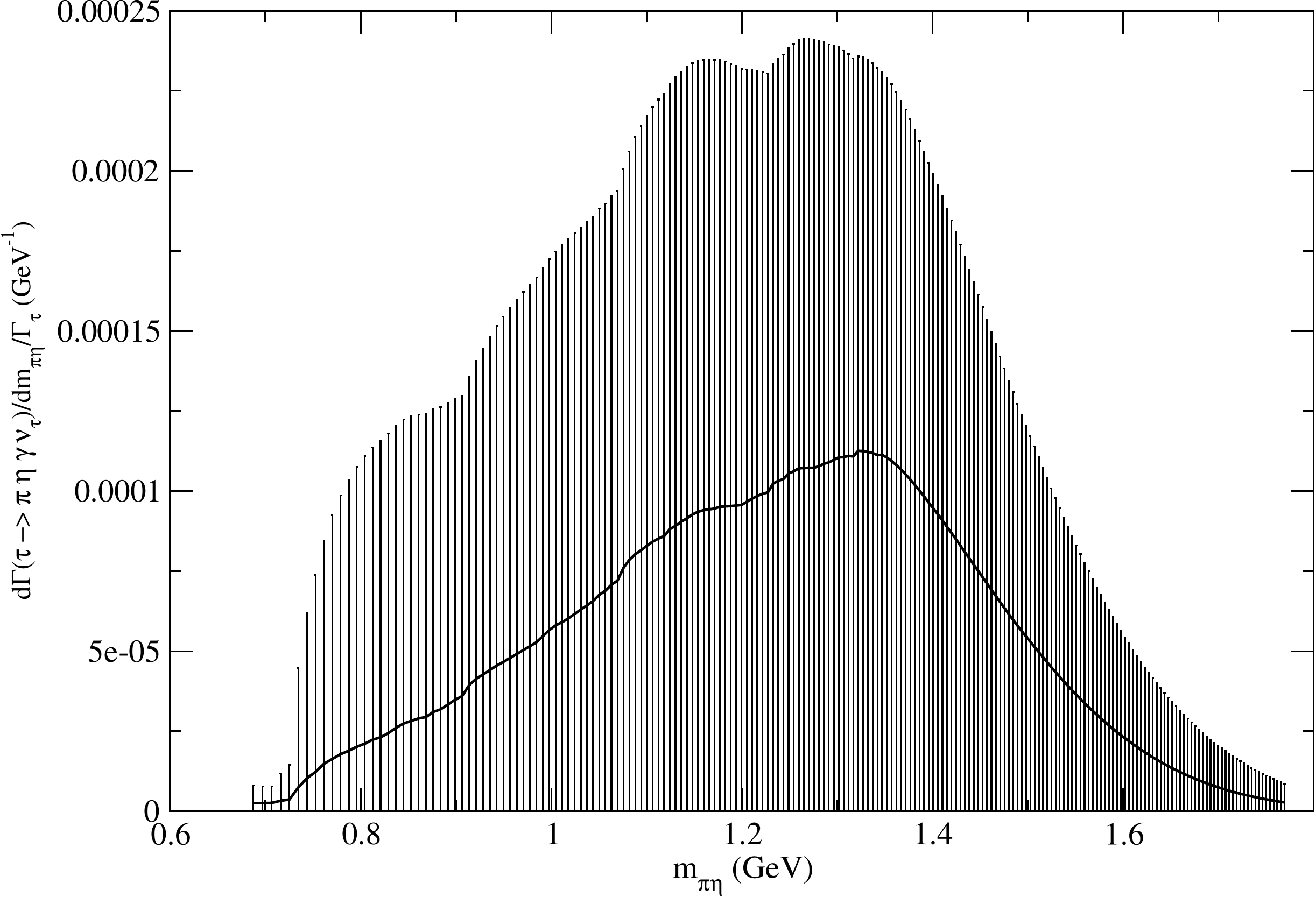}\includegraphics[scale=0.3,angle=-90]{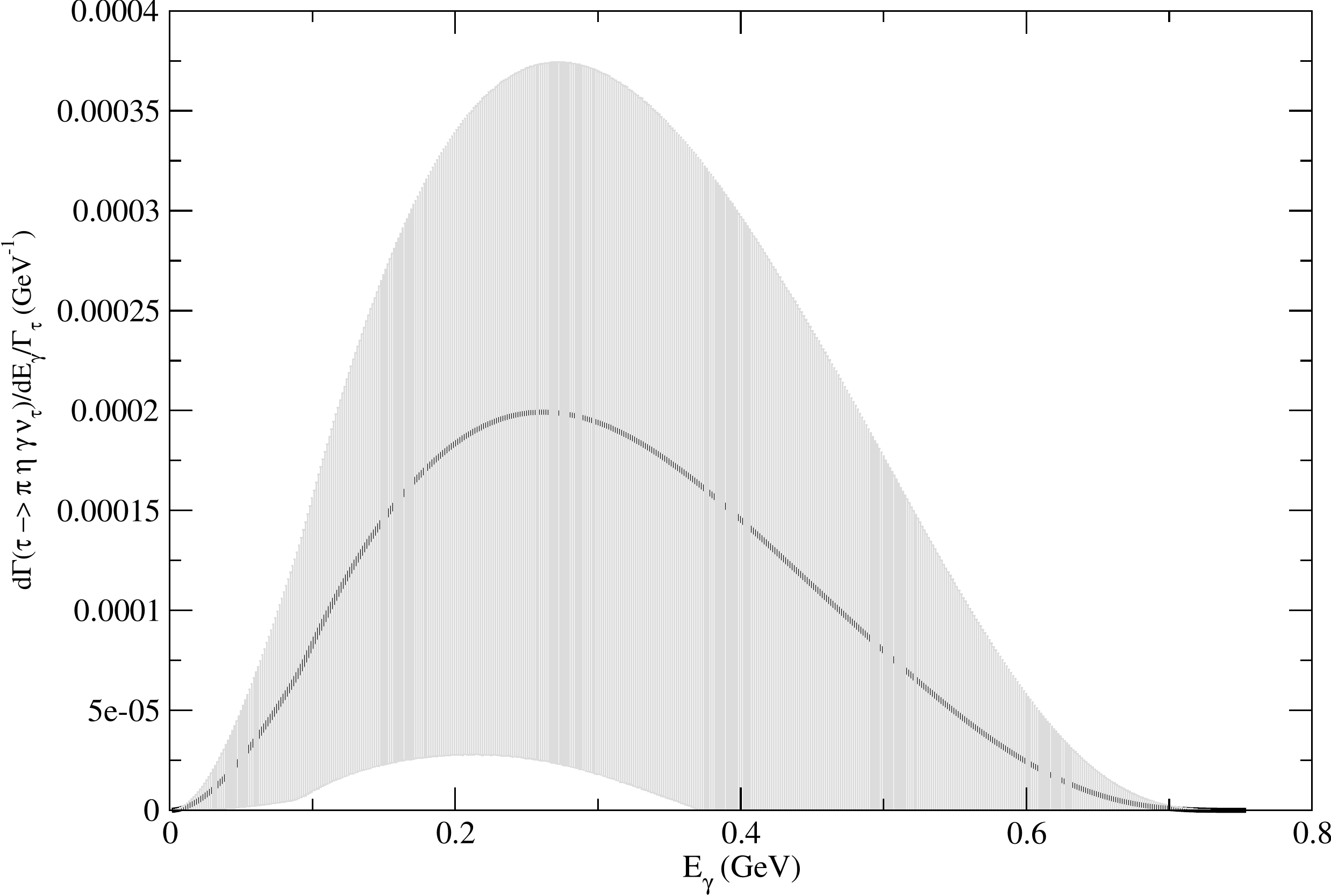}\caption{$\tau^-\to\pi^-\eta\gamma\nu_\tau$ 
  normalized spectra according to R$\chi$T in the invariant mass of the $\pi^-\eta$ system (left) and in the photon energy (right) are plotted.}\label{fig:RChTSpectra_eta}
 \end{figure}

 In fig \ref{fig:RChTSpectra_eta} we plot the normalized spectra in $\invm$ with 200 steps and $E_\gamma$ with 500 steps. Further analysis of the spectra 
 dependence on the statistics will be given in section \ref{sec:statistics}. Despite the dependence of the total decay width with respect to the statistics 
 of the error, an agreement is found with MDM respecting the (1.15,1.35) GeV $\invm$ region, this is, the enhancement in this region is also reproduced. 
 Also, it seems that the R$\chi$T for the $E_\gamma$ spectrum confirms our guess that a cut $\sim100\MeV$ will give a strong enough suppression. These features 
 do not seem to affect the analysis neglecting 2R diagrams. Further agreement with MDM is seen in the missing of any $a_0$ meson sign in the $\invm$ spectrum. \\
 
 \begin{figure}[!ht]
  \centering\includegraphics[scale=0.34,angle=-90]{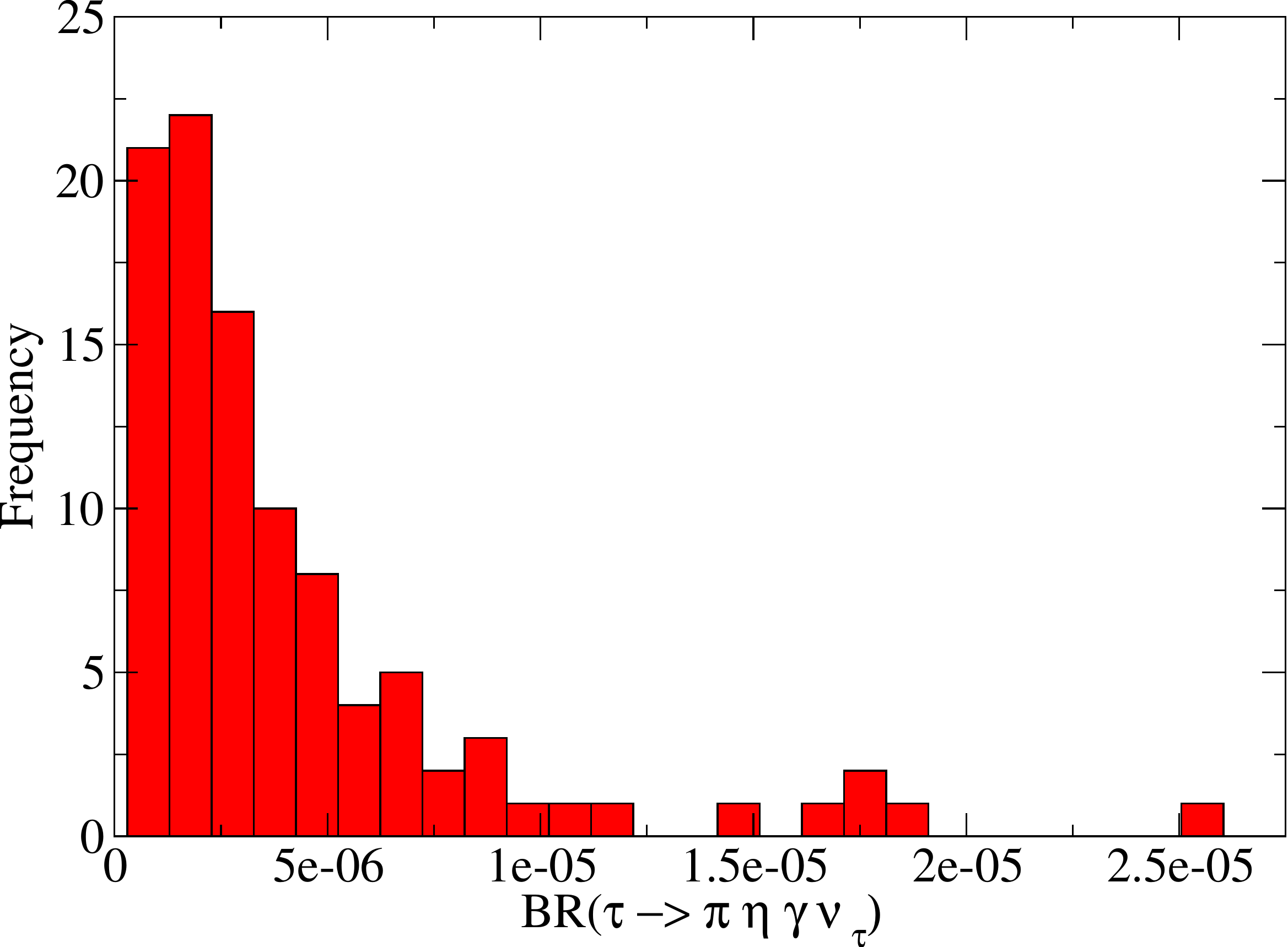}\caption{Histogram of $BR(\deceta)$ in R$\chi$T where photons with $E_\gamma>100$ MeV are rejected.}\label{fig:RChTcuts}
 \end{figure}

 In fig. \ref{fig:RChTcuts} we present the branching fraction for a cut in the photon energy of 100 MeV using 100 parameter space points. This yields a branching 
 ratio $(0.44\pm0.06)\times10^{-6}$. By neglecting photons with energies above 100 MeV the ratio of background events to non radiative decay event should be reduced 
 to 1/4. By neglecting 2R contributions the branching ratio changes to $(0.30\pm0.04)\times10^{-6}$, which lead us to the same conclusion.\\
  
 \begin{figure}[!ht]
  \centering
  \includegraphics[scale=0.3,angle=-90]{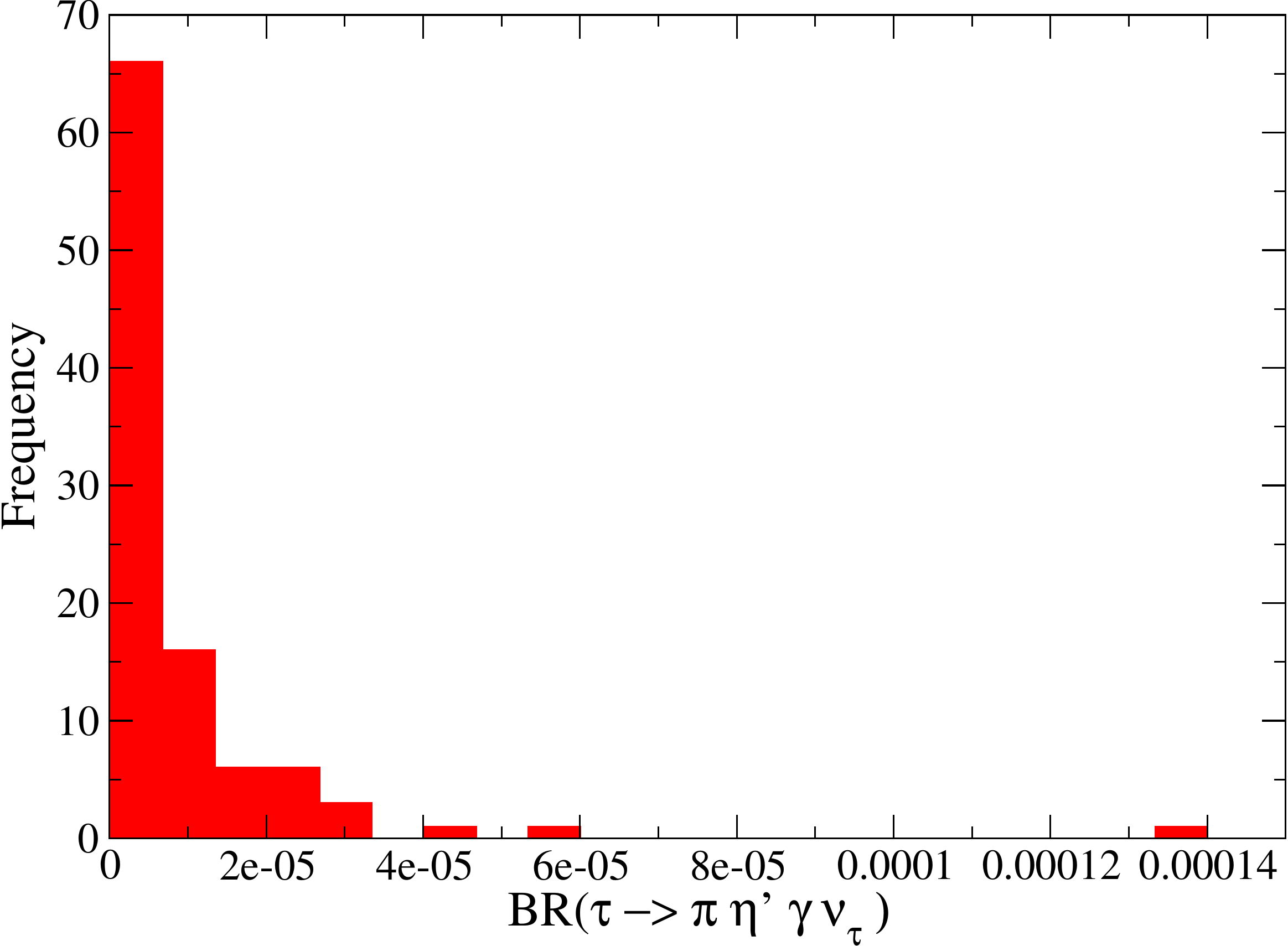}\includegraphics[scale=0.3,angle=-90]{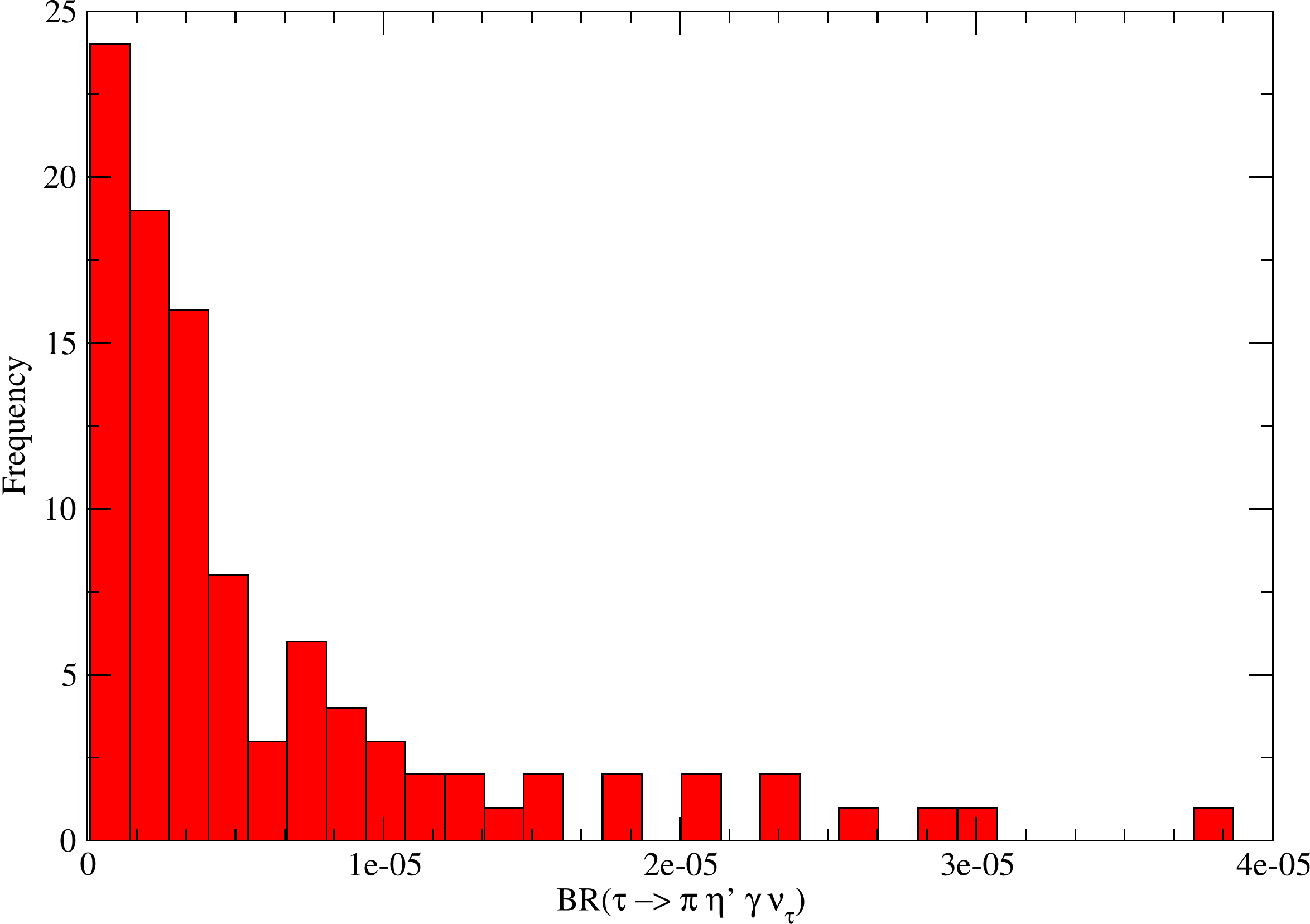}\caption{Histogram of $BR(\tau^-\to\pi^-\eta'\gamma\nu_\tau )$ 
  with a sample of $100$ R$\chi$T parameter space points for the complete (left) and neglecting 2R diagrams (right) branching fractions.}\label{fig:RChT100/1000p}
 \end{figure}

 Figure \ref{fig:RChT100/1000p} gives analogous plots to those in fig \ref{fig:RChT100/1000} for the $\eta'$ mode, this is, whole contribution (left) and 
 neglecting the 2R contributions (right). For 100 parameter space points the branching ratio 
 we get is $BR_{100}=(0.9\pm0.4)\times10^{-5}$, which is still larger than the non radiative process. As in the case for the $\eta$, as one takes an increasingly 
 larger region of the parameter space, the systematic theoretical error becomes the dominant uncertainty. Then, by taking the corresponding theoretical 
 uncertainty we get $BR_{1000}=(0.84\pm0.06)\times10^{-5}$ including all contributions, while neglecting 2R contributions we get $BR_{1000}=(0.65\pm0.05)\times10^{-6}$. 
 Again, the dependence on the statistics for the error is analyzed in section \ref{sec:statistics}. 
 
 \begin{figure}[!ht]
  \centering
    \includegraphics[scale=0.3,angle=-90]{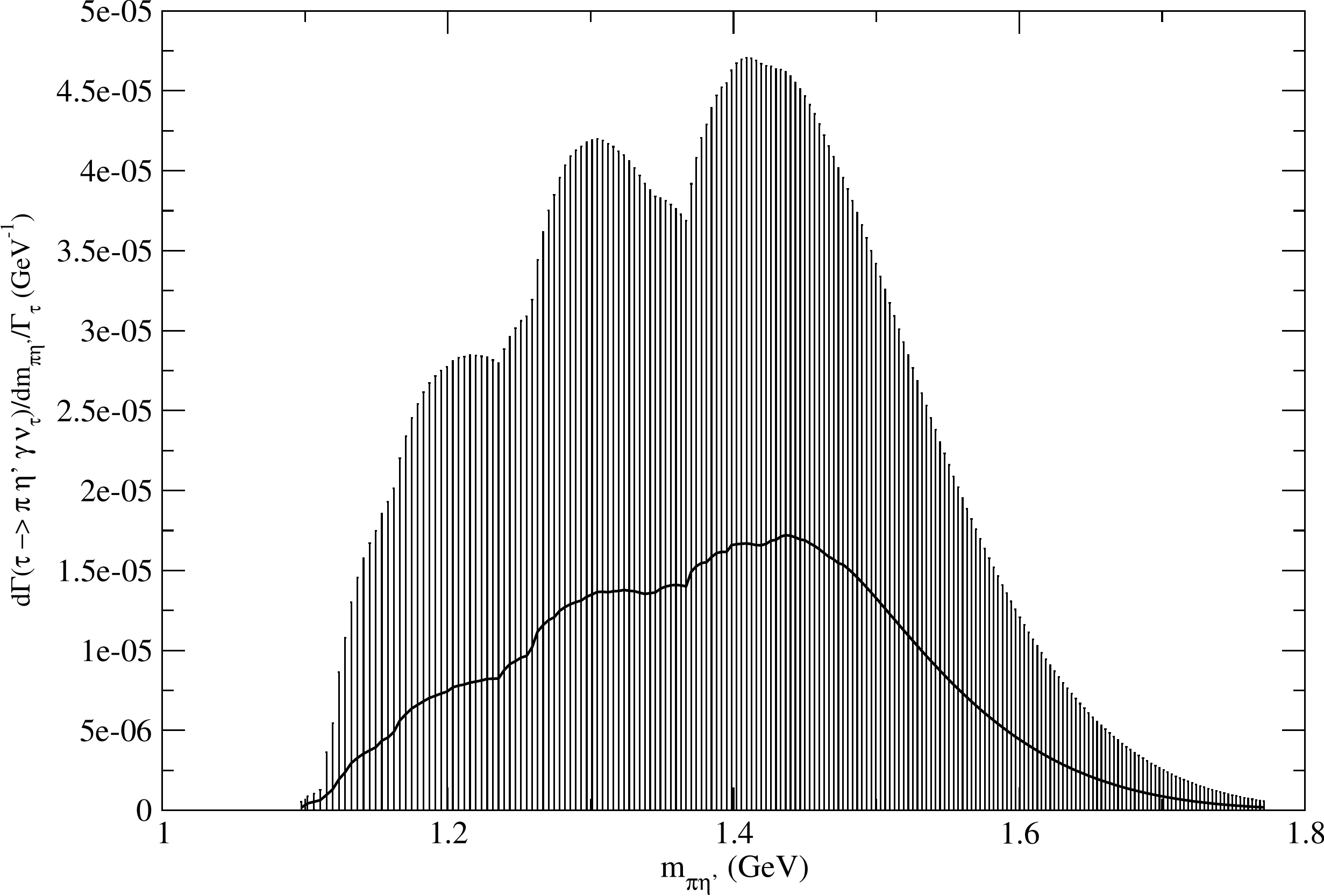}\includegraphics[scale=0.3,angle=-90]{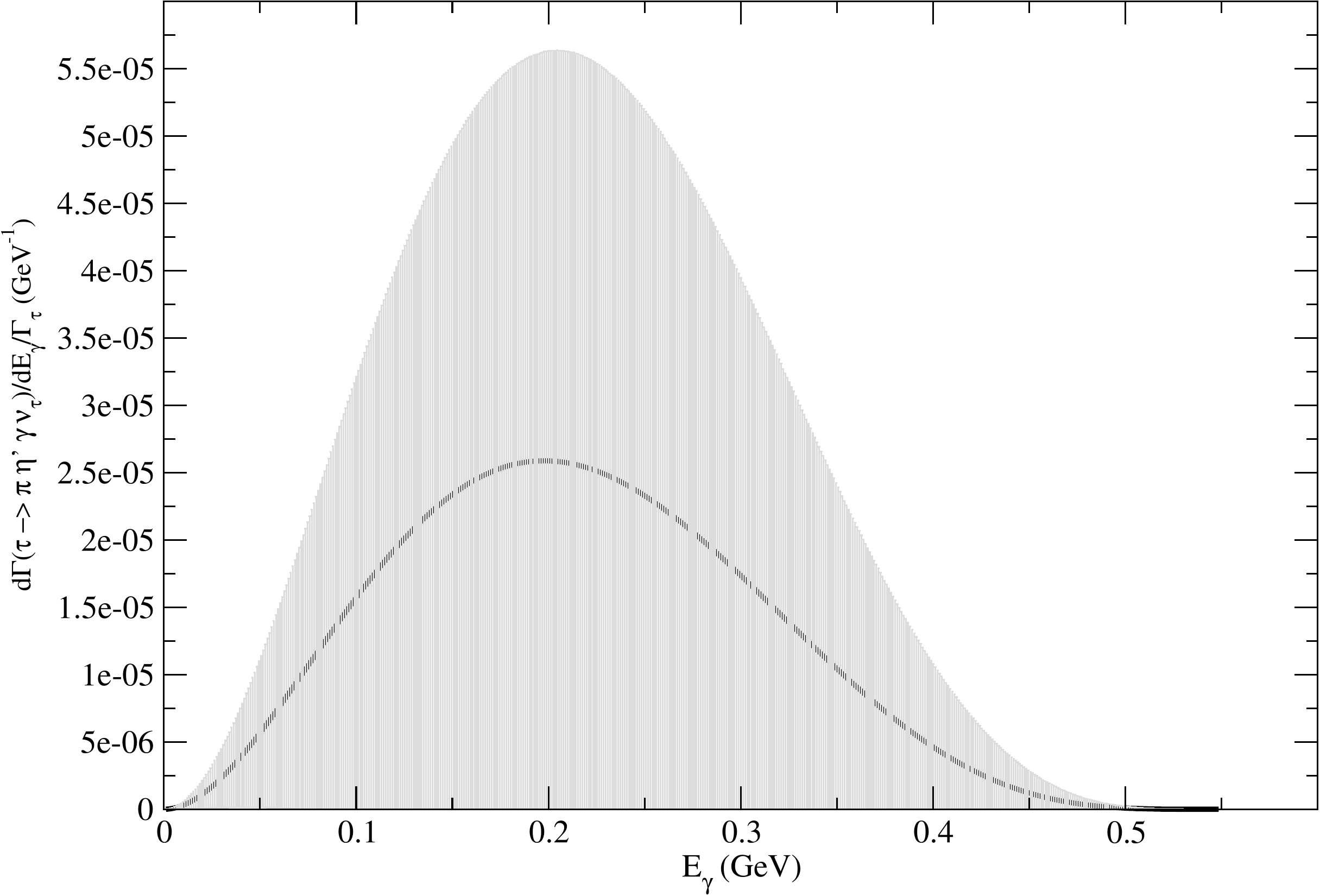}\caption{$\tau^-\to\pi^-\eta'\gamma\nu_\tau$ 
  normalized spectra according to R$\chi$T in the invariant mass of the $\pi^-\eta'$ system (left) and in the photon energy (right) are plotted.}\label{fig:RChTSpectra_etap}
 \end{figure}

 In fig. \ref{fig:RChTSpectra_etap} the normalized spectra in $\invm$ with 200 steps (left) and $E_\gamma$ with 500 steps (right) is shown. For the $\invmp$ spectrum, 
 a maximum is expected around the [1.30,1.45] GeV region. Also, the photon energy spectrum suggest a $\sim$ 100 MeV cut on the photon energy. The spectra barely 
 changes when neglecting the 2R contributions. 
  
 \begin{figure}[!ht]
  \centering\includegraphics[scale=0.34,angle=-90]{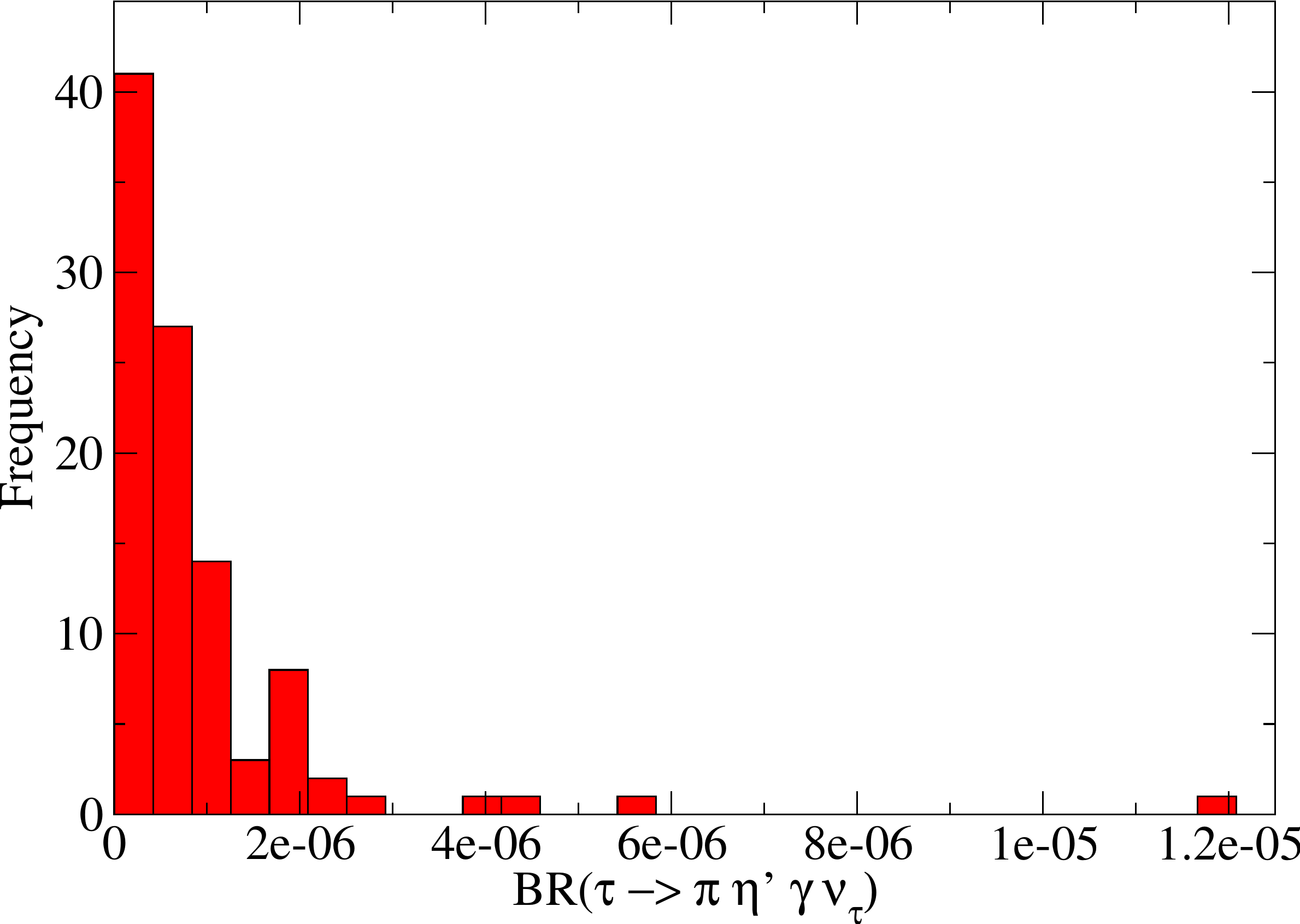}\caption{Histogram of $BR(\decetap)$ in R$\chi$T where photons with $E_\gamma>100$ MeV are rejected.}\label{fig:RChTcutsp}
 \end{figure}

 By applying the cut on the photon energy a $BR=(0.9\pm0.2)\times10^{-6}$, this result is shown in fig \ref{fig:RChTcutsp}. By neglecting 2R contributions one gets 
 $BR=(0.7\pm0.2)\times10^{-6}$.

 \section{Statistical error analysis}\label{sec:statistics}
 
 A different assignment of distribution for the parameters of both theories were done in the computation of these decays. In the case of MDM,
 the first consists in taking only the mean value for the couplings and letting them vary randomly within 1 sigma independently. For R$\chi$T the same 
 procedure is followed, but some of the couplings have rather large uncertainties due to the fact that those parameters cannot be fixed by nowadays 
 experimental data and we still do not know how to constrain them from short distance QCD. From here on, this will be called the
 1 sigma approach.\\
 
 In the previous sections, the coupling parameters were assigned as a Gaussian distribution around the mean value of the fit in the case of MDM. Since 
 almost all coupling parameters within this theory must be fitted phenomenologically from independent physical processes, the Gaussian behavior
 of the coupling constant gives a very good description of such parameters. \\
 
 What happens when we try to use the same Gaussian description for the R$\chi$T is rather startling at first, since the mean value of the 
 branching ratios seem to augment considerably (an order of magnitude in both channels). The problem here seems to rely on the fact that some of the 
 couplings take values mainly outside the 68\% around the mean value, leading to an artificial enhancement of the branching fractions. However, this 
 error analysis has a serious bias stemming from the fact that several of the known and unknown couplings must be related through short distance 
 constraints and should not be variated independently. The correct way to vary the parameters should be by means of constructing a parameter 
 vector and then assign a covariant matrix that would then correct correlation among the different parameters. The determination of the covariant matrix 
 would be such a formidable task that is far beyond the scope of this work. Therefore, for R$\chi$T the most reliable description of the decays 
 under study will be the 1 sigma approach. The corresponding values of the branching ratios are given in table \ref{Table:1sigma}.
 {\small\begin{table}[!h]
  \begin{center}
   \begin{tabular}{|c||c|c||c|c|}\hline
    &$\begin{array}{c}\text{complete}\\\deceta\end{array}$&$\begin{array}{c}\text{without 2R}\\\deceta\end{array}$&$\begin{array}{c}\text{complete}\\\decetap\end{array}$&$\begin{array}{c}\text{without 2R}\\\decetap\end{array}$\\\hline
   100 points&$(2.3\pm0.9)\cdot10^{-5}$& ----- &$(2.3\pm3.5)\cdot10^{-6}$&$(2.1\pm1.8)\cdot10^{-6}$ \\\hline
   1000 points&$(3.0\pm0.6)\cdot10^{-5}$&$(2.3\pm0.5)\cdot10^{-5}$&$(2.2\pm0.4)\cdot10^{-6}$&$(2.0\pm0.4)\cdot10^{-6}$\\\hline\hline
   $E_\gamma<100\MeV$&$(1.0\pm0.3)\cdot10^{-6}$&$(1.2\pm0.6)\cdot10^{-6}$&$(2\pm1)\cdot10^{-7}$&$(2\pm1)\cdot10^{-7}$\\\hline
   \end{tabular}
  \end{center}\caption{Branching fractions for different kinematical constraints and parameter space points.}\label{Table:1sigma}
 \end{table}}

 \section{Conclusions}
 
 The near start of Belle-II data taking brings us an excellent opportunity to search for second class currents, since the current limit on these 
 ($BR(\deceta)\lesssim9\times10^{-5}$ and $BR(\decetap)\lesssim7\times10^{-6}$)
 are very near to the expected predictions based on isospin breaking. Belle-II has become a very promising experiment to look for 
 SCC in $\tau$ decays due to its promised high precision.\\
 
 We have seen that a less restrictive error estimation might lead to an artificial enhancement of the observables computed within R$\chi$T 
 due to the dependence amidst known and/or unknown couplings. So, an appropriate description of the errors of the parameters 
 must be done, restricting them to lie within a certain error margin that can ensure the proper variation of parameters without introducing 
 bias due to this dependence.\\
 
\begin{table}[h!]
 \centering
   {\small\begin{tabular}{|c|c|c|c|c|} \hline
    SCC bkg & BR (no cuts) & BR ($E_\gamma^{\mathrm{cut}}=100$ MeV) & BR SCC signal & Bkg rejection\\ \hline
    $\tau^-\to\pi^-\eta\gamma\nu_\tau$ & $(3.0\pm0.6)\cdot10^{-5}$ & $(1.2\pm0.6)\cdot10^{-6}$ & $\sim1.7\cdot10^{-5}$ & Yes \\ \hline
    $\tau^-\to\pi^-\eta^\prime\gamma\nu_\tau$ & $(2.2\pm0.4)\cdot10^{-6}$ & $(2\pm1)\cdot10^{-7}$ & $[10^{-7},10^{-6}]$ & No \\ \hline
   \end{tabular}}
 \caption{The main conclusions of our analysis are summarized: Our predicted branching ratios for the $\tau^-\to\pi^-\eta^{(\prime)}\gamma\nu_\tau$ decays and the corresponding results when the cut $E_\gamma>100$ MeV is 
    applied. We also compare the latter results to the prediction for the corresponding non-radiative decay (SCC signal) according to ref.~\cite{Escribano:2016ntp} and conclude if this cut alone is able to get rid of the 
    corresponding background in SCC searches.}
   \label{resultsSCC}
\end{table}

 Using the appropriate error description for R$\chi$T we found the results given in table \ref{resultsSCC}, where we can guarantee background 
 rejection for $\deceta$, but not for the $\eta'$ channel.\\
 
  It is also interesting to note our finding that, within the $R\chi L$ frame, a simplified description of these decays neglecting the two-resonance mediated contributions is a good approximation for branching ratios and decay spectra, 
which will ease the coding of the corresponding form factors in the Monte Carlo generators.\\
 
 We have pointed out for the first time the importance of the process studied as an important background on the search for SCC. We also 
 found that {\GP} violation gives a suppression comparable to the $\alpha$, and thus neglect the bremsstrahlung contribution to the radiative 
 process by making a reasonable lower cut. Also, that by cutting photons with energies above 100 MeV will give the necessary suppression to 
 neglect the contribution from this process to the background in the search for SCC. (The cut is done leaving only small regions for the detection of $\pi^0$ and 
 $\etap$ through their decay into photons.)
  

\chapter{The $VV'P$ form factors in R$\chi$T and the $\pi-\eta-\eta'$ light-by-light contribution to the muon $g-2$}\label{g-2 chapter}

\section{Introduction}\label{IntroVVP}

The anomalous magnetic moment has been of such importance to physics that it lead the way in constructing the quantum theory of 
 the non-relativistic interactions of particles. Furthermore, once quantum field theory was constructed the anomalous magnetic 
 moment of the electron played a fundamental role in the understanding of renormalization and the perturbativity of QED. This 
 observable continues to be a very interesting one since currently there is a discrepancy of $\sim3.5\sigma$ between the latest 
 measurement of the anomalous magnetic moment of the muon and the prediction within the Standard Model. It has become more interesting 
 since, recently it has been announced that in the very near future the experimental accuracy will be improved by a factor of 4. 
 This makes mandatory for theoreticians to reduce the uncertainty in the predictions for this observable if one wants to explore 
 the plausibility of higher energy Beyond Standard Model effects that can be studied in this process.\\
 
 In section \ref{g} we introduce the anomalous 
 magnetic moment along with some historical development and the contributions to it. In section \ref{H amu} we introduce the 
 hadronic contributions to the anomalous magnetic moment of the muon and show the main contribution to the Hadronic Light by Light 
 scattering. In section \ref{TFFsec} we show our result for the pion transition form factor and how the couplings of the theory 
 are obtained. In section \ref{etaetap} we show the way to relate the pion transition form factor with the $\eta$ and $\eta'$ ones 
 and give our prediction for them. In section \ref{P amu} we give the contribution of the Goldstone exchange diagram. In section
 \ref{G-probe} we propose a new form of measuring the transition form factor when both photons are off-shell.

\section{The anomalous magnetic moment}\label{g}

   There is a fundamental property of particles that played a key role in the construction and understanding of Quantum Mechanics: the spin.
   The spin is one of the two Casimir invariants of the Poincar\'e group (the other being the mass) 
   upon which all particles in a theory invariant under Minkowski space-time isometries are classified. The quantum nature of spin as an 
   intrinsic angular momentum of fundamental particles was first discovered by Otto Stern and Walther Gerlach in 1922 by constructing a
   collimated beam of silver atoms passing through an inhomogeneous magnetic field \cite{Stern-Gerlach}. What they measured was a magnetic 
   dipole moment quantization due to a single electron in the outermost occupied shell, this is, the electron could have only two possible 
   magnetic moment values with the same magnitude. At the classical level, the magnetic moment is defined through the relation 
   \begin{equation}
    \boldsymbol{\mu}=\frac{q}{2m}\boldsymbol{L},
   \end{equation}
   where $q$ is the charge, $m$ the mass and $L$ the angular momentum. 
   At the quantum level, a massive particle with non-zero intrinsic angular momentum $\boldsymbol{s}$ must have a magnetic moment
   \begin{equation}
    {\boldsymbol{\mu}}=g\frac{q}{2m}\frac{\boldsymbol{s}}{2},
   \end{equation}
   where $\boldsymbol{s}$ is the spin operator, and $g$ is defined as the adimensional gyromagnetic factor.\\
   
\begin{figure}[ht!]
    \centering\includegraphics[scale=1]{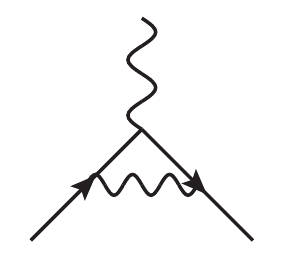}\caption{Next to leading order correction to the anomalous magnetic moment found by Schwinger.}\label{Schwinger}
   \end{figure}

   In Quantum Electrodynamics (QED), the way to obtain the magnetic moment of a fundamental particle is by making it interact with an 
   electromagnetic field. At leading order in QED one gets the Dirac result $g=2$. However, in 1947 Isidor Isaac Rabi and his team measured a 
   deviation from Dirac's prediction in hyperfine splitting of the ground state of hydrogen and deuterium \cite{Nafe}, which was
   \begin{equation}\label{DefAMM}
    a_e:=\delta\mu/\mu=\frac{g_e-2}{2}=0.00126\pm0.00019,
   \end{equation}
   where $a_\ell$ is defined as the anomalous magnetic moment of lepton $\ell$, and could be successfully explained by Julian Schwinger by computing a next 
   order term as a correction to the vertex \cite{Schwinger amu} as 
   shown in figure \ref{Schwinger}, where the correction is $\delta\mu/\mu=\alpha/2\pi=0.00116...$ in perfect agreement with the experimental value. 
   Ever since, more and more precise predictions and measurements of $a$ have been done. This is why it has become a suitable 
   observable for BSM phenomena.
   
   In their review of the anomalous magnetic moment of the muon $a_\mu$, Fred Jegerlehner and Andreas Nyffeler state that by the Appelquist-Carrazone theorem
   a heavier particle will have an enhanced sensibility to high energy scale phenomena \cite{Jegerlehner:2009ry}, since it would decouple from a lighter field. 
   The dependence on the mass of the particle and the scale of new phenomena is given by \cite{Jegerlehner:2009ry} 
   \begin{equation}
    \delta a_\ell\propto \frac{m_\ell^2}{\Lambda^2}\qquad \text{for}\qquad M\gg m_\ell
   \end{equation}
   This leads naturally to search deviations from the Standard Model in the anomalous magnetic moment of heavier particles. Since the 
   anomalous magnetic moment of the electron $a_e$ is very well known, the best option is the muon, which has a mass nearly 200 times greater 
   than the electron. This makes it more reliable to find effects that cannot be explained within the SM.\\
   
   Being nearly 17 times heavier than the muon, the $a_\tau$ should be the observable where to look for deviations from the SM. Nevertheless, the muon 
   is preferred instead of the $\tau$ since it's lifetime is so short ($\tau_\tau/\tau_\mu\sim10^{-7}$) that makes it really difficult 
   to measure its magnetic moment using spin precession techniques. Nowadays the experimental determination of the tau lepton anomalous magnetic moment $a_\tau$ is compatible with zero \cite{atau}.\\
 
   \begin{figure}[ht!]
    \centering\includegraphics[scale=0.9]{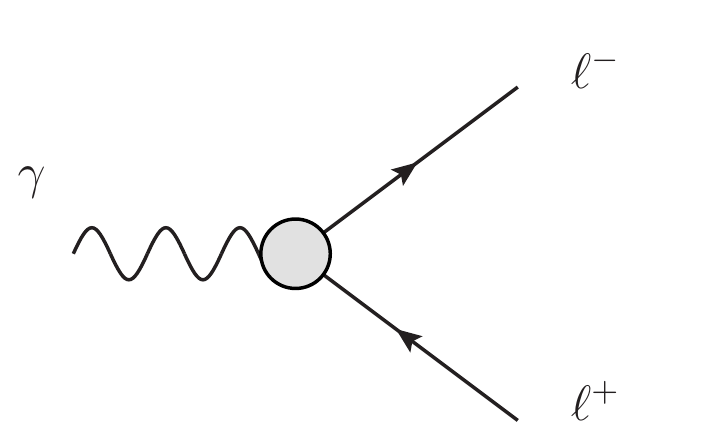}\caption{Feynman diagram of a fermion interaction with a classic electromagnetic field. The 
    blob represents all possible interactions that can happen in between.}\label{BLOB}
   \end{figure}

   Now, the most general vertex for magnetic moment is represented by the diagram in 
   figure \ref{BLOB} where all possible interactions are hidden in the blob (QED, EW and strong interactions)
   \begin{equation}
    a_{\ell}=a_{\ell}^{QED}+a_{\ell}^{EW}+a_{\ell}^{Had}.
   \end{equation}
   One can parametrize all the possible interactions in the blob into two form factors 
   by making use of the Gordon identity and separating the 
   fermion current into a vector ($\gamma_\mu$) and a tensor interaction ($\sigma_{\mu\nu}$) current. The tensor interaction will then have attached 
   the Pauli (or magnetic) form factor $F_M(q^2)$, which at $q^2=0$ gives the anomalous magnetic moment.\\
   
 \begin{table}[!ht]
 \centering
  \begin{tabular}{|c|c|c|}
\hline
{\footnotesize Type of contribution}&{\footnotesize Contribution to $a_\mu\times10^{11}$}&{\footnotesize Error$\times10^{11}$} \\
 \hline
 \small{QED} & \small{116'584,718.95}&\small{0.08}\\
\small{EW}&\small{153.6}&\small{1.0}\\
\small{Had} &\small{6930}&\small{$(42)_{HVP}(26)_{HLbL}$}\\\hline
\small{Total}&\small{116'591,803}&\small{(1)(42)(26)}\\\hline
\small{Exp}&\small{116'592,091}&\small{(54)(33)}\\\hline
   \end{tabular}\caption{Different types of contributions to the $a_\mu$. The hadronic contributions give the main theoretical uncertainty.}\label{Table:Contributions}
  \end{table}
   
\begin{figure}[!ht]
    \centering\includegraphics[scale=1]{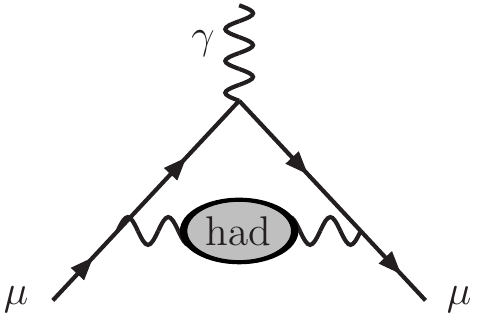}\caption{Hadronic Vacuum Polarization contribution to $a_\mu$, the blob stands for all possible 
    srong interaction processes.}\label{figHPV}
   \end{figure}

   The hadronic part has two contributions stemming from Hadronic Vacuum Polarization (HVP) diagrams (fig. \ref{figHPV}) and Hadronic Light by Light 
   (HLbL) (figs. \ref{lepLbL} and \ref{HLBL}). The former can be fully obtained by experimental data,
   while in the latter one has to rely on a model to predict its contribution. All the contributions to the $a_\mu$ are given in table 
   \ref{Table:Contributions}, where it can be seen that HVP gives a main source of error followed by HLbL. Nevertheless, the HVP error can be reduced by 
   using a better set of experimental data, while the HLbL error is given by the uncertainty from the theory and the fitting of the 
   parameters used to compute it.\\
   
   Here we discuss the contributions and errors from table \ref{Table:Contributions}: As can be seen in 
   the Particle Data Group Review of Particle Physics \cite{PDG16}, the anomalous magnetic 
   moment of the muon is almost fully determined by pure QED interaction, with an uncertainty $80\times10^{-14}$ \cite{Kinoshita alpha5}; the contribution arising 
   from weak boson exchange is very little, but non-negligible $(153.6\pm1.0)\times10^{-11}$ and the contributions stemming from strong interactions give the dominant 
   uncertainty to the anomalous magnetic moment of the muon $\sim 53\times10^{-11}$,
   which is comparable with the experimental error $\sim63\times10^{-11}$ \cite{PDG16}. The Fermilab \cite{Fermi g-2} and J-Parc \cite{J-PARC g-2} are planning 
   to reduce the present experimental error in a factor of 4, leading to an error $\sim16\times10^{-11}$. This is what makes mandatory from the theoretical 
   point to reduce the uncertainty in the prediction of the $a_\mu^{HLbL}$, since the error from $a_\mu^{HVP}$ can be reduced with further experimental data. \\

   \section{Hadronic contributions}\label{H amu}

   There is a contribution arising at order $\alpha^2$ where in the Schwinger correction the virtual photon polarizes the vacuum leading to a loop 
   correction in the propagator of the photon. The QED contribution when the fermions in the loop are leptons has been computed exactly 
   and is included in the QED correction \cite{lep VP}. 
   One of the hadronic contributions to the $a_\mu$, the Hadronic Vacuum Polarization (HPV), is obtained by changing the leptons in the loop by quarks, 
   where the strong interactions will come about 
   since the energy scale in the loop covers the region where quarks are confined in hadrons. The hadronic contribution is obtained by means of a dispersion 
   relation which can be connected by the optical theorem with the cross section $\sigma(e^+e^-\to\text{hadrons})$
   \begin{equation}\label{amuHVP}
    a_\mu^{HVP}=\left(\frac{\alpha m_\mu}{3\pi}\right)^2\int ds \frac{R(s)K(s)}{s^2}+\mathcal{O}(\alpha^3),
   \end{equation}
   where $R=\sigma(e^+e^-\to\text{hadrons})$ and the functional for of $K(s)$ can be found in ref \cite{Jegerlehner:2009ry}, thus obtained by 
   fitting data up to the charmonium region (5.2 GeV) and the bottomonium region ([9.46,13] GeV), the rest is obtained using pQCD. The greatest error in $a_\mu$
   comes from this contribution $\sim 42\times10^{-11}$, but this can be reduced by augmenting the quality and quantity of experimental data. In the
   expression for the $a_\mu^{HLbL}$ eq. (\ref{amuHVP}) it can be seen the dominance of low-energy contributions.\\
   
   \begin{figure}[ht!]
    \centering\includegraphics[scale=1]{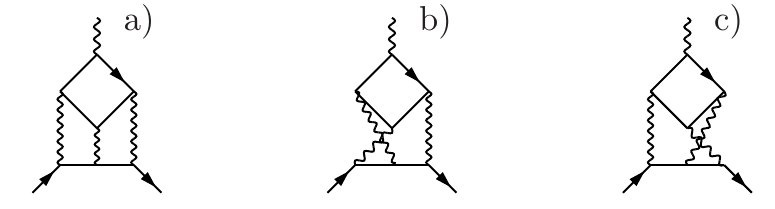}\caption{Light by light scattering insertions for a fermion loop.}\label{lepLbL}
   \end{figure}
   
   One can see that at the $\alpha^3$ order in the QED expansion a new kind of phenomena shows up, namely the {\it light by light} scattering (LbL) insertions. 
   Some typical diagrams for these contributions are shown in fig \ref{lepLbL}. This effect cannot be present in $\alpha^2$ contributions ($\gamma\gamma\to\gamma$), 
   since a closed fermion loop coupled to three photons would vanish due to Furry's theorem. These contributions can be separated into three categories, 
   one that gives a universal contribution to the $g-2$ where the particle in the loop is the same as the one under study to determine its magnetic moment. 
   One contribution comes from particles lighter than the one under study and the other from particles heavier than it. The latter ones are suppressed by 
   ratios of squared masses, the former are enhanced by logarithms of the mass ratio \cite{g-2 alpha3}. As in the HVP case, if we replace the leptons in the 
   loop by quarks we get the corresponding hadronic contribution, namely the Hadronic Light by Light scattering (HLbL), which cannot be obtained from 
   data and has to be theoretically predicted. Despite the fact that the Appelquist-Carrazone theorem cannot be applied to hadronic couplings, it can 
   give us a lead on which hadrons will contribute the most in the HLbL. \\

   \begin{figure}[ht!]
    \centering\includegraphics[scale=1.1]{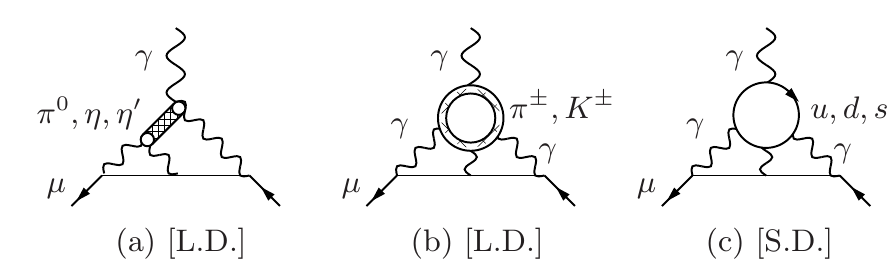}\caption{Contributions from Hadronic Light by Light scattering, $a\mu^{HLbL}$}\label{HLBL}
   \end{figure}
   
   Since HLbL cannot be fully obtained experimentally (See however recent progess in this direction in Refs. \cite{Hoferichter:2013ama, Colangelo:2014dfa, 
   Colangelo:2014qya,Colangelo:2014pva, Pauk:2014rfa, Colangelo:2015ama, Colangelo:2017fiz}), we will focus on this contribution to the $a_\mu$ trying to reduce the theoretical 
   uncertainty. There has been also a great advance in Lattice computations \cite{Blum:2014oka, Blum:2015gfa}. In R$\chi$T, the contribution comes from diagrams as those shown in fig. \ref{HLBL}, where to the Goldstone bosons exchange shown in the diagram
   one has to add resonances exchange. However by the argument of the previous paragraph one would expect these contributions 
   suppressed with respect to that of the Goldstone exchange. A previous computation of the resonances contribution showed that they are, in fact 
   suppressed with respect to the Goldstone ones \cite{DeRafael g-2}. The diagrams (b) and (c) of fig. \ref{HLBL} give a total contribution suppressed with respect 
   to the pseudoscalar exchange, therefore we will only focus in the main contribution to the $a_\mu^{HLbL}$\cite{DeRafael g-2} as shown in table \ref{Table:HLbL contributions}.
   It was at first assumed that the main contribution to HLbL would come from energy regions around the muon mass, but it was 
   noticed that some important contributions also come from the [0.5,1] GeV region\cite{HLbL not at mu mass1, HLbL not at mu mass1}, so that 
   one has to extend $\chi$PT to include resonances in order to include that energy region. \\

   \begin{table}[!ht]
   \begin{center}
    \begin{tabular}{|c|c|c|}\hline
     $a_\mu^{(a)}$&$a_\mu^{(b)}$&$a_\mu^{(c)}$\\\hline
     $(6.5\pm0.6)\times10^{-10}$&$ (-5.1\pm0.4)\times10^{-10}$&$(6.0\pm0.4)10^{-10}$\\\hline
    \end{tabular}
   \end{center}\caption{Contributions to $a_\mu$ from diagrams (a), (b) and (c) in fig \ref{HLBL} as given in ref. \cite{DeRafael g-2}.}\label{Table:HLbL contributions}
   \end{table}

\section{Transition Form Factor, TFF}\label{TFFsec}
   
\begin{figure}[ht!]
    \centering\includegraphics[scale=1.1]{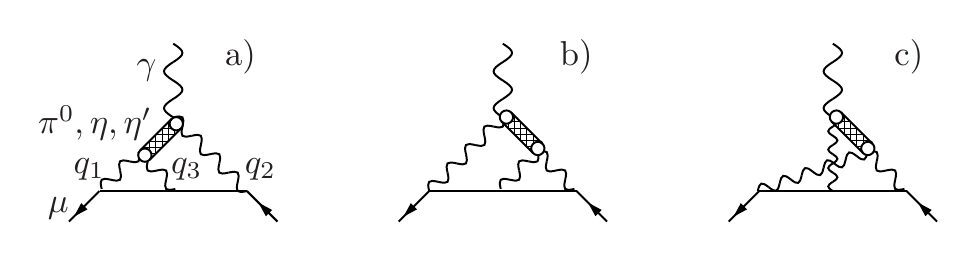}\caption{Main contribution to $a_\mu^{HLbL}$, internal photon lines include the $\rho-\gamma$ mixing}\label{pionpole}
   \end{figure}

   All the relevant diagrams to the pseudoscalar exchange are shown in figure \ref{pionpole}. Therefore, all we need to know to completely characterize 
   the Goldstone exchange contribution is the form factor of the effective vertex $\pi^*\gamma^*\gamma^*$.  Since the external photon must be taken with $q=0$ 
   and the form factor depends only on the squared of the momenta of the virtual photons, one finds that only three of the integrals in the 
   loop are non-trivial \cite{Knecht-Nyffeler}. The total pion exchange integrals to compute the contribution $a_\mu^{\pi^0,LbL}$ are given by the 
   expressions in appendix \ref{AppendixVVP}. 
   
   Previous computations of the $\pi$ Transition Form Factor ($\pi$TFF) were made using the naive simplification of taking the pion on-shell, 
   which might be seen as taking only the $\mu$ mass region ($m_\mu\approx m_\pi$) as told in the previous section. This 
   was shown to be an over-simplification of the problem \cite{HLbL not at mu mass1, HLbL not at mu mass2}. The following expression gives the 
   general $\pi$TFF when all particles are off-shell \cite{Kampf:2011ty, We 2013}
   
   \begin{multline}\label{pig*g* FF virtual pion}
   \mathcal{F}_{\pi^0\gamma^*\gamma^*}(p^2,q^2,r^2)=\frac{2r^2}{3F}\left[-\frac{N_C}{8\pi^2r^2}+4F_V^2\frac{d_3(p^2+q^2)}{(M_V^2-p^2)(M_V^2-q^2)r^2}
  \right.\\
   +\frac{4F_V^2d_{123}}{(M_V^2-p^2)(M_V^2-q^2)}+\frac{16F_V^2P_3}{(M_V^2-p^2)(M_V^2-q^2)(M_P^2-r^2)}\\
    \left.-\frac{2\sqrt{2}}{M_V^2-p^2}\left(\frac{F_V}{M_V}\frac{r^2c_{1235}-p^2c_{1256}+q^2c_{125}}{r^2}+\frac{8P_2F_V}{(M_P^2-r^2)}
  \right)+(q^2\leftrightarrow p^2)\right],
   \end{multline}
   where $p$, $q$ and $r$ are the photons and pion four-momenta, respectively. In the case where the pion is taken to be on-shell one gets
   \begin{eqnarray}\label{pig*g* FF real pion}
 \mathcal{F}_{\pi^0\gamma^*\gamma^*}(p^2,q^2,r^2=0) & = & \frac{2}{3F}\left[-\frac{N_C}{8\pi^2}+\frac{4F_V^2d_3(p^2+q^2)}{(M_V^2-p^2)(M_V^2-q^2)}+2\sqrt{2}\frac{F_V}{M_V}\frac{p^2c_{1256}-q^2c_{125}}{M_V^2-p^2}\right.\nonumber\\
& & \left. 
+2\sqrt{2}\frac{F_V}{M_V}\frac{q^2c_{1256}-p^2c_{125}}{M_V^2-q^2}\right]\,.
\end{eqnarray}

   One finds that the couplings in this form factor are all fixed from short distance constraints\cite{Roig:2013baa}. These couplings are given in the following 
   expression
   \begin{eqnarray}\label{SDVVP}
    F_V\,=\,\sqrt{3}F\,,\quad c_{125}\,=\,0\,,\quad c_{1256}\,=\,-\frac{N_C M_V}{32\sqrt{2}\pi^2F_V}\sim-3.26\cdot10^{-2}\,,\nonumber\\
    \quad c_{1235}\,=\,0\,,d_{123}\,=\,\frac{F^2}{8F_V^2}=\frac{1}{24}\,,\quad d_3\,=\,-\frac{N_C M_V^2}{64\pi^2F_V^2}\sim-0.112\,
   \end{eqnarray}
   where the condition we find for the coupling constant $d_3$ can be expressed in a different manner for convenience,
   \begin{equation}
    d_3\,=\,-\frac{N_C}{64\pi^2}\frac{M_V^2}{F_V^2}\,+\,\frac{F^2}{8F_V^2}\,+\,\frac{4\sqrt{2}P_2}{F_V}\,,
   \end{equation}
   provided the pseudoscalar resonance coupling $P_2\equiv d_m\kappa^{PV}_3$, where $\kappa^{PV}_3$ is the pseudoscalar vector coupling of 
   $\mathcal{O}^{PV}_3$ of ref \cite{Kampf:2011ty} fulfills the relation
   \begin{equation}
    P_2=-\frac{F^2}{32\sqrt{2}F_V}=-\frac{F}{32\sqrt{6}}\,,
   \end{equation}
   which belongs to the consistent set of short distance constraints \cite{Roig:2013baa}.\\
   
\begin{figure}[ht!]
    \centering\includegraphics[scale=0.5, angle=-90]{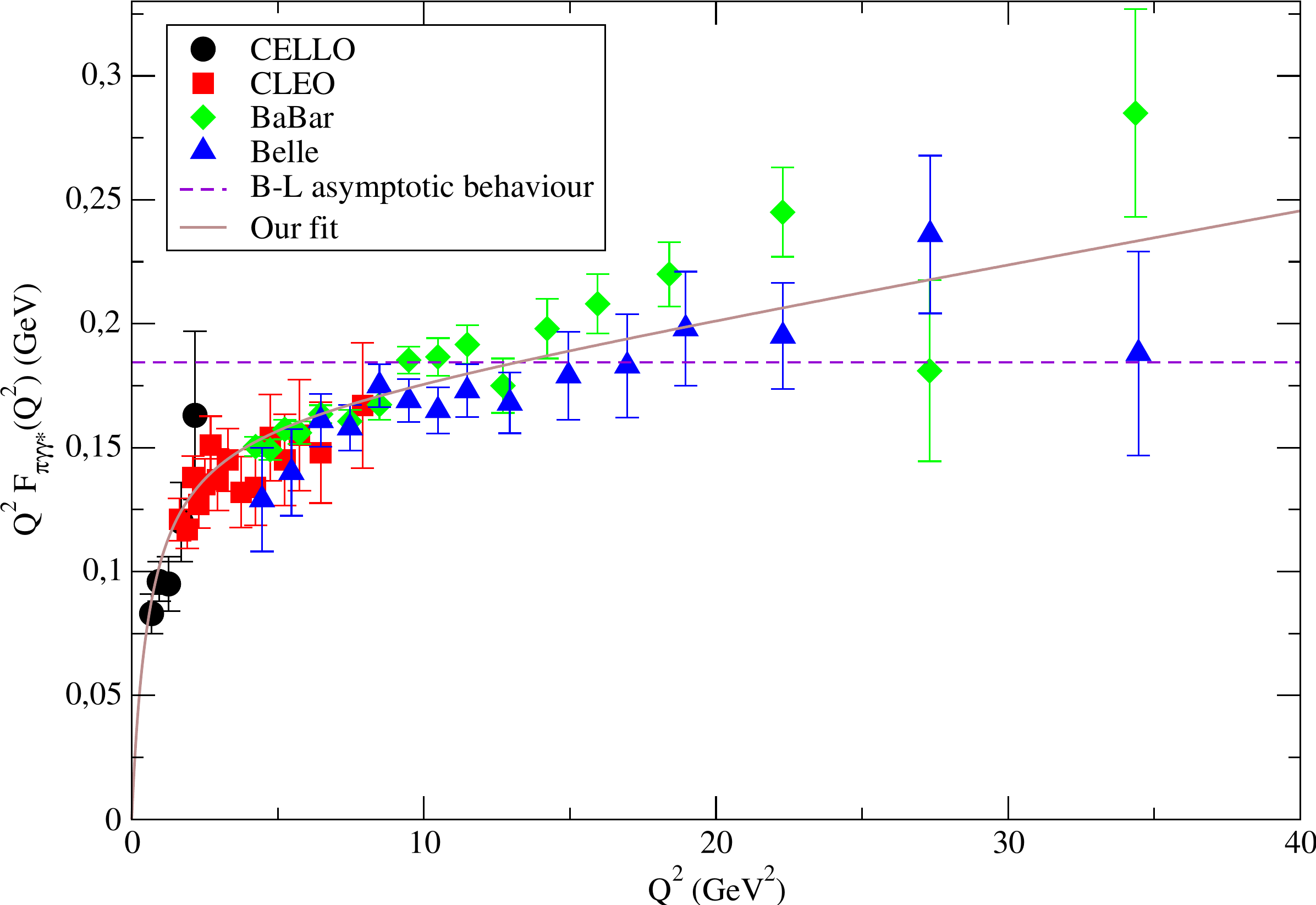}\caption{Our fit to the BaBar, Belle, CELLO and CLEO data compared to the Brodsky-Lepage behavior.}\label{PITFF}
   \end{figure}

   Contrary to ref. \cite{Kampf:2011ty}, we take the asymptotic value given by 
   the short distance constraints of $F_V=\sqrt{3}F$ and estimate the error varying it around a 10\% from the SD prediction. We rely on this 
   estimation since a fit done to BaBar data \cite{TAUOLA2} gives a variation around the 5\% of the asymptotic value. With this, we find the 
   behavior given in figure \ref{PITFF}. 
   The $P_3$ coupling is fixed phenomenologically from the combined analyses of the $\pi(1300)\to\gamma\gamma$ and $\pi(1300)\to\rho\gamma$
   \cite{Kampf:2011ty}, giving
   \begin{equation}
    P_3\,=\,-\left(1.2\pm0.3\right)\cdot10^{-2}\, \mathrm{\quad GeV}^2\,.
   \end{equation}
   
   Relying on the Brodsky-Lepage \cite{BrodskyLepage} behavior, which predicts the following limit of the form factor at high $q^2$ for one on-shell photon
   \begin{equation}
    \lim_{q^2\to\infty} \mathcal{F}_{\pi\gamma\gamma^*}(0,q^2,m_\pi^2) \sim \frac{2 F}{q^2},
   \end{equation}
   one finds a deviation from BaBar data \cite{Aubert:2009mc}. Since the high energy 
   description of the form factor is given mainly by the $P_2$ coupling we fix all the other couplings from equations (\ref{SDVVP})
   and fit $P_2$ to BaBar \cite{Aubert:2009mc}, Belle \cite{Uehara:2012ag}, CELLO \cite{Behrend:1990sr} and CLEO \cite{Gronberg:1997fj} data of the TFF. We, thus, find 
   that the best fit is given by
   \begin{equation}
    P_2=-(1.13\pm0.12)\times10^{-3} \quad\text{GeV},
   \end{equation}
   with the statistic test $\chi^2/\text{degrees of freedom}=1.01$. With this, we find that  
   the best fit for BaBar data is $\sim4\%$ away from the Brodsky-Lepage constraint.
   However, more accurate measurements of the $\pi$TFF at large $q^2$ and $p^2$ are needed to elucidate
   whether the Brodsky-Lepage-like asymptotic behavior (approached by Belle) or its violation (hinted by BaBar) describe the high-energy data. 
   Since a best fit of the form factor would be with the two photons being off-shell we propose a new observable from which to fit the TFF 
   (see section \ref{G-probe}).
   
\section{$\eta$- and $\eta'$- Transition Form Factor}\label{etaetap}

In this section we evaluate the contributions of the next lightest pseudoscalar mesons ($\eta$ and $\eta^\prime$) to $a_{\mu}^{HLbL}$. In order to do that we take advantage 
of the relation of the respective TFF with the $\pi$TFF. We will treat the $\eta$-$\eta^\prime$ mixing in the two-angle mixing scheme (consistent with the large-$N_C$ limit of QCD~\cite{KL}) and 
work in the quark flavor basis~\cite{twoanglemixing} where
\begin{equation}\label{diag_u}
 \mathrm{diag}(u)\,=\,\left( \frac{\pi^0+C_q\eta+C_{q^\prime}\eta^\prime}{\sqrt{2}},\; \frac{-\pi^0+C_q\eta+C_{q^\prime}\eta^\prime}{\sqrt{2}},\; -C_s\eta+C_{s^\prime}\eta^\prime\right)\,,
\end{equation}
in which
\begin{eqnarray}\label{definitions_Cq_Cs}
 C_q & \equiv & \frac{F}{\sqrt{3}\mathrm{cos}(\theta_8-\theta_0)}\left(\frac{\mathrm{cos}\theta_0}{f_8}-\frac{\sqrt{2}\mathrm{sin}\theta_8}{f_0}\right)\,,\nonumber\\ C_{q^\prime}&\equiv&\frac{F}{\sqrt{3}\mathrm{cos}(\theta_8-\theta_0)}\left(\frac{\sqrt{2}\mathrm{cos}\theta_8}{f_0}+\frac{\mathrm{sin}\theta_0}{f_8}\right)\,,\nonumber\\
 C_s & \equiv & \frac{F}{\sqrt{3}\mathrm{cos}(\theta_8-\theta_0)}\left(\frac{\sqrt{2}\mathrm{cos}\theta_0}{f_8}+\frac{\mathrm{sin}\theta_8}{f_0}\right)\,,\nonumber\\ C_{s^\prime}&\equiv&\frac{F}{\sqrt{3}\mathrm{cos}(\theta_8-\theta_0)}\left(\frac{\mathrm{cos}\theta_8}{f_0}-\frac{\sqrt{2}\mathrm{sin}\theta_0}{f_8}\right)\,.\nonumber\\
\end{eqnarray}
The values of the pairs of decay constants and mixing angles are~\cite{twoanglemixing} 
\begin{subequations}\label{values_mixingangless&decayscts}
\begin{align}
 &\theta_8\,=\,\left(-21.2\pm1.6\right)^\circ,\quad  \theta_0\,=\,\left(-9.2\pm1.7\right)^\circ,\quad\\ &f_8\,=\, \left(1.26\pm0.04\right)F,\quad f_0\,=\, \left(1.17\pm0.03\right)F\,.
 \end{align}
\end{subequations}
We will consider these errors as independent in the following.  

Within this mixing scheme, the $\eta$ and $\eta^\prime$ TFF can be easily related to the $\pi$TFF
\begin{eqnarray}\label{relation_etaTFF_piTFF}
 \mathcal{F}_{\eta\gamma\gamma}(p^2,\,q^2,\,r^2) & = & \left(\frac{5}{3}C_q-\frac{\sqrt{2}}{3}C_s\right)\mathcal{F}_{\pi\gamma\gamma}(p^2,\,q^2,\,r^2)\,,\nonumber\\
 \mathcal{F}_{\eta^\prime\gamma\gamma}(p^2,\,q^2,\,r^2) & = & \left(\frac{5}{3}C_{q^\prime}+\frac{\sqrt{2}}{3}C_{s^\prime}\right)\mathcal{F}_{\pi\gamma\gamma}(p^2,\,q^2,\,r^2)\,.
\end{eqnarray}
  
  \begin{figure}
   \centering\includegraphics[scale=0.32, angle=-90]{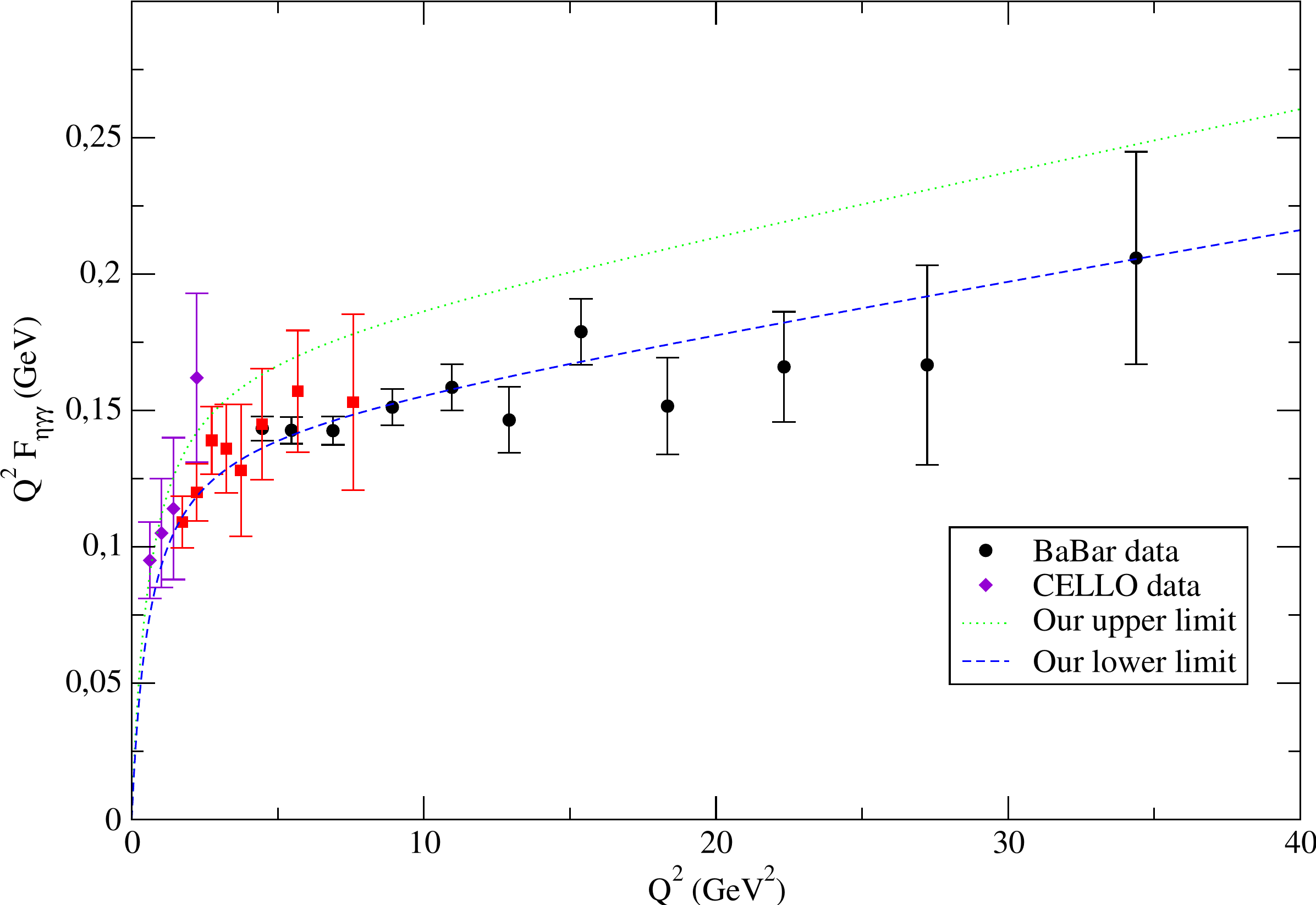}\includegraphics[scale=0.32,angle=-90]{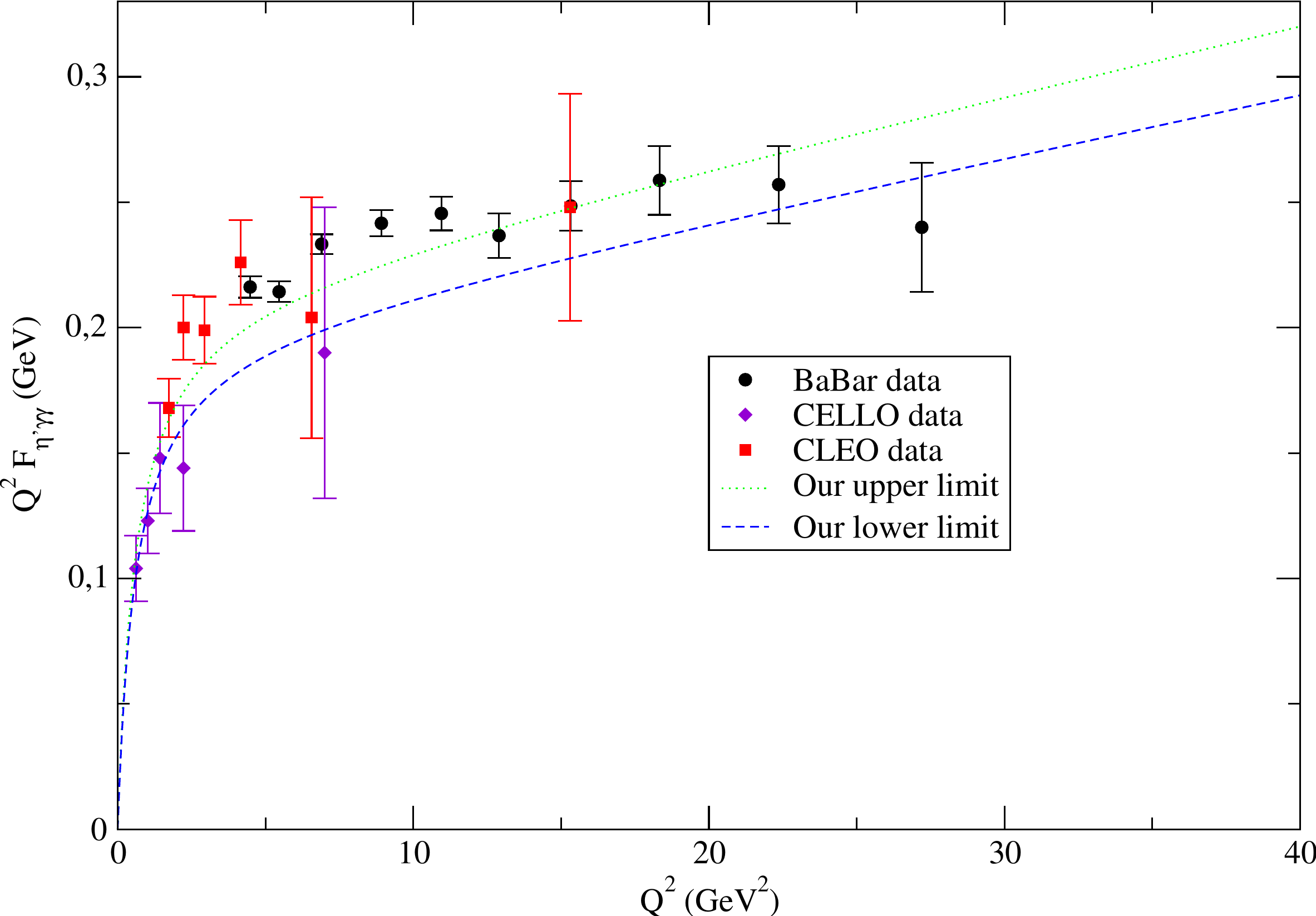}\caption{Our prediction for 
   the $\eta$ (left) and $\eta'$ (right) TFF cross section (left) using the couplings of eq. \ref{Couplings piggFF real pion} compared to 
   BaBar \cite{Aubert:2009mc}, CELLO \cite{Behrend:1990sr} and CLEO \cite{Gronberg:1997fj}.}\label{fig:etapTFF}
  \end{figure}

We have therefore predicted the $\eta$ and $\eta^\prime$ TFF using our results for the $\pi$TFF. 
The corresponding error is completely dominated by the $\eta$-$\eta^\prime$ 
mixing. In fig.~\ref{fig:etapTFF} we confront them to BaBar~\cite{Aubert:2009mc}, CELLO~\cite{Behrend:1990sr} and CLEO~\cite{Gronberg:1997fj} data. In the 
case of the $\eta$TFF good agreement can be seen, although BaBar data tend to lie in the border of our predicted lower limit. Even though data from different experiments on 
the $\eta^\prime$TFF show slight tension, the overall agreement of our prediction with them is quite good. We observe that our $\pi$TFF-based prediction tends to show a tiny 
larger slope than the $\eta$ and $\eta^\prime$ TFF data. This feature may be caused by BaBar data on the $\pi$TFF. It remains to be seen if new, more accurate, measurements 
of these TFF confirm this tendency or not.

\section{Pseudoscalar exchange contribution $a_\mu^{P,HLbL}$}\label{P amu}

   Using all the information in the previous sections and the integrals of appendix \ref{AppendixVVP} we can now compute the total 
   pseudoscalar exchange contribution to the $a_\mu$. However, before computing it, further analysis of the error in the $\pi$TFF can be done 
   to the classic WZW result in the chiral limit at very-low energy \cite{Jegerlehner:2009ry}. A small contribution to such value of the $\pi$TFF
   is included \cite{Ramsey-Musolf 2002,Engel:2012xb} by considering the correction at these limits stemming from the short distance constraints on $c_{1235}$ and $d_{123}$ and the 
   non-zero mass of the pion. The correction enters as follows,
   \begin{equation}
    \mathcal{F}_{\pi\gamma\gamma}(0)=-\frac{N_C}{4\pi^2F}(1-\Delta),
   \end{equation}
where the correction is 
   \begin{equation}
    \Delta=\frac{4\pi^2}{3}\frac{F^2}{M_V^2}\frac{m_\pi^2}{M_V^2}\sim5.9\times10^{-3}.
   \end{equation}

   Further corrections to the previous expression should be suppressed by further powers of $m_\pi^2/M_V^2$ and can be safely neglected.\\
   \begin{table}[ht!]
\begin{tabular}{|c|c|c|c|}
\hline 
$a_\mu^{\pi^0,HLbL}\cdot 10^{10}$ & Method and reference\\
\hline
$5.58\pm0.05$ & \small{Extended NJL Model \cite{NJL} (Bijnens, Pallante and Prades \cite{HLbL not at mu mass2})}\\
$5.56\pm0.01$ & \small{Naive VMD Model (Hayakawa, Kinoshita [and Sanda] \cite{HLbL not at mu mass1}\cite{Hayakawa-Kinoshita})}\\
$5.8\pm1.0$ & \small{Large-$N_C$ with two vector multiplets, $\pi$-pole contribution~\cite{Knecht-Nyffeler}}\\
$7.7\pm1.0$ &\small{Large-$N_C$ with two vector multiplets, $\pi$-pole contribution \cite{Melnikov:2003xd}}\\
$7.2\pm1.2$ & \small{$\pi$-exchange contribution corresponding to~\cite{Knecht-Nyffeler} evaluated in~\cite{Jegerlehner:2009ry} }\\
$6.9$ & \small{Holographic models of QCD~\cite{Hong:2009zw}}\\
$6.54\pm0.25$ & \small{Holographic models of QCD~\cite{Cappiello:2010uy}}\\
$6.58\pm0.12$ & \small{Lightest Pseudoscalar and Vector Resonance saturation~\cite{Kampf:2011ty}}\\
$6.49\pm0.56$ & \small{Rational Approximants \cite{Masjuan:2012qn}}\\
$5.0\pm0.4$ & \small{Non-local chiral quark model~\cite{Dorokhov:2012qa}}\\
\hline
$6.66\pm0.21$ & \small{This work, short-distance constraints of \cite{Kampf:2011ty} revisited and data set updated}\\
\hline
\end{tabular}
\caption{\label{Other_results} \small{Our result for $a_{\mu}^{\pi^0,HLbL}$ in eq.~(\ref{Result virtual pion}) is compared to other determinations. 
The method employed in each of them is also given. We specify those works that approximate $a_\mu^{\pi^0,HLbL}$ by the pion pole contribution. It is understood that all 
others consider the complete pion exchange contribution.}}
\end{table}

   Although the uncertainty stemming from this correction is negligible to the observables computed with the TFF (including those of section 
   \ref{G-probe}), it is not so for the $a_\mu^{P,HLbL}$, since the uncertainty in this contribution is essentially given by the low-energy 
   correction. Including this correction we find 
   \begin{equation}
    a_\mu^{\pi^0,HLbL}=(5.75\pm0.06)\times10^{-10}
   \end{equation}
   for the pion-pole simplification, {\it i. e.}, for on-shell pion and
   \begin{equation}\label{Result virtual pion}
    a_\mu^{{\pi^0}^*,HLbL}=(6.66\pm0.21)\times10^{-11}.
   \end{equation}
   We can see, as said above, that the pion-pole simplification underestimates the $a_\mu^{\pi^0,HLbL}$ by a 14\% and the error by a factor 4. 
   The off-shell result is shown in table \ref{Other_results}, where other results are quoted for comparison. For the 
   $\eta$ and $\eta'$ we made the same comparison between the pole estimation and the complete TFF, where we found 
   
   \begin{equation}\label{Result real eta}
    a_{\mu}^{\eta,HLbL}\,=\,\left(1.44\pm0.26\right)\cdot 10^{-10}\,,\quad a_{\mu}^{\eta^\prime,HLbL}\,=\,\left(1.08\pm0.09\right)\cdot 10^{-10}
   \end{equation}
   for the pole contribution and
   \begin{equation}\label{Result virtual eta}
    a_{\mu}^{\eta,HLbL}\,=\,\left(2.04\pm0.44\right)\cdot 10^{-10}\,,\quad a_{\mu}^{\eta^\prime,HLbL}\,=\,\left(1.77\pm0.23\right)\cdot 10^{-10}
   \end{equation}
   for the whole exchange contribution. 
   \begin{table*}[ht!]
\begin{center}
\begin{tabular}{|c|c|c|c|}
\hline 
$a_\mu^{P,HLbL}\cdot 10^{10}$ & Method and reference\\
\hline
$8.5\pm1.3$ & \small{Extended NJL Model \cite{NJL} (Bijnens, Pallante and Prades \cite{HLbL not at mu mass2})}\\
$8.27\pm0.64$ & \small{Naive VMD Model (Hayakawa, Kinoshita [and Sanda] \cite{HLbL not at mu mass1}\cite{Hayakawa-Kinoshita})}\\
$8.3\pm1.2$ & \small{Large-$N_C$ with two vector multiplets, $P$-pole contribution~\cite{Knecht-Nyffeler}}\\
$11.4\pm1.0$ &\small{Large-$N_C$ with two vector multiplets, $P$-pole contribution \cite{Melnikov:2003xd}}\\
$9.9\pm1.6$ & \small{$\pi$-exchange contribution corresponding to~\cite{Knecht-Nyffeler} evaluated in~\cite{Jegerlehner:2009ry}}\\
$10.7$ & \small{Holographic models of QCD~\cite{Hong:2009zw}}\\
$9.0\pm0.7$ & \small{Rational Approximants \cite{Escribano:2013kba} using half-width rule \cite{Masjuan:2012sk}, $P$-pole contribution}\\
$5.85\pm0.87$ & \small{Non-local chiral quark model~\cite{Dorokhov:2012qa}}\\
$11.4\pm1.3$ & \small{Average of various approaches (Prades, de Rafael and Vainshtein \cite{Prades:2009tw}}\\
\hline
$10.47\pm0.54$ & \small{This work, lightest Pseudoscalar and Vector Resonance saturation}\\
\hline
\end{tabular}
\caption{\label{Other_results_2} \small{Our result for $a_{\mu}^{P,HLbL}$ in eq.~(\ref{Result virtual P}) is compared to other determinations. 
The method employed in each of them is also given. We specify those works that approximate $a_\mu^{P,HLbL}$ by the pseudoscalar pole contribution. It is understood that all 
others consider the complete pseudoscalar exchange contribution.}}
\end{center}
\end{table*}

   As it happened in the $\pi^0$ case, the $\eta^{(\prime)}$-pole approximation underestimates clearly the $HLbL$ contribution, by $\sim 30(45)\%$, 
   and the error, by a factor of roughly two. This is confirmed by comparing our results in eq.~(\ref{Result real eta}) with those obtained in ref.~\cite{Escribano:2013kba}
   \begin{equation}\label{Result real eta Pere}
    a_{\mu}^{\eta,HLbL}\,=\,\left(1.38\pm0.16\right)\cdot 10^{-10}\,,\quad a_{\mu}^{\eta^\prime,HLbL}\,=\,\left(1.22\pm0.09\right)\cdot 10^{-10}\,
   \end{equation}
   which agree within errors.\\

   Taking into account our determinations of $a_{\mu}^{\pi^0,HLbL}$ (\ref{Result virtual pion}), $a_{\mu}^{\eta,HLbL}$ and $a_{\mu}^{\eta^\prime,HLbL}$(\ref{Result virtual eta}), 
   we obtain for the contribution of the three lightest pseudoscalars
   \begin{equation}\label{Result virtual P}
    a_{\mu}^{P,HLbL}\,=\,\left(10.47\pm0.54\right)\cdot 10^{-10}\,.
   \end{equation}
   This number is compared to other determinations in the literature in Table \ref{Other_results_2}. Again, the method employed in each determination is also given for reference.

\section{Genuine probe of the $\pi$TFF}\label{G-probe}

\begin{figure}[ht!]
    \centering\includegraphics[scale=0.5]{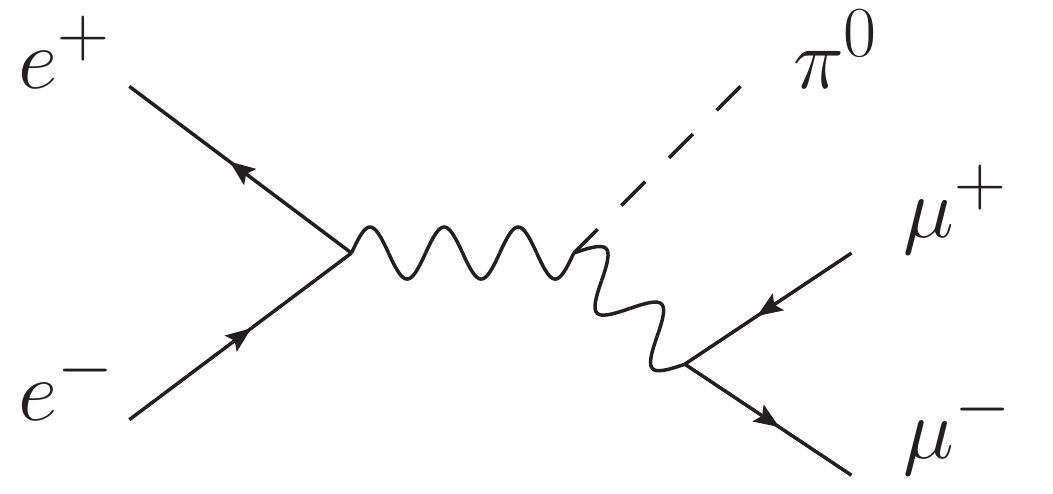}\caption{The $e^+e^-\to\mu^+\mu^-\pi^0$ scattering as a probe for $\pi$TFF with both photons off-shell.}\label{g-2 probe}
   \end{figure}

   Currently, there are no experimental data that can give us the behavior of the $\pi$TFF with both photons off-shell. Looking forward to fixing
   parameters from an observable with both photons off-shell we study the cross section $\sigma(e^+e^-\to\mu^+\mu^-\pi^0)$, which can be obtained 
   through the process shown in figure \ref{g-2 probe}.
   
   In this process the center of mass squared energy $p^2=s$ is the electron-positron invariant mass and the di-muon invariant mass will give the other photon 
   squared four-momenta $s_1=q^2$. Since now $s$ and $s_1$ are time-like, we must change the factors $1/(M_R^2-x)\to D_R(x)$ since now both four-momenta span the resonances 
   region and such effect must be taken into account. By using the values of the couplings as discussed in the previous section
   
   \begin{eqnarray}\label{Couplings piggFF real pion}
    F_V & = & \sqrt{3}F(1.0\pm0.1)\,,\quad c_{125}\,=\,0\,,\quad c_{1256}\,=\,-\frac{N_C M_V}{32\sqrt{2}\pi^2F_V}\,,\nonumber\\
    d_3 & = & -\frac{N_C M_V^2}{64\pi^2F_V^2}\,+\,\frac{F^2}{8F_V^2}\,+\,\frac{4\sqrt{2}P_2}{F_V}\,,\quad P_2 \, = \, -\left(1.13\pm0.12\right)\cdot 10^{-3}\,\mathrm{GeV}\,,
   \end{eqnarray}
   we can predict the dependence of the cross section with the center of mass energy, as well as the dependence on the cross section with 
   $s_1$ for a fixed center of mass energy. In terms of suitable invariants \cite{Kumar:1970cr},
\begin{equation}\label{invariants}
 s\equiv k^2\,,\quad s_1\equiv{k^\prime}^2\,,\quad t_0\equiv (q_+-p_\pi)^2\,,\quad t_1\equiv (q_+-p_+)^2\,,\quad u_1\equiv(k-p_+)^2\,,
\end{equation}
the corresponding spin-averaged and unpolarized squared matrix element reads
{\small\begin{eqnarray}\label{m.e. summed and averaged invariants}
  \sum \overline{\Big|\mathcal{M}\Big|^2} & = & \frac{512\alpha^4\pi^4}{s^2 s_1^2}\Big|\mathcal{F}_{\pi^0\gamma\gamma}(k^2,{k^\prime}^2)\Big|^2\left\lbrace -2 m_\mu^4 s^2\right.\nonumber\\
& & + m_\mu^2 s \left[m_\mu^4+m_\mu^2 \left(m_\pi^2+s+s_1-2t_0-4t_1+2u_1\right)\right.\nonumber\\
& & \left. \left. +m_\pi^4 +m_\pi^2 \left(-3 s+s_1-3 t_0-2 t_1+ u_1\right)+3 s^2-4 s s_1+5 s t_0+6 s t_1\right.\right.\nonumber\\
& & \left. - 3 s u_1+s_1^2-3 s_1 t_0-2 s_1 t_1+ s_1 u_1+3 t_0^2+4 t_0 t_1-2 t_0 u_1+4 t_1^2-4 t_1 u_1 + u_1^2\right]\nonumber\\
& &  +\frac{1}{4}\left[2 s \left(s_1-2 m_\mu^2\right) (s+t_1-u_1) \left(-m_\mu^2-m_\pi^2+s+t_0+t_1\right)\right.\nonumber\\
& & \left. \left. +4 \left(m_\mu^2-t_1\right) (s+t_1-u_1) \left(m_\mu^2+m_\pi^2-s-t_0-t_1\right)\right.\right.\nonumber\\
& & \left. \left. (s-s_1+t_0+t_1-u_1)+s \left(s_1-2 m_\mu^2\right) \left(-m_\mu^2-m_\pi^2+s+t_0+t_1\right)^2\right.\right.\nonumber\\
& & -2 (s+t_1-u_1)^2 \left(-m_\mu^2-m_\pi^2+s+t_0+t_1\right)^2-2 s^2 \left(s_1-2 m_\mu^2\right)^2\nonumber\\
& & -2 \left(m_\mu^2-t_1\right)^2 (s-s_1+t_0+t_1-u_1)^2+s \left(s_1-2 m_\mu^2\right) (s-s_1+t_0+t_1-u_1)^2\nonumber\\
& & \left. \left. +2 s \left(2 m_\mu^2-s_1\right) \left(m_\mu^2-t_1\right) (s-s_1+t_0+t_1-u_1)\right.\right.\nonumber\\
& & \left. \left. +s \left(s_1-2 m_\mu^2\right) (s+t_1-u_1)^2+s \left(s_1-2 m_\mu^2\right) \left(m_\mu^2-t_1\right)^2\right]\right\rbrace \,,
\end{eqnarray}}
where we have neglected the electron mass. Since the flavor facilities can measure this cross-section at very small values of $k^2$ --close to the threshold of 
$(2m_\mu+m_\pi)^2$-- we kept $m_\mu\neq0$ and $m_\pi\neq0$ in eq.~(\ref{m.e. summed and averaged invariants}) as we have done in the numerics.
The cross-section can be written \cite{Kumar:1970cr}
\begin{eqnarray}
\sigma&=\frac{1}{2^7\pi^4s^2}&\int_{4m_\mu^2}^{(\sqrt{s}-m_\pi)^2}\frac{\mathrm{d}s_1}{\lambda^{1/2}(s,s_1,m_\pi^2)}\int_{t_0^-}^{t_0^+}\frac{\mathrm{d}t_0}{\sqrt{1-\xi^2}}\nonumber\\
& &\qquad\int_{u_1^-}^{u_1^+}\frac{\mathrm{d}u_1}{\lambda^{1/2}(s,m_\mu^2,u_1)\sqrt{1-\eta^2}}\int_{t_1^-}^{t_1^+}\frac{\mathrm{d}t_1\overline{\Big|\mathcal{M}\Big|^2}}{\sqrt{1-\zeta^2}}\,,
\end{eqnarray}
with the definitions
{\small\begin{eqnarray}
 \zeta & =& (\omega-\xi\eta)\left[(1-\xi^2)(1-\eta^2)\right]^{-1/2}\,\\ \omega&=&(s-m_\mu^2-u_1+2t_1)\lambda^{-1/2}(s,m_\mu^2,u_1)\,,\\
 \eta & =& \left[2s s_1-(s+m_\mu^2-u_1)(s+s_1-m_\pi^2)\right]\lambda^{-1/2}(s,m_\mu^2,u_1)\lambda^{-1/2}(s,s_1,m_\pi^2)\,,\\\quad \xi&=&\frac{s-m_\pi^2-s_1+2t_0}{\lambda^{1/2}(s,s_1,m_\pi^2)}\,,\nonumber
\end{eqnarray}}
and the $t_0$, $u_1$ and $t_1$ integration limits
\begin{eqnarray}
 t_0^{\pm}& = & m_\pi^2-\frac{s+m_\pi^2-s_1}{2}\pm\frac{\lambda^{1/2}(s,m_\pi^2,s_1)}{2}\,,\nonumber\\
 \; u_1^{\pm}&=&s+m_\mu^2-\frac{s+s_1-m_\pi^2}{2}\pm\frac{\sqrt{s_1(s_1-4m_\mu^2)\lambda(s,s_1,m_\pi^2)}}{2s_1}\nonumber\\
 t_1^{\pm}& = & m_\mu^2-\frac{s+m_\mu^2-u_1}{2}+\frac{\lambda^{1/2}(s,m_\mu^2,u_1)}{2}\left[\xi\eta\pm\sqrt{(1-\xi^2)(1-\eta^2)}\right]\,.
 \end{eqnarray}
  
  \begin{figure}[ht!]
   \centering\includegraphics[scale=0.5, angle=-90]{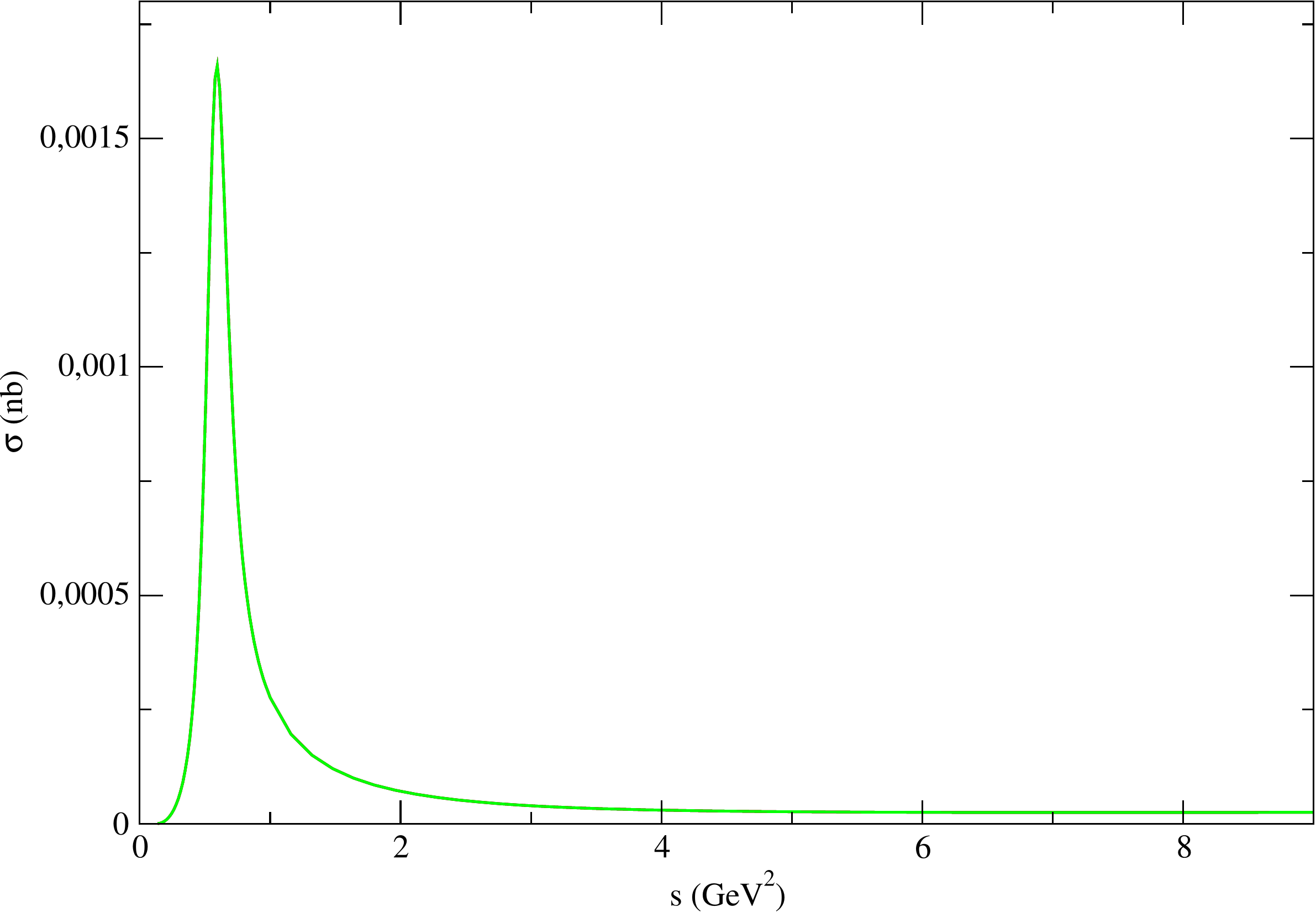}\caption{Our prediction for $\sigma(e^+e^-\to\mu^+\mu^-\pi^0)$ at different center of mass 
   energies using the couplings in eq. \ref{Couplings piggFF real pion}.}\label{sigmass}
  \end{figure}

\begin{figure}[ht!]
   \centering\includegraphics[scale=0.5, angle=-90]{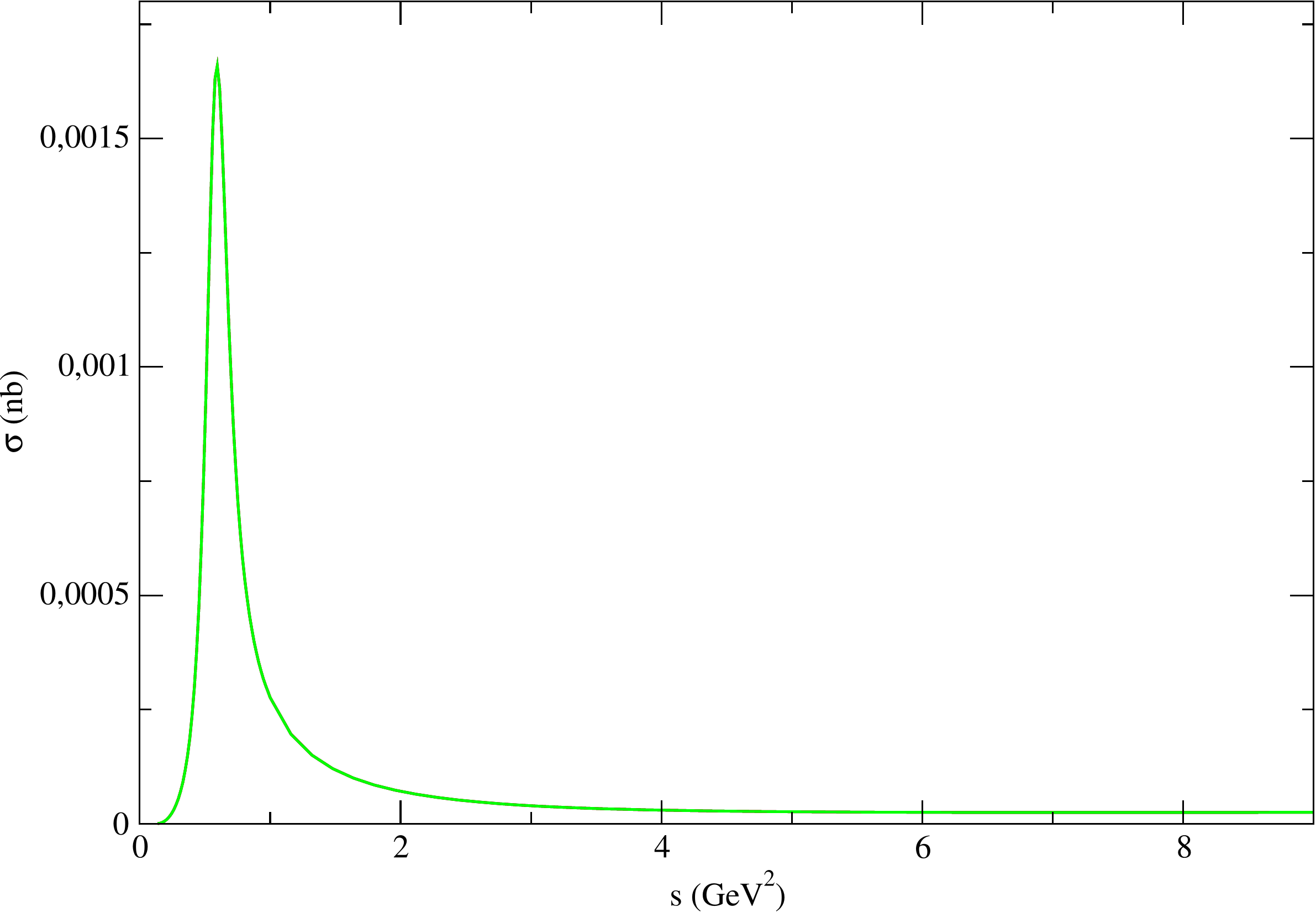}\caption{Our prediction for $\mu^+\mu^-$ distribution at $s=(1.02\text{ GeV})^2$  
   using the couplings in eq. \ref{Couplings piggFF real pion}.}\label{sigmas1}
  \end{figure}

  We therefore obtain the cross section for different center of mass energy shown in figure \ref{sigmass}, and the cross section dependence on $s_1$ with 
  $s=(1.02\text{ GeV})^2$ in fig. \ref{sigmas1}, since at that center of mass energy operates the KLOE experiment. By varying the parameters an upper 
  and a lower bound is obtained for both figures, however, the central value almost overlaps with both limits cannot be discerned in both figures.
  The $\rho(770)$ peak shows neatly and, at higher energies the cross section approaches a plateau. The excitations of the $\rho(770)$ and their associated 
  uncertainties are negligible. The profile of the $\frac{d\sigma}{d1}$ observable makes it appealing for its measurement at KLOE, this is why we show 
  its spectrum at $\sqrt{s}=M_\phi$ GeV. Although the plot for $M_{\Upsilon(4s)}$ is not shown, it would be very valuable to measure the behavior of 
  the TFF at high virtualities of both photons to check the predicted asymptotic behaviors. This process would provide complementary information to that 
  of BaBar and Belle, constraining the whole problematic mixed soft-hard regions needed to compute the internal vertex (the one which does not include 
  the real photon) of the diagrams in fig \ref{pionpole}. 

\begin{figure}
   \centering\includegraphics[scale=0.3, angle=-90]{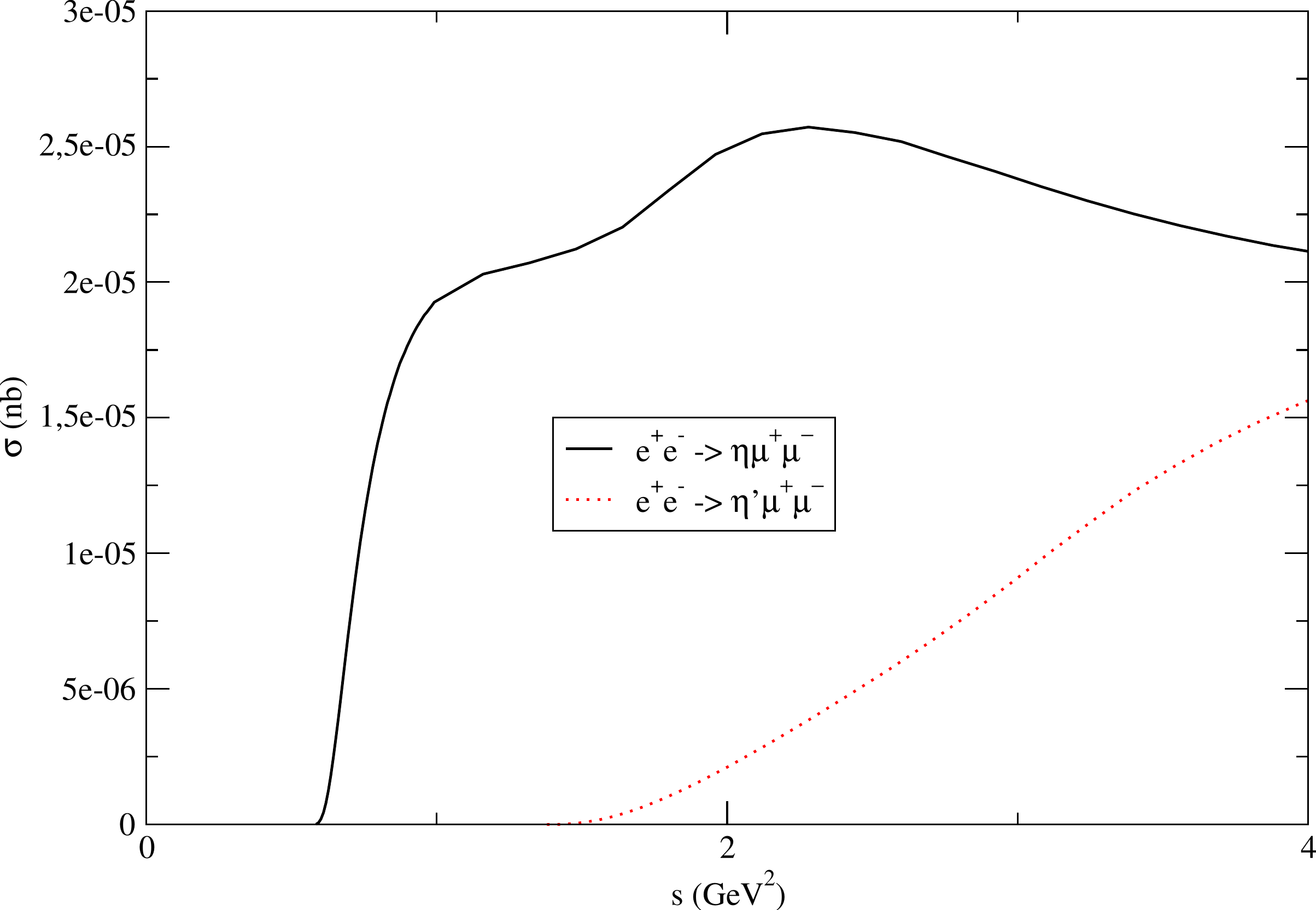}\includegraphics[scale=0.3,angle=-90]{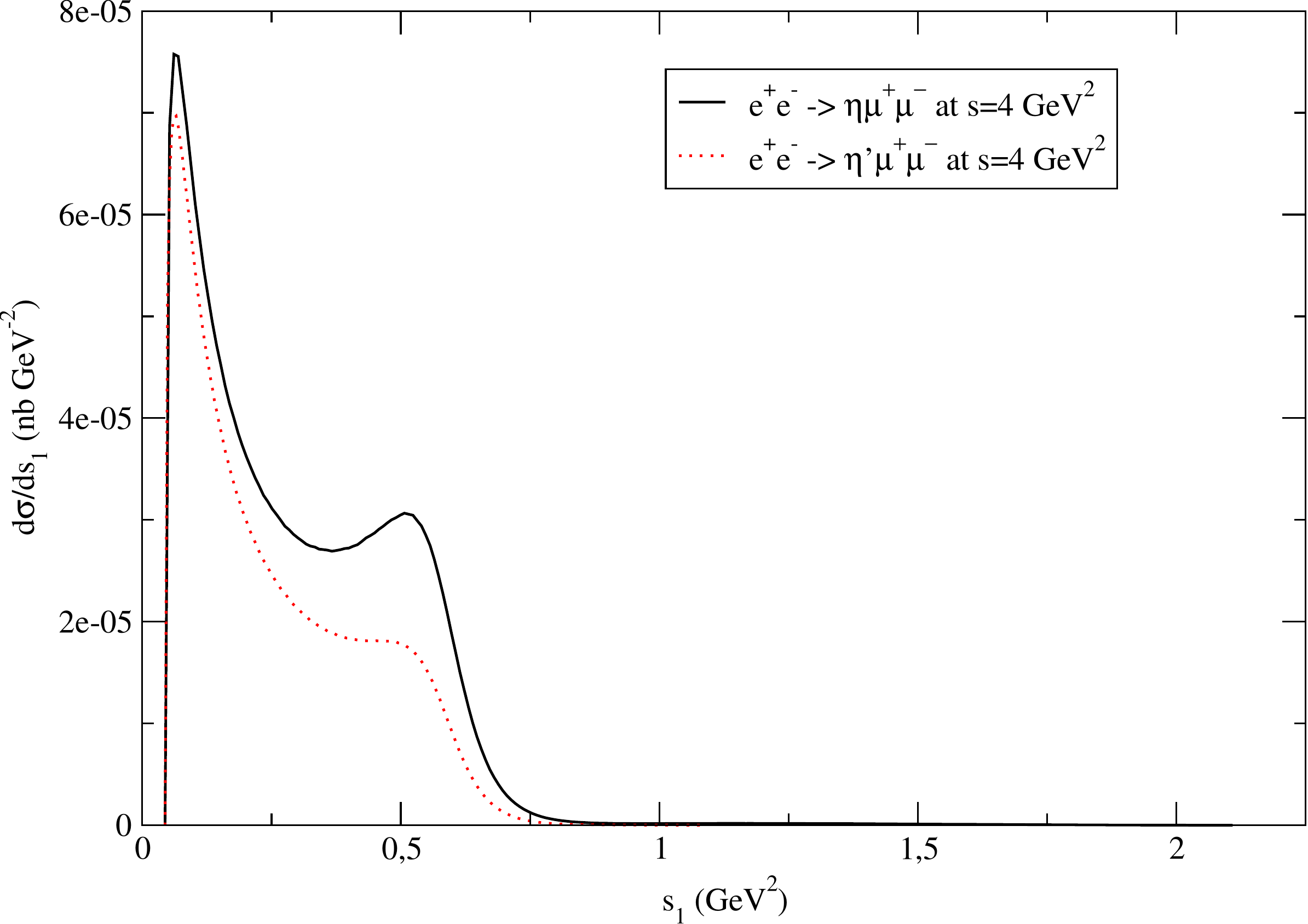}\caption{Our prediction for 
   the $\sigma(e^+e^-\to\mu^+\mu^-\eta)$ cross section (left) and $\mu^+\mu^-$ distribution at 4 GeV$^2$ (right).}\label{sigma eta}
  \end{figure}

  Using the relations between $\eta^{(\prime)}$TFF and the $\pi$TFF shown below in section \ref{etaetap}, we computed the $\sigma(e^+e^-\to\mu^+\mu^-\eta)$
  cross section in a completely analogous way to the pion case. The total cross section and $\frac{d\sigma}{ds_1}$ distribution are shown in fig. \ref{sigma eta}
  at $s=M_\phi^2$.
  The effect of the contribution of higher excited states is negligible in $\frac{d\sigma}{ds_1}$ and is at the same level induced by the 
  uncertainties on the $\eta-\eta'$ mixing in the cross-section plot. They are of order 30(20)\% for the $\eta(\eta^\prime)$ cases.
  The $\eta$ distribution at this energy will be less prominent and no hadronic structure will be appreciated since the available phase space is not 
  enough, while the $\eta'$ cannot even be produced.
  
  \section{Conclusions}
  
  A previous analysis of these TFF given by \cite{Kampf:2011ty} confirms the expressions obtained by us of the completely off-shell TFF, however 
  the main difference between our work \cite{We 2013} and the previous one comes from the more robust analysis of the error. This stems from the
  low energy behavior of the TFF, giving a correction to the classic WZW functional result which gives the dominant uncertainty in the $a_\mu^{\pi^0,HLbL}.$
  Also, we use a high energy constraint and Belle data, which appeared after the publication of ref \cite{Kampf:2011ty}.\\
  
  We have made a proposal of a new observable, namely the $e^+e^-\to \mu^+\mu^-\pi^0$ cross section and its dependence on the di-muon invariant mass, that 
  would give us relevant information of the $\pi$TFF in the whole problematic soft-hard regions needed to fully describe the TFF for two off-shell photons, 
  complementary to the $e^+e^-\to e^+e^-\pi^0$ data. These observables could be measured in Belle-II and KLOE.\\
  
  \begin{table}[ht!]
  \centering
   \begin{tabular}{|c|c|}\hline
    $a_\mu^{HLbL}\cdot10^{10}$ & Contributions\\\hline
    $11.6\pm4.0$&F. Jegerlehner y A. Nyffeler \cite{Jegerlehner:2009ry}\\
    $10.5\pm2.6$&Prades, De Rafael y Vainshtein \cite{Prades:2009tw}\\
    $13.7^{+2.7}_{-1.5}$&Erler and Toledo S\'anchez \cite{Erler:2006vu}\\
    $11.8\pm2.0$&Our contribution\cite{We 2013}\\\hline
   \end{tabular}\caption{Our contribution to the $a_\mu^{HLbL}$ compared to previous computations.}\label{wholePTFF}
 \end{table}

  We have obtained a total pseudoscalar exchange contribution $a_\mu^{P,HLbL}=(10.47\pm0.54)\times10^{-10}$ which is in good agreement with those 
  reported in previous analysis, namely $(9.9\pm1.6)\times10^{-10}$ \cite{Jegerlehner:2009ry} and $(11.4\pm1.3)\times10^{-10}$  
  \cite{Prades:2009tw}. The contribution for the whole HLbL is obtained by adding the rest of the contributions in fig \ref{HLBL}, which are obtained from ref 
  \cite{Jegerlehner:2009ry}, giving $a_\mu^{HLbL}=(11.8\pm2.0)\times10^{-10}$. The previous result is compared with previous reported analyses
  in table \ref{wholePTFF}.
  
  We can see that there is a very good agreement within errors with both results shown. Adding in quadrature the errors from the LO HVP and the 
  our HLbL contribution instead of that in ref. \cite{Nyffeler:2009tw}, we see that the uncertainty is $\pm5.1\times10^{-10}$, which is a 16\%
  smaller.
  
\chapter{Conclusions}

   We have computed several processes relevant for the search of BSM interactions due to the precision test that  
   will be done in high intensity frontier experiments in the very near future. Since most of the processes 
   studied in these facilities involve hadronic effects with an energy scale in which pQCD cannot give a reliable
   result, a better alternative to deal with this problem is the use of R$\chi$T as an effective field theory of QCD.\\

   By relying on the chiral symmetry of the fundamental field theory it is possible to construct an Effective Field Theory 
   that has been used to compute processes that otherwise is not possible to calculate. These theories have a 
   very wide range of applicability, meaning they can be used in any problem below certain energy scale so that 
   one can consistently compute amplitudes in a perturbative way, and even renormalize the theory to get further precision 
   in its predictions.\\
   
   In chapter 1 we showed the first computation of the $BR(\tau\to\pi\ell^+\ell^-\nu_\tau)$ decays, which gives us an
   effect that will be measurable at Belle-II. This process is a very important background for processes with lepton 
   number or lepton flavor violation. It may be an important background in the search for processes with lepton universality
   violation, since it gives different decay widths for different decay channels. \\
   
   This induced non-universality of leptons took us to try to explain the $R_K$ anomaly measured at LHCb, which comes 
   to be an observable free from hadronic-pollution in certain energy ranges, where we found that the energy range probed by LHCb for this 
   observable is free from hadronic pollution, giving a very clean window for the search of phenomena beyond the SM. 
   The different strong and weak phases led us to calculate a CP asymmetry that is large and in agreement with experimental data.\\
   
   Looking forward to improve the experimental constraints on new charged currents, we studied the decay $\tau\to\pi\eta\gamma\nu_\tau$, 
   which is a very important background for the search of this kind of currents. By cutting the energy of the photon and taking 
   values below that cut the background can be drastically reduced in one order of magnitude. Despite the fact that this would be enough
   for the $\eta$ channel, it is not so for the $\eta'$ channel since it cannot be determined due to the great uncertainty of the 
   non radiative process.\\
   
   We also found the important result that neglecting the contributions from two resonances exchange gives a very good estimation and 
   also, a significantly reduced uncertainty. The importance of this results rely in the fact that these form factors will be 
   inserted in the TAUOLA generator, so that this approximation will significantly simplify the codes (the computation time and 
   the uncertainty) to make them more efficient.\\
   
   Looking to provide a cleaner stage for the search of BSM effects, we computed the HLbL contributions to the $a_\mu$, where we 
   managed to reduce the uncertainty and at the same time, give a more robust estimation of the uncertainty. Since a full 
   description of the TFF will not be given by the current data fits, we proposed the measurement of (and predicted) the cross section
   $e^+e^-\to\mu^+\mu^-\pi^0$ and the invariant muon mass spectrum at a fixed center of mass energy, so that the TFF can be fitted 
   to an observable which involves both photons off-shell.
   
   \pagebreak
   
   {\huge Products derived from our research}\\
   
   We have published the following works: 
   
   \begin{itemize}
    \item {
  P. Roig, A. Guevara, G. L\'opez Castro, {\it Weak radiative pion vertex in $\tau^-\to\pi^-\nu_\tau\ell^+\ell^-$ decays}, Phys. Rev. D{\bf88} (2013) 033007
    }
    \item{
  P. Roig, A. Guevara, G. L\'opez Castro, {\it $VV'P$ form factors in resonance chiral theory and the $\pi-\eta-\eta'$
  light-by-light contribution to the muon g-2}, Phys. Rev. D{\bf89}  (2014) 073016    
    }
    \item{
  A. Guevara, G. L\'opez Castro, P. Roig and S. L. Tostado, {\it Long-distance weak annihilation contribution to the 
  $B^\pm\to(\pi^\pm,K^\pm)\ell^+\ell^-$ decays}, Phys. Rev. D{\bf92} (2015) 054035
    }
    \item{
  A. Guevara, G. L\'opez Castro, P Roig, {\it $\tau^-\to\eta^{(\prime)}\pi^-\nu_\tau\gamma$ decays as backgrounds in the search 
  for second class currents}, Phys. Rev. D{\bf95} (2017) 054015
    }
    \item{
  A. Guevara, {\it $W^*\gamma^*\pi$ form factors in the in $\tau^-\to\pi^-\nu_\tau\ell^+\ell^-$ decays}, 
  J. Phys. Conf. Ser. {\bf 761} (2016) 012088,\\ presented as a poster at the XIV Mexican Workshop on Particle Physics in November 2013
    }
    \item{
  A. Guevara, G. L\'opez Castro, P. Roig and S. L. Tostado, {\it A new long distance contribution to $B^\pm\to(\pi^\pm,K^\pm)\ell^+\ell^-$ decays}, Moriond QCD (2015) Conf. Proc. 107-110,\\ talk given by the first author 
  at the $50^\text{th}$ Rencontres de Moriond on QCD and High Energy Interactions in march 2015
    }
    \item{
  A. Guevara, {\it Hadronic light-by-light contribution to the muon g-2} J. Phys. Conf. Ser. {\bf 761} (2016) 012009,\\ talk given at the 30$^\text{th}$ Annual Meeting of the Particles and Fields division in may 2016
    }
   \end{itemize}

\addcontentsline{toc}{chapter}{Appendices}

\newcommand{\ko}{\boldsymbol{k}_1}
\newcommand{\ktw}{\boldsymbol{k}_2}
\newcommand{\kth}{\boldsymbol{k}_3}
\newcommand{\kot}{\boldsymbol{k}_{12}}
\newcommand{\ktt}{\boldsymbol{k}_{23}}
\newcommand{\Q}{\boldsymbol{Q}}
\newcommand{\QQ}{\boldsymbol{Q}^2}
\newcommand{\Qm}{\|\boldsymbol{Q}\|}

 \section*{Appendix A: Kinematics for cross section $\sigma(k k'\to k_1 k_2 k_3)$}
 
    \subsection*{Introduction}
    
     As part of an almost finished work with the Ph.D. student Bryan Larios (student of Dr. Lorenzo D\'iaz Cruz), the author computed the phase space for 
     a cross section of two-to-three particles as a function of two-particle 
     invariant masses. This was done to compute the cross section of $e^-(k)e^+(k')\to \tilde{G}(k_1)\tilde{G}(k_2)\gamma(k_3)$, where $\tilde{G}$ is a 
     gravitino as a function of the mass of the gravitino, of a virtual neutralino and the energy spectrum of the photon. 
     This is a simplification to the case where the amplitude does not depend on the angle between $\ktw$ and $\boldsymbol{k}$.
     
     \subsection*{Differential cross section}
     
     The general expression for the differential cross section can be obtained from Quantum Field Theory books, in the following we will use 
     Peskin and Schroeder definition \cite{Peskin}
     \begin{equation}\label{dsigma}
      d\sigma=\frac{1}{2E2E'|\boldsymbol{v}-\boldsymbol{v}'|}\left(\prod_f^3\frac{d^3k_f}{2E_f}\right)|\mathcal{M}|^2 (2\pi)^4\delta^{(4)}(k+k'-\Sigma k_f),
     \end{equation}

     where $k$ and $k'$ are the electron and positron four-momentum and $k_f$ are the four-momenta of the final state particles with mass $m_f$ and 
     energy $E_X$. The term in the denominator can be expressed as a function of the center of mass energy and the masses of the initial state particles 
     using the relation $\boldsymbol{v}_X=\boldsymbol{k}_X/E_X$, with $\boldsymbol{k}$ the three-momentum,
     \begin{multline}\label{speed}
      2E\cdot2E'|\boldsymbol{v}-\boldsymbol{v}'|=4EE'\left|\frac{\boldsymbol{k}}{E}-\frac{\boldsymbol{k}'}{E'}\right|=4EE'\left|\frac{\boldsymbol{k}}{E}+
      \frac{\boldsymbol{k}}{E'}\right|\\
      \hspace*{32ex}=4|\boldsymbol{k}|(E+E')=4E_{CM}|\boldsymbol{k}|,
      \\
     \end{multline}
     where $E_{CM}$ is the center of mass energy. For initial state particles with same mass, the energy of both must be equal in the center of 
     mass reference frame (CM), since from the dispersion relation one has $|\boldsymbol{k}|=\sqrt{E^2-m^2}=\sqrt{E'^2-m'^2}=$ and $E+E'=E_{CM}$. Now,
     if initial state particles have equal masses $|\boldsymbol{k}|$ can be expressed as a function of the center of mass energy and the mass of 
     initial state particles
     \begin{equation}
      |\boldsymbol{k}|=\sqrt{E^2-m^2}=\frac{1}{2}\sqrt{E_{CM}^2-4m^2}=\frac{E_{CM}}{2}\sqrt{1-\frac{4m^2}{E_{CM}^2}}=\frac{E_{CM}}{2}\sigma_m(E_{CM}^2)
     \end{equation}
      where $\sigma_X(p^2)=\sqrt{1-\frac{4X^2}{p^2}}$, being $m_X$ the mass of particle $X$. And thus, the factor can be expressed as 
     \begin{equation}\label{fator}
      \boxed{2E\cdot2E'|\boldsymbol{v}-\boldsymbol{v}'|=2Q^2\sigma_m(Q^2)},
     \end{equation}
     where $Q=k+k'$, {\it i.e.}, $Q^2=E_{CM}^2$.\\
     

     To express the cross section as a function of the invariant masses of particles with four-momenta $k_1-k_2$ and $k_2-k_3$, we introduce the variables
     $k_{12}$ and $k_{23}$, and the invariant masses $s=(k_1+k_2)^2$ and $t=(k_2+k_3)^2$ by using integrals which are trivially equal to 1,
     \begin{subequations}
      \begin{align}
       &\int \delta^{(4)}(k_{12}-k_1-k_2)d^4k_{12},\\
       &\int \delta^{(4)}(k_{23}-k_2-k_3)d^4k_{23},\\
       &\int \delta\left[s-(Q-k_3)^2\right]ds,\\
       &\int \delta\left[t-(Q-k_1)^2\right]dt,
      \end{align}
     \end{subequations}

     So that the differential cross section takes the form 
     \begin{equation}
      d\sigma(Q^2)=\frac{|\mathcal{M}|^2}{2Q^2\sigma_m(Q^2)}\frac{1}{(2\pi)^5}Idsdt,
     \end{equation}
     where $I=I_1I_2$, and
     \begin{subequations}
     \begin{align}
      I_1&=\int d^4k_{12}\frac{d^3k_3}{2E_3}\delta\left[s-(Q-k_3)^2\right]\delta^{(4)}(Q-k_{12}-k_3),\\
      I_2&=\int d^4k_{23}dsdt\frac{d^3k_1d^3k_2}{4E_1E_2}
      \delta^{(4)}(k_{12}-k_1-k_2)\delta^{(4)}(k_{23}-k_2-k_3)\delta\left[t-(Q-k_1)^2\right].
     \end{align}
     \end{subequations}

     Since $\frac{d^3k_3}{2E_3}=\int d^4k\delta(k_3^2-m_3^2)$, where $m_3$ is the mass of the particle with $k_3$ four-momentum with positive time component,
     $I_1$ is a Lorentz scalar and can be computed in any reference frame. By choosing the CM, one gets
     \begin{multline}\label{I1}
      I_1=\int\frac{d^3k_3}{2E_3}dk^0_{12}d^3k_{12}\delta\left(\sqrt{Q^2}-k_{12}^0-E_3\right)\delta^{(3)}(\kot+\kth)\delta(s-{k_{12}^0}^2-\kot^2)\hspace*{5ex}\\
      =\int \frac{d^3k_3}{2E_3}\frac{dk^0_{12}d^3k_{12}}{2\sqrt{\kot^2+s}}\delta\left(\sqrt{Q^2}-\sqrt{\kot^2+s}-E_3\right)\delta^{(3)}(\kot+\kth)
      \delta\left(k_{12}^0-\sqrt{\kot^2+s}\right),\hspace*{8ex}\\
      =\int \frac{d^3k_3}{2\sqrt{\kth^2+m_3^2}}\frac{\|\kot\|^2d\|\kot\|d\Omega_{12}}{2\sqrt{\kot^2+s}}\delta\left(\sqrt{Q^2}-\sqrt{\kot^2+s}-\sqrt{\kth^2+m_3^2}\right)
      \delta^{(3)}(\kot+\kth),\\
      =\int\frac{4\pi}{2\sqrt{\kot^2+m_3^2}}\frac{\|\kot\|^2d\|\kot\|}{2\sqrt{\kot^2+s}}\delta\left(\sqrt{Q^2}-\sqrt{\kot^2+s}-\sqrt{\kot^2+m_3^2}\right),\hspace*{24ex}
     \end{multline}
     In the second line we used the delta function property $\delta[f(x)]=\sum_i\frac{1}{|f(x_i)|}\delta(x-x_i)$, where $x_i\in$ ker $f$. In the third line 
     $d\Omega_{12}$ is the differential solid angle subtended by $d\|\kot\|$.For the delta function in the 
     last expression we get the following expression
     \begin{equation}\label{whatever}
      \delta\left(\sqrt{Q^2}-\sqrt{\kot^2+s}-\sqrt{\kot^2+m_3^2}\right)=\left(\frac{\|\kot\|}{\sqrt{\kot^2+s}}+\frac{\|\kot\|}{\sqrt{\kot^2+m_3^2}}\right)^{-1}
      \delta\left(\|\kot\|-\|\kot\|_0\right),
     \end{equation}
     where $\|\kot\|_0$ is found as the root of the function of $\|\kot\|$ in the delta function. This root is found analytically in the following way
     \begin{multline}\label{bloodyroots}
     \sqrt{Q^2}-\sqrt{\kot^2+s}-\sqrt{\kot^2+m_3^2}=0\hspace*{18ex}\\
     \left(\sqrt{Q^2}-\sqrt{\kot^2+s}\right)^2=\kot^2+m_3^2\hspace*{28ex}\\
     Q^2-2\sqrt{Q^2}\sqrt{\kot^2+s}+ \kot^2 +s=\kot^2+m_3\hspace*{36ex}\\
     Q^2+s-m_3^2=2\sqrt{Q^2}\sqrt{\kot^2+s}\hspace*{11ex}\\
     \left(\frac{Q^2+s-m_3^2}{2\sqrt{Q^2}}\right)^2=\|\kot\|_0^2+s\frac{4Q^2}{4Q^2}\hspace*{17ex}\\
     \hspace*{27ex}\|\kot\|_0^2=\frac{Q^4+s^2+m_3^4+2Q^2s-2Q^2m_3^2-2sm_3^2-4Q^2s}{4Q^2}\\
     =\frac{\lambda(Q^2,s,m_3^2)}{4Q^2}\\
     \|\kot\|_0=\frac{\lambda^{1/2}(Q^2,s,m_3^2)}{2\sqrt{Q^2}},\hspace*{25ex}
     \end{multline}
     where the positive root has been chosen, since otherwise $\|\cdot\|$ would not be a norm. Here $\lambda(a,b,c):=a^2+b^2+c^2-2ab-2ac-2bc$ is the 
     K\"allen function. Now, the factor in parenthesis in eq (\ref{whatever}) can be expressed in a more simple manner
     \begin{equation}
      \left(\frac{\|\kot\|}{\sqrt{\kot^2+s}}+\frac{\|\kot\|}{\sqrt{\kot^2+m_3^2}}\right)^{-1}=
      \frac{\frac{1}{\|\kot\|}\sqrt{\kot^2+s}\sqrt{\kot^2+m_3^2}}{\sqrt{\kot^2+m_3^2}+\sqrt{\kot^2+s}}=
      \frac{\sqrt{\kot^2+s}\sqrt{\kot^2+m_3^2}}{\sqrt{Q^2}\|\kot\|},
     \end{equation}
     where, in the last line we made use of the fact that $\|\kot\|$ must be a root of the argument of the delta function in the lhs of eq (\ref{whatever}). 
     Thus, by inserting the previous expression in eq. (\ref{I1}), substituing the root $\|\kot\|_0$ and trivially integrating over $d\|\kot\|$ we get
     
     \begin{equation}
      \boxed{I_1=\frac{\pi}{2}\frac{\lambda^{1/2}(Q^2,s,m_3^2)}{Q^2}}
     \end{equation}\\
     
     By the same argument used for $I_1$, it is showed that $I_2$ is Lorentz invariant, so that we can choose a reference frame to integrate it. 
     For the sake of simplicity we used the reference frame in which $\kot=0$, so that $k_{12}^2=(k_{12}^0,\boldsymbol{0})^2=s$ and therefore
     $k_{12}=(\sqrt{s},\boldsymbol{0})$. In order to compute the integrals of $I_2$ we need to obtain an expression for $\Qm$ and $E=\sqrt{Q^2+\QQ}$ 
     in this reference frame. Since $\boldsymbol{Q}=\gamma m\boldsymbol{v}$ is a three-momentum, $E=\gamma m$ is the time component of the 
     four-momentum vector $Q=(E,\boldsymbol{Q})$, where $m=\sqrt{Q^2}$ and $\gamma$ is the Lorentz factor associated with $\boldsymbol{v}$, the relative 
     velocity between this frame and the CM one. Therefore, the relative speed is
     \begin{equation}\label{veso}
      v:=\|\boldsymbol{v}\|=\frac{\Qm}{E}.
     \end{equation}
     By squaring $k_3=Q-k_{12}$ one can find that $Q^2+s-2Ek_{12}^0=m_3^2$, and by clearing for $E$ we find
     \begin{equation}\label{Eso}
      E=\frac{Q^2+s-m_3^2}{2\sqrt{2}},
     \end{equation}
     and since $Q^2=E^2-\QQ$, we get
     \begin{equation}\label{Queso}
      \Qm=\frac{\lambda^{1/2}(Q^2,s,m_3^2)}{2\sqrt{s}},
     \end{equation}
     so that $Q^T=\Qm(0,0,-1)$, since we have chosen $\kot$ to be parallel to the third direction in the CM reference frame. This last expression will 
     help us later to express contractions of four-momenta in terms of Lorentz scalars. Thus, now we can follow a development for $I_2$ similar to that 
     of $I_1$,
     \begin{multline}
      I_2=\int\frac{d^4k_{23}d^3k_1d^3k_2}{4E_1E_2}\delta^{(4)}(k_{12}-k_1-k_2)\delta^{(4)}(k_{23}-k_{12}-Q+2k_1)\delta(t-k_{23}^2)\\
         =\int\frac{d^4k_{23}d^3k_1d^3k_2\delta^{(3)}(\ko+\ktw)}{4E_1\sqrt{\ktw^2+m_2^2}}\delta\left(\sqrt{s}-E_1-\sqrt{\ktw^2+m_2^2}\right)\delta^{(4)}(\cdots)
         \delta(t-k_{23}^2)\\
         =\int\frac{d^4k_{23}d^3k_1}{4\sqrt{\ko^2+m_1^2}\sqrt{\ko^2+m_2^2}}\delta\left(\sqrt{s}-E_1-\sqrt{\ko^2+m_2^2}\right)\delta^{(4)}(\cdots)
         \delta(t-k_{23}^2).\hspace*{1ex}
     \end{multline}
     Following the procedure used in eqs. (\ref{whatever}) and (\ref{bloodyroots}), we find that 
     \begin{equation}\label{root1}
      \delta\left(\sqrt{s}-\sqrt{\ko^2+m_1^2}-\sqrt{\ktw^2+m_2^2}\right)=\frac{\sqrt{\ko^2+m_1^2}\sqrt{\ko^2+m_2^2}}{\sqrt{s}\|\ko\|}
      \delta\left(\|\ko\|-\frac{\lambda^{1/2}(s,m_1^2,m_2^2)}{2\sqrt{s}}\right).
     \end{equation}
     Substituing the delta function by the expressions found in the previous equation in the $I_2$ integrals we get
     \begin{multline}
     I_2=\int d^4k_{23}\frac{\|\ko\|d\|\ko\|d\Omega_1}{4\sqrt{s}}\delta(\|\ko\|-\|\ko\|_0)\delta(t-k_{23}^2)\delta^{(4)}(k_{23}-k_{12}-Q+2k_1)\\
     =\int d^4k_{23}d\Omega_1\frac{\|\ko\|_0}{4\sqrt{s}}\delta(t-k_{23}^2)\delta^{(4)}(k_{23}-k_{12}-Q+2k_1)\hspace*{22ex}\\
     =\frac{\lambda^{1/2}(s,m_1^2,m_2^2)}{8s}\int \frac{dk^0_{23}d^3_{23}}{2\sqrt{\ktt^2+t}}d\Omega_1\delta\left(k_{23}^0-\sqrt{\ktt^2+t}\right)\times\hspace*{18ex}\\
     \delta\left(k_{23}^0-\sqrt{s}-E+2\frac{s+m_1^2+m_2^2}{2\sqrt{s}}\right)\delta^{(3)}(\ktt-\Q+2\ko)\\
     =\frac{\lambda^{1/2}(s,m_1^2,m_2^2)}{16s}\int\frac{d^3k_{23}}{\sqrt{\ktt^2+t}}d\Omega_1\delta\left[\sqrt{\ktt^2+t}-\sqrt{s}-E+
     \frac{s+m_1^2-m_2^2}{\sqrt{s}}\right]\times\\
     \hspace*{52ex}\delta^{(3)}(\ktt-\Q+2\ko)\\
     =\frac{\lambda^{1/2}(s,m_1^2,m_2^2)}{16s}\int d\Omega_1\frac{\delta\left[\sqrt{(\Q-2\ko)^2+t}-\sqrt{s}-E+
     \frac{s+m_1^2-m_2^2}{\sqrt{s}}\right]}{\sqrt{(\Q-2\ko)^2+t}}\hspace*{8ex}\\
     \end{multline}
     In the third line we used the relation $E_1=\sqrt{\ko^2+m_1^2}$ along with the $\|\ko\|=\|\ko\|_0$ from eq. (\ref{root1}). By realizing the expansion
     $(\Q-2\ko)^2=4\ko^2+\Q^2-4\|\ko\|\Qm\cos(\theta_1)$ in order to obtain the $\cos(\theta_1)$ dependance explicitly, 
     the derivative $\frac{d(\Q-2\ko)^2}{d\cos(\theta_1)}=-4\|\ko\|\Qm$ was obtained which will be needed to integrate the delta function. So, the delta 
     function can be expressed as 
     \begin{multline}
      \delta\left[\sqrt{(\Q-2\ko)^2+t}-\sqrt{s}-E+\frac{s+m_1^2-m_2^2}{\sqrt{s}}\right]\\
      =\delta\left[\sqrt{4\ko^2+\QQ-4\|\ko\|\Qm\cos(\theta_1)+t}-E-\frac{s+m_1^2-m_2^2}{\sqrt{s}}\right]\hspace*{25ex}\\
      =\left|\frac{-\|\ko\|\Qm}{\sqrt{(\Q-2\ko)^2+t}}\right|^{-1}\delta\left\{\cos(\theta_1)-\frac{\left[(Q^2+3s-m_3^2)\lambda^{1/2}(s,m_1^2,m_2^2)
      -s(2s+sQ^2-m_3^2)\right]}{4\Qm\|\ko\|s}
      \right\}\\
      =\frac{\sqrt{(\Q-2\ko)^2+t}}{4\Qm\|\ko\|}\delta\left[(\cos(\theta_1)-\cos(\theta_1)_0\right]\hspace*{42ex}\\
     \end{multline}
     Finally, substituing the delta function in the expression for $I_2$ and integrating trivially the azimuthal angle $\phi_1$ we get
     \begin{multline}
     I_2= \frac{\lambda^{1/2}(s,m_1^2,m_2^2)}{16s}\frac{2\pi}{\sqrt{(\Q-2\ko)^2+t}}\frac{\sqrt{(\Q-2\ko)^2+t}}{4\|\ko\|\Qm}\int d\cos(\theta_1)
     \delta\left[(\cos(\theta_1)-\cos(\theta_1)_0\right]\\
     =2\pi\frac{\lambda^{1/2}(s,m_1^2,m_2^2)}{16s}\frac{4s}{4\lambda^{1/2}(Q^2,s,m_3^2)\lambda^{1/2}(s,m_1^2,m_2^2)}\hspace*{23ex}
     \end{multline}
     And thus we get
     \begin{equation}
      \boxed{I_2=\frac{\pi}{8\lambda^{1/2}(Q^2,s,m_3^2)}}.
     \end{equation}
     Since $I=I_2I_2$, we get by using the boxed expressions for each integral
     \begin{equation}
      I=\frac{\pi^2}{16Q^2},
     \end{equation}
     and sustituing the value of $I$ and that of equation \ref{fator} we find the expression of the differential cross section as a function of 
     $s$ and $t$,
     \begin{equation}
      \boxed{\frac{d\sigma}{dtds}=\frac{|\mathcal{M}|^2}{1024\pi^3Q^4\sigma_m(Q^2)}}
     \end{equation}
     In order to be able to compute the total cross section for any process, the range of $s$ and $t$ must be specified. If we take $s$ as the 
     last-to-be-integrated variable we find its minimum $s_-$ by taking $\ko=\ktw=0$ and its maximum $s_+$ by taking $\kth=0$
     \begin{subequations}
      \begin{align}
       s_-=&(k_1+k_2)^2=(m_1+m_2)^2,\\
       s_+=&(Q-k_3)^2=(\sqrt{Q^2}-m_3)^2.
      \end{align}
     \end{subequations}
     We compute $t$ in the $\kot=0$ frame, so that
     \begin{equation}
     t=(Q-k_1)^2=(E-E_1)^2 -(\Q-\ko)^2,
     \end{equation}
     where
     \begin{equation}
     \left(\begin{array}{c}\max\\\min\end{array}\right)\left[(\Q-\ko)^2\right]=\QQ+\|\ko\|^2\pm2\Qm\|\ko\|=(\Qm\pm\|\ko\|)^2,
     \end{equation}
     Now, substituing the values for $E$, $E_1$, $\Qm$ and $\|\ko\|$ from eqs. (\ref{Eso}), (\ref{Queso}) and (\ref{root1}) in the expression for $t$ 
     we find
     \begin{multline}
     t_\pm=(E-E_1)^2-(\Qm\mp\|\ko\|)^2\\
     =\left(\frac{Q^2+s-m_3^2}{2\sqrt{s}}-\frac{s+m_1^2+m_2^2}{2\sqrt{s}}\right)^2-\left(\frac{\lambda^{1/2}(Q^2,s,m_3^2)}{2\sqrt{s}}-
     \frac{\lambda^{1/2}(s,m_1^2,m_2^2)}{2\sqrt{s}}\right)^2.
     \end{multline}
     So that the range of both variables are determined to be
     \begin{subequations}
      \begin{align}
      &\hspace*{19ex}\boxed{(m_1+m_2)^2\le s\le(\sqrt{Q^2}-m_3)^2}\\
      &\boxed{t_\pm=\frac{1}{4s}\left\{(Q^2+m_1^2-m_2^2-m_3^3)^2-\left[\lambda^{1/2}(Q^2,s,m_3^2)\mp\lambda^{1/2}(s,m_1^2m_2^2)\right]^2\right\}}
      \end{align}
     \end{subequations}
     The scalar products among final-state particles four-momenta are easily find from the definitions of $s$, $t$ and the relation $Q^2=(k_1+k_2+k_3)^2$.
     Since $s=(k_1+k_2)^2$,
     \begin{equation}
      k_1\cdot k_2=\frac{1}{2}(s-m_1^2-m_2^2).
     \end{equation}
     Analogously, from $t=(k_2+k_3)^2$
     \begin{equation}
      k_2\cdot k_3=\frac{1}{2}(t-m_2^2-m_3^2).
     \end{equation}
     By using the previous equations in $Q^2$ we can clear the remaining scalar product
     \begin{equation}
      k_1\cdot k_3=\frac{1}{2}(Q^2-s-t+m_2^2).
     \end{equation}
     To compute the scalar products between the initial state particles momenta and the final state paticles we must make a Lorentz boost to the $\kot=0$ 
     frame. First note that $Q^2=k^2+k'^2+2k\cdot k'$, so that $k\cdot Q=k'\cdot Q=Q^2/2$. The expressions for $v$, $E$ and $\Qm$ are given in eqs. (\ref{veso}), 
     (\ref{Eso}) and (\ref{Queso}) respectively. So that by expressing $k$ as perpendicular to $\kot$ in the CM reference frame and boosting it 
     to the $\kot=0$ frame we get
     \begin{equation}
      k^{CM\to12}\overset{\cdot}{=}\left(\begin{array}{cccc}\gamma&0&0&-\beta\gamma\\0&\hspace*{1ex}1\hspace*{1ex}&0&0\\0&0&\hspace*{1ex}1\hspace*{1ex}&0\\
      -\beta\gamma&0&0&\gamma\end{array}\right)\left(\begin{array}{c}\frac{1}{2}\sqrt{Q ^2}\\k^1\\k^2\\0\end{array}\right)=
      \left(\begin{array}{c}\frac{1}{2}\sqrt{Q ^2}\gamma\\k^1\\k^2\\\frac{1}{2}\sqrt{Q^2}\beta\gamma\end{array}\right),
     \end{equation}
     where $\beta$ is $v$ in eq. (\ref{veso}), and thus we must substitute the values of $E$ and $\Q$ en eqs. (\ref{Eso}) and (\ref{Queso}). Therefore,
     \begin{equation}
     \beta=\frac{\Qm}{E}=\frac{\lambda^{1/2}(Q^2,s,m_3^2)}{Q^2+s-m_3^2}=\left(1-\frac{4Q^2s}{(Q^2+s-m_3^2)^2}\right)^{1/2},
     \end{equation}
     from which we can obtain the Lorentz factor $\gamma=(1-\beta^2)^{-1/2}$
     \begin{equation}
      \gamma=\left[1-\left(1-\frac{4Q^2s}{(Q^2+s-m_3^2)^2}\right)\right]^{-1/2}=\frac{Q^2+s-m_3^2}{2\sqrt{Q^2s}}.
     \end{equation}
     And thus, the time component of the four momenta $k$ can be expressed in the following way
     \begin{equation}
      k^0=\frac{1}{2}\sqrt{Q^2}\frac{Q^2+s-m_3^2}{2\sqrt{Q^2s}}=\frac{Q^2+s-m_3^2}{4\sqrt{s}}.
     \end{equation}
     With this last expression, we find that
     \begin{equation}
      \boxed{k\cdot k_3=k\cdot Q-k\cdot k_{12}=\frac{1}{2}Q^2-2k^0\sqrt{s}=\frac{Q^2-s+m_3^2}{4}}.
     \end{equation}
     On the other hand, we will also be needing the expressions for $v$, $E$ and $\Qm$ in the $\ktt=0$ reference frame. So, in a completely analogous way 
     we find that 
     \begin{subequations}
      \begin{align}
       \Qm^*&=\frac{\lambda^{1/2}(Q^2+t-m_1^2)}{2\sqrt{t}},\\
       E^*&=\frac{Q^2+t-m_1^2}{2\sqrt{t}},\\
       \beta^*&=\left(1-\frac{4Q^2t}{(Q^2+t-m_1^2)^2}\right)^{1/2},\\
       \gamma^*&=\frac{Q^2+t-m_1^2}{2\sqrt{Q^2t}},
      \end{align}
     \end{subequations}
     so that the time component of the $k$ four-vector is
     \begin{equation}
      {k^0}^*=\frac{Q^2+t-m_1^2}{4\sqrt{t}}.
     \end{equation}
     Therefore,
     \begin{equation}
      \boxed{k\cdot k_1=k\cdot Q-k\cdot k_{23}=Q^2-2{k^0}^*\sqrt{t}=\frac{Q^2-t+m_1^2}{4}}.
     \end{equation}
     Finally, the last Lorentz invariant remaining is obtained by contracting the $k$ four-momentum with $Q$ and using the two previous contractions
     \begin{equation}
      \boxed{k\cdot k_2=k\cdot(Q-k_1-k_3)=\frac{1}{2}Q^2-\frac{Q^2-t+m_1^2}{4}-\frac{Q^2-s+m_3^2}{4}=\frac{s+t-m_1^2-m_2^2}{4}}.
     \end{equation}
     To obtain the cross section as a function of the energy of the final state particle with four-momentum $k_3$ and a mass invariant pair we first obtain 
     the dependence of $s$ in $E_3$,
     \begin{equation}\label{ese}
      s=(Q-k_3)^2=Q^2-2Q\cdot k_3+m_3^2=Q^2-2\sqrt{Q^2}E_3+m_3^2,
     \end{equation}
     so that $\frac{ds}{dE_3}=-2\sqrt{Q^2}$, with which we obtain
     \begin{equation}
      \boxed{\frac{d\sigma}{dEdt}=\left|\frac{ds}{dE_3}\right|\frac{d\sigma}{dsdt}=\frac{\sqrt{Q^2}\left|\mathcal{M}\right|^2}{512\pi^3Q^4\sigma_m(Q^2)}}.
     \end{equation}
     The kinematical limits for the energy $E_3$ are obtained by taking the two extremal cases where $\kth=0$ and when $\kot=0$. In the latter, eq. (\ref{ese})
     can be used by subsituing $s\to (m_1+m_2)^2$; in the former we just make use of the dispersion relation $E_3=\sqrt{\kth^2+m_3^2}$. Thus,
     \begin{equation}
      m_3\le E_3\le \frac{Q^2+m_3^2-(m_1+m_2)^2}{2\sqrt{Q^2}}.
     \end{equation}

 \section*{Appendix B: Contributions to the $\tau\to\pi\ell^+\ell^-\nu_\tau$ decay amplitude}\label{LongExpressions}
We collect in this appendix the results of summing over polarizations and averaging over that of the tau the different contributions to the squared matrix element. We refrain 
from writing the lengthy outcome of the contraction of the indices which was used in our programs.
\begin{eqnarray}\label{averaged IB}
& & \overline{\Big|\mathcal{M}_{IB}\Big|^2} = 16 G_F^2 |V_{ud}|^2 \frac{e^4}{k^4}F_\pi^2 M_\tau^2 \ell_{\mu\nu} \left[\frac{-\tau^{\mu\nu} k^2}
{\left(k^2-2 k\cdot p_\tau\right)^2}+\frac{4 p^{\mu} q^{\nu} k\cdot p_\tau}{\left(k^2+2 k\cdot p\right) \left(k^2-2 k\cdot p_\tau\right)}\right.\nonumber\\
& & \left. +\frac{4 p_\tau^{\mu} q^\nu k\cdot p_\tau}{\left(k^2-2 k\cdot p_\tau\right)^2}-\frac{2 g^{\mu \nu} k\cdot p_\tau k\cdot q}{\left(k^2-2 k\cdot p_\tau\right)^2}-\frac{4 p^{\mu} p_\tau^{\nu} k\cdot q}{\left(k^2+2 k\cdot p\right) \left(k^2-2 k\cdot p_\tau\right)}\right.\nonumber\\
& & \left. -\frac{4 p_\tau^{\mu} p_\tau^{\nu} k\cdot q}{\left(k^2-2 k\cdot p_\tau\right)^2}+\frac{8 p^{\mu} p_\tau^{\nu} p_\tau\cdot q}{\left(k^2+2 k\cdot p\right) \left(k^2-2 k\cdot p_\tau\right)}\right.\nonumber\\
& & \left. +\frac{4 p^{\mu} p^{\nu} p_\tau\cdot q}{\left(k^2+2 k\cdot p\right)^2}\,+\frac{4 p_\tau^{\mu} p_\tau^{\nu} p_\tau\cdot q}{\left(k^2-2 k\cdot p_\tau\right)^2}\right]\,,\nonumber\\
\end{eqnarray}
\begin{eqnarray}\label{averaged IBV}
& & \overline{2 \Re e\left[\mathcal{M}_{IB}\mathcal{M}_V^*\right]} = - 32  G_F^2|V_{ud}|^2 \frac{e^4}{k^4}F_\pi M_\tau^2 \Im m \left\lbrace F_V^*(p\cdot k,k^2)\ell^\mu_{\nu^\prime} \epsilon^{\mu^\prime \nu^\prime \rho^\prime \sigma^\prime}k_{\rho^\prime}p_{\sigma^\prime} \mathcal{V}_{\mu\mu^\prime}\right\rbrace\,,\nonumber
\end{eqnarray}
\begin{eqnarray}\label{averaged IBA}
& &  \overline{2 \Re e\left[\mathcal{M}_{IB}\mathcal{M}_A^*\right]} = -64 G_F^2 |V_{ud}|^2 \frac{e^4}{k^4}F_\pi M_\tau^2 \ell_\mu^{\nu^\prime} \Re e\left[ \mathcal{A}^*_{\mu^\prime\nu^\prime} \mathcal{V}^{\mu\mu^\prime}\right]\,,\nonumber
\end{eqnarray}
\begin{eqnarray}\label{averaged V}
& & \overline{ \Big|\mathcal{M}_{V}\Big|^2} = 16 G_F^2 |V_{ud}|^2 \frac{e^4}{k^4} \Big|F_V(p\cdot k,k^2)\Big|^2\epsilon_{\mu^\prime\nu^\prime\rho^\prime\sigma^\prime}\epsilon_{\mu\nu\rho\sigma}
k^\rho p^\sigma k^{\rho\prime} p^{\sigma\prime} \ell^{\nu{\nu^\prime}}\tau^{\mu{\mu^\prime}}\,,\nonumber
\end{eqnarray}
\begin{eqnarray}\label{averaged A}
& & \overline{\Big|\mathcal{M}_{A}\Big|^2} = 64 G_F^2 |V_{ud}|^2 \frac{e^4}{k^4}\ell_{\nu{\nu^\prime}}\tau_{\mu{\mu^\prime}}\mathcal{A}^{\mu\nu}{\mathcal{A}^{\mu^\prime\nu^\prime}}^*\,,\nonumber
\end{eqnarray}
\begin{eqnarray}\label{averaged VA}
& &  \overline{2 \Re e\left[\mathcal{M}_{V}\mathcal{M}_A^*\right]} = -64  G_F^2 |V_{ud}|^2 \frac{e^4}{k^4}\Im m\left[ F_V(p\cdot k,k^2)\epsilon_{\mu\nu\rho\sigma}k^\rho p^\sigma \ell_{\nu^\prime}^\mu \tau^{\mu\mu^\prime}{\mathcal{A}_{\mu^\prime}^{\nu^\prime}}^*\right]\nonumber\,,
\end{eqnarray} 
where we have defined
\begin{eqnarray}
 \ell^{\mu\nu} & = & p_-^\mu p_+^\nu+p_-^\nu p_+^\mu-g^{\mu\nu}(m_\ell^2+p_-\cdot p_+)\, \nonumber \\  \tau^{\mu\nu}& =& p_\tau^\mu q^\nu+p_\tau^\nu q^\mu-g^{\mu\nu}p_\tau \cdot q\,,\\
 \mathcal{A}^{\mu\nu}& = & F_A(p\cdot k,k^2)\left[(k^2+p\cdot k)g^{\mu\nu}-k^\mu p^\nu\right]+B(k^2) k^2 \left[g^{\mu\nu}-\frac{(p+k)^\mu p^\nu}{k^2+2p\cdot k}\right]\,,\nonumber\\
 \mathcal{V}_{\mu\nu} & = & \frac{2p_\mu q_{\nu}}{2k\cdot p+k^2}+\frac{-g_{\mu\nu}k\cdot q+2q_{\nu}p_{\tau\,\mu} -i\epsilon_{\mu\nu\rho\sigma}k^\rho q^\sigma+k_{\nu}q_\mu}{k^2-2k\cdot p_\tau}\, ,\nonumber
\end{eqnarray}\label{definitions}
and used the conservation of the electromagnetic currents implying $k_\mu\ell^{\mu\nu}=0=\ell^{\mu\nu}k_\nu$.



 \section*{Appendix C: Form factors results according to Resonance Chiral Lagrangians in $\tau^\pm\to\pi^\pm\eta\gamma\nu_\tau$ decays}\label{AppSCC}
\footnotesize{
In this appendix we include the different contributions to the (axial-)vector form factors obtained using $R\chi L$. Only the anomalous contribution was included in section \ref{sec:FFsRChT}. Here we explicitly quote the analytic expressions for the 
model-dependent (resonant-mediated) contributions to these form factors following the order in the figures. We start with fig.~\ref{fig:1R-FA}, giving rise to $a_{i=1,2,3,4}^{1R}$ in $R\chi L$:\\

\begin{multline}\label{a1-1R-RChL}
 a_1^{1R}=
   -\frac{4C_q}{F^2M_Vm_\rho^2D_\rho[(p_0+k)^2]}\left(k\cdot p_0 \left(\left(F_V-2 G_V\right) \left(-\left(8 c_3+2
   c_5+3 c_7\right) m_{\eta }^2+8 c_3 m_{\pi }^2+2 \left(c_5+c_7\right)
   p\cdot p_0\right)\right.\right.\\\left.
   +m_{\rho }^2 \left(\left(c_7+c_{1256}\right) F_V-2
   c_7 G_V\right)\right)-\frac{1}{2} \left(F_V-2 G_V\right) k\cdot p
   \left(4 c_5 k\cdot p_0+2 \left(c_5-4 c_3\right) m_{\eta }^2-\left(2
   c_5+c_{1256}\right) m_{\rho }^2+8 c_3 m_{\pi }^2\right)\\
   -2 c_7
   \left(F_V-2 G_V\right) \left(k\cdot p_0\right){}^2+\frac{1}{2}
   m_{\rho }^2 \left(m_{\eta }^2 \left(\left(2 \left(8
   c_3+c_5+c_7\right)+c_{1256}\right) F_V-4 \left(4 c_3+c_5+c_7\right)
   G_V\right)\right.\\\left.\left.
   +16 c_3 m_{\pi }^2 \left(G_V-F_V\right)-\left(2
   \left(c_5+c_7\right)-c_{1256}\right) p\cdot p_0 \left(F_V-2
   G_V\right)\right)\right.\\\left.+\left(\left(4 c_3+c_5+c_7\right) m_{\eta }^2-4 c_3
   m_{\pi }^2\right) \left(F_V-2 G_V\right) \left(p\cdot p_0-m_{\eta
   }^2\right)\right)\\
 +\frac{4F_A}{F^2D_{a_1}[(p+p_0+k)^2]}\left(-C_q \left(-\left(\kappa _5^A+\kappa _6^A-\kappa
   _7^A+\kappa _{16}^A\right) \left(k\cdot p+m_\pi^2\right)-\left(\kappa
   _3^A+2 \kappa _8^A+\kappa _{15}^A\right) k\cdot p_0\right.\right.\\\left.
+2 m_{\pi }^2
   \left(2 \kappa _9^A-\kappa _{10}^A+2 \kappa _{11}^A+\kappa
   _{14}^A\right)+\left(-3 \kappa _3^A+4 \kappa _4^A+\kappa _5^A+\kappa
   _6^A+\kappa _7^A-2 \kappa _8^A-\kappa _{15}^A\right) p\cdot
   p_0\right)\\
   -C_q \left(-\left(\kappa _5^A+\kappa _6^A+\kappa
   _7^A\right) k\cdot p_0+2 m_{\pi }^2 \left(\kappa _{14}^A+2
   \left(\kappa _9^A+\kappa _{11}^A\right)\right)+m_\eta^2 \left(\kappa
   _3^A-\kappa _5^A-\kappa _6^A-\kappa _7^A+2 \kappa _8^A+\kappa
   _{15}^A\right)\right.\\\left.\left.+\left(-2 \kappa _3^A+4 \kappa _4^A+\kappa _5^A+\kappa
   _6^A-\kappa _7^A+\kappa _{16}^A\right) p\cdot p_0\right)-2 \sqrt{2}
   \kappa _9^A C_s \left(m_{\pi }^2-2 m_K^2\right)+\sqrt{2} \left(2
   \kappa _9^A-\kappa _{10}^A\right) C_s \left(2 m_K^2-m_{\pi
   }^2\right)\right)\\
   -\frac{4F_A}{F^2m_{a_1}^2D_{a_1}[(p+k)^2]}\left(\left(m_{a_1}^2-k\cdot p\right) \left(C_q \left(\kappa
   _{16}^A \left(2 k\cdot p+m_\pi^2\right)+\kappa _{16}^A \left(2
   \left(k\cdot p_0+p\cdot p_0\right)+m_\eta^2\right)\right.\right.\right.\\\left.\left.-\left(\kappa _3^A+2
   \kappa _8^A+\kappa _{15}^A\right) \left(k\cdot p_0+m_\eta^2+p\cdot
   p_0\right)-2 m_{\pi }^2 \left(2 \kappa _9^A+\kappa _{10}^A+2 \kappa
   _{11}^A+\kappa _{12}^A\right)\right)+\sqrt{2} \left(2 \kappa
   _9^A+\kappa _{10}^A\right) C_s \left(2 m_K^2-m_{\pi
   }^2\right)\right)\\
   +\sqrt{2} C_s \left(2 m_K^2-m_{\pi }^2\right)
   \left(-2 \kappa _9^A \left(-m_{a_1}^2+2 k\cdot p+m_\pi^2+p\cdot
   p_0\right)-\kappa _{10}^A p\cdot p_0\right)\\
   +C_q \left(p\cdot p_0
   \left(-m_{a_1}^2 \left(\kappa _3^A+\kappa
   _{15}^A\right)+\left(\kappa _3^A+\kappa _{15}^A-\kappa
   _{16}^A\right) \left(2 k\cdot p+m_\pi^2\right)+2 m_{\pi }^2 \left(2
   \kappa _9^A+\kappa _{10}^A+2 \kappa _{11}^A+\kappa
   _{12}^A\right)\right.\right.\\\left.
   +m_\eta^2 \left(\kappa _3^A+2 \kappa _8^A+\kappa
   _{15}^A-\kappa _{16}^A\right)\right)-\left(4 m_{\pi }^2 \left(\kappa
   _9^A+\kappa _{11}^A\right)+m_\eta^2 \left(\kappa _3^A+\kappa
   _{15}^A\right)\right) \left(m_{a_1}^2-2 k\cdot p-m_\pi^2\right)\\\left.\left.
   +k\cdot
   p_0 \left(\left(\kappa _3^A+\kappa _{15}^A\right) \left(-m_{a_1}^2+2
   k\cdot p+m_\pi^2\right)+\left(\kappa _3^A+2 \kappa _8^A+\kappa _{15}^A-2
   \kappa _{16}^A\right) p\cdot p_0\right)+\left(\kappa _3^A+2 \kappa
   _8^A+\kappa _{15}^A-2 \kappa _{16}^A\right) \left(p\cdot
   p_0\right){}^2\right)\right)\\
  -\frac{4F_VC_q}{F^2 m_\rho^2}\left(\frac{2 \sqrt{2} C_s \left(2 m_K^2-m_{\pi }^2\right)
   \kappa _{13}^V}{C_q}+\frac{2 \sqrt{2} C_s \left(2 m_K^2-m_{\pi
   }^2\right) \left(\kappa _{13}^V+\kappa
   _{18}^V\right)}{C_q}\right.\\
   -\left(\kappa _1^V+\kappa _2^V+\kappa
   _3^V+\kappa _6^V+\kappa _7^V+\kappa _8^V\right) k\cdot
   p_0-\left(\kappa _6^V+\kappa _8^V+2 \kappa _{12}^V+\kappa _{16}^V-2
   \kappa _{17}^V\right) k\cdot p_0\\
   +\left(-\kappa _1^V+\kappa
   _2^V-\kappa _3^V+\kappa _7^V+\kappa _{17}^V\right) k\cdot p-m_{\eta
   }^2 \kappa _1^V-m_{\pi }^2 \left(\kappa _1^V+2 \left(-\kappa
   _4^V+\kappa _{10}^V+2 \left(\kappa _{13}^V+\kappa _{14}^V+\kappa
   _{15}^V\right)\right)\right)\\
   -2 m_{\pi }^2 \left(-\kappa _4^V+2
   \kappa _9^V+\kappa _{10}^V+2 \left(\kappa _{13}^V+\kappa
   _{14}^V+\kappa _{15}^V+\kappa _{18}^V\right)\right)+m_\pi^2 \left(\kappa
   _2^V-\kappa _3^V+\kappa _7^V\right)-m_\eta^2 \left(\kappa _2^V+\kappa
   _3^V+\kappa _6^V+\kappa _7^V+\kappa _8^V\right)\\
   -m_\eta^2 \left(\kappa
   _6^V+\kappa _8^V+2 \kappa _{12}^V+\kappa _{16}^V-\kappa
   _{17}^V\right)+p\cdot p_0 \kappa _1^V+p\cdot p_0 \left(\kappa
   _1^V+\kappa _2^V+\kappa _3^V-4 \kappa _5^V-\kappa _6^V-\kappa
   _7^V-\kappa _8^V\right)\\\left.
   -p\cdot p_0 \left(-3 \kappa _2^V+3 \kappa
   _3^V+4 \kappa _5^V+\kappa _6^V-\kappa _7^V+\kappa _8^V+2 \kappa
   _{12}^V+\kappa _{16}^V-\kappa _{17}^V\right)\frac{}{}\right)\,,
   \\
\end{multline}
where $c_{1256}\equiv c_1-c_2-c_5+2c_6$ was used. Its value is fixed by eqs.~(─\ref{eq: Consistent set of relations}).\\

\begin{multline}\label{a2-1R-RChL}
 a_2^{1R}=
  \frac{4C_q\left(F_V-2 G_V\right)}{F^2M_Vm_\rho^2D_\rho[(p_0+k)^2]} \left(c_7 k\cdot p_0+\left(4
   c_3+c_5+c_7\right) m_{\eta }^2-4 c_3 m_{\pi }^2\right) \left(-2
   k\cdot p_0-m_{\eta }^2+m_{\rho }^2\right)\\
 +\frac{4F_A}{F^2D_{a_1}[(p+p_0+k)^2]}\left(C_q \left(-\left(\kappa _5^A+\kappa _6^A+\kappa
   _7^A\right) k\cdot p_0+2 m_{\pi }^2 \left(\kappa _{14}^A+2
   \left(\kappa _9^A+\kappa _{11}^A\right)\right)\right.\right.\\\left.\left.+m_\eta^2 \left(\kappa
   _3^A-\kappa _5^A-\kappa _6^A-\kappa _7^A+2 \kappa _8^A+\kappa
   _{15}^A\right)+\left(-2 \kappa _3^A+4 \kappa _4^A+\kappa _5^A+\kappa
   _6^A-\kappa _7^A+\kappa _{16}^A\right) p\cdot p_0\right)+2 \sqrt{2}
   \kappa _9^A C_s \left(m_{\pi }^2-2 m_K^2\right)\right)\\
   +\frac{4F_A}{F^2m_{a_1}^2D_{a_1}[(p+k)^2]}\left(\sqrt{2} C_s \left(2 m_K^2-m_{\pi }^2\right) \left(2 \kappa _9^A \left(m_{a_1}^2-2 k\cdot
   p-m_\pi^2\right)-\left(2 \kappa _9^A+\kappa _{10}^A\right) p\cdot p_0\right)\right.\\
   +C_q \left(p\cdot p_0 \left(-m_{a_1}^2
   \left(\kappa _3^A+\kappa _{15}^A\right)+\left(\kappa _3^A+\kappa _{15}^A-\kappa _{16}^A\right) \left(2 k\cdot
   p+m_\pi^2\right)+2 m_{\pi }^2 \left(2 \kappa _9^A+\kappa _{10}^A+2 \kappa _{11}^A+\kappa _{12}^A\right)\right.\right.\\\left.
   +m_\eta^2 \left(\kappa
   _3^A+2 \kappa _8^A+\kappa _{15}^A-\kappa _{16}^A\right)\right)-\left(4 m_{\pi }^2 \left(\kappa _9^A+\kappa
   _{11}^A\right)+m_\eta^2 \left(\kappa _3^A+\kappa _{15}^A\right)\right) \left(m_{a_1}^2-2 k\cdot p-m_\pi^2\right)\\\left.\left.
   +k\cdot p_0
   \left(\left(\kappa _3^A+2 \kappa _8^A+\kappa _{15}^A-2 \kappa _{16}^A\right) p\cdot p_0-\left(\kappa _3^A+\kappa
   _{15}^A\right) \left(m_{a_1}^2-2 k\cdot p-m_\pi^2\right)\right)+\left(\kappa _3^A+2 \kappa _8^A+\kappa _{15}^A-2 \kappa
   _{16}^A\right) \left(p\cdot p_0\right){}^2\right)\right)\\
  +\frac{4F_VC_q}{F^2 m_\rho^2}\left(\frac{2 \sqrt{2} C_s \left(2 m_K^2-m_{\pi }^2\right)
   \left(\kappa _{13}^V+\kappa _{18}^V\right)}{C_q}-\left(\kappa
   _1^V+\kappa _2^V+\kappa _3^V+\kappa _6^V+\kappa _7^V+\kappa
   _8^V\right) k\cdot p_0-m_{\eta }^2 \kappa _1^V\right.\\
   -2 m_{\pi }^2
   \left(-\kappa _4^V+2 \kappa _9^V+\kappa _{10}^V+2 \left(\kappa
   _{13}^V+\kappa _{14}^V+\kappa _{15}^V+\kappa
   _{18}^V\right)\right)-m_\eta^2 \left(\kappa _2^V+\kappa _3^V+\kappa
   _6^V+\kappa _7^V+\kappa _8^V\right)\\\left.
   +p\cdot p_0 \left(\kappa
   _1^V+\kappa _2^V+\kappa _3^V-4 \kappa _5^V-\kappa _6^V-\kappa
   _7^V-\kappa _8^V\right)\frac{}{}\right)\,,
  \\
\end{multline}

\begin{multline}\label{a3-1R-RChL}
 a_3^{1R}=
 \frac{16G_VC_q}{F^2M_V[m_\eta^2+2(k\cdot p + k\cdot p_0 + p\cdot p_0)]D_\rho[(p_0+k)^2]}\left[-(c_{1256}+8c_3)\frac{m_\eta^2}{2}+c_{1256} k\cdot p_0
 +4c_3m_\pi^2\right]\\
 +\frac{4F_A}{F^2m^2_{a_1}D_{a_1}[(p+p_0+k)^2]} \left(\kappa _{10}^A \left(\sqrt{2} C_s \left(2 m_K^2-m_{\pi
   }^2\right)-2 m_{\pi }^2 C_q\right)-m_{a_1}^2 \left(\kappa
   _5^A+\kappa _6^A-\kappa _7^A+\kappa _{16}^A\right)
   C_q\right)\\
   +\frac{4F_A}{F^2m_{a_1}^2D_{a_1}[(p+k)^2]}\left(C_q \left(\kappa _{16}^A \left(m_{a_1}^2+2 k\cdot
   p_0+m_\eta^2+2 p\cdot p_0\right)-\left(\kappa _3^A+2 \kappa _8^A+\kappa
   _{15}^A\right) \left(k\cdot p_0+m_\eta^2+p\cdot p_0\right)\right.\right.\\\left.\left.
   -2 m_{\pi }^2
   \left(2 \kappa _9^A+\kappa _{10}^A+2 \kappa _{11}^A+\kappa
   _{12}^A\right)\right)+\sqrt{2} \left(2 \kappa _9^A+\kappa
   _{10}^A\right) C_s \left(2 m_K^2-m_{\pi }^2\right)\right)\\
  +\frac{4F_VC_q}{F^2 m_\rho^2}\left(3 \kappa _1^V-3 \kappa _2^V+3 \kappa _3^V-\kappa
   _6^V+\kappa _7^V-\kappa _8^V\right)\,,
   \\
\end{multline}

\begin{multline}\label{a4-1R-RChL}
 a_4^{1R}= 
  \frac{4C_q\left(F_V-2 G_V\right) }{F^2M_Vm_\rho^2D_\rho[(p_0+k)^2]}\left(\left(c_5+c_7\right) \left(2
   k\cdot p_0+m_{\eta }^2\right)+4 c_3 \left(m_{\eta }^2-m_{\pi }^2\right)+\left(-c_5-c_7+\frac{c_{1256}}{2}\right) m_{\rho }^2\right)
   \\
 -\frac{4F_A C_q}{F^2D_{a_1}[(p+p_0+k)^2]}\left(\kappa _3^A-2 \kappa _5^A-2 \kappa _6^A+2 \kappa
   _8^A+\kappa _{15}^A-\kappa _{16}^A\right)\\
   -\frac{4F_A}{F^2m_{a_1}^2D_{a_1}[(p+k)^2]}\left(C_q \left(\kappa _{16}^A \left(m_{a_1}^2+2 k\cdot
   p_0+m_\eta^2+2 p\cdot p_0\right)-\left(\kappa _3^A+2 \kappa _8^A+\kappa
   _{15}^A\right) \left(k\cdot p_0+m_\eta^2+p\cdot p_0\right)\right.\right.\\\left.\left.-2 m_{\pi }^2
   \left(2 \kappa _9^A+\kappa _{10}^A+2 \kappa _{11}^A+\kappa
   _{12}^A\right)\right)+\sqrt{2} \left(2 \kappa _9^A+\kappa
   _{10}^A\right) C_s \left(2 m_K^2-m_{\pi }^2\right)-C_q \left(\kappa _{16}^A-2 \kappa _8^A\right)
   \left(m_{a_1}^2-2 k\cdot p-m_\pi^2\right)\right)\\
  -\frac{4F_VC_q}{F^2 m_\rho^2}\left(2 \kappa _1^V-4 \kappa _2^V+6 \kappa _3^V-2 \kappa _6^V-2
   \kappa _8^V+2 \kappa _{12}^V-\kappa _{16}^V-\kappa _{17}^V\right)\,.\\
\end{multline}

The two-resonance mediated contributions to the axial-vector form factors, corresponding to figs.~\ref{fig:2R-FA}, are given in the following:\\

\begin{multline}\label{a1-2R-RChL}
  a_1^{2R}=
   -\frac{8F_AC_q}{F^2M_Vm_\rho^2D_{a_1}[(p+p_0+k)^2]D_\rho[(p_0+k)^2]}\\
   \left(-2 \sqrt{2} (k\cdot p)^2 \left(\frac{1}{2} \left(2
   c_5+c_{1256}\right) m_{\rho }^2-c_5 \left(m_{\eta }^2+2 k\cdot
   p_0\right)+4 c_3 \left(m_{\eta }^2-m_{\pi }^2\right)\right) \lambda
   ^{\prime\prime}\right.\\
   -\left(\left(c_5+c_7\right) m_{\eta }^2+k\cdot p_0 c_7-4
   c_3 \left(m_{\pi }^2+m_{\eta }^2\right)\right) \left(m_{\eta
   }^2-m_{\rho }^2+2 k\cdot p_0\right) \left(-2 \sqrt{2} \left(k\cdot
   p+p\cdot p_0\right) \lambda ^{\prime\prime}-2 m_{\pi }^2 \left(2 \lambda
   _1+\lambda _2\right)\right.\\\left.
   +\left(m_{\eta }^2+2 k\cdot p_0\right) \lambda
   _4\right)-\frac{1}{2} k\cdot p \left(-4 \sqrt{2} p\cdot p_0 \left(8
   c_3 m_{\pi }^2-c_7 m_{\eta }^2+\left(c_7-c_{1256}\right) m_{\rho
   }^2\right) \lambda ^{\prime\prime}\right.\\
   +2 c_5 \left(4 \lambda _4 \left(k\cdot
   p_0\right){}^2+\left(m_{\eta }^2-m_{\rho }^2\right) \left(m_{\eta }^2
   \lambda _4-2 m_{\pi }^2 \left(2 \lambda _1+\lambda
   _2\right)\right)\right)\\
   -8 c_3 \left(m_{\pi }^2-m_{\eta }^2\right)
   \left(2 \left(2 \lambda _1+\lambda _2\right) m_{\pi }^2-m_{\eta }^2
   \lambda _4+m_{\rho }^2 \left(\lambda _3-2 \lambda
   _5\right)\right)+c_{1256} m_{\rho }^2 \left(2 \left(2 \lambda
   _1+\lambda _2\right) m_{\pi }^2+m_{\eta }^2 \left(\lambda _3-\lambda
   _4-2 \lambda _5\right)\right)\\
   +2 k\cdot p_0 \left(4 \sqrt{2} p\cdot
   p_0 c_7 \lambda ^{\prime\prime}+2 c_5 \left(\left(2 m_{\eta }^2-m_{\rho
   }^2\right) \lambda _4-2 m_{\pi }^2 \left(2 \lambda _1+\lambda
   _2\right)\right)+8 c_3 \left(\left(4 \lambda _1+\lambda _4\right)
   m_{\pi }^2+m_{\rho }^2 \lambda _3-2 m_{\eta }^2 \lambda
   _4\right)\right.\\\left.\left.
   +c_{1256} m_{\rho }^2 \left(\lambda _3-\lambda _4-2 \lambda
   _5\right)\right)\right)+p\cdot p_0 \left(4 c_3 \left(m_{\pi
   }^2+m_{\eta }^2\right) \left(2 \left(2 \lambda _1+\lambda _2\right)
   m_{\pi }^2-m_{\eta }^2 \lambda _4+p\cdot p_0 \left(2 \lambda
   _2-\lambda _4-2 \lambda _5\right)\right)\right.\\\left.
   -\left(c_5+c_7\right)
   \left(-4 \lambda _4 \left(k\cdot p_0\right){}^2-m_{\eta }^2 \left(-2
   \left(2 \lambda _1+\lambda _2\right) m_{\pi }^2+m_{\eta }^2 \lambda
   _4+p\cdot p_0 \left(-2 \lambda _2+\lambda _4+2 \lambda
   _5\right)\right)\right)\right)\\
   +k\cdot p_0 \left(\left(-c_{1256}
   \lambda _3 m_{\rho }^2-4 p\cdot p_0 \left(c_5+c_7\right) \lambda
   _2\right) m_{\pi }^2\right.\\
   +p\cdot p_0 \left(-4 \sqrt{2} p\cdot p_0
   \left(c_5+c_7\right) \lambda ^{\prime\prime}+4 \left(-4
   c_3+c_5+c_7\right) m_{\eta }^2 \lambda _4+8 m_{\pi }^2
   \left(-\left(-4 c_3+c_5+c_7\right) \lambda _1-c_3 \lambda
   _4\right)\right)\\\left.
   -m_{\rho }^2 \left(c_{1256} \left(4 \lambda _1 m_{\pi
   }^2+p\cdot p_0 \left(\lambda _3-\lambda _4-2 \lambda
   _5\right)\right)+2 p\cdot p_0 \left(\left(c_5+c_7\right) \lambda _4+4
   c_3 \left(\lambda _3-2 \left(\lambda _4+\lambda
   _5\right)\right)\right)\right)\right)\\
   -m_{\rho }^2 \left(-\sqrt{2}
   \left(p\cdot p_0\right){}^2 \left(2
   \left(c_5+c_7\right)-c_{1256}\right) \lambda ^{\prime\prime}+4 m_{\pi
   }^2 \left(\frac{1}{2} \left(8 c_3+c_{1256}\right) m_{\eta }^2-4 c_3
   m_{\pi }^2\right) \lambda _1\right.\\
   +\frac{1}{2} m_{\pi }^2 \left(2 p\cdot
   p_0 \left(c_{1256}-2 \left(c_5+c_7\right)\right) \lambda
   _2+\left(\left(8 c_3+c_{1256}\right) m_{\eta }^2-8 c_3 m_{\pi
   }^2\right) \lambda _3\right)\\
   +p\cdot p_0 \left(\frac{1}{2} m_{\eta }^2
   \left(2 \left(c_5+c_7\right) \lambda _4+c_{1256} \left(\lambda
   _3-\lambda _4-2 \lambda _5\right)+8 c_3 \left(\lambda _3-2
   \left(\lambda _4+\lambda _5\right)\right)\right)\right.\\\left.\left.\left.
   \left(\left(c_5+c_7-\frac{c_{1256}}{2}\right) \lambda _1+c_3
   \left(\lambda _3-2 \lambda _5\right)\right)\right)\right)\frac{}{}\right)\\
 +\frac{8 F_A C_q \kappa _1^{SA} \left(c_d p\cdot
    p_0+m_{\pi }^2 c_m\right)}{F^2D_{a_1}[(p+p_0+k)^2]D_{a_0}[(p+p_0)^2]}\,,
  \\
\end{multline}

\begin{multline}\label{a2-2R-RChL}
 a_2^{2R}=
  -\frac{8F_AC_q}{F^2M_Vm_\rho^2D_{a_1}[(p+p_0+k)^2]D_\rho[(p_0+k)^2]}\\
   \left(c_7 k\cdot p_0+\left(c_5+c_7\right) m_{\eta }^2-4 c_3
   \left(m_{\eta }^2+m_{\pi }^2\right)\right) \left(2 k\cdot p_0+m_{\eta
   }^2-m_{\rho }^2\right)\\
   \left(\lambda _4 \left(2 k\cdot p_0+m_{\eta
   }^2\right)-2 \sqrt{2} \lambda ^{\prime\prime} k\cdot p-2 \left(2 \lambda
   _1+\lambda _2\right) m_{\pi}^2-2 \sqrt{2} p\cdot p_0 \lambda
   ^{\prime\prime}\right)\\
   -\frac{4 F_A C_q \kappa _1^{SA} \left(c_d p\cdot
   p_0+m_{\pi }^2 c_m\right)}{F^2D_{a_1}[(p+p_0+k)^2]D_{a_0}[(p+p_0)^2]}\,,
   \\
\end{multline}

\begin{multline}\label{a3-2R-RChL}
 a_3^{2R}=
   -\frac{4\sqrt{2}F_AC_q\left(\lambda ^\prime+\lambda^{\prime\prime}\right)}{F^2M_VD_{a_1}[(p+p_0+k)^2]D_\rho[(p_0+k)^2]}
    \left(2 \left(8 c_3+c_{1256}\right) k\cdot
   p_0+\left(8 c_3+c_{1256}\right) m_{\eta }^2-8 c_3 m_{\pi }^2\right)\,,
 \\
\end{multline}

\begin{multline}\label{a4-2R-RChL}
 a_4^{2R}=
 -\frac{8F_AC_q}{F^2M_Vm_\rho^2D_{a_1}[(p+p_0+k)^2]D_\rho[(p_0+k)^2]}\left(\frac{1}{2} \left(2 \lambda _2-\lambda _3\right)
 m_{\rho }^2 \left(c_{1256} \left(2 k\cdot p_0+m_{\eta
   }^2\right)+8 c_3 \left(m_{\eta }^2-m_{\pi }^2\right)\right)\right.\\
   -2 \sqrt{2} \lambda ^{\prime\prime} k\cdot p
   \left(-\left(c_5+c_7\right) \left(2 k\cdot p_0+m_{\eta }^2\right)+4 c_3 \left(m_{\eta }^2+m_{\pi }^2\right)+\frac{1}{2}
   \left(2 \left(c_5+c_7\right)-c_{1256}\right) m_{\rho }^2\right)\\
   +2 k\cdot p_0 \left(\left(c_5+c_7\right) \left(\lambda _4
   \left(m_{\rho }^2-2 m_{\eta }^2\right)+2 \left(2 \lambda _1+\lambda _2\right) m_{\pi }^2\right)+4 c_3 \left(2 \lambda _4
   m_{\eta }^2+\left(\lambda _4-4 \lambda _1\right) m_{\pi }^2\right)+2 \sqrt{2} \left(c_5+c_7\right) p\cdot p_0 \lambda
   ^{\prime\prime}\right)\\
   -4 \left(c_5+c_7\right) \lambda _4 \left(k\cdot p_0\right){}^2-\left(4 c_3 \left(m_{\eta }^2+m_{\pi
   }^2\right)-\left(c_5+c_7\right) m_{\eta }^2\right) \left(-\lambda _4 m_{\eta }^2+2 \left(2 \lambda _1+\lambda _2\right)
   m_{\pi }^2+\left(2 \lambda _2-\lambda _4-2 \lambda _5\right) p\cdot p_0\right)\\\left.
   -m_{\rho }^2 \left(\lambda _4 \left(4 c_3
   \left(m_{\eta }^2+m_{\pi }^2\right)-\left(c_5+c_7\right) m_{\eta }^2\right)+\left(2 \left(c_5+c_7\right)-c_{1256}\right)
   \left(2 \lambda _1+\lambda _2\right) m_{\pi }^2+\sqrt{2} \left(2 \left(c_5+c_7\right)-c_{1256}\right) p\cdot p_0 \lambda
   ^{\prime\prime}\right)\right).
 \\
\end{multline}

We will display separately the contributions from the last diagram in the first line of figure \ref{fig:2R-FA}, due to the length of the corresponding expressions.\\

\begin{multline}
 a_1^{W^-\to (a_1^-)\eta\to\pi^-(\rho^0)\eta\to\pi^-\gamma\eta}=   
  +\frac{8F_V}{F^2m_{a_1}^2m_\rho^2D_{a_1}[(p+k)^2]}\left(8 m_\pi^2 C_q m_{\pi }^2 m_\eta^2 \lambda _1 \kappa
   _3^A+16 k\cdot p C_q m_{\pi }^2 m_\eta^2 \lambda _1 \kappa _3^A\right.\\
   -8 C_q
   m_{\pi }^2 m_{a_1}^2 m_\eta^2 \lambda _1 \kappa _3^A-2 m_\pi^2 C_q
   m_{a_1}^2 m_\eta^2 \lambda _2 \kappa _3^A+2 k\cdot p C_q m_{a_1}^2
   m_\eta^2 \lambda _2 \kappa _3^A+2 p^4 C_q m_\eta^2 \lambda _2 \kappa
   _3^A-4 (k\cdot p)^2 C_q m_\eta^2 \lambda _2 \kappa _3^A\\
   +2 m_\pi^2 k\cdot p
   C_q m_\eta^2 \lambda _2 \kappa _3^A+2 k\cdot p C_q m_{a_1}^2 m_\eta^2
   \lambda _4 \kappa _3^A-4 (k\cdot p)^2 C_q m_\eta^2 \lambda _4 \kappa
   _3^A-2 m_\pi^2 k\cdot p C_q m_\eta^2 \lambda _4 \kappa _3^A
   +4 k\cdot p C_q m_{a_1}^2 m_\eta^2 \lambda _5 \kappa _3^A\\
   -8 (k\cdot p)^2 C_q m_\eta^2
   \lambda _5 \kappa _3^A-4 m_\pi^2 k\cdot p C_q m_\eta^2 \lambda _5 \kappa
   _3^A+8 C_q m_{\pi }^2 p_0^4 \kappa _8^A \lambda _1+8 k\cdot p C_q
   m_{\pi }^2 m_\eta^2 \kappa _8^A \lambda _1
   -8 C_q m_{\pi }^2 m_{a_1}^2 m_\eta^2 \kappa _8^A \lambda _1\\
   +32 m_\pi^2 C_q m_{\pi }^4 \kappa _9^A
   \lambda _1+64 k\cdot p C_q m_{\pi }^4 \kappa _9^A \lambda _1+16
   \sqrt{2} m_\pi^2 C_s m_{\pi }^4 \kappa _9^A \lambda _1+32 \sqrt{2}
   k\cdot p C_s m_{\pi }^4 \kappa _9^A \lambda _1
   -32 \sqrt{2} m_\pi^2 C_s
   m_K^2 m_{\pi }^2 \kappa _9^A \lambda _1\\
   -64 \sqrt{2} k\cdot p C_s
   m_K^2 m_{\pi }^2 \kappa _9^A \lambda _1-32 C_q m_{\pi }^4 m_{a_1}^2
   \kappa _9^A \lambda _1-16 \sqrt{2} C_s m_{\pi }^4 m_{a_1}^2 \kappa
   _9^A \lambda _1
   +32 \sqrt{2} C_s m_K^2 m_{\pi }^2 m_{a_1}^2 \kappa
   _9^A \lambda _1+8 k\cdot p C_q m_{\pi }^4 \kappa _{10}^A \lambda
   _1\\
   +4 \sqrt{2} k\cdot p C_s m_{\pi }^4 \kappa _{10}^A \lambda _1-8
   \sqrt{2} k\cdot p C_s m_K^2 m_{\pi }^2 \kappa _{10}^A \lambda _1
   -8
   C_q m_{\pi }^4 m_{a_1}^2 \kappa _{10}^A \lambda _1-4 \sqrt{2} C_s
   m_{\pi }^4 m_{a_1}^2 \kappa _{10}^A \lambda _1
   +8 \sqrt{2} C_s m_K^2
   m_{\pi }^2 m_{a_1}^2 \kappa _{10}^A \lambda _1\\
   +8 C_q m_{\pi }^4
   m_\eta^2 \kappa _{10}^A \lambda _1+4 \sqrt{2} C_s m_{\pi }^4 m_\eta^2
   \kappa _{10}^A \lambda _1
   -8 \sqrt{2} C_s m_K^2 m_{\pi }^2 m_\eta^2
   \kappa _{10}^A \lambda _1+32 m_\pi^2 C_q m_{\pi }^4 \kappa _{11}^A
   \lambda _1+64 k\cdot p C_q m_{\pi }^4 \kappa _{11}^A \lambda _1\\
   -32
   C_q m_{\pi }^4 m_{a_1}^2 \kappa _{11}^A \lambda _1+8 k\cdot p C_q
   m_{\pi }^4 \kappa _{12}^A \lambda _1
   -8 C_q m_{\pi }^4 m_{a_1}^2
   \kappa _{12}^A \lambda _1+8 C_q m_{\pi }^4 m_\eta^2 \kappa _{12}^A
   \lambda _1+8 m_\pi^2 C_q m_{\pi }^2 m_\eta^2 \kappa _{15}^A \lambda _1\\
   +16
   k\cdot p C_q m_{\pi }^2 m_\eta^2 \kappa _{15}^A \lambda _1-8 C_q m_{\pi
   }^2 m_{a_1}^2 m_\eta^2 \kappa _{15}^A \lambda _1
   -4 C_q m_{\pi }^2 p_0^4
   \kappa _{16}^A \lambda _1-8 (k\cdot p)^2 C_q m_{\pi }^2 \kappa
   _{16}^A \lambda _1-4 m_\pi^2 k\cdot p C_q m_{\pi }^2 \kappa _{16}^A
   \lambda _1\\
   +4 m_\pi^2 C_q m_{\pi }^2 m_{a_1}^2 \kappa _{16}^A \lambda
   _1+8 k\cdot p C_q m_{\pi }^2 m_{a_1}^2 \kappa _{16}^A \lambda _1
   -4
   m_\pi^2 C_q m_{\pi }^2 m_\eta^2 \kappa _{16}^A \lambda _1-12 k\cdot p C_q
   m_{\pi }^2 m_\eta^2 \kappa _{16}^A \lambda _1+4 C_q m_{\pi }^2
   m_{a_1}^2 m_\eta^2 \kappa _{16}^A \lambda _1\\
   -4 k\cdot p C_q p_0^4
   \kappa _8^A \lambda _2-4 m_\pi^2 C_q m_{a_1}^2 m_\eta^2 \kappa _8^A \lambda
   _2
   -4 (k\cdot p)^2 C_q m_\eta^2 \kappa _8^A \lambda _2-8 \sqrt{2} p^4
   C_s m_K^2 \kappa _9^A \lambda _2+16 \sqrt{2} (k\cdot p)^2 C_s m_K^2
   \kappa _9^A \lambda _2\\
   -8 \sqrt{2} m_\pi^2 k\cdot p C_s m_K^2 \kappa _9^A
   \lambda _2+8 p^4 C_q m_{\pi }^2 \kappa _9^A \lambda _2
   -16 (k\cdot
   p)^2 C_q m_{\pi }^2 \kappa _9^A \lambda _2+8 m_\pi^2 k\cdot p C_q m_{\pi
   }^2 \kappa _9^A \lambda _2+4 \sqrt{2} p^4 C_s m_{\pi }^2 \kappa _9^A
   \lambda _2\\
   -8 \sqrt{2} (k\cdot p)^2 C_s m_{\pi }^2 \kappa _9^A
   \lambda _2+4 \sqrt{2} m_\pi^2 k\cdot p C_s m_{\pi }^2 \kappa _9^A
   \lambda _2
   +8 \sqrt{2} m_\pi^2 C_s m_K^2 m_{a_1}^2 \kappa _9^A \lambda
   _2-8 \sqrt{2} k\cdot p C_s m_K^2 m_{a_1}^2 \kappa _9^A \lambda _2\\
   -8
   m_\pi^2 C_q m_{\pi }^2 m_{a_1}^2 \kappa _9^A \lambda _2+8 k\cdot p C_q
   m_{\pi }^2 m_{a_1}^2 \kappa _9^A \lambda _2
   -4 \sqrt{2} m_\pi^2 C_s
   m_{\pi }^2 m_{a_1}^2 \kappa _9^A \lambda _2+4 \sqrt{2} k\cdot p C_s
   m_{\pi }^2 m_{a_1}^2 \kappa _9^A \lambda _2\\
   +4 \sqrt{2} (k\cdot p)^2
   C_s m_K^2 \kappa _{10}^A \lambda _2-4 (k\cdot p)^2 C_q m_{\pi }^2
   \kappa _{10}^A \lambda _2
   -2 \sqrt{2} (k\cdot p)^2 C_s m_{\pi }^2
   \kappa _{10}^A \lambda _2+4 \sqrt{2} m_\pi^2 C_s m_K^2 m_{a_1}^2 \kappa
   _{10}^A \lambda _2\\
   -4 m_\pi^2 C_q m_{\pi }^2 m_{a_1}^2 \kappa _{10}^A
   \lambda _2-2 \sqrt{2} m_\pi^2 C_s m_{\pi }^2 m_{a_1}^2 \kappa _{10}^A
   \lambda _2
   +4 \sqrt{2} k\cdot p C_s m_K^2 m_\eta^2 \kappa _{10}^A
   \lambda _2-4 k\cdot p C_q m_{\pi }^2 m_\eta^2 \kappa _{10}^A \lambda
   _2+8 p^4 C_q m_{\pi }^2 \kappa _{11}^A \lambda _2\\
   -2 \sqrt{2} k\cdot p C_s m_{\pi }^2 m_\eta^2 \kappa _{10}^A \lambda
   _2
   -16 (k\cdot p)^2
   C_q m_{\pi }^2 \kappa _{11}^A \lambda _2+8 m_\pi^2 k\cdot p C_q m_{\pi
   }^2 \kappa _{11}^A \lambda _2-8 m_\pi^2 C_q m_{\pi }^2 m_{a_1}^2 \kappa
   _{11}^A \lambda _2+8 k\cdot p C_q m_{\pi }^2 m_{a_1}^2 \kappa
   _{11}^A \lambda _2\\
   -4 (k\cdot p)^2 C_q m_{\pi }^2 \kappa _{12}^A
   \lambda _2
   -4 m_\pi^2 C_q m_{\pi }^2 m_{a_1}^2 \kappa _{12}^A \lambda
   _2-4 k\cdot p C_q m_{\pi }^2 m_\eta^2 \kappa _{12}^A \lambda _2-2 m_\pi^2
   C_q m_{a_1}^2 m_\eta^2 \kappa _{15}^A \lambda _2+2 k\cdot p C_q
   m_{a_1}^2 m_\eta^2 \kappa _{15}^A \lambda _2\\
   +2 p^4 C_q m_\eta^2 \kappa
   _{15}^A \lambda _2
   -4 (k\cdot p)^2 C_q m_\eta^2 \kappa _{15}^A \lambda
   _2+2 m_\pi^2 k\cdot p C_q m_\eta^2 \kappa _{15}^A \lambda _2+2 k\cdot p C_q
   p_0^4 \kappa _{16}^A \lambda _2+2 p^4 C_q m_{a_1}^2 \kappa _{16}^A
   \lambda _2+4 m_\pi^2 k\cdot p C_q m_{a_1}^2 \kappa _{16}^A \lambda _2\\
   +2
   m_\pi^2 C_q m_{a_1}^2 m_\eta^2 \kappa _{16}^A \lambda _2
   +6 (k\cdot p)^2 C_q
   m_\eta^2 \kappa _{16}^A \lambda _2+2 m_\pi^2 k\cdot p C_q m_\eta^2 \kappa
   _{16}^A \lambda _2+4 (k\cdot p)^3 C_q \kappa _{16}^A \lambda _2+2
   m_\pi^2 (k\cdot p)^2 C_q \kappa _{16}^A \lambda _2\\
   -2 \left(p\cdot p_0\right){}^2 \left(\sqrt{2} C_s \left(2
   m_K^2-m_{\pi }^2\right) \kappa _{10}^A+C_q
   \left(\left(m_\pi^2-m_{a_1}^2+2 k\cdot p\right) \kappa _3^A+2
   \left(m_\pi^2-m_{a_1}^2-m_\eta^2+k\cdot p\right) \kappa _8^A-2 m_{\pi }^2
   \kappa _{10}^A\right.\right.\\\left.\left.
   -2 m_{\pi }^2 \kappa _{12}^A+m_\pi^2 \kappa
   _{15}^A-m_{a_1}^2 \kappa _{15}^A+2 k\cdot p \kappa _{15}^A-m_\pi^2
   \kappa _{16}^A+2 m_{a_1}^2 \kappa _{16}^A+m_\eta^2 \kappa
   _{16}^A\right)\right) \lambda _2-2 k\cdot p C_q p_0^4 \kappa _8^A
   \lambda _4+2 k\cdot p C_q m_{a_1}^2 m_\eta^2 \kappa _8^A \lambda _4\\
   -2
   (k\cdot p)^2 C_q m_\eta^2 \kappa _8^A \lambda _4+16 \sqrt{2} (k\cdot
   p)^2 C_s m_K^2 \kappa _9^A \lambda _4+8 \sqrt{2} m_\pi^2 k\cdot p C_s
   m_K^2 \kappa _9^A \lambda _4-16 (k\cdot p)^2 C_q m_{\pi }^2 \kappa
   _9^A \lambda _4\\
   -8 m_\pi^2 k\cdot p C_q m_{\pi }^2 \kappa _9^A \lambda
   _4-8 \sqrt{2} (k\cdot p)^2 C_s m_{\pi }^2 \kappa _9^A \lambda _4-4
   \sqrt{2} m_\pi^2 k\cdot p C_s m_{\pi }^2 \kappa _9^A \lambda _4-8
   \sqrt{2} k\cdot p C_s m_K^2 m_{a_1}^2 \kappa _9^A \lambda _4\\
   +8
   k\cdot p C_q m_{\pi }^2 m_{a_1}^2 \kappa _9^A \lambda _4+4 \sqrt{2}
   k\cdot p C_s m_{\pi }^2 m_{a_1}^2 \kappa _9^A \lambda _4+2 \sqrt{2}
   (k\cdot p)^2 C_s m_K^2 \kappa _{10}^A \lambda _4-2 (k\cdot p)^2 C_q
   m_{\pi }^2 \kappa _{10}^A \lambda _4\\
   -\sqrt{2} (k\cdot p)^2 C_s
   m_{\pi }^2 \kappa _{10}^A \lambda _4-2 \sqrt{2} k\cdot p C_s m_K^2
   m_{a_1}^2 \kappa _{10}^A \lambda _4+2 k\cdot p C_q m_{\pi }^2
   m_{a_1}^2 \kappa _{10}^A \lambda _4+\sqrt{2} k\cdot p C_s m_{\pi }^2
   m_{a_1}^2 \kappa _{10}^A \lambda _4\\
   +2 \sqrt{2} k\cdot p C_s m_K^2
   m_\eta^2 \kappa _{10}^A \lambda _4-2 k\cdot p C_q m_{\pi }^2 m_\eta^2
   \kappa _{10}^A \lambda _4-\sqrt{2} k\cdot p C_s m_{\pi }^2 m_\eta^2
   \kappa _{10}^A \lambda _4-16 (k\cdot p)^2 C_q m_{\pi }^2 \kappa
   _{11}^A \lambda _4-8 m_\pi^2 k\cdot p C_q m_{\pi }^2 \kappa _{11}^A
   \lambda _4\\
   +8 k\cdot p C_q m_{\pi }^2 m_{a_1}^2 \kappa _{11}^A
   \lambda _4-2 (k\cdot p)^2 C_q m_{\pi }^2 \kappa _{12}^A \lambda _4+2
   k\cdot p C_q m_{\pi }^2 m_{a_1}^2 \kappa _{12}^A \lambda _4-2 k\cdot
   p C_q m_{\pi }^2 m_\eta^2 \kappa _{12}^A \lambda _4   +4 \left(p\cdot
   p_0\right){}^3 C_q \left(\kappa _8^A-\kappa _{16}^A\right) \lambda
   _2\\
   +2 k\cdot p C_q
   m_{a_1}^2 m_\eta^2 \kappa _{15}^A \lambda _4-4 (k\cdot p)^2 C_q m_\eta^2
   \kappa _{15}^A \lambda _4-2 m_\pi^2 k\cdot p C_q m_\eta^2 \kappa _{15}^A
   \lambda _4+k\cdot p C_q p_0^4 \kappa _{16}^A \lambda _4-2 (k\cdot
   p)^2 C_q m_{a_1}^2 \kappa _{16}^A \lambda _4
   \nonumber
   \end{multline}
   
   \begin{multline}
   -m_\pi^2 k\cdot p C_q
   m_{a_1}^2 \kappa _{16}^A \lambda _4-k\cdot p C_q m_{a_1}^2 m_\eta^2
   \kappa _{16}^A \lambda _4+3 (k\cdot p)^2 C_q m_\eta^2 \kappa _{16}^A
   \lambda _4+m_\pi^2 k\cdot p C_q m_\eta^2 \kappa _{16}^A \lambda _4+2
   (k\cdot p)^3 C_q \kappa _{16}^A \lambda _4\\
   +m_\pi^2 (k\cdot p)^2 C_q
   \kappa _{16}^A \lambda _4-4 k\cdot p C_q p_0^4 \kappa _8^A \lambda
   _5+4 k\cdot p C_q m_{a_1}^2 m_\eta^2 \kappa _8^A \lambda _5-4 (k\cdot
   p)^2 C_q m_\eta^2 \kappa _8^A \lambda _5+32 \sqrt{2} (k\cdot p)^2 C_s
   m_K^2 \kappa _9^A \lambda _5\\
   +16 \sqrt{2} m_\pi^2 k\cdot p C_s m_K^2
   \kappa _9^A \lambda _5-32 (k\cdot p)^2 C_q m_{\pi }^2 \kappa _9^A
   \lambda _5-16 m_\pi^2 k\cdot p C_q m_{\pi }^2 \kappa _9^A \lambda _5-16
   \sqrt{2} (k\cdot p)^2 C_s m_{\pi }^2 \kappa _9^A \lambda _5\\
   -8
   \sqrt{2} m_\pi^2 k\cdot p C_s m_{\pi }^2 \kappa _9^A \lambda _5-16
   \sqrt{2} k\cdot p C_s m_K^2 m_{a_1}^2 \kappa _9^A \lambda _5+16
   k\cdot p C_q m_{\pi }^2 m_{a_1}^2 \kappa _9^A \lambda _5+8 \sqrt{2}
   k\cdot p C_s m_{\pi }^2 m_{a_1}^2 \kappa _9^A \lambda _5\\
   +4 \sqrt{2}
   (k\cdot p)^2 C_s m_K^2 \kappa _{10}^A \lambda _5-4 (k\cdot p)^2 C_q
   m_{\pi }^2 \kappa _{10}^A \lambda _5-2 \sqrt{2} (k\cdot p)^2 C_s
   m_{\pi }^2 \kappa _{10}^A \lambda _5-4 \sqrt{2} k\cdot p C_s m_K^2
   m_{a_1}^2 \kappa _{10}^A \lambda _5\\
   +4 k\cdot p C_q m_{\pi }^2
   m_{a_1}^2 \kappa _{10}^A \lambda _5+2 \sqrt{2} k\cdot p C_s m_{\pi
   }^2 m_{a_1}^2 \kappa _{10}^A \lambda _5+4 \sqrt{2} k\cdot p C_s
   m_K^2 m_\eta^2 \kappa _{10}^A \lambda _5-4 k\cdot p C_q m_{\pi }^2
   m_\eta^2 \kappa _{10}^A \lambda _5\\
   -2 \sqrt{2} k\cdot p C_s m_{\pi }^2
   m_\eta^2 \kappa _{10}^A \lambda _5-32 (k\cdot p)^2 C_q m_{\pi }^2
   \kappa _{11}^A \lambda _5-16 m_\pi^2 k\cdot p C_q m_{\pi }^2 \kappa
   _{11}^A \lambda _5+16 k\cdot p C_q m_{\pi }^2 m_{a_1}^2 \kappa
   _{11}^A \lambda _5\\
   -4 (k\cdot p)^2 C_q m_{\pi }^2 \kappa _{12}^A
   \lambda _5+4 k\cdot p C_q m_{\pi }^2 m_{a_1}^2 \kappa _{12}^A
   \lambda _5-4 k\cdot p C_q m_{\pi }^2 m_\eta^2 \kappa _{12}^A \lambda
   _5+4 k\cdot p C_q m_{a_1}^2 m_\eta^2 \kappa _{15}^A \lambda _5-8
   (k\cdot p)^2 C_q m_\eta^2 \kappa _{15}^A \lambda _5\\
   -4 m_\pi^2 k\cdot p C_q
   m_\eta^2 \kappa _{15}^A \lambda _5+2 k\cdot p C_q p_0^4 \kappa _{16}^A
   \lambda _5-4 (k\cdot p)^2 C_q m_{a_1}^2 \kappa _{16}^A \lambda _5-2
   m_\pi^2 k\cdot p C_q m_{a_1}^2 \kappa _{16}^A \lambda _5-2 k\cdot p C_q
   m_{a_1}^2 m_\eta^2 \kappa _{16}^A \lambda _5\\
   +6 (k\cdot p)^2 C_q m_\eta^2
   \kappa _{16}^A \lambda _5+2 m_\pi^2 k\cdot p C_q m_\eta^2 \kappa _{16}^A
   \lambda _5+4 (k\cdot p)^3 C_q \kappa _{16}^A \lambda _5+2 m_\pi^2
   (k\cdot p)^2 C_q \kappa _{16}^A \lambda _5
   \\
   -2 \left(k\cdot
   p_0\right){}^2 C_q \left(\kappa _8^A-\kappa _{16}^A\right) \left(4
   \lambda _1 m_{\pi }^2+2 \left(m_\pi^2-p\cdot p_0\right) \lambda
   _2-k\cdot p \left(\lambda _4+2 \lambda _5\right)\right)\\
   +2 p\cdot p_0
   \left(\sqrt{2} C_s \left(2 m_K^2-m_{\pi }^2\right) \left(2
   \left(m_\pi^2-m_{a_1}^2+2 k\cdot p\right) \kappa
   _9^A+\left(m_\pi^2-m_{a_1}^2+k\cdot p\right) \kappa _{10}^A\right)
   \lambda _2\right.\\
   +C_q \left(\left(m_\pi^2-m_{a_1}^2+2 k\cdot p\right)
   \left(\left(m_\pi^2-m_\eta^2-k\cdot p\right) \lambda _2-k\cdot p
   \left(\lambda _4+2 \lambda _5\right)\right) \kappa _3^A-2 (k\cdot
   p)^2 \kappa _8^A \lambda _2-2 m_\pi^2 m_{a_1}^2 \kappa _8^A \lambda _2\right.\\
   +2 m_{a_1}^2 m_\eta^2 \kappa _8^A
   \lambda _2-4 k\cdot p m_\eta^2 \kappa _8^A \lambda _2+p^4 \kappa
   _{15}^A \lambda _2-2 (k\cdot p)^2 \kappa _{15}^A \lambda _2-m_\pi^2
   m_{a_1}^2 \kappa _{15}^A \lambda _2+k\cdot p m_{a_1}^2 \kappa
   _{15}^A \lambda _2-m_\pi^2 m_\eta^2 \kappa _{15}^A \lambda _2-2
   m_\pi^2 m_\eta^2 \kappa _8^A \lambda _2\\
   +m_{a_1}^2
   m_\eta^2 \kappa _{15}^A \lambda _2-2 k\cdot p m_\eta^2 \kappa _{15}^A
   \lambda _2+m_\pi^2 k\cdot p \kappa _{15}^A \lambda _2+p^4 \kappa _{16}^A
   \lambda _2+4 (k\cdot p)^2 \kappa _{16}^A \lambda _2+m_\pi^2 m_{a_1}^2
   \kappa _{16}^A \lambda _2-2 k\cdot p m_{a_1}^2 \kappa _{16}^A
   \lambda _2\\
   +m_\pi^2 m_\eta^2 \kappa _{16}^A \lambda _2-m_{a_1}^2 m_\eta^2
   \kappa _{16}^A \lambda _2+3 k\cdot p m_\eta^2 \kappa _{16}^A \lambda
   _2+3 m_\pi^2 k\cdot p \kappa _{16}^A \lambda _2+2 m_{\pi }^2 \left(2
   \left(m_\pi^2-m_{a_1}^2+2 k\cdot p\right) \lambda _1 \kappa _3^A-2
   m_{a_1}^2 \kappa _8^A \lambda _1\right.\\
   +2 m_\eta^2 \kappa _8^A \lambda _1+2
   m_\pi^2 \kappa _{15}^A \lambda _1-2 m_{a_1}^2 \kappa _{15}^A \lambda
   _1+2 (m_{a_1}^2-2 m_\eta^2) \kappa _{16}^A
   \lambda _1-2 m_\pi^2 \kappa _9^A \lambda _2+2 m_{a_1}^2 \kappa _9^A
   \lambda _2-m_\pi^2 \kappa _{10}^A \lambda _2+m_{a_1}^2 \kappa _{10}^A
   \lambda _2-2 m_\pi^2 \kappa _{11}^A \lambda _2\\\left.
   +2 m_{a_1}^2 \kappa
   _{11}^A \lambda _2-m_\pi^2 \kappa _{12}^A \lambda _2+m_{a_1}^2 \kappa
   _{12}^A \lambda _2+k\cdot p \left(2 \lambda _1 \kappa _8^A+4 \kappa
   _{15}^A \lambda _1-2 \kappa _{16}^A \lambda _1-4 \kappa _9^A \lambda
   _2-\kappa _{10}^A \lambda _2-4 \kappa _{11}^A \lambda _2-\kappa
   _{12}^A \lambda _2\right)\right)\\
   -(k\cdot p)^2 \kappa _8^A \lambda
   _4+k\cdot p m_{a_1}^2 \kappa _8^A \lambda _4-k\cdot p m_\eta^2 \kappa
   _8^A \lambda _4-2 (k\cdot p)^2 \kappa _{15}^A \lambda _4+k\cdot p
   m_{a_1}^2 \kappa _{15}^A \lambda _4-m_\pi^2 k\cdot p \kappa _{15}^A
   \lambda _4+(k\cdot p)^2 \kappa _{16}^A \lambda _4\\
   -k\cdot p m_{a_1}^2
   \kappa _{16}^A \lambda _4+k\cdot p m_\eta^2 \kappa _{16}^A \lambda _4-2
   (k\cdot p)^2 \kappa _8^A \lambda _5+2 k\cdot p m_{a_1}^2 \kappa _8^A
   \lambda _5-2 k\cdot p m_\eta^2 \kappa _8^A \lambda _5-4 (k\cdot p)^2
   \kappa _{15}^A \lambda _5\\\left.\left.
   +2 k\cdot p m_{a_1}^2 \kappa _{15}^A
   \lambda _5-2 m_\pi^2 k\cdot p \kappa _{15}^A \lambda _5+2 (k\cdot p)^2
   \kappa _{16}^A \lambda _5-2 k\cdot p m_{a_1}^2 \kappa _{16}^A
   \lambda _5+2 k\cdot p m_\eta^2 \kappa _{16}^A \lambda
   _5\right)\right)\\
   +k\cdot p_0 \left(\sqrt{2} C_s \left(2 m_K^2-m_{\pi
   }^2\right) \left(4 \lambda _1 m_{\pi }^2+2 \left(m_\pi^2-p\cdot
   p_0\right) \lambda _2-k\cdot p \left(\lambda _4+2 \lambda
   _5\right)\right) \kappa _{10}^A\right.\\
   +C_q \left(2 \left(m_\pi^2-m_{a_1}^2+2
   k\cdot p\right) \left(\left(m_\pi^2-k\cdot p-p\cdot p_0\right) \lambda
   _2-k\cdot p \left(\lambda _4+2 \lambda _5\right)\right) \kappa
   _3^A-8 m_{\pi }^4 \left(\kappa _{10}^A+\kappa _{12}^A\right) \lambda
   _1-4 (k\cdot p)^2 \kappa _8^A \lambda _2\right.\\
   \lambda_2\left\{4\kappa _8^A\left[2(p\cdot p_0)^2-(k\cdot p)^2-m_\pi^2(m_{a_1}^2
   +m_\eta^2+2p\cdot p_0)+p\cdot p_0 m_{a_1}^2+m_\eta^2(p\cdot p_0-k\cdot p)-p\cdot p_0 k\cdot p\right]\right.\\
   +2\kappa_{15}^A\left[m_\pi^2\left(m_\pi^2-m_{a_1}^2+2k\cdot p-p\cdot p_0\right)-(k\cdot p)(k\cdot p+2p\cdot p_0-m_{a_1}^2)+2p\cdot p_0 m_{a_1}^2\right]
   +2\kappa_{16}^A\left[m_\pi^2(m_\pi^2+2m_{a_1}^2\right.\\\left.+m_\eta^2+2k\cdot p+3p\cdot p_0)+m_\eta^2(2k\cdot p-p\cdot p_0)\right]
   +\lambda_4\{\kappa_8^A[-2 (k\cdot p)^2 +2
   k\cdot p m_{a_1}^2 +2 k\cdot p p\cdot p_0]-2\kappa_{15}^A[m_\pi^2k\cdot p+2(k\cdot p)^2
   -k\cdot pm_{a_1}^2]\\-\kappa_{16}^Ak\cdot p[m_{a_1}^2+m_\pi^2-m_\eta^2+p\cdot p_0]\}
   +2\lambda_5k\cdot p\{2\kappa_8^A[m_{a_1}^2-k\cdot p+p\cdot p_0]+2\kappa_{15}^A[m_{a_1}^2-m_\pi^2-k\cdot p]+\kappa_{16}^A[m_\eta^2-m_\pi^2-m_{a_1}^2-p\cdot p_0]\}\\
   +2 m_{\pi }^2 \left(4
   \left(m_\pi^2-m_{a_1}^2+2 k\cdot p\right) \lambda _1 \kappa _3^A+2
   \left(-m_\pi^2 \lambda _2 \kappa _{10}^A+2 m_\pi^2 \kappa _{15}^A \lambda
   _1+ (m_\pi^2-m_\eta^2) \kappa _{16}^A \lambda _1-2
   m_{a_1}^2 \left(\kappa _8^A+\kappa _{15}^A-\kappa _{16}^A\right)
   \lambda _1\right.\right.\\\left.\left.\left.\left.\left.
   -m_\pi^2 \kappa _{12}^A \lambda _2+p\cdot p_0 \left(-2 \lambda
   _1 \kappa _8^A+2 \kappa _{16}^A \lambda _1+\left(\kappa
   _{10}^A+\kappa _{12}^A\right) \lambda _2\right)\right)+k\cdot p
   \left(4 \lambda _1 \kappa _8^A+8 \kappa _{15}^A \lambda
   _1+\left(\kappa _{10}^A+\kappa _{12}^A\right) \left(\lambda _4+2
   \lambda _5\right)\right)\right)\right)\right)\right),\nonumber
\end{multline}

\begin{multline}
 a_2^{W^-\to (a_1^-)\eta\to\pi^-(\rho^0)\eta\to\pi^-\gamma\eta}=   
  +\frac{8F_V}{F^2m_{a_1}^2m_\rho^2D_{a_1}[(p+k)^2]}\left(2 \left(C_q m_\eta^2 \kappa _3^A-4 \sqrt{2} C_s m_K^2 \kappa _9^A
  +4 C_q m_{\pi }^2 \kappa _9^A+2
   \sqrt{2} C_s m_{\pi }^2 \kappa _9^A\right.\right.\\\left.
   +4 C_q m_{\pi }^2 \kappa _{11}^A+C_q m_\eta^2 \kappa _{15}^A-C_q m_\eta^2 \kappa
   _{16}^A+k\cdot p_0 C_q \left(\kappa _3^A+\kappa _{15}^A\right)+p\cdot p_0 C_q \left(\kappa _3^A+\kappa _{15}^A-\kappa
   _{16}^A\right)\right) \left(2 \lambda _2+\lambda _4+2 \lambda _5\right) (k\cdot p)^2\\
   +\left(-8 p\cdot p_0 C_q m_{\pi }^2
   \lambda _1 \kappa _3^A-8 C_q m_{\pi }^2 m_\eta^2 \lambda _1 \kappa _3^A-2 p\cdot p_0 C_q m_{a_1}^2 \lambda _2 \kappa _3^A-2
   C_q m_{a_1}^2 m_\eta^2 \lambda _2 \kappa _3^A+4 p\cdot p_0 C_q m_\eta^2 \lambda _2 \kappa _3^A\right.\\
   +4 \left(p\cdot p_0\right){}^2 C_q
   \lambda _2 \kappa _3^A-p\cdot p_0 C_q m_{a_1}^2 \lambda _4 \kappa _3^A-C_q m_{a_1}^2 m_\eta^2 \lambda _4 \kappa _3^A-2 p\cdot
   p_0 C_q m_{a_1}^2 \lambda _5 \kappa _3^A-2 C_q m_{a_1}^2 m_\eta^2 \lambda _5 \kappa _3^A\\
   -32 C_q m_{\pi }^4 \kappa _9^A \lambda
   _1-16 \sqrt{2} C_s m_{\pi }^4 \kappa _9^A \lambda _1+32 \sqrt{2} C_s m_K^2 m_{\pi }^2 \kappa _9^A \lambda _1-32 C_q m_{\pi
   }^4 \kappa _{11}^A \lambda _1-8 p\cdot p_0 C_q m_{\pi }^2 \kappa _{15}^A \lambda _1\\
   -8 C_q m_{\pi }^2 m_\eta^2 \kappa _{15}^A
   \lambda _1+8 p\cdot p_0 C_q m_{\pi }^2 \kappa _{16}^A \lambda _1+8 C_q m_{\pi }^2 m_\eta^2 \kappa _{16}^A \lambda _1+4 C_q
   p_0^4 \kappa _8^A \lambda _2+8 p\cdot p_0 C_q m_\eta^2 \kappa _8^A \lambda _2+4 \left(p\cdot p_0\right){}^2 C_q \kappa _8^A
   \lambda _2\\
   -16 \sqrt{2} p\cdot p_0 C_s m_K^2 \kappa _9^A \lambda _2+16 p\cdot p_0 C_q m_{\pi }^2 \kappa _9^A \lambda _2+8
   \sqrt{2} p\cdot p_0 C_s m_{\pi }^2 \kappa _9^A \lambda _2+8 \sqrt{2} C_s m_K^2 m_{a_1}^2 \kappa _9^A \lambda _2-8 C_q
   m_{\pi }^2 m_{a_1}^2 \kappa _9^A \lambda _2\\
   -4 \sqrt{2} C_s m_{\pi }^2 m_{a_1}^2 \kappa _9^A \lambda _2-4 \sqrt{2} p\cdot
   p_0 C_s m_K^2 \kappa _{10}^A \lambda _2+4 p\cdot p_0 C_q m_{\pi }^2 \kappa _{10}^A \lambda _2+2 \sqrt{2} p\cdot p_0 C_s
   m_{\pi }^2 \kappa _{10}^A \lambda _2-4 \sqrt{2} C_s m_K^2 m_\eta^2 \kappa _{10}^A \lambda _2\\
   +4 C_q m_{\pi }^2 m_\eta^2 \kappa
   _{10}^A \lambda _2+2 \sqrt{2} C_s m_{\pi }^2 m_\eta^2 \kappa _{10}^A \lambda _2+16 p\cdot p_0 C_q m_{\pi }^2 \kappa _{11}^A
   \lambda _2-8 C_q m_{\pi }^2 m_{a_1}^2 \kappa _{11}^A \lambda _2+4 p\cdot p_0 C_q m_{\pi }^2 \kappa _{12}^A \lambda _2+4 C_q
   m_{\pi }^2 m_\eta^2 \kappa _{12}^A \lambda _2\\
   -2 p\cdot p_0 C_q m_{a_1}^2 \kappa _{15}^A \lambda _2-2 C_q m_{a_1}^2 m_\eta^2
   \kappa _{15}^A \lambda _2+4 p\cdot p_0 C_q m_\eta^2 \kappa _{15}^A \lambda _2+4 \left(p\cdot p_0\right){}^2 C_q \kappa _{15}^A
   \lambda _2-2 C_q p_0^4 \kappa _{16}^A \lambda _2\\
   +4 p\cdot p_0 C_q m_{a_1}^2 \kappa _{16}^A \lambda _2-6 p\cdot p_0 C_q
   m_\eta^2 \kappa _{16}^A \lambda _2+2 C_q p_0^4 \kappa _8^A \lambda _4+4 p\cdot p_0 C_q m_\eta^2 \kappa _8^A \lambda _4+2
   \left(p\cdot p_0\right){}^2 C_q \kappa _8^A \lambda _4\\
   +4 \sqrt{2} C_s m_K^2 m_{a_1}^2 \kappa _9^A \lambda _4-4 C_q m_{\pi
   }^2 m_{a_1}^2 \kappa _9^A \lambda _4-2 \sqrt{2} C_s m_{\pi }^2 m_{a_1}^2 \kappa _9^A \lambda _4-2 \sqrt{2} p\cdot p_0 C_s
   m_K^2 \kappa _{10}^A \lambda _4+2 p\cdot p_0 C_q m_{\pi }^2 \kappa _{10}^A \lambda _4\\
   +\sqrt{2} p\cdot p_0 C_s m_{\pi }^2
   \kappa _{10}^A \lambda _4-2 \sqrt{2} C_s m_K^2 m_\eta^2 \kappa _{10}^A \lambda _4+2 C_q m_{\pi }^2 m_\eta^2 \kappa _{10}^A
   \lambda _4+\sqrt{2} C_s m_{\pi }^2 m_\eta^2 \kappa _{10}^A \lambda _4-4 C_q m_{\pi }^2 m_{a_1}^2 \kappa _{11}^A \lambda _4\\
   +2
   p\cdot p_0 C_q m_{\pi }^2 \kappa _{12}^A \lambda _4+2 C_q m_{\pi }^2 m_\eta^2 \kappa _{12}^A \lambda _4-p\cdot p_0 C_q
   m_{a_1}^2 \kappa _{15}^A \lambda _4-C_q m_{a_1}^2 m_\eta^2 \kappa _{15}^A \lambda _4-C_q p_0^4 \kappa _{16}^A \lambda _4-3
   p\cdot p_0 C_q m_\eta^2 \kappa _{16}^A \lambda _4\\
   -2 \left(p\cdot p_0\right){}^2 C_q \kappa _{16}^A \lambda _4+4 C_q p_0^4
   \kappa _8^A \lambda _5+8 p\cdot p_0 C_q m_\eta^2 \kappa _8^A \lambda _5+4 \left(p\cdot p_0\right){}^2 C_q \kappa _8^A \lambda
   _5+8 \sqrt{2} C_s m_K^2 m_{a_1}^2 \kappa _9^A \lambda _5\\
   -8 C_q m_{\pi }^2 m_{a_1}^2 \kappa _9^A \lambda _5-4 \sqrt{2} C_s
   m_{\pi }^2 m_{a_1}^2 \kappa _9^A \lambda _5-4 \sqrt{2} p\cdot p_0 C_s m_K^2 \kappa _{10}^A \lambda _5+4 p\cdot p_0 C_q
   m_{\pi }^2 \kappa _{10}^A \lambda _5+2 \sqrt{2} p\cdot p_0 C_s m_{\pi }^2 \kappa _{10}^A \lambda _5\\
   -4 \sqrt{2} C_s m_K^2
   m_\eta^2 \kappa _{10}^A \lambda _5+4 C_q m_{\pi }^2 m_\eta^2 \kappa _{10}^A \lambda _5+2 \sqrt{2} C_s m_{\pi }^2 m_\eta^2 \kappa
   _{10}^A \lambda _5-8 C_q m_{\pi }^2 m_{a_1}^2 \kappa _{11}^A \lambda _5+4 p\cdot p_0 C_q m_{\pi }^2 \kappa _{12}^A \lambda
   _5\\
   +4 C_q m_{\pi }^2 m_\eta^2 \kappa _{12}^A \lambda _5-2 p\cdot p_0 C_q m_{a_1}^2 \kappa _{15}^A \lambda _5-2 C_q m_{a_1}^2
   m_\eta^2 \kappa _{15}^A \lambda _5-2 C_q p_0^4 \kappa _{16}^A \lambda _5-6 p\cdot p_0 C_q m_\eta^2 \kappa _{16}^A \lambda _5-4
   \left(p\cdot p_0\right){}^2 C_q \kappa _{16}^A \lambda _5\\
   +m_\pi^2 \left(-4 \sqrt{2} C_s m_K^2 \kappa _9^A
   +2 (\sqrt{2} C_s +2 C_q) m_{\pi }^2 \kappa _9^A+4 C_q m_{\pi }^2 \kappa _{11}^A+C_q m_\eta^2 \left(\kappa _3^A+\kappa
   _{15}^A-\kappa _{16}^A\right)+p\cdot p_0 C_q \left(\kappa _3^A+\kappa _{15}^A-\kappa _{16}^A\right)\right) \\
   \left(2 \lambda
   _2+\lambda _4+2 \lambda _5\right)+k\cdot p_0 C_q \left(-8 m_{\pi }^2 \lambda _1 (\kappa _3^A+\kappa_{15}^A)-(2\lambda_2+\lambda_4+2\lambda_5)
   \left[(m_{a_1}^2-m_\pi^2)(\kappa_3^A+\kappa_{15}^A)+2m_\eta^2(\kappa_{16}^A-\kappa_8^A)\right]\right.
   \\\left.\left.
   +2 p\cdot p_0 \left[2\lambda_2(\kappa_3^A+\kappa_8^A+\kappa_{15}^A)+(\lambda_4+2\lambda_5)(\kappa_8-\kappa_{16})
   \right]\right)\right) k\cdot p\\
   +2 \left(
   C_qm_{a_1}^2(p\cdot p_0+m_\eta^2)(m_\pi^2\lambda_1+p\cdot p_0\lambda_2)\kappa_3^A
   +2m_\pi^2\lambda_1\left\{-2C_q\kappa_8^A\left[m_\eta^2+p\cdot p_0\right]^2-\left[\sqrt{2}C_s(2m_K^2-m_\pi^2)-m_\pi^2C_q\right]\right.\right.\\
   \left.\left[2m_{a_1}^2\kappa_9^A-(p\cdot p_0+m_\eta^2)\kappa_{10}^A\right]+C_q\left[(p\cdot p_0+m_\eta^2)(m_{a_1}^2\kappa_{15}^A
   +(m_\eta^2+2p\cdot p_0)\kappa_{16}^A-m_\pi^2\kappa_{12}^A)\right]\right\}\\
   +p\cdot p_0 \lambda_2\left\{-C_q(m_{a_1}^2+p\cdot p_0)\left[2\kappa_8^A(m_\eta^2+p\cdot p_0)+2m_\pi^2\kappa_{12}^A
   -\kappa_{16}^A(2p\cdot p_0+m_\eta^2)\right]-C_qm_{a_1}^2\left[4m_\pi^2\kappa_{11}^A+\kappa_{15}^A(p\cdot p_0)\right]\right.\\
   \left.+\left[\sqrt{2}C_s(2m_K^2-m_\pi^2)-m_\pi^2C_q\right]\left[m_{a_1}^2\kappa_9^A+(p\cdot p_0+m_{a_1}^2)\kappa_{10}^A\right]\right\}
      +2 \left(k\cdot p_0\right){}^2 p\cdot p_0 C_q \left(\kappa _{16}^A-\kappa _8^A\right) \lambda _2\\
   +m_\pi^2 \left(C_q
   \left(\kappa _3^A+\kappa _{15}^A+\kappa _{16}^A\right) \lambda _2 \left(p\cdot p_0\right){}^2+\left(2 \sqrt{2} C_s
   \left(m_{\pi }^2-2 m_K^2\right) \lambda _2 \kappa _9^A+C_q m_\eta^2 \left(\kappa _3^A+\kappa _{15}^A\right) \lambda _2\right.\right.\\\left.
   +C_q
   \left(m_{a_1}^2 \lambda _2 \kappa _{16}^A+m_{\pi }^2 \left(-2 \lambda _1 \kappa _3^A-2 \kappa _{15}^A \lambda _1+2 \kappa
   _{16}^A \lambda _1+4 \kappa _9^A \lambda _2+4 \kappa _{11}^A \lambda _2\right)\right)\right) p\cdot p_0\\\left.
   -2 m_{\pi }^2
   \left(2 \sqrt{2} C_s \left(m_{\pi }^2-2 m_K^2\right) \kappa _9^A+4 C_q m_{\pi }^2 \left(\kappa _9^A+\kappa
   _{11}^A\right)+C_q m_\eta^2 \left(\kappa _3^A+\kappa _{15}^A-\kappa _{16}^A\right)\right) \lambda _1\right)   \nonumber
   \end{multline}
   
   \begin{multline}
   +k\cdot p_0
   \left(C_q \left(p\cdot p_0 \left(\kappa _3^A+\kappa _{15}^A+\kappa _{16}^A\right) \lambda _2-2 m_{\pi }^2 \left(\kappa
   _3^A+\kappa _{15}^A\right) \lambda _1\right) m_\pi^2+2 C_q m_{\pi }^2 \left(\left(\kappa _3^A+\kappa _{15}^A\right) m_{a_1}^2+2
   m_\eta^2 \left(\kappa _{16}^A-\kappa _8^A\right)\right) \lambda _1\right.\\
   +4 \left(p\cdot p_0\right){}^2 C_q \left(\kappa
   _{16}^A-\kappa _8^A\right) \lambda _2+p\cdot p_0 \left(\sqrt{2} C_s \left(2 m_K^2-m_{\pi }^2\right) \lambda _2 \kappa
   _{10}^A+C_q m_\eta^2 \left(\kappa _{16}^A-2 \kappa _8^A\right) \lambda _2\right.\\\left.\left.\left.\left.
   +C_q \left(-2 \left(2 \lambda _1 \kappa _8^A-2 \kappa
   _{16}^A \lambda _1+\left(\kappa _{10}^A+\kappa _{12}^A\right) \lambda _2\right) m_{\pi }^2-m_{a_1}^2 \left(\kappa _3^A+2
   \kappa _8^A+\kappa _{15}^A-2 \kappa _{16}^A\right) \lambda _2\right)\right)\right)\right)\right)\,,
\end{multline}

\begin{multline}
 a_3^{W^-\to (a_1^-)\eta\to\pi^-(\rho^0)\eta\to\pi^-\gamma\eta}=   
  +\frac{8F_V}{F^2m_{a_1}^2m_\rho^2D_{a_1}[(p+k)^2]}\left(2 C_q m_{a_1}^2 m_\eta^2 \lambda _2 \kappa
   _3^A-2 m_\pi^2 C_q m_\eta^2 \lambda _2 \kappa _3^A-4 k\cdot p C_q m_\eta^2
   \lambda _2 \kappa _3^A\right.\\
   +8 C_q m_{\pi }^4 \kappa _{10}^A \lambda _1+4 \sqrt{2} C_s m_{\pi
   }^4 \kappa _{10}^A \lambda _1-8 \sqrt{2} C_s m_K^2 m_{\pi }^2 \kappa
   _{10}^A \lambda _1+8 C_q m_{\pi }^4 \kappa _{12}^A \lambda _1-4 C_q
   m_{\pi }^2 m_{a_1}^2 \kappa _{16}^A \lambda _1-4 C_q m_{\pi }^2
   m_\eta^2 \kappa _{16}^A \lambda _1\\
   +4 C_q m_{a_1}^2 m_\eta^2 \kappa _8^A
   \lambda _2-4 k\cdot p C_q m_\eta^2 \kappa _8^A \lambda _2+8 \sqrt{2}
   m_\pi^2 C_s m_K^2 \kappa _9^A \lambda _2+16 \sqrt{2} k\cdot p C_s m_K^2
   \kappa _9^A \lambda _2-8 m_\pi^2 C_q m_{\pi }^2 \kappa _9^A \lambda
   _2\\
   -16 k\cdot p C_q m_{\pi }^2 \kappa _9^A \lambda _2-4 \sqrt{2} m_\pi^2
   C_s m_{\pi }^2 \kappa _9^A \lambda _2-8 \sqrt{2} k\cdot p C_s m_{\pi
   }^2 \kappa _9^A \lambda _2-8 \sqrt{2} C_s m_K^2 m_{a_1}^2 \kappa
   _9^A \lambda _2+8 C_q m_{\pi }^2 m_{a_1}^2 \kappa _9^A \lambda _2\\
   +4
   \sqrt{2} C_s m_{\pi }^2 m_{a_1}^2 \kappa _9^A \lambda _2+4 \sqrt{2}
   k\cdot p C_s m_K^2 \kappa _{10}^A \lambda _2-4 k\cdot p C_q m_{\pi
   }^2 \kappa _{10}^A \lambda _2-2 \sqrt{2} k\cdot p C_s m_{\pi }^2
   \kappa _{10}^A \lambda _2-4 \sqrt{2} C_s m_K^2 m_{a_1}^2 \kappa
   _{10}^A \lambda _2\\
   +4 C_q m_{\pi }^2 m_{a_1}^2 \kappa _{10}^A \lambda
   _2+2 \sqrt{2} C_s m_{\pi }^2 m_{a_1}^2 \kappa _{10}^A \lambda _2-8
   m_\pi^2 C_q m_{\pi }^2 \kappa _{11}^A \lambda _2-16 k\cdot p C_q m_{\pi
   }^2 \kappa _{11}^A \lambda _2+8 C_q m_{\pi }^2 m_{a_1}^2 \kappa
   _{11}^A \lambda _2\\
   -4 k\cdot p C_q m_{\pi }^2 \kappa _{12}^A \lambda
   _2+4 C_q m_{\pi }^2 m_{a_1}^2 \kappa _{12}^A \lambda _2+2 C_q
   m_{a_1}^2 m_\eta^2 \kappa _{15}^A \lambda _2-2 m_\pi^2 C_q m_\eta^2 \kappa
   _{15}^A \lambda _2-4 k\cdot p C_q m_\eta^2 \kappa _{15}^A \lambda _2-2
   m_\pi^2 C_q m_{a_1}^2 \kappa _{16}^A \lambda _2\\
   -2 k\cdot p C_q m_{a_1}^2
   \kappa _{16}^A \lambda _2-2 C_q m_{a_1}^2 m_\eta^2 \kappa _{16}^A
   \lambda _2+2 k\cdot p C_q m_\eta^2 \kappa _{16}^A \lambda _2+4
   \left(k\cdot p_0\right){}^2 C_q \left(\kappa _8^A-\kappa
   _{16}^A\right) \lambda _2\\
   +4 \left(p\cdot p_0\right){}^2 C_q
   \left(\kappa _8^A-\kappa _{16}^A\right) \lambda _2-2 k\cdot p C_q
   m_\eta^2 \kappa _8^A \lambda _4+2 \sqrt{2} k\cdot p C_s m_K^2 \kappa
   _{10}^A \lambda _4-2 k\cdot p C_q m_{\pi }^2 \kappa _{10}^A \lambda
   _4-\sqrt{2} k\cdot p C_s m_{\pi }^2 \kappa _{10}^A \lambda _4\\
   -2
   k\cdot p C_q m_{\pi }^2 \kappa _{12}^A \lambda _4+k\cdot p C_q
   m_{a_1}^2 \kappa _{16}^A \lambda _4+k\cdot p C_q m_\eta^2 \kappa
   _{16}^A \lambda _4-4 k\cdot p C_q m_\eta^2 \kappa _8^A \lambda _5+4
   \sqrt{2} k\cdot p C_s m_K^2 \kappa _{10}^A \lambda _5\\
   -4 k\cdot p C_q
   m_{\pi }^2 \kappa _{10}^A \lambda _5-2 \sqrt{2} k\cdot p C_s m_{\pi
   }^2 \kappa _{10}^A \lambda _5-4 k\cdot p C_q m_{\pi }^2 \kappa
   _{12}^A \lambda _5+2 k\cdot p C_q m_{a_1}^2 \kappa _{16}^A \lambda
   _5+2 k\cdot p C_q m_\eta^2 \kappa _{16}^A \lambda _5\\
   +8 C_q m_{\pi }^2 m_\eta^2 \kappa _8^A \lambda
   _1+2 p\cdot p_0
   \left(C_q \left(-\left(m_\pi^2-m_{a_1}^2+2 k\cdot p\right) \lambda _2
   \kappa _3^A+2 m_{a_1}^2 \kappa _8^A \lambda _2+2 m_\eta^2 \kappa _8^A
   \lambda _2\right.\right.\\
   -2 k\cdot p \kappa _8^A \lambda _2-m_\pi^2 \kappa _{15}^A
   \lambda _2+m_{a_1}^2 \kappa _{15}^A \lambda _2-2 k\cdot p \kappa
   _{15}^A \lambda _2-m_\pi^2 \kappa _{16}^A \lambda _2-2 m_{a_1}^2 \kappa
   _{16}^A \lambda _2-m_\eta^2 \kappa _{16}^A \lambda _2\\\left.
   +2 m_{\pi }^2
   \left(2 \lambda _1 \kappa _8^A-2 \kappa _{16}^A \lambda
   _1+\left(\kappa _{10}^A+\kappa _{12}^A\right) \lambda
   _2\right)-k\cdot p \kappa _8^A \lambda _4+k\cdot p \kappa _{16}^A
   \lambda _4-2 k\cdot p \kappa _8^A \lambda _5+2 k\cdot p \kappa
   _{16}^A \lambda _5\right)\\\left.
   -\sqrt{2} C_s \left(2 m_K^2-m_{\pi
   }^2\right) \kappa _{10}^A \lambda _2\right)+2 k\cdot p_0 \left(C_q
   \left(-\left(m_\pi^2-m_{a_1}^2+2 k\cdot p\right) \lambda _2 \kappa
   _3^A+2 m_{a_1}^2 \kappa _8^A \lambda _2+2 m_\eta^2 \kappa _8^A \lambda
   _2-2 k\cdot p \kappa _8^A \lambda _2\right.\right.\\
   +4 p\cdot p_0 \kappa _8^A
   \lambda _2-m_\pi^2 \kappa _{15}^A \lambda _2+m_{a_1}^2 \kappa _{15}^A
   \lambda _2-2 k\cdot p \kappa _{15}^A \lambda _2-m_\pi^2 \kappa _{16}^A
   \lambda _2-2 m_{a_1}^2 \kappa _{16}^A \lambda _2-m_\eta^2 \kappa
   _{16}^A \lambda _2-4 p\cdot p_0 \kappa _{16}^A \lambda _2\\\left.
   +2 m_{\pi
   }^2 \left(2 \lambda _1 \kappa _8^A-2 \kappa _{16}^A \lambda
   _1+\left(\kappa _{10}^A+\kappa _{12}^A\right) \lambda
   _2\right)-k\cdot p \kappa _8^A \lambda _4+k\cdot p \kappa _{16}^A
   \lambda _4-2 k\cdot p \kappa _8^A \lambda _5+2 k\cdot p \kappa
   _{16}^A \lambda _5\right)\\\left.\left.
   -\sqrt{2} C_s \left(2 m_K^2-m_{\pi
   }^2\right) \kappa _{10}^A \lambda _2\right)\right)\,,
\end{multline}

\begin{multline}
 a_4^{W^-\to (a_1^-)\eta\to\pi^-(\rho^0)\eta\to\pi^-\gamma\eta}=   
  +\frac{16F_V}{F^2m_{a_1}^2m_\rho^2D_{a_1}[(p+k)^2]}\left(C_q \left(-\left(p\cdot p_0\right) m_{a_1}^2
   \lambda _2 \kappa _3^A+m_\pi^2 m_\eta^2 \lambda _2 \kappa _3^A-m_{a_1}^2
   m_\eta^2 \lambda _2 \kappa _3^A\right.\right.\\
   +2 k\cdot p m_\eta^2 \lambda _2 \kappa
   _3^A+m_\pi^2 p\cdot p_0 \lambda _2 \kappa _3^A+2 k\cdot p p\cdot p_0
   \lambda _2 \kappa _3^A+4 (k\cdot p)^2 \kappa _8^A \lambda _2-2
   \left(p\cdot p_0\right){}^2 \kappa _8^A \lambda _2-2 k\cdot p
   m_{a_1}^2 \kappa _8^A \lambda _2\\
   -2 p\cdot p_0 m_{a_1}^2 \kappa _8^A
   \lambda _2-2 m_{a_1}^2 m_\eta^2 \kappa _8^A \lambda _2-2 p\cdot p_0
   m_\eta^2 \kappa _8^A \lambda _2+2 m_\pi^2 k\cdot p \kappa _8^A \lambda
   _2-p\cdot p_0 m_{a_1}^2 \kappa _{15}^A \lambda _2+m_\pi^2 m_\eta^2 \kappa
   _{15}^A \lambda _2\\
   -m_{a_1}^2 m_\eta^2 \kappa _{15}^A \lambda _2+2
   k\cdot p m_\eta^2 \kappa _{15}^A \lambda _2+m_\pi^2 p\cdot p_0 \kappa
   _{15}^A \lambda _2+2 k\cdot p p\cdot p_0 \kappa _{15}^A \lambda _2+2
   \left(p\cdot p_0\right){}^2 \kappa _{16}^A \lambda _2+m_\pi^2 m_{a_1}^2
   \kappa _{16}^A \lambda _2+2 k\cdot p m_{a_1}^2 \kappa _{16}^A
   \lambda _2\\
   +2 p\cdot p_0 m_{a_1}^2 \kappa _{16}^A \lambda
   _2+m_{a_1}^2 m_\eta^2 \kappa _{16}^A \lambda _2+p\cdot p_0 m_\eta^2 \kappa
   _{16}^A \lambda _2+m_\pi^2 p\cdot p_0 \kappa _{16}^A \lambda _2+2 k\cdot
   p p\cdot p_0 \kappa _{16}^A \lambda _2-2 \left(k\cdot p_0\right){}^2
   \left(\kappa _8^A-\kappa _{16}^A\right) \lambda _2\\
   +k\cdot p_0
   \left(\left(m_\pi^2-m_{a_1}^2+2 k\cdot p\right) \kappa _3^A-2 m_{a_1}^2
   \kappa _8^A-2 m_\eta^2 \kappa _8^A+m_\pi^2 \kappa _{15}^A-m_{a_1}^2 \kappa
   _{15}^A+2 k\cdot p \kappa _{15}^A+m_\pi^2 \kappa _{16}^A+2 m_{a_1}^2
   \kappa _{16}^A+m_\eta^2 \kappa _{16}^A\right.\\\left.
   +2 k\cdot p \kappa _{16}^A-4
   p\cdot p_0 \left(\kappa _8^A-\kappa _{16}^A\right)\right) \lambda
   _2-2 m_{\pi }^2 \left(2 \left(m_\pi^2-m_{a_1}^2+2 k\cdot p\right) \kappa
   _8^A \lambda _1-\left(2 \left(m_\pi^2-m_{a_1}^2+2 k\cdot p\right) \kappa
   _9^A-m_{a_1}^2 \kappa _{10}^A\right.\right.\\\left.\left.
   -p\cdot p_0 \kappa _{10}^A+2 m_\pi^2 \kappa
   _{11}^A-2 m_{a_1}^2 \kappa _{11}^A+4 k\cdot p \kappa
   _{11}^A-m_{a_1}^2 \kappa _{12}^A-p\cdot p_0 \kappa _{12}^A-k\cdot
   p_0 \left(\kappa _{10}^A+\kappa _{12}^A\right)\right) \lambda
   _2\right)+2 (k\cdot p)^2 \kappa _8^A \lambda _4\\\left.
   -k\cdot p m_{a_1}^2
   \kappa _8^A \lambda _4+m_\pi^2 k\cdot p \kappa _8^A \lambda _4+4 (k\cdot
   p)^2 \kappa _8^A \lambda _5-2 k\cdot p m_{a_1}^2 \kappa _8^A \lambda
   _5+2 m_\pi^2 k\cdot p \kappa _8^A \lambda _5\right)\\\left.
   -\sqrt{2} C_s \left(2
   m_K^2-m_{\pi }^2\right) \left(2 \left(m_\pi^2-m_{a_1}^2+2 k\cdot
   p\right) \kappa _9^A-\left(m_{a_1}^2+k\cdot p_0+p\cdot p_0\right)
   \kappa _{10}^A\right) \lambda _2\right)\,.
\end{multline}

We turn now to the vector factors, with the one-resonance exchange contributions (fig.~\ref{fig:1R-FV}) listed next:\\

\begin{multline}\label{v1-1R-RChL}
 v_1^{1R} =
 \frac{4\sqrt{2}C_q}{3F^2M_V^2m_\rho^2D_\rho[(p+k)^2]}\left(2 k\cdot p \left(4 \left(c_5+c_7\right) \left(m_{\rho }^2-2 m_{\pi }^2\right) 
 \left(\left(c_5+c_7\right) \left(k\cdot p_0+m_{\eta
   }^2\right)+4 c_3 \left(m_{\pi }^2-m_{\eta }^2\right)\right)\right.\right.\\\left.
   +p\cdot p_0 \left(8 \left(c_5+c_7\right) \left(2 c_3 m_{\eta }^2-\left(2
   c_3+c_5+c_7\right) m_{\pi }^2\right)+\left(4 \left(c_5+c_7\right){}^2+c_{1256}^2\right) m_{\rho }^2\right)\right)\\
   +4
   \left(c_5+c_7\right) m_{\pi }^2 \left(m_{\rho }^2-m_{\pi }^2\right) \left(\left(c_5+c_7\right) \left(k\cdot p_0+m_{\eta }^2\right)+4
   c_3 \left(m_{\pi }^2-m_{\eta }^2\right)\right)\\
   -16 \left(c_5+c_7\right) (k\cdot p)^2 \left(\left(c_5+c_7\right) \left(k\cdot p_0+m_{\eta
   }^2+p\cdot p_0\right)+4 c_3 \left(m_{\pi }^2-m_{\eta }^2\right)\right)\\\left.
   +p\cdot p_0 \left(-8 c_3 \left(m_{\pi }^2-m_{\eta }^2\right)
   \left(\left(c_{1256}-2 \left(c_5+c_7\right)\right) m_{\rho }^2+2 \left(c_5+c_7\right) m_{\pi }^2\right)+\left(4
   \left(c_5+c_7\right){}^2(m_\rho^2-m_\pi^2)+c_{1256}^2 m_{\rho }^2\right) m_{\pi }^2
   \right)\right)\\
  +\frac{4\sqrt{2}C_q}{3F^2M_V^2m_\omega^2D_\omega[(p_0+k)^2]}\left(16 c_3 m_{\pi }^2 k\cdot p_0 \left(\left(c_5+c_7\right) \left(2 k\cdot
   p_0+m_{\eta }^2\right)+4 c_3 \left(m_{\eta }^2-m_{\pi
   }^2\right)\right)\right.\\
   -m_{\omega }^2 \left(\left(c_{1256} \left(8 c_3+3
   c_{1256}\right) m_{\eta }^2+16 c_3 \left(c_5+c_7\right) m_{\pi
   }^2\right) k\cdot p_0+2 c_{1256}^2 \left(k\cdot p_0\right){}^2+64
   c_3^2 m_{\pi }^2 m_{\eta }^2\right.\\\left.\left.
   +c_{1256} \left(8 c_3+c_{1256}\right)
   m_{\eta }^4-64 c_3^2 m_{\pi }^4\right)\right)\\
 -\frac{8\sqrt{2}C_q\lambda_{15}^{S}}{3F^2D_{a_0}[(p+p_0)^2]}\left(c_mm_{\pi }^2 +c_d p\cdot p_0\right)\\
  -\frac{2C_qF_V}{3F^2m_\omega^2}\left(-\left(\lambda _{13}^V-\lambda _{14}^V-\lambda
   _{15}^V\right) \left(k\cdot p_0+m_\eta^2+p\cdot p_0\right)+4 m_{\pi }^2
   \lambda _6^V+2 p\cdot p_0 \left(\lambda _{11}^V+\lambda
   _{12}^V\right)\right)\,,
   \\
\end{multline}

\begin{multline}\label{v2-1R-RChL}
 v_2^{1R} =
 \frac{4\sqrt{2}C_q}{3F^2M_V^2m_\rho^2D_\rho[(p+k)^2]}\left(m_{\rho }^2 \left(-\left(\left(3 c_{1256}^2 m_{\pi }^2-8 c_3 \left(2 \left(c_5+c_7\right)+c_{1256}\right) 
 \left(m_{\pi }^2-m_{\eta
   }^2\right)\right) k\cdot p+2 c_{1256}^2 (k\cdot p)^2\right.\right.\right.\\\left.\left.\left.
   +c_{1256} m_{\pi }^2 \left(8 c_3 m_{\eta }^2+\left(c_{1256}-8 c_3\right) m_{\pi
   }^2\right)\right)\right)-16 c_3 \left(c_5+c_7\right) \left(m_{\pi }^2
   -m_{\eta }^2\right) k\cdot p \left(2 k\cdot p+m_{\pi }^2\right)\right)\\
  -\frac{16\sqrt{2}C_q}{3F^2M_V^2m_\omega^2D_\omega[(p_0+k)^2]}\left(-\frac{1}{2} k\cdot p_0 \left(p\cdot p_0 \left(8
   \left(c_5+c_7\right) \left(4 c_3 m_{\pi }^2-\left(2
   c_3+c_5+c_7\right) m_{\eta }^2\right)\right.\right.\right.\\\left.\left.
   +\left(4
   \left(c_5+c_7\right){}^2+c_{1256}^2\right) m_{\omega }^2\right)-4
   \left(4 c_3-c_5-c_7\right) m_{\pi }^2 \left(-2 \left(2
   c_3+c_5+c_7\right) m_{\eta }^2+\left(c_5+c_7\right) m_{\omega }^2+4
   c_3 m_{\pi }^2\right)\right)\\
   +\left(c_5+c_7\right) k\cdot p
   \left(\left(c_5+c_7\right) \left(2 k\cdot p_0+m_{\eta }^2\right)+4
   c_3 \left(m_{\eta }^2-m_{\pi }^2\right)\right) \left(2 k\cdot
   p_0+m_{\eta }^2-m_{\omega }^2\right)\\\left.+4 \left(c_5+c_7\right)
   \left(k\cdot p_0\right){}^2 \left(\left(c_5+c_7\right) \left(m_{\pi
   }^2+p\cdot p_0\right)-4 c_3 m_{\pi }^2\right)+\frac{1}{4} m_{\omega
   }^2 \left(16 c_3 m_{\pi }^2 \left(\left(c_5+c_7\right) \left(m_{\pi
   }^2+2 p\cdot p_0\right)-4 c_3 m_{\pi }^2\right)\right.\right.\\\left.
   -m_{\eta }^2 \left(4
   \left(\left(c_5+c_7\right){}^2-16 c_3^2\right) m_{\pi
   }^2+\left(c_{1256}^2+8 c_3 c_{1256}+4 \left(c_5+c_7\right) \left(4
   c_3+c_5+c_7\right)\right) p\cdot p_0\right)\right)\\\left.
   +\left(4 c_3 m_{\pi
   }^2-\left(4 c_3+c_5+c_7\right) m_{\eta }^2\right) \left(4 c_3 m_{\pi
   }^2 p\cdot p_0-m_{\eta }^2 \left(\left(c_5+c_7\right) \left(m_{\pi
   }^2+p\cdot p_0\right)-4 c_3 m_{\pi }^2\right)\right)\right)\\
+ \frac{8\sqrt{2}C_q\lambda_{15}^{S}}{3F^2D_{a_0}[(p+p_0)^2]}\left(c_mm_{\pi }^2 +c_d p\cdot p_0\right)\\
  -\frac{2C_qF_V}{3F^2m_\omega^2}\left(-\left(\lambda _{13}^V-\lambda _{14}^V-\lambda
   _{15}^V\right) \left(k\cdot p_0+m_\eta^2+p\cdot p_0\right)+4 m_{\pi }^2
   \lambda _6^V+2 p\cdot p_0 \left(\lambda _{11}^V+\lambda
   _{12}^V\right)\right)\,,
 \\
\end{multline}

\begin{multline}\label{v3-1R-RChL}
 v_3^{1R} = 
 -\frac{4\sqrt{2}C_q}{3F^2M_V^2m_\rho^2D_\rho[(p+k)^2]}\left(16 c_3 \left(c_5+c_7\right) \left(m_{\pi }^2-m_{\eta }^2\right)
   \left(2 k\cdot p+m_{\pi }^2\right)-m_{\rho }^2 \left(2 c_{1256}^2 k\cdot p\right.\right.\\\left.\left.-8 c_3 \left(2 \left(c_2+4 c_3+c_7\right)-c_{1256}\right)
   m_{\eta }^2+\left(c_{1256}^2+8 c_3 \left(2 \left(c_5+c_7\right)-c_{1256}\right)\right) m_{\pi }^2\right.\right.\\\left.\left.
   +2 \left(4 c_3-c_5\right) \left(2 c_5-2 c_6+c_{1256}\right) p\cdot p_0\right)\right)\\
  +\frac{16\sqrt{2}C_q}{3F^2M_V^2m_\omega^2D_\omega[(p_0+k)^2]}\left(c_5+c_7\right) \left(\left(c_5+c_7\right) \left(2 k\cdot p_0+m_{\eta
   }^2\right)+4 c_3 \left(m_{\eta }^2-m_{\pi }^2\right)\right) \left(2 k\cdot p_0+m_{\eta }^2-m_{\omega }^2\right)\\
  -\frac{2C_qF_V}{3F^2m_\omega^2} \left(\lambda _{13}^V-\lambda _{14}^V-\lambda_{15}^V\right)\,,
   \\
\end{multline}

\begin{multline}\label{v4-1R-RChL}
 v_4^{1R} = 
 -\frac{4\sqrt{2}C_q}{3F^2M_V^2m_\rho^2D_\rho[(p+k)^2]}4 \left(c_5+c_7\right){}^2 \left(2 k\cdot p+m_{\pi }^2\right) \left(2
   k\cdot p-m_{\rho }^2+m_{\pi }^2\right)\\
  -\frac{4\sqrt{2}C_q}{3F^2M_V^2m_\omega^2D_\omega[(p_0+k)^2]}\left(m_{\omega }^2 \left(c_{1256} \left(c_{1256} \left(2
   k\cdot p_0+m_{\eta }^2\right)+8 c_3 m_{\eta }^2\right)-16 c_3
   \left(c_5+c_7\right) m_{\pi }^2\right)\right.\\\left.
   +16 c_3 m_{\pi }^2
   \left(\left(c_5+c_7\right) \left(2 k\cdot p_0+m_{\eta }^2\right)+4
   c_3 \left(m_{\eta }^2-m_{\pi }^2\right)\right)\right)\\
  +\frac{2C_qF_V}{3F^2m_\omega^2} \left(\lambda _{13}^V-\lambda _{14}^V-\lambda_{15}^V\right)\,.
   \\
\end{multline}

Finally, we will give the two-resonance mediated contributions to the vector form factors (figure \ref{fig:2R-FV}):\\

\begin{multline}\label{v1-2R-RChL}
 v_1^{2R}=
  \frac{8F_VC_q}{3F^2M_Vm_\omega^2D_\rho[(p+p_0+k)^2]D_\omega[(p_0+k)^2]}\left(m_{\omega }^2 \left(k\cdot p_0 \left(2 \left(8
   c_3+3 c_{1256}\right) d_3 m_{\eta }^2\right.\right.\right.\\\left.
   +m_{\pi }^2 \left(-\left(16 c_3
   d_3+\left(2 \left(c_5+c_7\right)+c_{1256}\right)
   d_{12}\right)\right)+2 \left(2 \left(c_5+c_7\right)+c_{1256}\right)
   d_3 p\cdot p_0\right)+4 c_{1256} d_3 \left(k\cdot
   p_0\right){}^2\\\left.
   +\left(8 c_3 m_{\pi }^2-\left(8 c_3+c_{1256}\right)
   m_{\eta }^2\right) \left(d_{12} m_{\pi }^2-2 d_3 \left(m_{\eta
   }^2+p\cdot p_0\right)\right)\right)\\
   +2 d_3 k\cdot p \left(m_{\omega
   }^2 \left(\left(2 \left(c_5+c_7\right)+c_{1256}\right) k\cdot
   p_0+\left(8 c_3+c_{1256}\right) m_{\eta }^2-8 c_3 m_{\pi }^2\right)\right.\\\left.
   +2
   k\cdot p_0 \left(4 c_3 \left(m_{\pi }^2-m_{\eta
   }^2\right)-\left(c_5+c_7\right) \left(2 k\cdot p_0+m_{\eta
   }^2\right)\right)\right)\\\left.
   -2 k\cdot p_0 \left(2 d_3 p\cdot p_0-d_{12}
   m_{\pi }^2\right) \left(\left(c_5+c_7\right) \left(2 k\cdot
   p_0+m_{\eta }^2\right)+4 c_3 \left(m_{\eta }^2-m_{\pi
   }^2\right)\right)\right)\\
  -\frac{8F_VC_q}{3F^2M_Vm_\rho^2D_\rho[(p+p_0+k)^2]D_\rho[(p+k)^2]}\left(-16 c_5 d_2 m_{\pi }^6\right.\\
  +16 c_5
   d_2 m_{\rho }^2 m_{\pi }^4
   +16 c_7 d_2 m_{\rho }^2 m_{\pi }^4-16
   p\cdot p_0 c_5 d_2 m_{\pi }^4-16 p\cdot p_0 c_7 d_2 m_{\pi }^4-2
   p\cdot p_0 c_5 d_3 m_{\pi }^4-2 p\cdot p_0 c_7 d_3 m_{\pi }^4-2
   p\cdot p_0 c_5 d_4 m_{\pi }^4\\
   -2 p\cdot p_0 c_7 d_4 m_{\pi }^4+16
   p\cdot p_0 c_5 d_2 m_{\rho }^2 m_{\pi }^2+16 p\cdot p_0 c_7 d_2
   m_{\rho }^2 m_{\pi }^2-8 p\cdot p_0 c_{1256} d_2 m_{\rho }^2 m_{\pi
   }^2+2 p\cdot p_0 c_5 d_3 m_{\rho }^2 m_{\pi }^2\\
   +2 p\cdot p_0 c_7 d_3
   m_{\rho }^2 m_{\pi }^2-2 p\cdot p_0 c_{1256} d_3 m_{\rho }^2 m_{\pi
   }^2+2 p\cdot p_0 c_5 d_4 m_{\rho }^2 m_{\pi }^2+2 p\cdot p_0 c_7 d_4
   m_{\rho }^2 m_{\pi }^2-4 \left(p\cdot p_0\right){}^2 c_5 d_3 m_{\pi
   }^2-16 c_7 d_2 m_{\pi }^6\\
   -4 \left(p\cdot p_0\right){}^2 c_7 d_3 m_{\pi }^2+4 \left(p\cdot
   p_0\right){}^2 c_5 d_3 m_{\rho }^2+4 \left(p\cdot p_0\right){}^2 c_7
   d_3 m_{\rho }^2-2 \left(p\cdot p_0\right){}^2 c_{1256} d_3 m_{\rho
   }^2\\
   -8 (k\cdot p)^2 \left(c_5+c_7\right)
   \left(\left(-d_3+d_4+d_{12}\right) m_{\eta }^2+\left(k\cdot
   p_0+p\cdot p_0\right) \left(d_3+d_4\right)+8 d_2 \left(m_{\pi
   }^2-m_{\eta }^2\right)\right)\\
   +2 k\cdot p_0 \left(\left(c_5+c_7\right)
   \left(d_3+d_4\right) \left(m_{\rho }^2-m_{\pi }^2\right) m_{\pi
   }^2+p\cdot p_0 d_3 \left(\left(2 \left(c_5+c_7\right)-c_{1256}\right)
   m_{\rho }^2-2 \left(c_5+c_7\right) m_{\pi }^2\right)\right)\\
   -m_{\eta
   }^2 \left(2 \left(c_5+c_7\right) \left(8 d_2+d_3-d_4-d_{12}\right)
   m_{\pi }^2 \left(m_{\rho }^2-m_{\pi }^2\right)-p\cdot p_0 \left(8
   d_2-d_{12}\right) \left(2 \left(c_5+c_7\right) m_{\pi
   }^2+\left(c_{1256}-2 \left(c_5+c_7\right)\right) m_{\rho
   }^2\right)\right)\\
   +4 k\cdot p \left(-2 \left(c_5+c_7\right) d_3
   \left(p\cdot p_0\right){}^2+\left(\left(\left(c_5+c_7-c_{1256}\right)
   d_3\right.\right.\right.\\\left.\left.
   +\left(c_5+c_7\right) d_4\right) m_{\rho }^2+\left(c_5+c_7\right)
   \left(\left(8 d_2-d_{12}\right) m_{\eta }^2-2 \left(4
   d_2+d_3+d_4\right) m_{\pi }^2\right)\right) p\cdot
   p_0\\\left.\left.
   -\left(c_5+c_7\right) \left(8 d_2 m_{\pi }^2+\left(-8
   d_2-d_3+d_4+d_{12}\right) m_{\eta }^2\right) \left(2 m_{\pi
   }^2-m_{\rho }^2\right)-k\cdot p_0 \left(c_5+c_7\right) \left(2 p\cdot
   p_0 d_3+\left(d_3+d_4\right) \left(2 m_{\pi }^2-m_{\rho
   }^2\right)\right)\right)\right)\\
 +\frac{8F_VC_q\lambda^{SV}_3}{3F^2D_\rho[(p + p_0 + k)^2]D_{a_0}[(p+p_0)^2]}\left(c_d p\cdot p_0+c_m m_{\pi
   }^2\right)\\
  -\frac{16F_VC_q}{3F^2M_Vm_\rho^2 m_\omega^2D_\rho[(p+k)^2]}\left(k\cdot p \left(2 \left(\left(c_5+c_7\right)
   k\cdot p_0+\left(c_5+c_7\right) m_{\eta }^2+4 c_3 \left(m_{\pi
   }^2-m_{\eta }^2\right)\right)\right.\right.\\
   \left(m_{\pi }^2 (d_1+8
   d_2+d_3+2 d_4)-d_4 m_{\rho }^2\right)+p\cdot
   p_0 \left(2 c_5 \left(m_{\pi }^2 (d_1+8 d_2+d_3+2
   d_4)-d_4 m_{\rho }^2\right)+2 c_7 d_1 m_{\pi }^2+16
   c_7 d_2 m_{\pi }^2\right.\\\left.\left.
   +c_{1256} d_3 m_{\rho }^2+2 c_7
   d_3 m_{\pi }^2+8 c_3 d_4 \left(m_{\pi }^2-m_{\eta
   }^2\right)-2 c_7 d_4 m_{\rho }^2+4 c_7 d_4 m_{\pi
   }^2\right)\right)\\
   +m_{\pi }^2 \left(m_{\pi }^2-m_{\rho }^2\right)
   (d_1+8 d_2+d_3+d_4)
   \left(\left(c_5+c_7\right) k\cdot p_0+\left(c_5+c_7\right) m_{\eta
   }^2+4 c_3 \left(m_{\pi }^2-m_{\eta }^2\right)\right)\\
   +\frac{1}{2}
   p\cdot p_0 \left(8 c_3 \left(m_{\pi }^2-m_{\eta }^2\right)
   \left(m_{\pi }^2 (d_1+8
   d_2+d_3+d_4)-(d_3+d_4) m_{\rho
   }^2\right)+m_{\pi }^2 \left(2 c_5 \left(m_{\pi }^2-m_{\rho }^2\right)
   (d_1+8 d_2+d_3+d_4)\right.\right.\\\left.\left.
   +2 c_7 \left(m_{\pi
   }^2-m_{\rho }^2\right) (d_1+8
   d_2+d_3+d_4)+c_{1256} (-(d_1+8 d_2))
   m_{\rho }^2\right)\right)\\\left.
   +4 d_4 (k\cdot p)^2
   \left(\left(c_5+c_7\right) k\cdot p_0-4 c_3 m_{\eta }^2+c_5 m_{\eta
   }^2+c_7 m_{\eta }^2+4 c_3 m_{\pi }^2+\left(c_5+c_7\right) p\cdot
   p_0\right)\right)\\   
  +\frac{\sqrt{2}F_V^2C_q}{3F^2m_\rho^2m_\omega^2D_\rho[(p+p_0+k)^2]}\left(2 k\cdot p_0+m_{\eta }^2+m_{\rho
   }^2+2 p\cdot p_0\right) \\
   \left(\left(\lambda _3^{\text{VV}}+\lambda
   _4^{\text{VV}}+2 \lambda _5^{\text{VV}}\right) \left(k\cdot
   p_0+m_{\eta }^2+p\cdot p_0\right)+4 m_{\pi }^2 \lambda
   _6^{\text{VV}}+4 p\cdot p_0 \left(\lambda _1^{\text{VV}}+\lambda
   _2^{\text{VV}}\right)\right)\,, 
 \\
\end{multline}
where $d_{12}\equiv d_1+8d_2$ is fixed by the short-distance constraints (\ref{eq: Consistent set of relations}).\\

\begin{multline}\label{v2-2R-RChL}
 v_2^{2R}=
  \frac{8F_VC_q}{3F^2M_Vm_\omega^2D_\rho[(p+p_0+k)^2]D_\omega[(p_0+k)^2]}\left(8 c_3 d_3 m_{\omega }^2 m_{\pi }^4-8 c_3 d_4
   m_{\omega }^2 m_{\pi }^4+8 c_3 d_{12} m_{\omega }^2 m_{\pi }^4-8
   p\cdot p_0 c_3 d_{12} m_{\pi }^4\right.\\
   +64 p\cdot p_0 c_3 d_2 m_{\omega }^2
   m_{\pi }^2-16 p\cdot p_0 c_5 d_2 m_{\omega }^2 m_{\pi }^2+8 p\cdot
   p_0 c_3 d_3 m_{\omega }^2 m_{\pi }^2-8 p\cdot p_0 c_3 d_4 m_{\omega
   }^2 m_{\pi }^2-8 p\cdot p_0 c_3 d_{12} m_{\omega }^2 m_{\pi }^2\\
   -2
   p\cdot p_0 c_7 d_{12} m_{\omega }^2 m_{\pi }^2+p\cdot p_0 c_{1256}
   d_{12} m_{\omega }^2 m_{\pi }^2+8 p\cdot p_0 c_3 \left(d_{12}-8
   d_2\right) m_{\omega }^2 m_{\pi }^2-2 p\cdot p_0 c_5 \left(d_{12}-8
   d_2\right) m_{\omega }^2 m_{\pi }^2\\
   +16 \left(p\cdot p_0\right){}^2
   c_3 d_3 m_{\pi }^2-2 \left(4 c_3+c_5+c_7\right) \left(p\cdot p_0
   \left(d_3+d_4\right)-\left(d_3-d_4+d_{12}\right) m_{\pi }^2\right)
   m_{\eta }^4+4 \left(p\cdot p_0\right){}^2 c_5 d_3 m_{\omega }^2\\
   +4
   \left(p\cdot p_0\right){}^2 c_7 d_3 m_{\omega }^2-2 \left(p\cdot
   p_0\right){}^2 c_{1256} d_3 m_{\omega }^2+8 \left(k\cdot
   p_0\right){}^2 \left(c_5+c_7\right) \left(\left(d_3-d_4+d_{12}\right)
   m_{\pi }^2-p\cdot p_0 \left(d_3+d_4\right)\right)\\
   +2 m_{\eta }^2
   \left(-2 \left(4 c_3+c_5+c_7\right) d_3 \left(p\cdot
   p_0\right){}^2+\left(\left(\left(c_5+c_7\right) d_{12}+4 c_3
   \left(d_3+d_4+d_{12}\right)\right) m_{\pi }^2+\left(\left(-4
   c_3+c_5+c_7-c_{1256}\right) d_3\right.\right.\right.\\\left.\left.\left.
   +\left(4 c_3+c_5+c_7\right) d_4\right)
   m_{\omega }^2\right) p\cdot p_0-\left(d_3-d_4+d_{12}\right) m_{\pi
   }^2 \left(4 c_3 m_{\pi }^2+\left(4 c_3+c_5+c_7\right) m_{\omega
   }^2\right)\right)\\
   +4 k\cdot p_0 \left(-2 \left(c_5+c_7\right) d_3
   \left(p\cdot p_0\right){}^2+\left(\left(\left(c_5+c_7-c_{1256}\right)
   d_3+\left(c_5+c_7\right) d_4\right) m_{\omega }^2+4 c_3
   \left(d_3+d_4\right) \left(m_{\pi }^2-m_{\eta
   }^2\right)\right.\right.\\\left.\left.
   +\left(c_5+c_7\right) \left(d_{12} m_{\pi }^2-2
   \left(d_3+d_4\right) m_{\eta }^2\right)\right) p\cdot
   p_0+\left(d_3-d_4+d_{12}\right) m_{\pi }^2 \left(4 c_3 \left(m_{\eta
   }^2-m_{\pi }^2\right)+\left(c_5+c_7\right) \left(2 m_{\eta
   }^2-m_{\omega }^2\right)\right)\right)\\
   +2 k\cdot p \left(-4
   \left(c_5+c_7\right) \left(d_3+d_4\right) \left(k\cdot
   p_0\right){}^2+2 \left(\left(d_3+d_4\right) \left(4 c_3 m_{\pi }^2-2
   \left(2 c_3+c_5+c_7\right) m_{\eta }^2+\left(c_5+c_7\right) m_{\omega
   }^2\right)\right.\right.\\
   \left.-2 p\cdot p_0 \left(c_5+c_7\right) d_3\right) k\cdot
   p_0+\left(d_3+d_4\right) \left(4 c_3 m_{\pi }^2-\left(4
   c_3+c_5+c_7\right) m_{\eta }^2\right) \left(m_{\eta }^2-m_{\omega
   }^2\right)\\
   \left.\left.
   +p\cdot p_0 d_3 \left(8 c_3 m_{\pi }^2-2 \left(4
   c_3+c_5+c_7\right) m_{\eta }^2+\left(2
   \left(c_5+c_7\right)-c_{1256}\right) m_{\omega
   }^2\right)\right)\right)\\
   -\frac{8F_VC_q}{3F^2M_Vm_\rho^2D_\rho[(p+p_0+k)^2]D_\rho[(p+k)^2]}\\
   \left(4 (k\cdot p)^2 \left(c_{1256} d_3 m_{\rho
    }^2-\left(c_5+c_7\right) \left(2 d_3 \left(k\cdot p_0+p\cdot
    p_0\right)+\left(d_{12}-8 d_2\right) m_{\eta }^2+8 d_2 m_{\pi
    }^2\right)\right)\right.\\
    +k\cdot p \left(m_{\rho }^2 \left(2 \left(2
    \left(c_5+c_7\right)+c_{1256}\right) d_3 \left(k\cdot p_0+p\cdot
    p_0\right)-\left(2 \left(c_5+c_7\right)+c_{1256}\right) \left(8
    d_2-d_{12}\right) m_{\eta }^2\right.\right.\\
     \left.\left.
    +2 m_{\pi }^2 \left(4 \left(2
    \left(c_5+c_7\right)+c_{1256}\right) d_2+3 c_{1256}
    d_3\right)\right)-2 \left(c_5+c_7\right) m_{\pi }^2 \left(2 d_3
    \left(k\cdot p_0+p\cdot p_0\right)+\left(d_{12}-8 d_2\right) m_{\eta
    }^2+8 d_2 m_{\pi }^2\right)\right)\\\left.
    +c_{1256} m_{\pi }^2 m_{\rho }^2
    \left(2 d_3 \left(k\cdot p_0+m_{\pi }^2+p\cdot
    p_0\right)+\left(d_{12}-8 d_2\right) m_{\eta }^2+8 d_2 m_{\pi
    }^2\right)\right)\\   
  +\frac{8F_VC_q\lambda^{SV}_3}{3F^2D_\rho[(p + p_0 + k)^2]D_{a_0}[(p+p_0)^2]}\left(c_d p\cdot p_0+c_m m_{\pi
    }^2\right)\\
   -\frac{8F_VC_q}{3F^2M_V m_\rho^2 m_\omega^2D_\rho[(p+k)^2]}\left(k\cdot p \left(8 c_3 \left(m_{\pi
    }^2-m_{\eta }^2\right) \left(m_{\pi }^2 (d_1+8
    d_2+d_3+d_4)+(d_3-d_4) m_{\rho
    }^2\right)\right.\right.\\\left.
    +c_{1256} m_{\pi }^2 m_{\rho }^2 (d_1+8 d_2-2
    d_3)\right)+m_{\pi }^2 (d_1+8 d_2) m_{\rho }^2
    \left(c_{1256} m_{\pi }^2-8 c_3 \left(m_{\pi }^2-m_{\eta
    }^2\right)\right)\\\left.
    +2 (k\cdot p)^2 \left(8 c_3 d_4 \left(m_{\pi
    }^2-m_{\eta }^2\right)-c_{1256} d_3 m_{\rho }^2\right)\right)\\
   +\frac{\sqrt{2}F_V^2C_q}{3F^2m_\rho^2m_\omega^2D_\rho[(p+p_0+k)^2]}\left(2 k\cdot p_0+m_{\eta }^2+m_{\rho
    }^2+2 p\cdot p_0\right)\\
    \left(\left(\lambda _3^{\text{VV}}+\lambda
    _4^{\text{VV}}+2 \lambda _5^{\text{VV}}\right) \left(k\cdot p+m_{\pi
    }^2+p\cdot p_0\right)+4 m_{\pi }^2 \lambda _6^{\text{VV}}+4 p\cdot
    p_0 \left(\lambda _1^{\text{VV}}+\lambda _2^{\text{VV}}\right)\right)\,,
    \\
\end{multline}

\begin{multline}\label{v3-2R-RChL}
 v_3^{2R}=
  -\frac{16F_VC_q(d_3-d_4)}{3F^2M_Vm_\omega^2D_\rho[(p+p_0+k)^2]D_\omega[(p_0+k)^2]}\left(4c_3(m_\pi^2-m_\eta^2)-(c_5+c_5)(2k\cdot p_0+m_\eta^2)\right)
  (m_\omega^2-m_\eta^2-2k\cdot p_0)\\
  +\frac{16F_VC_q}{3F^2M_Vm_\rho^2D_\rho[(p+p_0+k)^2]D_\rho[(p+k)^2]}\left(\frac{m_\rho^2}{2}\left(2 d_3 \left(c_{1256}
   \left(2 k\cdot p+k\cdot p_0+p\cdot p_0\right)\right.\right.\right.\\\left.\left.
   -2 \left(c_5+c_7\right)
   \left(k\cdot p_0+p\cdot p_0\right)\right)+\left(2
   \left(c_5+c_7\right)-c_{1256}\right) \left(8 d_2-d_{12}\right)
   m_{\eta }^2+2 m_{\pi }^2 \left(4 \left(c_{1256}-2
   \left(c_5+c_7\right)\right) d_2+c_{1256}
   d_3\right)\right)\\\left.
   +\left(c_5+c_7\right) \left(2 k\cdot p+m_{\pi
   }^2\right) \left(2 d_3 \left(k\cdot p_0+p\cdot
   p_0\right)+\left(d_{12}-8 d_2\right) m_{\eta }^2+8 d_2 m_{\pi
   }^2\right)\frac{}{}\right)\\
-\frac{8F_VC_q}{3F^2M_V m_\rho^2 m_\omega^2D_\rho[(p+k)^2]} \left(8 c_3 \left(m_{\pi }^2-m_{\eta }^2\right)
   \left(m_{\pi }^2 (d_1+8 d_2+d_3+d_4)-(d_3+d_4) m_{\rho }^2\right)\right.\\
   \left.+c_{1256} m_{\pi }^2
   (-(d_1+8 d_2)) m_{\rho }^2+2 k\cdot p \left(c_{1256} d_3 m_{\rho }^2+8 c_3 d_4 \left(m_{\pi }^2-m_{\eta
   }^2\right)\right)\right) \\
  -\frac{\sqrt{2}F_V^2C_q}{3F^2m_\rho^2m_\omega^2D_\rho[(p+p_0+k)^2]}\left(\lambda _3^{\text{VV}}+\lambda _4^{\text{VV}}+2 \lambda _5^{\text{VV}}\right) 
   \left(2 k\cdot p_0+m_{\eta }^2+m_{\rho }^2+2 p\cdot p_0\right)\,,
 \\
\end{multline}

\begin{multline}\label{v4-2R-RChL}
 v_4^{2R}=
  \frac{16F_VC_q(d_3-d_4)}{3F^2M_Vm_\omega^2D_\rho[(p+p_0+k)^2]D_\omega[(p_0+k)^2]}
  \left(\frac{1}{2} m_{\omega }^2 \left(4 c_{1256} d_3 k\cdot
   p_0+2 \left(c_{1256}-2 \left(c_5+c_7\right)\right) d_3 k\cdot p\right.\right.\\\left.
   +2
   \left(8 c_3+c_{1256}\right) d_3 m_{\eta }^2-16 c_3 d_3 m_{\pi }^2+2
   c_5 d_{12} m_{\pi }^2+2 c_7 d_{12} m_{\pi }^2-c_{1256} d_{12} m_{\pi
   }^2+2 \left(c_{1256}-2 \left(c_5+c_7\right)\right) d_3 p\cdot
   p_0\right)\\\left.
   +\left(2 d_3 \left(k\cdot p+p\cdot p_0\right)-d_{12} m_{\pi
   }^2\right) \left(\left(c_5+c_7\right) \left(2 k\cdot p_0+m_{\eta
   }^2\right)+4 c_3 \left(m_{\eta }^2-m_{\pi }^2\right)\right)\frac{}{}\right)\\
  -\frac{16F_VC_q\left(c_5+c_7\right) \left(d_3-d_4\right)}{3F^2M_Vm_\rho^2D_\rho[(p+p_0+k)^2]D_\rho[(p+k)^2]} 
   \left(2 k\cdot p+m_{\pi }^2\right) \left(2 k\cdot p-m_{\rho }^2+m_{\pi }^2\right)\\
  +\frac{16F_VC_q\left(c_5+c_7\right)}{3F^2M_V m_\rho^2 m_\omega^2D_\rho[(p+k)^2]}\left(-2 k\cdot p+m_{\rho
   }^2-m_{\pi }^2\right) \left(m_{\pi }^2 (d_1+8
   d_2+d_3+d_4)+2 d_4 k\cdot p\right)\\
  +\frac{\sqrt{2}F_V^2C_q}{3F^2m_\rho^2m_\omega^2D_\rho[(p+p_0+k)^2]}\left(\lambda _3^{\text{VV}}+\lambda _4^{\text{VV}}+2 \lambda _5^{\text{VV}}\right) 
   \left(2 k\cdot p_0+m_{\eta }^2+m_{\rho }^2+2 p\cdot p_0\right)\,.
 \\
\end{multline}
}
\pagebreak

\section*{Appendix D: Off-shell width of meson resonances}\label{OffShellWidthApp}
For completeness we explain in this appendix the expressions that we have used for the off-shell width of meson resonances relevant to our study. The $\rho(770)$ width is basically driven by Chiral Perturbation Theory results
\begin{equation}
 \Gamma_\rho(s)\,=\,\frac{s M_\rho}{96\pi F^2}\left[\sigma_\pi^{3/2}(s)\theta(s-4m_\pi^2)+\frac{1}{2}\sigma_K^{3/2}(s)\theta(s-4m_K^2)\right]\,,
\end{equation}
where $\sigma_P(s)=\sqrt{1-4\frac{m_P^2}{s}}$. We note that the definition of the vector meson width is independent of the realization of the spin-one fields \cite{GomezDumm:2000fz}. Given the narrow character of the $\omega(782)$ resonance the 
off-shellness of its width can be neglected. A similar comment would apply to the $\phi(1020)$ meson, although it does not contribute to the considered processes in the ideal-mixing scheme for the $\omega-\phi$ mesons that we are following.\\

The $a_1(1260)$ meson energy-dependent width was derived in Ref.~\cite{Dumm:2009va} applying the Cutkosky rules to the analytical results for the form factors into $3\pi$~\cite{Dumm:2009va} and $KK\pi$ channels~\cite{Dumm:2009kj} that are the main 
contributions to this width. Since its computation requires the time-consuming numerical calculation of the corresponding correlator over phase-space, we computed $\Gamma_{a_1}(s)$ at 800 values of $s$ and use linear interpolation to obtain the 
width function at intermediate values.\\

Finally, the $a_0(980)$ meson is also needed as an input in the analyses. We have used the functional dependence advocated in eqs.~(19) and (20) of Ref.~\cite{Escribano:2016ntp} which take into account the main absorptive parts given by the 
$\pi\eta$, $K\bar{K}$ and $\pi\eta^\prime$ cuts. The very low-energy (G-parity violating) $\pi\pi$ cut has been neglected.\\

We point out that we are considering only the imaginary parts of the meson-meson loop functions giving rise to the resonance widths. On the contrary, we are disregarding the corresponding real parts. Although this procedure violates analyticity 
at NNLO in the chiral expansion, the numerical impact of this violation is negligible (see e.g. Ref.~\cite{Boito:2008fq}) and, for simplicity, we take this simplified approach in our study.\\


 \section*{Appendix E: Integration formulas needed to compute the $a_\mu^{P,HLbL}$}\label{AppendixVVP}

This appendix collects some formulae used for the evaluation of the pion pole/exchange contribution to the hadronic light-by-light muon anomalous magnetic moment in Chapter.~\ref{g-2 chapter}. We will follow 
the notation of ref.~\cite{Knecht-Nyffeler}, where angular integrations of the relevant two-loop integrals were first performed analytically using the method of Gegenbauer 
polynomials. The remaining two-dimensional integrations can be readily performed numerically provided the $\pi$TFF can be written
\begin{equation}\label{decomposition KN}
 \mathcal{F}_{\pi^0\gamma\gamma}(q_1^2,q_2^2)\,=\,\frac{F}{3}\left[f(q_1^2)-\sum_{M_{V_i}}\frac{1}{q_2^2-M_{V_i}^2}g_{M_{V_i}}(q_1^2)\right]\,.
\end{equation}
Then, the hadronic light-by-light contribution to $a_\mu$ reads
\begin{equation}\label{on-shell pi LbL 1}
 a_\mu^{\pi^0,\,HLbL}\,=\,\left(\frac{\alpha}{\pi}\right)^3\left[a_\mu^{\pi^0(1),\,HLbL}+a_\mu^{\pi^0(2),\,HLbL}\right]\,,
\end{equation}
with
\begin{equation}\label{on-shell pi LbL 2}
 a_\mu^{\pi^0(1),\,HLbL}\,=\,\int_0^\infty dQ_1 \int_0^\infty dQ_2 \left[w_{f_1}(Q_1,Q_2)\,f^{(1)}(Q_1^2,Q_2^2)+\sum_{M_{V_i}}w_{g_1}(M_{V_i},Q_1,Q_2)\,g^{(1)}_{M_{V_i}}(Q_1^2,Q_2^2)\right]\,,
\end{equation}
and
\begin{equation}\label{on-shell pi LbL 3}
 a_\mu^{LbL,\pi^0(2),\,HLbL}\,=\,\int_0^\infty dQ_1 \int_0^\infty dQ_2 \sum_{M=m_\pi,M_{V_i}}w_{g_2}(M,Q_1,Q_2)\,g^{(2)}_M(Q_1^2,Q_2^2)\,.
\end{equation}
In the previous equation, $w_{{\left\lbrace f/g\right\rbrace}_i}(q_1^2,q_2^2)$ are weight factors, whose expressions can be found in ref.~\cite{Knecht-Nyffeler}. 
${\left\lbrace f/g\right\rbrace}^{(i)}$ are generalized form factors given by
\begin{eqnarray}\label{GFFs}
& & f^{(1)}(Q_1^2,Q_2^2)\,=\,\frac{F}{3}f(-Q_1^2)\,\mathcal{F}_{\pi^0\gamma\gamma}(-Q_2^2,0)\,,\quad g^{(1)}_{M_{V_i}}(Q_1^2,Q_2^2)\,=\,\frac{F}{3}\frac{g_{M_{V_i}}(-Q_1^2)}{M_{V_i}^2}\,\mathcal{F}_{\pi^0\gamma\gamma}(-Q_2^2,0)\,,\nonumber\\
& & g^{(2)}_{m_\pi}(Q_1^2,Q_2^2)\,=\,\frac{F}{3}\,\mathcal{F}_{\pi^0\gamma\gamma}(-Q_1^2,-Q_2^2)\left[f(0)+\sum_{M_{V_i}}\frac{g_{M_{V_i}}(0)}{M_{V_i}^2-m_\pi^2}\right]\,,\nonumber\\
& & g^{(2)}_{M_{V_i}}(Q_1^2,Q_2^2)\,=\,\frac{F}{3}\,\mathcal{F}_{\pi^0\gamma\gamma}(-Q_1^2,-Q_2^2)\frac{g_{M_{V_i}(0)}}{m_\pi^2-M_{V_i}^2}\,.
\end{eqnarray}
Our expressions for the $\pi$TFF in the case of virtual (\ref{pig*g* FF virtual pion}) and real pion (\ref{pig*g* FF real pion}) can indeed be written 
according to eq.~(\ref{decomposition KN})
:
\begin{eqnarray}\label{KN on-shell}
 f(q^2) & = & \frac{2}{F^2}\left[\frac{-2\sqrt{2}c_{1256}F_V(M_V^2-2q^2)}{M_V(M_V^2-q^2)}-\frac{N_C}{8\pi^2}-\frac{4d_3F_V^2}{M_V^2-q^2}
\right]\,,\nonumber\\
 g_{M_V}(q^2) & = & \frac{2}{F^2}\left[2\sqrt{2}c_{1256}F_VM_V+4d_3F_V^2\frac{M_V^2+q^2}{M_V^2-q^2}
\right]\,,
\end{eqnarray}
for on-shell pion, and the additional contributions
\begin{eqnarray}\label{KN off-shell}
 \Delta f(q^2,r^2) & = & \frac{2r^2}{F^2}
\frac{-16\sqrt{2}P_2F_V}{(M_V^2-q^2)(M_P^2-r^2)}
\,,\\
 \Delta g_{M_V}(q^2,r^2) & = & \frac{2r^2}{F^2}\left\lbrace
 \frac{4 d_{123}F_V^2}{M_V^2-q^2}-\frac{16\sqrt{2}P_2 F_V}{M_P^2-r^2}+\frac{16F_V^2P_3}{(M_V^2-q^2)(M_P^2-r^2)}\right\rbrace\nonumber
\end{eqnarray}
for the general situation in which the pion is off its mass-shell. The predicted vanishing of the $c_{1235}$, $c_{125}$ and $P_1$ couplings according to asymptotic constraints 
has already been taken into account to simplify eqs.~(\ref{KN on-shell}) and (\ref{KN off-shell}).

In the latter case, eqs.~(\ref{on-shell pi LbL 1})-(\ref{on-shell pi LbL 3}) should be replaced by \cite{Jegerlehner:2009ry}
\begin{eqnarray}\label{off-shell pi LbL 1}
  a_\mu^{\pi^0,\,HLbL} & = & -\frac{2\alpha^3}{3\pi^2}\int_0^\infty dQ_1 \int_0^\infty dQ_2\int_{-1}^{+1} dt \sqrt{1-t^2} Q_1^3 Q_2^3 \left[\frac{F_1(Q_1^2,Q_2^2,t)}{Q_2^2+m_\pi^2}I_1(Q_1,Q_2,t)\right.\nonumber\\
& & \left. +\frac{F_2(Q_1^2,Q_2^2,t)}{Q_3^2+m_\pi^2}I_2(Q_1,Q_2,t)\right]\,,
\end{eqnarray}
where $Q_3=(Q_1+Q_2)$, $t=$cos$(\widehat{Q_1,Q_2})$,
\begin{eqnarray}\label{off-shell pi LbL 2}
 F_1(Q_1^2,Q_2^2,t)\,=\,\mathcal{F}_{\pi\gamma\gamma}(-Q_1^2,-Q_3^2,-Q_2^2)\,\mathcal{F}_{\pi\gamma\gamma}(-Q_2^2,0,-Q_2^2)\nonumber\\
 F_2(Q_1^2,Q_2^2,t)\,=\,\mathcal{F}_{\pi\gamma\gamma}(-Q_1^2,-Q_2^2,-Q_3^2)\,\mathcal{F}_{\pi\gamma\gamma}(-Q_3^2,0,-Q_3^2)\,,
\end{eqnarray}
and the integration kernels $I_1(Q_1,Q_2,t)$ and $I_2(Q_1,Q_2,t)$ can be found in ref.~\cite{Jegerlehner:2009ry}.


\renewcommand\bibname{References}

\end{document}